\documentclass[nofootinbib,preprintnumbers]{revtex4}

\usepackage{graphicx}
\usepackage{dcolumn}
\usepackage{amsmath,amssymb,epsfig}
\usepackage{paralist}
\usepackage{comment}
\usepackage{graphicx}
\usepackage{multirow} 
\usepackage{wrapfig}

\def\OO#1{{\cal O}(c^{-#1})}
\def\ve#1{{\bf #1}}

\def\vw{\ve{w}}
\def\vx{\ve{x}}
\def\vv{\ve{v}}

\def\vS{\ve{S}}

\def\vQ{\ve{Q}}
\def\vX{\ve{X}}

\def\vOmega{\ve{\Omega}}

\def\cO{{\cal O}}

\usepackage{wasysym}

\usepackage{hyperref}

\renewcommand{\vec}[1]{\boldsymbol{\mathrm{#1}}}

\newcommand{\editnote}[2]{}

\allowdisplaybreaks

\begin{document}

\title{High-Precision Relativistic Time Scales for Cislunar Navigation}

\author{Slava G. Turyshev$^1$}

\affiliation{\vskip 3pt$^1$Jet Propulsion Laboratory, California Institute of Technology,\\
4800 Oak Grove Drive, Pasadena, CA 91109-0899, USA}

\date{\today}

\begin{abstract}

We present a unified post-Newtonian framework for relativistic timing and coordinate transformations that covers three reference systems, including the Barycentric, Geocentric, and Lunicentric Celestial Reference Systems ({\tt BCRS}, {\tt GCRS}, {\tt LCRS}, respectively), and six time scales, including Barycentric Coordinate Time ({\tt TCB}), Geocentric Coordinate Time ({\tt TCG}),
Terrestrial Time ({\tt TT}), Barycentric Dynamical Time ({\tt TDB}), Lunicentric Coordinate Time ({\tt TCL}), and Lunar Time ({\tt TL}). Following IAU Resolution II (2024), which adopts an {\tt LCRS} with time coordinate {\tt TCL} constructed by the same techniques as the {\tt GCRS} and related to {\tt TCB} via the IAU B1.5 prescription with lunar quantities replacing terrestrial ones, we provide an explicit implementation of the {\tt LCRS} metric and the associated time and
coordinate transformations. We construct the {\tt LCRS} metric tensor so as to retain all contributions above a fractional threshold of $5\times10^{-18}$ and timing terms larger than $0.1$ ps by expanding the lunar gravity field to spherical-harmonic degree $\ell=9$ with Love-number variations and including external tidal and inertial multipoles to the octupole. We derive closed-form mappings among {\tt TCB}, {\tt TCG}, {\tt TT}, {\tt TDB}, {\tt TCL}, and
{\tt TL}, yielding proper-to-coordinate time transformations and two-way time-transfer corrections at sub-ps accuracy. We evaluate secular rate constants and periodic perturbations arising
from kinematic dilation, lunar monopole and multipoles, Earth tides, and gravitomagnetic effects for clocks on the lunar surface, in very low and low lunar orbits (vLLO/LLO), in elliptical lunar frozen
orbits (ELFOs), at the Earth--Moon L1 point, and in near-rectilinear halo orbits (NRHOs). Our analysis demonstrates that harmonics through $\ell=9$ and tides through $\ell=8$ are sufficient to
achieve $5\times10^{-18}$ fractional stability for deep cis-lunar regimes (e.g., NRHO, Earth--Moon L1), supporting sub-ps clock synchronization and cm-level navigation,
while near-surface and very low lunar orbit realizations generally require a much higher spherical-harmonic degree, $\ell_{\max}\gtrsim 300$, to meet the same stability goal. This framework
provides a worked-out realization and accuracy assessment of the IAU {\tt LCRS}/{\tt TCL} prescription and supports high-precision time and frequency transfer, relativistic geodesy, quantum communication links, and fundamental-physics experiments beyond low Earth orbit.

\end{abstract}


\maketitle

\tableofcontents

\section{Introduction}
\label{sec:intro}

The era of sustained lunar activity, including crewed outposts, robotic landers and rovers, and future quantum-enabled time-transfer networks, places stringent requirements on navigation and
timing systems. In the Earth--Moon system, contributions from the Moon's multipolar gravitational field, Earth and solar tidal potentials, spacecraft orbital dynamics, and relativistic frame-dragging produce coordinate-time offsets and frequency shifts ranging from the microsecond ($\mu$s) to the sub-picosecond (ps) level. The IAU-endorsed Barycentric and Geocentric Celestial Reference Systems ({\tt BCRS} and {\tt GCRS}) establish a consistent framework for relativistic timing in the solar system \cite{Soffel-etal:2003,Kaplan:2005}, and IAU Resolution II (2024)
\cite{IAU:2024-Resol2} has now extended this framework by adopting a Lunicentric Celestial Reference System ({\tt LCRS}) with time coordinate {\tt TCL} constructed by the same post-Newtonian techniques as the {\tt GCRS} and related to {\tt TCB} via the IAU B1.5 prescription with lunar quantities substituted for terrestrial ones. However, the resolution itself does not give an explicit {\tt LCRS} metric, does not quantify the magnitude of individual terms for different orbital regimes, and does not provide guidance on which contributions may be safely neglected for specific timing and navigation requirements. In particular, practical cis-lunar PNT
architectures must know how far the IAU prescriptions can be truncated in order to reach the $5\times10^{-18}$ fractional-frequency and $0.1$ ps timing levels relevant to modern optical
clocks and centimeter-level ranging.

To address this need, we construct an explicit post-Newtonian realization of the {\tt LCRS} and its associated time scales that is fully consistent with the IAU resolutions but optimized for
cis-lunar applications. We implement the {\tt LCRS} metric to a well-defined accuracy threshold by carrying the lunar gravitational potential to spherical-harmonic degree $\ell=9$ (including
time-dependent Love-number variations) and by incorporating external multipoles through the octupole. We then derive analytic time and coordinate transformations among the six time scales
{\tt TCB}, {\tt TCG}, {\tt TT}, {\tt TDB}, {\tt TCL}, and {\tt TL}, retaining all metric and potential contributions above a fractional threshold of $5\times10^{-18}$ and timing terms larger than $0.1$ ps. By evaluating these transformations for clocks on the lunar surface, in very low and low lunar orbits (vLLO and LLO), in ELFOs, at the Earth--Moon L1 point, and in NRHOs, we provide orbit-specific error budgets that indicate which terms must be retained to meet specified accuracy goals. 

Throughout the paper our emphasis is implementation-oriented: we translate the general relativistic prescriptions of the IAU framework into explicit, orbit-dependent formulas and truncation rules that can be implemented directly in cis-lunar navigation and timing software. Our work can therefore be viewed as an explicit, worked-out realization of the IAU prescription for the {\tt LCRS}/{\tt TCL}: we write down the {\tt LCRS} metric, truncate it at a controlled level, and
quantify the magnitude of all retained terms for a set of  representative cis-lunar orbits.

A complementary line of work has focused on the definition and realization of {\tt TCL} and lunar time scales. Reference~\cite{Kopeikin-Kaplan-2024} derives a general-relativistic definition of {\tt TCL} within the IAU 2000 framework and constructs the {\tt TCL}--{\tt TCG} relation by specializing the {\tt BCRS}/{\tt GCRS} metric potentials to the lunar case, yielding a surface
clock transformation with several-nanosecond accuracy in the Earth's Hill sphere. Reference~\cite{Bourgoin-Defraigne-Meynadier-2025} provides an order-of-magnitude analysis of
relativistic frequency shifts for lunar clocks, decomposing monopole, multipole, tidal, and velocity-dependent contributions and discussing candidate lunar reference timescales and their
traceability to UTC. Reference~\cite{Defraigne-Meynadier-Bourgoin-2025} compares surface, orbital,
and ephemeris-based realizations of lunar time and concludes that {\tt TCL} is a natural choice for an operational reference timescale when long-term stability and traceability are required, while
Ref.~\cite{Lu-Yang-Xie-2025} develops the numerical Lunar Time Ephemeris LTE440 implementing the IAU {\tt TCL} definition and providing sub-nanosecond {\tt TCL}--{\tt TCB} and {\tt TCL}--{\tt TDB} transformations.

On the reference-frame side, Ref.~\cite{Fienga-Rambaux-Sosnica-2024} reviews lunar reference systems, frames, and time scales in the context of the ESA Moonlight programme, emphasizing the coupling between dynamical and kinematic realizations of the lunar center-of-mass frame and their connection to terrestrial reference frames and time scales. Ref.~\cite{Sosnica-etal-2025-ILRF} defines and realizes an International Lunar Reference Frame (ILRF) tied to the lunar center of mass by combining several high-precision lunar ephemerides and quantifies the origin, scale, and
orientation uncertainties that must be accounted for in any high-precision {\tt LCRS}/{\tt TCL}-based timing scheme.

Reference~\cite{Turyshev:2025} develops a post-Newtonian model for relativistic time transformations between the solar-system barycenter, Earth, and Moon, deriving the {\tt TL}--{\tt TT} relation with explicit secular and periodic terms and computing the associated {\tt TT}-compatible spatial scales and Lorentz contraction of Moon-centered coordinates. In the present paper we adopt that formalism and notation for the Earth--Moon sector as our starting point for the combined {\tt BCRS}/{\tt GCRS}/{\tt LCRS} description and extend it by (i) including additional cis-lunar orbital regimes (vLLO, LLO, ELFO, Earth--Moon L1, NRHO) and (ii) providing explicit truncation criteria and error budgets for time-scale transformations at the $5\times10^{-18}$ fractional-
frequency and $0.1$~ps timing levels. In this way our results complement the existing work on {\tt TCL}, lunar time ephemerides, and lunar reference frames by supplying a unified analytic
framework and orbit-specific accuracy requirements for cis-lunar timing and navigation.

This paper is organized as follows. In Section~\ref{sec:time-E} we review the chain of post-Newtonian time and position transformations among {\tt TT}, {\tt TCG}, {\tt TCB}, and
{\tt TDB} within the Earth system. Section~\ref{sec:time-M} extends this framework to the Moon, defines the Lunicentric Coordinate Times ({\tt TCL}, {\tt TL}), and quantifies the tidal and inertial contributions to both time and spatial mappings. In Section~\ref{sec:imp} we present a practical implementation algorithm, showing how to apply these relativistic corrections to raw
timing observables in the {\tt BCRS}. Section~\ref{sec:pt-TT-tau} derives the proper-time relation for cis-lunar spacecraft clocks relative to {\tt TT}, combining Earth- and Moon-based models to
capture cumulative gravitational and kinematic shifts. Finally, Section~\ref{sec:conc} summarizes our main findings, highlights the dominant perturbations, and offers recommendations for deploying
high-precision PNT services throughout the Earth--Moon system. Technical derivations are relegated to two appendices: Appendix~\ref{sec:BCRS-GCRS} reviews the IAU definitions of the {\tt BCRS} and {\tt GCRS}, their metric tensors and potentials, and the explicit post-Newtonian coordinate transformations between them; Appendix~\ref{sec:time-M-coord} constructs the {\tt LCRS} metric and its mapping to the {\tt BCRS}, including lunar self-potentials and external tides.

\section{Time and Position Transformations for the Earth system}
\label{sec:time-E}

For practical purposes, one needs a chain of time transformations from {\tt TT} to Geocentric Coordinate Time ({\tt {\tt TCG}}) in the {\tt GCRS},  to  Barycentric Coordinate Time ({\tt TCB}) in the {\tt BCRS}, and to {\tt TDB} in the {\tt SSB} frame.  
For that purpose, IAU Resolution B1.3 \cite{Kaplan:2005,Soffel-etal:2003} defines two harmonic, post‑Newtonian frames—the {\tt BCRS} and {\tt GCRS}—with metrics\footnote{The notational conventions employed in this paper are those used in \cite{Landau-Lifshitz:1988,Moyer:2003}. Letters from the second half of the Latin alphabet, $m,n,\ldots=0...3$, denote space-time indices, and Greek letters $\alpha,\beta,\ldots=1...3$ denote spatial indices. The metric $\gamma_{mn}$ is that of Minkowski space-time with $\gamma_{mn}=\mathrm{diag}(+1,-1,-1,-1)$ in the Cartesian representation. This differs by an overall sign from the convention adopted in the IAU 2000 resolutions, which are written in the $(-,+,+,+)$ signature. Our expressions for the scalar and vector potentials $(w,w^{\alpha})$ and for the time-scale transformations are algebraically equivalent to the IAU formulae after this global sign change; no observable predictions are affected. We employ the Einstein summation convention with indices being lowered or raised using $\gamma_{mn}$. We use powers of $G$ and negative powers of $c$ as bookkeeping devices for post-Newtonian orders.}
 \(g_{mn}(t,\mathbf x)\) and \(G_{mn}(T,\mathbf X)\) specified to \(O(c^{-4})\) by potentials \((w,w^\alpha)\) and \((W,W^\alpha)\). It also derives the \(O(c^{-4})\) coordinate transformation \((t,\mathbf x)\to(T,\mathbf X)\), including the external tidal potential \(w_{\rm ext}\). 

In Appendix~\ref{sec:BCRS-GCRS}, we review the  definitions for {\tt BCRS} and {\tt LCRS} to show that many terms in the recommended expressions lie below the resolution of current and near future instruments. For that, we computed the magnitude of each term under realistic mission scenarios and truncate the series by retaining only those contributions exceeding a fractional frequency contribution of \(5\times10^{-18}\) and  timing accuracy of 0.1 ps. The resulting expressions capture all physically measurable proper‐time effects while eliminating negligible terms.  

In particular, in Appendix~\ref{sec:BCRS} we discuss {\tt BCRS} which is defined with metric tensor $g_{mn}(t,\vec x)$ and coordinates $(ct, x^\alpha) = x^m$, where $t$ is defined as Barycentric Coordinate Time ($\tt TCB$), or $t\equiv {\tt TCB}$. We also derive Eqs.~(\ref{eq:BCRS-metric-00})--(\ref{eq:BCRS-metric-ab}) that establish the practically-relevant form of the metric tensor $g_{mn}(t,\vec x)$ of the  {\tt BCRS}. 

In this Section, we review the time transformation models specifically developed for the Earth system. This review is essential, as our method for introducing the {\tt LCRS} in Sec.~\ref{sec:time-M} will closely parallel the approach used for the {\tt GCRS}. 

\subsection{GCRS: the practical form}
\label{sec:GCRS}

We discuss the definition of the {\tt GCRS} in 
Appendix~\ref{sec:B-GCRS}. According to IAU, the  {\tt GCRS},  is  defined by the geocentric metric tensor \(G_{mn}\) with coordinates \((T,\vec X)\), where $T$ is the Geocentric Coordinate Time ($\tt TCG$) or $T\equiv \tt TCG$. In the from sufficient to modern timing applications in the solar system\footnote{Notation:
Bold symbols denote spatial vectors; $(\cdot)$ is the Euclidean dot product. 
BCRS positions/velocities of body $B$ are $x_B(t),\,v_B(t)$; $r_{BE}\equiv x_E-x_B$, $r_{BM}\equiv x_M-x_B$, $R_{12}\equiv \|x_2-x_1\|$. 
GCRS vectors are $X$ and LCRS vectors are $\mathbf{X}$; when unambiguous we drop boldface. 
Coordinate times are $t\equiv\mathrm{TCB}$, $T\equiv\mathrm{TCG}$, $\mathrm{TDB}$, $\mathrm{TT}$, $\mathrm{TCL}$, and $\mathrm{TL}$. 
We use $c^{-n}$ to indicate post‑Newtonian order; fractional‑frequency thresholds $<5\times 10^{-18}$ or timing amplitudes $<0.1$\,ps are neglected.}, \(G_{mn}\)  is given  by \eqref{eq:G00tr}–\eqref{eq:Gabtr}. 

The coordinate transformations between the {\tt GCRS} ($T = \tt TCG$, $\vec X$) and the {\tt BCRS} ($t = \tt TCB$, $ \vec x $) that are sufficient for modern high-precision PNT applications are given by (\ref{eq:coord-tr-T1-rec})--(\ref{eq:coord-tr-Xrec}) and repeated here for convenience\footnote{In the notation $O(c^{-n};\,\epsilon_{f};\,\epsilon_{t})$, the first term specifies the post-Newtonian order $n$, the second gives the bound $\epsilon_{f}$ on the fractional frequency (rate) contribution, and the third gives the bound $\epsilon_{t}$ on the corresponding timing effect.}:
{}
\begin{eqnarray}
T &=& t-c^{-2}\Big\{\int^{t}_{t_{0}}\Big({\textstyle{1\over 2}} v_{\tt E}^{2} +
\sum_{{\tt B\not= E}} {G M_{\tt B} \over r_{\tt BE}} \Big)dt + (\vec v_{\tt E}  \cdot \vec r_{\tt E} ) \Big\}-
\nonumber \\ && \hskip 6pt -\, 
   c^{-4}\Big\{\int^t_{t_0}\Big({\textstyle{1 \over 8}}
v_{\tt E} ^4  
+ {\textstyle{3 \over 2}} v_{\tt E} ^2 \sum_{{\tt B\not= E}}{G M_{\tt B} \over r_{\tt BE}}- 
{\textstyle{\textstyle{1 \over 2}}}\Big[\sum_{{\tt B\not= E}} {G M_{\tt B} \over r_{\tt BE}}\Big]^2\Big)dt
+
\Big({\textstyle{\textstyle{1 \over 2}}}v_{\tt E} ^2 +3\sum_{\tt B \not =E}{G M_{\tt B} \over r_{\tt BE}}  \Big)(\vec v_{\tt E}  \cdot \vec r_{\tt E} ) 
\Big\} +\nonumber\\
  && \hskip 6pt +\, 
 {\cal O}\Big(c^{-5};\, 2.14\times 10^{-19}(t-t_0);\, 
 1.91\times 10^{-16}\,{\rm s}\Big),
 \label{eq:coord-tr-T1-rec-sum}\\[4pt]
\vec X &=& \vec r_{\tt E}  + c^{-2} \Big\{{\textstyle\frac{1}{2}}( \vec v_{\tt E}  \cdot\vec r_{\tt E} )\vec v_{\tt E}  +\sum_{{\tt B\not= E}}{G M_{\tt B} \over r_{\tt BE}} \vec r_{\tt E}  +(\vec a_{\tt E} \cdot \vec r_{\tt E} )\vec r_{\tt E} - {\textstyle\frac{1}{2}}r^2_{\tt E} \vec a_{\tt E} \Big\}  +
 {\cal O}\Big(c^{-4};\, 1.28\times 10^{-12}~{\rm m}\Big),
\label{eq:coord-tr-Xrec-sum}
\end{eqnarray}
where $\vec r_{\tt E} \equiv \vec x - \vec x_{\tt E}(t)$ with $\vec x_{\tt E}$  and $\vec v_{\tt E}=d\vec x_{\tt E}/dt$ being the Earth's position and velocity vectors  in the {\tt BCRS} and where the error bounds for secular 
\(\mathcal O(2.1\times10^{-19}(t-t_{0}))\), periodic 
\(\mathcal O(1.9\times10^{-16}\,\mathrm s)\), and positional 
\(\mathcal O(1.3\times10^{-12}\,\mathrm m)\) terms 
arise from omitted external vector‐potentials (\ref{eq:coord-tr-BtE}) and (\ref{eq:coord-tr-A-BiE}), and solar \(J_2\) contributions (\ref{w_ext-mono}), respectively.   

Note that the $c^{-4}$-terms included in (\ref{eq:coord-tr-T1-rec-sum}) are evaluated to contribute up to  $c^{-4}\big\{\frac{1}{8}v_{\tt E}^2+\frac{3}{2}v^2_{\tt E}G M_{\tt S}/r_{\tt E}-\frac{1}{2}(G M_{\tt S}/r_{\tt E})^2\big\}\lesssim 1.10\times 10^{-16}=9.50$~ps/d. 
Also, the  acceleration-dependent terms present in the spatial transformation (\ref{eq:coord-tr-Xrec-sum}), when evaluated at the Earth's surface contribute  $c^{-2}\big((\vec a_{\tt E}\cdot \vec r_{\tt E})\vec r_{\tt E}- \frac{1}{2}r^2_{\tt E}\vec a_{\tt E}\big)\simeq 1.34\times 10^{-6}$~m. Even at the lunar distance, this term is only $\sim4.87 \times 10^{-3} \, \text{m}$, which is negligible for our purposes and may be omitted. 

As a result, (\ref{eq:coord-tr-T1-rec-sum})--(\ref{eq:coord-tr-Xrec-sum})  provide the highest‐precision relativistic coordinate transformations, retaining all contributions down to \(\sim5\times10^{-18}\); these are essential for deep‐space navigation, time transfer, and fundamental‐physics research. 

\subsection{Relativistic time scales at {\tt GCRS}}
\label{sec:GCRS-time}

\subsubsection{Relating TT and TCG}
\label{sec:TT}

We first consider  the relationship between $\tt TT$  and $\tt TCG$. Time $\tt TT$ was defined by IAU Resolution A4 (1991) \cite{Guinot:1992} as: a time scale differing from $\tt TCG$ by a constant rate, with the unit of measurement of $\tt TT$ chosen so that it matches the SI second on the geoid. With  the {\tt GCRS}  metric tensor \(G_{mn}\)   in the form of (\ref{eq:G00tr})--(\ref{eq:Gabtr}), to sufficient accuracy, the  transformation between the proper time of a clock, $\tau$, and the coordinate time of the {\tt GCRS}, ${\rm  T}\equiv {\tt TCG}$, given as 
{}
\begin{eqnarray}
\frac{d\tau}{d{ T}}&=&
1-\frac{1}{c^2}\Big\{{\textstyle\frac{1}{2}}V^2+U_{\tt E}(T,{\bf X}) + U_{\rm tid}(T,{\bf X})\Big\}+{\cal O}\Big(c^{-4};\, 2.42\times 10^{-19}\Big),
\label{eq:prop-coord-time-GCRS-TT}
\end{eqnarray}
where $U_{\tt E}(T,{\bf X})$ and $U_{\rm tid}(T,{\bf X})$  are the Newtonian Earth gravity  and tidal potentials, correspondingly, which are obtained by truncating their post-Newtonian definitions (see Sec.~\ref{sec:BCRS-GCRS}): $W_{\tt E}(T,{\bf X})=U_{\tt E}(T,{\bf X}) +{\cal O}(c^{-2})$ and $W_{\rm tid}(T,{\bf X})=U_{\rm tid}(T,{\bf X}) +{\cal O}(c^{-2})$. Also, ${\vec V}=d{\vec X}/dT$ and $V=|\vec V|$ is the velocity of the clock, as observed from within the {\tt GCRS}. The error bound in (\ref{eq:prop-coord-time-GCRS-TT}) is due to omitted $c^{-4}{\textstyle\frac{1}{2}}U^2_{\tt E}$ term that on the  Earth's surface may have a contribution of up to $c^{-4}{\textstyle\frac{1}{2}}(GM_{\tt E}/R_{\tt E})^2 \lesssim 2.42 \times 10^{-19}$, with other terms being much smaller \cite{Turyshev-Toth:2023-grav-phase}. 

Considering a clock is situated at a ground station on the surface of the Earth. In this case, the first two terms in (\ref{eq:prop-coord-time-GCRS-TT}) are due to the  geocentric velocity of the ground station and the Newtonian potential at its location. Assuming a uniform diurnal rotation of the Earth, so that ${\textstyle\frac{1}{2}}{V}_{\tt }^2={\textstyle\frac{1}{2}}\omega_{\tt E}^2 R^2_{\tt C}(\theta)\sin^2\theta$, we evaluate the magnitudes of the largest contributions produced by these terms, evaluated at the Earth's equator $R_{\tt C}(\frac{\pi}{2})=R_{\tt E}$:
{}
\begin{eqnarray}
c^{-2}{\textstyle\frac{1}{2}}{V}_{\tt }^2&=& \frac{1}{2c^2} \omega_{\tt E}^2 R^2_{\tt E}
\lesssim
1.20 \times 10^{-12},
\label{eq:tau-vel}
\qquad
c^{-2}U_{\tt E} = \frac{1}{c^2}\frac{GM_{\tt E}}{R_{\tt E}} \lesssim
6.95 \times 10^{-10}.
\label{eq:tau-U}
\end{eqnarray}
Thus, both of these terms are very large and must be kept in the model. In addition, as we will see below, one would have to account for several terms in the spherical harmonics expansion of the Earth gravity potential.

The last $c^{-2}$-term in (\ref{eq:prop-coord-time-GCRS-TT}) is the sum of the Newtonian tides due to other bodies (mainly the Sun and the Moon) at the clock location ${\vec X}_{\tt C}$. Using their explicit from (\ref{W-tidal}), the quadrupole tides $(\ell =2)$ contribute at the following level
{}
\begin{equation}
c^{-2}U_{\rm tid[2]}^{\tt (M)}
\simeq  \frac{GM_{\tt M} R_{\tt E}^2}{c^2r^3_{\tt EM}}P_2(\vec{n}_{\tt EM}\cdot\vec{n}_{\tt C})\lesssim
3.91 \times 10^{-17}, \qquad
c^{-2}U_{\rm tid[2]}^{\tt (S)}
\simeq \frac{GM_{\tt S} R_{\tt E}^2}{c^2{\rm AU}^3} P_2(\vec{n}_{\tt SE}\cdot\vec{n}_{\tt C})\lesssim
1.79 \times 10^{-17}.
\label{eq:tidalTS}
\end{equation}
Thus, both quadrupole tides are larger than our accuracy threshold and must be kept in the model.  The octuple $\ell=3$ tides for the Moon and the Sun are at $6.48 \times10^{-19}$ and $7.65 \times10^{-22}$, correspondingly, and, thus, may be omitted. 

Averaging readings of many clock on the Earth's surface, one can form a notion of the {\tt TT}. Denoting $\left<...\right>$ to be the long time averaging procedure, the constant rate between $\tt TCG$ and  $\tt TT$ is expressed as 
{}
\begin{equation}
\Big<\frac{d{\tt TT}}{ d {\tt TCG}}\Big> 
= 1 - \frac{1}{c^2}\big<U_{\tt gE} \big>
= 1 - L_{\tt G},
\label{eq:LG}
\end{equation}
where $U_{\tt gE}$ is the combined long-time averages of the rotational, gravitational and tidal potentials at the geoid, determined as $U_{\tt gE}=(62636856.0\pm0.5)~{\rm m}^2{\rm s}^{-2}$ \cite{Groten:2004}. The IAU value for $L_{\tt G}$ is $6.969\,290\,134 \times 10^{-10}\approx 60.2147$ microseconds/day $(\mu{\rm s/d})$,  a defining constant as set by IAU 2000 Resolution B1.9, Table 1.1 in \cite{Petit-Luzum:2010}. 

The constant $L_{\tt G}$ may be formally defined on the geoid and, with the help of (\ref{eq:prop-coord-time-GCRS-TT}), it may be written as below
\begin{eqnarray}
\label{eq:LG-geoid}
L_{\tt G} \equiv \frac{1}{c^2}\big<U_{\tt gE}\big>
=\frac{1}{c^2}\Big\{{\textstyle\frac{1}{2}}\omega_{\tt E}^2 R^2_{\tt E}+\big<U_{\tt E}({ T},\vec { X}) \big>
+\frac{GM_{\tt M} R_{\tt E}^2}{4a^3_{\tt EM}}\Big\}+
{\cal O}\Big(c^{-4};\, 4.49\times 10^{-18}\Big),
\end{eqnarray}
where the last term is the contribution of the lunar $\ell=2$ tide $c^{-2}\big<U_{\rm tid[2]}^{\tt (M)}\big>=c^{-2}{GM_{\tt M}R_{\tt E}^2}/({4a^3_{\tt EM}})\simeq 9.78\times 10^{-18}$ and the error term is set by the omitted $\ell=2$ solar tide evaluated to be  $c^{-2}\big<U_{\rm tid[2]}^{\tt (S)}\big>=c^{-2}{GM_{\tt S}R_{\tt E}^2}/({4a^3_{\tt SE}})\simeq 4.49\times 10^{-18}$.

Note that, to reach the accuracy of $5\times 10^{-18}$, the Earth gravity field must be known to a similar level. Thus, keeping only the leading terms with gravitational harmonics $J_\ell, C_{\ell k}$ and $S_{\ell k}$  up to $\ell=8$ order, (\ref{eq:prop-coord-time-GCRS-TT}) takes the form \cite{Turyshev-Toth:2023-grav-phase}:
{}
\begin{eqnarray}
L_{\tt G} \equiv \frac{1}{c^2}\big<U_{\tt g E}\big>=
\frac{1}{c^2}\Big\{{\textstyle\frac{1}{2}}\omega_{\tt E}^2 R^2_{\tt E} &+&
\frac{GM_{\tt E}}{R_{\tt E}}\Big(1+{\textstyle\frac{1}{2}}J_2-{\textstyle\frac{3}{8}}J_4+{\textstyle\frac{5}{16}}J_6-{\textstyle\frac{35}{128}}J_8+P_{22}(0)\big(C_{22}\cos 2\phi+S_{22}\sin 2\phi\big)+\nonumber\\
&+&
\sum_{\ell=3}^{8}\sum_{k=1}^{+\ell}P_{\ell k}(0)(C_{\ell k}\cos k\phi+S_{\ell k}\sin k\phi)\Big)
\Big\}+{\cal O}(c^{-4}; 5.83\times 10^{-17}),
\label{eq:prop-coord-time-J2J8-LG}
\end{eqnarray}
where the error bound is set by the omitted contribution from $J_{10}$ and some low-order tesseral harmonics. In fact, not only many more terms are needed to reach the accuracy of  $5\times 10^{-18}$ level, but all the physical parameters involved (i.e., $GM_{\tt E}$, $R_{\tt E}$, $C_{\ell k}, S_{\ell k}$, etc.) must also be known to the stated level of accuracy, which currently is not the case.

Recognizing the challenges involved  in defining relativistic geoid (e.g., \cite{Philipp:2020}), the constant $L_{\tt G}$ was turned into a defining constant with its value fixed to $6.969\,290\,134 \times 10^{-10}$ (2000 IAU Resolution B1.9) \cite{Soffel-etal:2003,Kaplan:2005}.
The conversion from {\tt TT} to Geocentric Coordinate Time ({\tt TCG}), on average, involves a rate change  (\ref{eq:LG})
{}
\begin{equation}
\frac{	d {\tt {\tt TCG}}}{d {\tt TT}} = \frac{1}{1  -  L_{\tt G}} = 1 + \frac{L_{\tt G}}{1 -  L_{\tt G}},
\label{eq:(4)}
\end{equation}
which may be used to introduce the following  relationship between  {\tt TCG} and {\tt TT}, starting at time ${\tt T}_0$:
{}
\begin{equation}
 {\tt TCG}-{\tt TT}  = \frac{L_{\tt G}}{1 -  L_{\tt G}}({\tt TT}-{\tt T}_0).
\label{eq:(4)in}
\end{equation}

For convenience, the defining constants and adopted values used throughout this paper 
(\emph{e.g.}, $L_{\tt G}$, $L_{\tt C}$, $L_{\tt B}$, $T_0$, $\tt TDB_0$) are summarized in Table~\ref{tab:defs}.

\begin{table*}[t]
\centering
\caption{Defining constants and adopted values used in this work.}
\label{tab:defs}
\begin{tabular}{lclll}
\hline
Quantity & Symbol & Value & Drift (ms/d)  & Notes \\
\hline
Geocentric scaling (defining) & $L_{\tt G}$ & $6.969\,290\,134\times 10^{-10}$ & 60.2147 $(\mu{\rm s/d})$ & IAU 2000 B1.9, \cite{Petit-Luzum:2010} \\
{\tt TCG}--{\tt TCB} mean rate & $L_{\tt C}$ & $1.480\,826\,854\,55\times 10^{-8}$ & $1.2794344$ & long-term average \\
{\tt TDB} scaling (defining) & $L_{\tt B}$ & $1.550\,519\,768\times 10^{-8}$ & $1.339650$ & IAU 2006 B3, \cite{Kaplan:2005,Petit-Luzum:2010,Luzum-etal:2011}. \\
{\tt TCL}--{\tt TCB} mean rate & $L_{\tt H}$ & $1.482\,536\,24\times10^{-8}$ & $1.280913$ & from Eq.~(\ref{eq:LH}) \\
Lunar surface scaling & $L_{\tt L}$ & $3.13905\times 10^{-11}$ & $0.0027121$ & selenoid‑anchored; see Sec.~\ref{sec:tm-tdb} \\
{\tt TL}--{\tt TCB} mean rate & $L_{\tt M}$ & $1.485\,675\,294\times10^{-8}$ & $1.283620$ & via Eq.~(\ref{eq:consTM}) \\
Epoch & $T_0$ & JD $2443144.5003725$ & & 1977-01-01 00:00:32.184 TAI \\
Offset & ${\tt TDB}_0$ & $-65.5~\mu$s & & DE405 convention \\
\hline
\end{tabular}
\end{table*}

As shown in Table~\ref{tab:defs}. we adopt the IAU 2000/2006 conventions for $L_{\tt G}$, $L_{\tt B}$, $T_0$, ${\tt TDB}_0$; a conventional $L_{\tt L}$ as above;  and evaluate $L_{\tt C}$, $L_{\tt H}$, $L_{\tt M}$, $L_{\tt EM}$ from long-time averages per Eqs.~(\ref{eq:coord-tr-LC}), (\ref{eq:LH}), (\ref{eq:consTM}), and (\ref{eq:expRR1+}).  All path delays (Sec.~\ref{sec:lighttime}) are modeled in the {\tt BCRS} with station vectors transformed from the {\tt GCRS} (Sec.~\ref{sec:GCRS}).

According, the scaling of spatial coordinates and mass factors is designed to maintain the invariance of the speed of light and the equations of motion in the {\tt GCRS} \cite{Klioner:2008}, applicable to the Moon's or Earth’s artificial satellites, during the transformation from {\tt TCG} to {\tt TT}. This transformation, which includes the scaling of temporal and spatial coordinates and mass factors, ensures the invariance of the metric (up to a constant factor)
{}
\begin{equation}
(ds^2)_{\tt TT} = (1- L_{\tt G})^2ds^2_{\tt TCG} ,
\label{eq:interv}
\end{equation}
where $(ds^2)_{\tt TT}$ maintains the same form in terms of {\tt TT}, $\vec X_{\tt TT}$, $(G M )_{\tt TT}$ as  (\ref{eq:G00tr})--(\ref{eq:Gabtr}) do in terms of $T$, $\vec X$, $(GM)_{\tt TCG}$. 

As a result, instead of coordinate time ${\tt T} = {\tt TCG}$, spatial coordinates $\vec X$ and mass factors $(G M)_{\tt TCG}$ related to {\tt GCRS}, the following scaling of these quantiles is used \cite{Brumberg-Groten:2001}
{}
\begin{equation}
{\tt TT} ={\tt TCG}- L_{\tt G}({\tt TCG}-{\tt T}_0), \qquad
\vec X_{\tt TT} = (1-L_{\tt G})\vec X_{\tt TCG}, \qquad
(GM)_{\tt TT} = (1-L_{\tt G})(GM)_{\tt TCG}.
\label{eq:TCGT}
\end{equation}

\subsubsection{Relating TCG and TCB}
\label{sec:TCG-TCB}

Another constant, $L_{\tt C}$, removes the average rate between ${\tt TCG}$ and ${\tt TCB}$. It is determined as the long time average of the rate computed from  transformation (\ref{eq:coord-tr-T1-rec-sum}) given as below
{}
\begin{eqnarray}
{\tt TCG} - {\tt TCB}  &=& -\frac{1}{c^2}\Big\{\int^{t}_{t_{0}}\Big({\textstyle{1\over 2}} v_{\tt E}^{2} +
\sum_{{\tt B\not= E}} {G M_{\tt B} \over r_{\tt BE}} \Big)d{\tt TCB} + (\vec v_{\tt E}  \cdot \vec r_{\tt E} ) \Big\}_{\tt TCB}-
\nonumber \\ && \hskip 0pt -\, 
  \frac{1}{c^4}\Big\{\int^t_{t_0}\Big({\textstyle{1 \over 8}}
v_{\tt E} ^4  
+ {\textstyle{3 \over 2}} v_{\tt E} ^2 \sum_{{\tt B\not= E}}{G M_{\tt B} \over r_{\tt BE}}- 
{\textstyle{\textstyle{1 \over 2}}}\Big[\sum_{{\tt B\not= E}} {G M_{\tt B} \over r_{\tt BE}}\Big]^2\Big)d{\tt TCB}
+
\Big({\textstyle{\textstyle{1 \over 2}}}v_{\tt E} ^2 +3\sum_{\tt B \not =E}{G M_{\tt B} \over r_{\tt BE}}  \Big)(\vec v_{\tt E}  \cdot \vec r_{\tt E} ) 
\Big\}_{\tt TCB} +\nonumber\\
  && \hskip 0pt +\, 
 {\cal O}\Big(c^{-5};\, 2.14\times 10^{-19}(t-t_0);\, 
 1.91\times 10^{-16}\,{\rm s}\Big),
\label{eq:(12)}
\end{eqnarray}
where the subscript $\{...\}_{\tt TCB}$  are used to identify ${\tt TCB}$-compatible quantities.  
Although the integrals in (\ref{eq:(12)}) may be calculated by a numerical integration (see details in \cite{Fukushima:1995,Irwin-Fukushima:1999}), there are analytic formulations available (e.g., \cite{Fairhead-Bretagnon:1990,Harada-Fukushima:2003}). For that, expression for the the total Earth's energy of its orbital motion may be given as below:
{}
\begin{eqnarray}
\frac{1}{c^2}\Big({\textstyle{1\over 2}} v_{\tt E}^{2} +
\sum_{{\tt B\not= E}} {G M_{\tt B} \over r_{\tt BE}} \Big)+
\frac{1}{c^4}\Big({\textstyle{1 \over 8}}
v_{\tt E} ^4  
+ {\textstyle{3 \over 2}} v_{\tt E} ^2 \sum_{{\tt B\not= E}}{G M_{\tt B} \over r_{\tt BE}}- 
{\textstyle{\textstyle{1 \over 2}}}\Big[\sum_{{\tt B\not= E}} {G M_{\tt B} \over r_{\tt BE}}\Big]^2\Big)
=L_{\tt C}+ \dot P(t)+ {\cal O}\Big(c^{-5};\, 2.14\times 10^{-19}\Big), 
 \label{eq:coord-tr-QQ}
\end{eqnarray}
where $L_{\tt C}$ and $\dot P(t)$ are given below
\begin{eqnarray}
L_{\tt C} &=&
\frac{1}{c^2}\Big<{\textstyle{1\over 2}} v_{\tt E}^{2} +
\sum_{{\tt B\not= E}} {G M_{\tt B} \over r_{\tt BE}} \Big>+
\frac{1}{c^4}\Big<{\textstyle{1 \over 8}}
v_{\tt E} ^4  
+ {\textstyle{3 \over 2}} v_{\tt E} ^2 \sum_{{\tt B\not= E}}{G M_{\tt B} \over r_{\tt BE}}- 
{\textstyle{\textstyle{1 \over 2}}}\Big[\sum_{{\tt B\not= E}} {G M_{\tt B} \over r_{\tt BE}}\Big]^2\Big>+\, 
 {\cal O}\Big(c^{-5};\, 2.14\times 10^{-19}\Big),
 \label{eq:coord-tr-LC}\\
\dot P(t)&=&
\frac{1}{c^2}\Big({\textstyle{1\over 2}} v_{\tt E}^{2} +
\sum_{{\tt B\not= E}} {G M_{\tt B} \over r_{\tt BE}} \Big)+
\frac{1}{c^4}\Big({\textstyle{1 \over 8}}
v_{\tt E} ^4  
+ {\textstyle{3 \over 2}} v_{\tt E} ^2 \sum_{{\tt B\not= E}}{G M_{\tt B} \over r_{\tt BE}}- 
{\textstyle{\textstyle{1 \over 2}}}\Big[\sum_{{\tt B\not= E}} {G M_{\tt B} \over r_{\tt BE}}\Big]^2\Big)-L_{\tt C}.
 \label{eq:coord-tr-Pt}
\end{eqnarray}
Thus, the constant \( L_{\tt C}\) is derived from long-term averaging of Earth's total orbital energy, as expressed in (\ref{eq:coord-tr-LC}), yielding \( L_{\tt C} = 1.480\,826\,854\,55 \times 10^{-8} \approx 1.279\,434\,4~\text{ms/d} \) (milliseconds per day). The term \( P(t) \) in (\ref{eq:coord-tr-Pt}) represents a series of periodic components, as detailed in Refs.~\cite{Fairhead-Bretagnon:1990,Irwin-Fukushima:1999}. 

As a result, Eq.~(\ref{eq:(12)}) may be used to determine mean rate between ${\tt TCG}$ and ${\tt TCB}$:
{}
\begin{equation}
\Big<\frac{d{\tt TCG}}{d{\tt TCB}}\Big> =1-L_{\tt C}.
\label{eq:constTCGTCB}
\end{equation}

\subsubsection{Relating TCB  and TDB}
\label{sec:TCB-TDB}

Similar to (\ref{eq:LG}), we can formally relate {\tt TCB} and {\tt TDB}. The  IAU 2006 Resolution B3 for {\tt TDB}  \cite{Luzum-etal:2011} defines the relationship between \( {\tt TDB} \) and \( {\tt TCB} \) using the constant \( L_{\tt B} \) while ensuring there is no rate difference between \( {\tt TDB} \) and \( {\tt TT} \):
{}
\begin{equation}
\Big<\frac{d{\tt TDB}}{d{\tt TCB}}\Big>= 1-L_{\tt B} \qquad {\rm and}\qquad 
\frac{d{\tt TDB}}{d{\tt TT}}=1.
\label{eq:const=}
\end{equation}
Using these expressions together with (\ref{eq:LG}) and (\ref{eq:constTCGTCB}), we have
{}
\begin{equation}
\Big<\frac{d{\tt TDB}}{d{\tt TCB}}\Big>=\Big(\frac{d{\tt TDB}}{d{\tt TT}}\Big)\Big<\frac{d{\tt TT}}{d{\tt TCG}}\Big>\Big<\frac{d{\tt TCG}}{d{\tt TCB}}\Big> \qquad
\Rightarrow \qquad 1-L_{\tt B}=(1-L_{\tt G})(1-L_{\tt C}),
\label{eq:constLBLCLG}
\end{equation}
where $L_{\tt B}$ is determined as $L_{\tt B} = L_{\tt G}+L_{\tt C}-L_{\tt G}L_{\tt C}=
1.550\,519\,768 \times 10^{ -8}\pm 
2\times 10^{ - 17}
\approx 1.339\,65~{\rm ms/d}\pm 1.7$~ps/d,  an IAU defining constant \cite{Kaplan:2005,Petit-Luzum:2010,Luzum-etal:2011}. 

As a result, {\tt TDB} is a timescale rescaled from {\tt TCB}, as defined by IAU 2006 Resolution B3 and  IAU 2009 Resolution 3 \cite{IAU2009ResB3,Petit-Luzum:2010}, given by the following set of expressions:
{}
\begin{equation}
{\tt TDB} = {\tt TCB} - L_{\tt B}({\tt TCB}-{\tt T}_0)+{\tt TDB}_0, \qquad
\vec x_{\tt TDB} = (1-L_{\tt B})\vec x_{\tt TCB}, \qquad
(GM)_{\tt TDB} = (1-L_{\tt B})(GM)_{\tt TCB},
\label{eq:TDBC}
\end{equation}
were, the defining constants
$L_{\tt B} = 1.550519768\times10^{-8},
{\tt T}_0 = 2443144.5003725\,\mathrm{JD}$, and $
{\tt TDB}_0 = -65.5\,\mu\mathrm{s},$
match those used in the JPL DE405 ephemeris \cite{Standish:DE405-1998}. This ensures that \(\tt TDB\) advances at the same rate as \(\tt TT\) at the geocenter. The offset \({\tt TDB}_0\) is chosen to align with the standard $(\tt TDB - TT)$ relation \cite{Fairhead-Bretagnon:1990}, which implies that \({\tt TDB}\) is not synchronized with \({\tt TT}\), \({\tt TCG}\), or \({\tt TCB}\) at 1977‑01‑01\,00:00:32.184\,TAI, at the geocenter (see discussion in \cite{Turyshev:2025}). 

\subsection{Transformation {\tt TT} vs {\tt TDB}} 
\label{sec:tdb-tai}

To establish relationships between  {\tt TT} and {\tt TDB} as a function of {\tt TDB}, we use the chain of  transformations: ${\tt TT}  -  {\tt TDB}=({\tt TT}  -  {\tt TCG})+({\tt TCG}  -  {\tt TCB})+({\tt TCB}  -  {\tt TDB})$, 
with  expressions  given by (\ref{eq:TCGT}),  (\ref{eq:(12)}), and (\ref{eq:TDBC}).  As a result, we have: 
{}
\begin{align}
  {\tt TT} -{\tt TDB} &=
  \frac{L_{\tt B}-L_{\tt G}}{1-L_{\tt B}}({\tt TDB}-{\tt T}_0)-
  \frac{1-L_{\tt G}}{1-L_{\tt B}}\Big\{{\tt TDB}_0+\frac{1}{c^2}\int_{{\tt T_0+TDB_0}}^{\tt TDB} \Big({\textstyle{1\over 2}} v_{\tt E}^{2} +
\sum_{{\tt B\not= E}} {G M_{\tt B} \over r_{\tt BE}} \Big)d{\tt TDB}+\frac{1}{c^2}(\vec v_{\tt E}\cdot \vec {r_{\tt E}}_{\tt TDB})  +\nonumber\\
  & \hskip 15pt +\, 
  \frac{1}{c^4}\int_{{\tt T_0+TDB_0}}^{\tt TDB} \Big({\textstyle{1 \over 8}}
v_{\tt E} ^4  
+ {\textstyle{3 \over 2}} v_{\tt E} ^2 \sum_{{\tt B\not= E}}{G M_{\tt B} \over r_{\tt BE}}- 
{\textstyle{\textstyle{1 \over 2}}}\Big[\sum_{{\tt B\not= E}} {G M_{\tt B} \over r_{\tt BE}}\Big]^2\Big)d{\tt TDB}
+
\frac{1}{c^4}\Big({\textstyle{\textstyle{1 \over 2}}}v_{\tt E} ^2 +3\sum_{\tt B \not =E}{G M_{\tt B} \over r_{\tt BE}}  \Big)(\vec v_{\tt E}\cdot \vec {r_{\tt E}}_{\tt TDB})
\Big\}+\nonumber\\
  & \hskip 15pt +\, 
 {\cal O}\Big(c^{-5};\, 2.14\times 10^{-19}({\tt TDB -T_0- TDB_0});\, 1.91\times 10^{-16}\,{\rm s}\Big).
\label{eq:(nonN1-TCB)}
\end{align}

The constant rate of  $ (L_{\tt B}-L_{\tt G})/(1-L_{\tt B}) =1.480\,826\,878 \times 10^{-8}\simeq1.279~{\rm ms/d}$ is removed by taking the integral in (\ref{eq:(nonN1-TCB)}) with the help of (\ref{eq:coord-tr-QQ}). As a result,  we have the following expression for  {\tt TT} as a function of  {\tt TDB}
{}
\begin{align}
 {\tt TT} -{\tt TDB} = -  {\tt TDB}_0 & -\Big\{P({\tt TDB})-P({\tt T}_0+{\tt TDB}_0)+\frac{1}{c^2} (\vec v_{\tt E} \cdot \vec {r_{\tt E}}_{\tt TDB}) + \frac{1}{c^4}\Big({\textstyle{\textstyle{1 \over 2}}}v_{\tt E} ^2 +3\sum_{\tt B \not =E}{G M_{\tt B} \over r_{\tt BE}}  \Big)(\vec v_{\tt E}\cdot \vec {r_{\tt E}}_{\tt TDB})\Big\}+
  \nonumber\\
 &
 +
 {\cal O}\Big(c^{-5};\, 2.14\times 10^{-19}({\tt TDB -T_0- TDB_0});\, 
 1.91\times 10^{-16}\,{\rm s}\Big).
\label{eq:(nonN1TCB)}
\end{align}

Thus, there is no secular rate difference between \texttt{TT} and \texttt{TDB}; only small periodic variations $\propto P({\tt TDB})$ remain (cf.\ \eqref{eq:(nonN1TCB)}). The resulting  relation for {\tt TT}  achieves fractional‐frequency accuracy of  \(\lesssim 2.14\times10^{-19}\) and its position-dependent periodic terms are accurate to $1.91\times 10^{-16}\,{\rm s}$, meeting our accuracy thresholds.

\section{Time and Position Transformations for the Moon system}
\label{sec:time-M}

In the Moon's vicinity, we require a coordinate system suitable for both surface observers and lunar-orbiting spacecraft, each with its own proper time to be used for PNT applications. Following
IAU Resolution II (2024) \cite{IAU:2024-Resol2}, we adopt a {\tt LCRS} constructed by the same post-Newtonian techniques as the {\tt GCRS}, with Lunicentric Coordinate Time {\tt TCL} as its time coordinate and with {\tt TCL} related to {\tt TCB} via the IAU B1.5 relations specialized to lunar quantities. By paralleling the ${\tt TT}\rightarrow{\tt TCG}\rightarrow{\tt TCB}\rightarrow{\tt TDB}$ time-scale chain and the {\tt GCRS} construction, we work out an explicit {\tt LCRS} metric and derive relations describing time transformations between {\tt LCRS} and {\tt BCRS}.

\subsection{Lunicentric Coordinate Reference System (LCRS)}
\label{sec:LCRS}

The {\tt LCRS} is defined by the lunicentric metric tensor ${\cal G}_{mn}$ with lunicentric coordinates $({\cal T},\vec{\cal X})$, where ${\cal T}$ is the Lunicentric Coordinate Time {\tt TCL}, or ${\cal T}\equiv{\tt TCL}$ \cite{Soffel-etal:2003,Turyshev:2025,IAU:2024-Resol2}. Analogous to the {\tt GCRS} metric construction (see Eqs.~(\ref{eq:G00tr})--(\ref{eq:Gabtr})), in Appendix~\ref{sec:time-M-coord} we derive an explicit {\tt LCRS} metric tensor that is consistent with the IAU prescription but truncated so as to retain only those terms whose omission would exceed our adopted $5\times10^{-18}$ fractional-frequency or $0.1$ ps timing threshold. The resulting practical metric is given by Eqs.~(\ref{eq:G00tr-MM})--(\ref{eq:Gabtr-MM}); this truncation eliminates all sub-threshold contributions from the full {\tt LCRS} metric (\ref{LCRS_metric}), retaining only the monopole, tidal, and inertial components that produce measurable proper-time effects.

In addition, also in Appendix \ref{sec:time-M-coord}, we  derived the coordinate transformations between the {\tt LCRS} (${\cal T} = \tt TCL$, $\vec {\cal X}$) and the {\tt BCRS} ($t = \tt TCB$, $ \vec x $)  (see (\ref{eq:coord-tr-T1-rec-M})--(\ref{eq:coord-tr-Xrec-M})) that retain  terms that are sufficient for modern high-precision PNT applications in cislunar space. These transformations are repeated here for convenience: 
{}
\begin{eqnarray}
{\cal T} &=& t-c^{-2}\Big\{\int^{t}_{t_{0}}\Big({\textstyle{1\over 2}} v_{\tt M}^{2} +
\sum_{{\tt B\not= M}} {G M_{\tt B} \over r_{\tt BM}} \Big)dt + (\vec v_{\tt M}  \cdot \vec r_{\tt M} ) \Big\}-
\nonumber \\ && \hskip 6pt -\, 
   c^{-4}\Big\{\int^t_{t_0}\Big({\textstyle{1 \over 8}}
v_{\tt M} ^4  
+ {\textstyle{3 \over 2}} v_{\tt M} ^2 \sum_{{\tt B\not= M}}{G M_{\tt B} \over r_{\tt BM}}- 
{\textstyle{\textstyle{1 \over 2}}}\Big[\sum_{{\tt B\not= M}} {G M_{\tt B} \over r_{\tt BM}}\Big]^2\Big)dt
+
\Big({\textstyle{\textstyle{1 \over 2}}}v_{\tt M} ^2 +3\sum_{\tt B \not =M}{G M_{\tt B} \over r_{\tt BM}}  \Big)(\vec v_{\tt M}  \cdot \vec r_{\tt M} ) 
\Big\} +\nonumber\\
  && \hskip 6pt +\, 
 {\cal O}\Big(c^{-5};\, 6.86\times 10^{-19}\,(t-t_0);\, 
 1.37\times 10^{-15}\,{\rm s}\Big),
 \label{eq:coord-time-TL}\\[4pt]
\vec {\cal X} &=& \vec r_{\tt M}  + c^{-2} \Big\{{\textstyle\frac{1}{2}}( \vec v_{\tt M}  \cdot\vec r_{\tt M} )\vec v_{\tt M}  +\sum_{{\tt B\not= M}}{G M_{\tt B} \over r_{\tt BM}} \vec r_{\tt M}  +(\vec a_{\tt M} \cdot \vec r_{\tt M} )\vec r_{\tt M} - {\textstyle\frac{1}{2}}r^2_{\tt M} \vec a_{\tt M} \Big\}  +
 {\cal O}\Big(c^{-4};\, 2.94\times 10^{-12}~{\rm m}\Big),
\label{eq:coord-space-TL}
\end{eqnarray}
where $\vec r_{\tt M} \equiv \vec x - \vec x_{\tt M}(t)$ with $\vec x_{\tt M}$  and $\vec v_{\tt M}=d\vec x_{\tt M}/dt$ being the Moon's position and velocity vectors  in the {\tt BCRS}. The error in the time transformation is set by the omitted contribution of the external vector potential 
$4\sum_{\tt B \not =M} (G M_{\tt B}/ r_{\tt BM}) (\vec v_{\tt M} \cdot \vec v_{\tt B})\sim 6.86\times10^{-19}$, as established by  (\ref{eq:coord-tr-BtE-M});  the error in the position transformation is due to omitted contribution of the solar quadrupole moment $J_2 = 2.25 \times 10^{-7}$ \cite{Park:2017,MecheriMeftah:2021},  
 estimated even at the Earth-Moon Lagrange point L1 at which is the distance $a_{\tt L1}\simeq 5.80\times10^{7}\,\mathrm{m}$ from the Moon (see Sec.~\ref{sec:LCRS-L1}) contributing only $c^{-2}w^*_{2,{\tt S}}(t,\vx) \vec r_{\tt M}  \simeq c^{-2} (GM_{\tt S} J_2R_{\tt S}^2/{\rm AU}^3)a_{\tt L1}\sim  2.94 \times 10^{-12}$\,m
to (\ref{w_ext-mono-M}), which is clearly impractical for our purposes. 

Results~(\ref{eq:coord-time-TL}) and~(\ref{eq:coord-space-TL}) specify the relativistic time‐ and space‐coordinate transformations required for modern high‐precision cis‐lunar applications: deep‐space navigation, time transfer, and fundamental‐physics tests. Eqs.~(\ref{eq:coord-time-TL})--(\ref{eq:coord-space-TL}) are the practically relevant forms of the BCRS$\leftrightarrow$LCRS transformations, retaining all contributions above the $5\times 10^{-18}$/0.1\,ps thresholds, analogous to Eqs.~\eqref{eq:coord-tr-T1-rec-sum}--\eqref{eq:coord-tr-Xrec-sum} for the Earth system.

\subsection{Relativistic time scales at {\tt LCRS}}

\subsubsection{TCL vs TCB}
\label{sec:TCL}

It is straightforward to establish the relationship between the Luni-centric Coordinate Time  ({\tt TCL}) vs {\tt TCB}, thus, we will start with that. 
The transformation from {\tt TCL} to {\tt TCB} is analogous to Eq.~(\ref{eq:(12)}) and from (\ref{eq:coord-time-TL}) is determined to be
{}
\begin{eqnarray}
{\tt TCL} -{\tt TCB}  &=& -\frac{1}{c^2}\Big\{\int \Big( {\textstyle\frac{1}{2}}v_{\tt M}^2 + 
\sum_{{\tt B\not= M}} {G M_{\tt B} \over r_{\tt BM}} \Big)d{\tt TCB}+\big(\vec v_{\tt M} \cdot \vec {r_{\tt M}}\big)\Big\}_{\tt TCB}  -
\nonumber \\ && \hskip 0pt -\, 
   c^{-4}\Big\{\int^t_{t_0}\Big({\textstyle{1 \over 8}}
v_{\tt M} ^4  
+ {\textstyle{3 \over 2}} v_{\tt M} ^2 \sum_{{\tt B\not= M}}{G M_{\tt B} \over r_{\tt BM}}- 
{\textstyle{\textstyle{1 \over 2}}}\Big[\sum_{{\tt B\not= M}} {G M_{\tt B} \over r_{\tt BM}}\Big]^2\Big)d{\tt TCB}
+
\Big({\textstyle{\textstyle{1 \over 2}}}v_{\tt M} ^2 +3\sum_{\tt B \not =M}{G M_{\tt B} \over r_{\tt BM}}  \Big)(\vec v_{\tt M}  \cdot \vec r_{\tt M} ) 
\Big\}_{\tt TCB} +\nonumber\\
  && \hskip 0pt +\, 
 {\cal O}\Big(c^{-5};\, 6.86\times 10^{-19}\,(t-t_0);\, 
 1.37\times 10^{-15}\,{\rm s}\Big),
\label{eq:(30-n)}
\end{eqnarray}
where $\vec v_{\tt M}$ is the solar system barycentric velocity vector of the Moon, and $\vec r_{\tt M}=\vec x-\vec x_{\tt M}$ is the {\tt BCRS} vector from the center of the Moon to the surface site. The potential and kinetic energy use the Moon centered reference frame.  The dot product annually reaches $\pm0.58~\mu$s with smaller variations  of $\pm 21$~ps at Moon's sidereal period of $t_{\tt M}=27.32166~{\rm d}$.

Similar to (\ref{eq:coord-tr-QQ}), ${\tt TCB} - {\tt TCL}$ from (\ref{eq:(30-n)})  has a mean rate given by constant $L_{\tt H}$  
{}
\begin{eqnarray}
\frac{1}{c^2}\Big( {\textstyle\frac{1}{2}}v_{\tt M}^2 + 
\sum_{{\tt B\not= M}} {G M_{\tt B} \over r_{\tt BM}} \Big)+\frac{1}{c^4}\Big({\textstyle{1 \over 8}}
v_{\tt M} ^4  
+ {\textstyle{3 \over 2}} v_{\tt M} ^2 \sum_{{\tt B\not= M}}{G M_{\tt B} \over r_{\tt BM}}- 
{\textstyle{\textstyle{1 \over 2}}}\Big[\sum_{{\tt B\not= M}} {G M_{\tt B} \over r_{\tt BM}}\Big]^2\Big)=L_{\tt H}+
\dot P_{\tt H}(t)+ {\cal O}\Big(c^{-5};\, 6.86\times 10^{-19}\Big),
 \label{eq:coord-tr-QQM}
\end{eqnarray}
where the constant $L_{\tt H}$ and periodic terms $\dot P_{\tt H}(t)$ are given as below \cite{Turyshev:2025}
{}
\begin{eqnarray}
L_{\tt H} &=&
\frac{1}{c^2}\Big< {\textstyle\frac{1}{2}}v_{\tt M}^2 + 
\sum_{{\tt B\not= M}} {G M_{\tt B} \over r_{\tt BM}} \Big>+\frac{1}{c^4}\Big<{\textstyle{1 \over 8}}
v_{\tt M} ^4  
+ {\textstyle{3 \over 2}} v_{\tt M} ^2 \sum_{{\tt B\not= M}}{G M_{\tt B} \over r_{\tt BM}}- 
{\textstyle{\textstyle{1 \over 2}}}\Big[\sum_{{\tt B\not= M}} {G M_{\tt B} \over r_{\tt BM}}\Big]^2\Big>+ {\cal O}\Big(c^{-5};\, 6.86\times 10^{-19}\Big),
 \label{eq:LH}\\
\dot P_{\tt H}(t) &=&
\frac{1}{c^2}\Big( {\textstyle\frac{1}{2}}v_{\tt M}^2 + 
\sum_{{\tt B\not= M}} {G M_{\tt B} \over r_{\tt BM}} \Big)+\frac{1}{c^4}\Big({\textstyle{1 \over 8}}
v_{\tt M} ^4  
+ {\textstyle{3 \over 2}} v_{\tt M} ^2 \sum_{{\tt B\not= M}}{G M_{\tt B} \over r_{\tt BM}}- 
{\textstyle{\textstyle{1 \over 2}}}\Big[\sum_{{\tt B\not= M}} {G M_{\tt B} \over r_{\tt BM}}\Big]^2\Big)- L_{\tt H},
\end{eqnarray}
where constant $L_{\tt H}$ results from the long-time averaging of the Moon's total orbital energy in the {\tt BCRS} determined as  $L_{\tt H}=1.482\,536\,24 \times 10^{-8} \approx 1.280\,913\,2$~ms/d, and $P_{\tt H}(t)$ represents a series of small periodic terms. If needed,  the term $P_{\tt H}(t)$ can be developed semi-analytically in the same manner as the time-series $P_{\tt }(t)$ for the Earth, e.g. \citep{Fairhead-Bretagnon:1990,Fukushima:1995,Irwin-Fukushima:1999,Fukushima:2010}. 

Eq.~(\ref{eq:(30-n)}) together with (\ref{eq:coord-tr-QQM}) gives the mean rate between ${\tt TCL}$ and ${\tt TCB}$:
{}
\begin{equation}
\Big<\frac{d{\tt TCL}}{d{\tt TCB}}\Big> =1-L_{\tt H}.
\label{eq:constTCGTCBH}
\end{equation}

\subsubsection{TL vs TCL}
\label{sec:tm-tdb}

The definition of the Lunar Time ({\tt TL}) is a bit trickier. In analogy with {\tt TT} (see Sec.~\ref{sec:TT}), may want to define time $\tt TL$ as a time scale  at or near the Moon’s surface that differs from $\tt TCL$ by a constant rate, with the unit of measurement of $\tt TL$ chosen at a well-justified reference surface on the Moon. Then, a lunar surface time $\tt TL$ may be defined as a time scale differing from {\tt TCL} by a constant rate, $ L_{\tt L}$, with an appropriately chosen unit of measurement.

To develop the relevant expression,  we consider  the transformation between proper and coordinate time in the {\tt LCRS} given by (\ref{eq:tau-LCRS-lim}) that is given in a form suitable for modern timekeeping applications in cislunar space:
{}
\begin{eqnarray}
\label{eq:prop-coord-time-LCRS-TL}
\frac{d\tau}{d{\cal T}}
&=&1
-\frac{1}{c^2}\Big\{\tfrac12\,{\cal V}^2+U_{\tt M}({\cal T},\vec {\cal X}) 
+U^*_{\rm tid}({\cal T},\vec {\cal X}) \Big\}+
{\cal O}\Big(c^{-4};\, 1.46\times 10^{-21}\Big),
\end{eqnarray}
where $U_{\tt M}({\cal T},\vec {\cal X})$ and $U^*_{\rm tid}({\cal T},\vec {\cal X})$ are is the Newtonian lunar gravitational and tidal potentials, correspondingly; $\vec {\cal V}=d\vec {\cal X}/d{\cal T}$ with $\cal V=|\vec{\cal V}|$ is the clock' velocity in the {\tt LCRS}. 
As we shall see in Sec.~\ref{sec:proper-time-LCRS}, the error bound in (\ref{eq:prop-coord-time-LCRS-TL}) is due to the largest $c^{-4}$-term omitted in (\ref{eq:tau-LCRS}) evaluated to contribute  \(c^{-4}\tfrac32{\cal V}_{\tt vLLO}^2U_{\tt M}\simeq c^{-4}\tfrac32{\cal V}_{\tt vLLO}^2\bigl(GM_{\tt M}/r_{\tt vLLO}\bigr)\simeq 1.46 \times 10^{-21}\).

Similar to  (\ref{eq:LG}), the transformation from the {\tt TL} to {\tt TCL} involves a rate change:
{}
\begin{equation}
\Big<\frac{	d{\tt TL}}{d{\tt TCL}}\Big> = 1 - \frac{1}{c^2}\big<U_{\tt gM}\big> \equiv 1  -  L_{\tt L},
\qquad {\rm or} \qquad\frac{d{\tt TCL}}{d{\tt TL}} = \frac{1}{1  -  L_{\tt L}} = 1 +  \frac{L_{\tt L}}{1  -  L_{\tt L}},
\label{eq:(26)}
\end{equation}
where $U_{\tt gM}$ is the combined rotational, gravitational, and tidal potential at a yet-to-be-defined surface or location. 

In practice, the reference value \(L_{\mathrm{L}}\) remains ambiguous. For consistency, it is most natural to anchor \(L_{\mathrm{L}}\) on the lunar selenoid (the Moon’s geoid). However, deploying and interconnecting a network of high‑precision clocks across the lunar surface—analogous to the terrestrial realization of \texttt{TT} via \texttt{TAI}~\cite{Turyshev:2025}—is not foreseen in the near term. Rather, current lunar exploration efforts envisage one or two primary frequency standards located near the lunar South Pole. 

To consider both of the plausible locations, using (\ref{eq:prop-coord-time-LCRS-TL}), we introduce $L_{\tt L}$ as below
{}
\begin{equation}
L_{\tt L} \equiv \frac{1}{c^2}\Big<U_{\tt gM}\Big>=\frac{1}{c^2}\Big<\tfrac12\,{\cal V}^2+U_{\tt M}({\cal T},\vec {\cal X}) 
+U^*_{\rm tid}({\cal T},\vec {\cal X})\Big>+{\cal O}\Big(c^{-4};\, 1.46\times 10^{-21}\Big),
\label{eq:LL-selen)}
\end{equation}
where $U_{\tt gM}$ is the reference surface of the selenopotential at a yet to be specified location either on the selenoid or at a particular  location neat the South Pole. Below, we examine both these possibilities.

\paragraph{Selenopotential:}

Although, it is natural to define the selenopotential (and, thus, the constant $L_{\tt L}$) based on (\ref{eq:prop-coord-time-LCRS-TL}), there is significant uncertainty in  determining the reference level surface of the selenopotential, \( U_{\tt gM} \). Reported \( U_{\tt gM} \) values vary widely, from \(2825390~\mathrm{m}^2\mathrm{s}^{-2} \) \citep{Bursa-Sima:1980}, derived from gravity measurements at the Apollo 12 landing site, to \( 2821713.3~\mathrm{m}^2\mathrm{s}^{-2} \) \citep{Martinec-Pec:1988}, based on a lunar gravity model \citep{Ferrari-etal:1980} utilizing Doppler tracking data from Lunar Orbiter 4 and Lunar Laser Ranging (LLR) data, adjusted for lunar topography. More recently, \( U_{\tt gM} = (2822336.927 \pm 23)~\mathrm{m}^2\mathrm{s}^{-2} \) \citep{Ardalan-Karimi:2014} was determined using pre-GRAIL global gravity models (GGMs), incorporating topographic bias corrections on geoidal heights.

We begin by considering a clock on the lunar reference radius for gravity of $R_{\tt MQ} = 1738.00\;\mathrm{km}$. Because the Moon is in synchronous rotation, the clock’s velocity in the {\tt LCRS} frame is purely due to that rotation, thus 
${\cal V} = \omega_{\tt M}\,R_{\tt MQ}\sin\theta_{\tt M} \simeq 4.62\,{\rm m/s}$, with $\omega_{\tt M}
= {2\pi}/{T_{\rm sid}}
\approx2.66\times10^{-6}\;\mathrm{s^{-1}},
$ and $T_{\rm sid}\approx27.32\;\mathrm{d}.$
For  $\theta_{\tt M}=\tfrac\pi2$, one finds
$c^{-2}\,{\textstyle\frac{1}{2}}{\cal V}^2
= \frac12\omega_{\tt M}^2\,R_{\tt MQ}^2/c^2
\approx 1.19\times10^{-16}$,
which exceeds our retention threshold of $5\times10^{-18}$ and thus must be kept.

The next contribution is from the Moon’s gravitational potential evaluated at the surface,
$ c^{-2}\,U_{\tt M}
=c^{-2}{GM_{\tt M}}/{R_{\tt MQ}}
\approx3.14\times10^{-11}$,
a term that dominates all others in magnitude.  Given the values of the lunar gravitational spherical harmonics (Table~\ref{tab:sp-harmonics-moon}), achieving our target accuracy requires including a large number of additional harmonics.

The  tidal quadrupole perturbations  due to the Earth and the Sun are evaluated to be 
\begin{equation}
\label{eq:LCRS-surf-tide}
c^{-2}\,U_{\rm tid[2]}^{*\tt (E)}
\;\lesssim\;\frac{GM_{\tt E}\,R_{\tt MQ}^2}{c^2r_{\tt EM}^3}
P_2(\vec n_{\tt EM}\!\cdot\!\widehat{\vec {\cal X}})\approx 2.36\times10^{-16}
\qquad
c^{-2}\,U_{\rm tid[2]}^{*\tt (S)}\;\lesssim\;\frac{GM_{\tt S}\,R_{\tt MQ}^2}{c^2{\rm AU}^3}P_2(\vec n_{\tt SM}\!\cdot\!\widehat{\vec {\cal X}})
\lesssim 1.33\times10^{-18},
\end{equation}
which, after averaging, yield values $\tfrac14$ smaller, i.e., $5.90 \times 10^{-17}$ for the Earth $(\ell=2)$ and $3.33 \times 10^{-19}$ for the solar $(\ell=2)$ tide, correspondingly.  With $(\ell=3)$ tides averaging out to zero, the contributions of $(\ell=4)$ tides are negligible.

 Hence, retaining only terms larger than $5\times10^{-18}$, the selenoid-based definition of $L_{\tt L}$ becomes
\begin{equation}
\label{eq:LL-selen}
L^{\rm sel}_{\tt L} \simeq
\frac{1}{c^2}\Big\{
   \tfrac12\,\omega_{\tt M}^2\,R_{\tt MQ}^2
 + \big<U_{\tt M}({\cal T},\vec {\cal X})\big> \big|_{\rm sel}
 + \frac{GM_{\tt E}R_{\tt MQ}^2}{4r^3_{\tt EM}}
\Big\}
+ {\cal O}\Bigl(c^{-4};\,3.33\times10^{-19}\Bigr),
\end{equation}
where  the last term in the averaged value of the Earth's tidal quadrupole potential from (\ref{eq:LCRS-surf-tide}), contributing to the rate $c^{-2}\big<U_{\rm tid[2]}^{*\tt (E)}\big>=c^{-2}{GM_{\tt E}R_{\tt MQ}^2}/({4r^3_{\tt EM}})\simeq 5.90\times 10^{-17}$. The error bound here due to the $\tfrac14$ part of the solar quadrupole tide shown in (\ref{eq:LCRS-surf-tide}).  

Limiting in (\ref{eq:LL-selen}), the lunar gravity potential (\ref{eq:pot_w_0sh}) only to  quadrupole, $J_{\tt 2M}$,  one obtains a less accurate value   
{}
\begin{equation}
L^{\rm sel}_{\tt L} 
\simeq \frac{1}{c^2}\Big\{{\textstyle\frac{1}{2}}  \omega_{\tt M}^2R_{\tt MQ}^2 + \frac{GM_{\tt M}}{R_{\tt MQ}} \big( 1 + {\textstyle\frac{1}{2}}J_{2\tt M}\big) \Big\}+{\cal O}\Big(c^{-4};\, 2.11 \times 10^{-15}\Big),
\label{eq:(29)}
\end{equation}
where to estimate $L_{\tt L}$, we adopted: lunar reference radius for gravity is $R_{\tt MQ} = 1738.0$~km, which is larger than the mean radius of $R_{\tt M} = 1737.1513$~km \citep{Smith-etal:2017}, the lunar gravitational constant $GM_{\tt M} = 4902.800118~{\rm km}^3/{\rm s}^2$ (DE440, \citep{Park-etal:2021}), gravity harmonic $J_{2 \tt M}$ is $2.033\times 10^{-4}$ \citep{Williams-etal:2014}, and $\omega_{\tt M} = 2\pi/(27.321\,661~{\rm d} \times 86400~{\rm s/d})=2.6616996\times 10^{-6}~{\rm s}^{-1}$. The value of $L_{\tt L}$, with the larger radius, $R_{\tt MQ} $, is then estimated to be $L_{\tt L}^{\rm sel}=
3.139\,05 \times 10^{-11}\simeq 2.7121 \,\mu$s/d. Also, the error bound in (\ref{eq:(29)}) is set by the omitted term with the tesseral harmonics $C_{22}=2.242\,615\times 10^{-5}$ of the Moon's gravity field \citep{Konopliv-etal:2013}. Note that, if the smaller value  for the lunar radius $R_{\tt M}$ is used in (\ref{eq:(29)}), the result is $L_{\tt L}^{\rm sel}=
3.140\,59\times 10^{-11}\simeq 
2.7135\,\mu$s/d.

\paragraph{Lunar South Pole:} Evaluating (\ref{eq:LL-selen)}) at the lunar South Pole, we recognize that for $\theta_{\tt M}\simeq0$, we have no kinetic contribution. In addition, contribution from lunar gravity spherical harmonics will be different, yielding:
\begin{equation}
\label{eq:LL-SP}
L_{\tt L}^{\rm pole} \simeq
\frac{1}{c^2}\Big\{
 \big<U_{\tt M}({\cal T},\vec {\cal X})\big> \big|_{\rm pole}
 + \frac{GM_{\tt E}R_{\tt MQ}^2}{4r^3_{\tt EM}}
\Big\}
+ {\cal O}\Bigl(c^{-4};\,3.33\times10^{-19}\Bigr),
\end{equation}
Again, truncating lunar gravity potential (\ref{eq:pot_w_0sh}) at the quadrupole level, we have 
{}
\begin{equation}
L_{\tt L}^{\rm pole} 
\simeq \frac{1}{c^2}\Big\{\frac{GM_{\tt M}}{R_{\tt MQ}} \big( 1 - J_{2\tt M}\big)\Big\}+{\cal O}\Big(c^{-4};\,2.11 \times 10^{-15}\Big),
\label{eq:(29)-sp}
\end{equation}
with the estimated value of $L_{\tt L}^{\rm pole}=
3.138\,09 \times 10^{-11}\simeq 2.7113 \,\mu$s/d. With the smaller value of the lunar radius the value is $L_{\tt L}^{\rm pole}=
3.139\,62 \times 10^{-11}\simeq 2.7126 \,\mu$s/d. Thus, the difference between the two possible definitions of the constant $L_{\tt L}$ is small, and, depending on the chosen lunar radius, it is either $\delta L_{\tt L}\simeq2.17$\,ns/d for $R_{\tt MQ}$ or $\delta L_{\tt L}\simeq0.84$\,ns/d for $R_{\tt M}$, with both differences potentially measurable at the current sensitivity of timing instruments.

Note that, by \eqref{eq:(29)} and  \eqref{eq:(29)-sp}, the constant \(L_{\tt L}\) is determined only to \(\mathcal O(2.11\times10^{-15})\), a factor of $\sim 10^3$ less precise than our chosen accuracy threshold of $5\times 10^{-18}\simeq 0.4$\,ps/d. Achieving higher precision would require including many higher‐degree terms in the lunar gravity potential—an impractical task given the logistical challenges of deploying and synchronizing multiple high‐stability clocks on the lunar surface. In practice, only one or two clocks are likely to operate at the lunar bases, which may be insufficient to refine \(L_{\tt L}\) beyond its current uncertainty. Therefore, analogous to the IAU decision for \(L_{\tt G}\) in the {\tt GCRS}, the constant $L_{\tt L}$ may also become a defining constant for the {\tt LCRS}.

Ultimately, the constant  $L_{\tt L}$ allows us to establish the scaling of  coordinates and mass factors to maintain the invariance of the speed of light and the equations of motion in the {\tt LCRS},  for the transformation from {\tt TCL} to {\tt TL}. Similarly to (\ref{eq:interv}), this transformation, which includes the scaling of temporal and spatial coordinates and mass factors, ensures the invariance of the metric (up to a constant factor) and has the form:  
{}
\begin{equation}
(ds^2)_{\tt TL} = (1- L_{\tt L})^2ds^2_{\tt TCL} ,
\label{eq:interv-M}
\end{equation}
where $(ds^2)_{\tt TL}$ maintains the same form in terms of {\tt TL}, $\vec {\cal X}_{\tt TL}$, $(G M )_{\tt TL}$ as  (\ref{eq:G00tr-MM})--(\ref{eq:Gabtr-MM}) do in terms of $\cal T$, $\vec {\cal X}$, $(GM)_{\tt TCL}$. 

As a result, instead of using coordinate time ${\cal T} = {\tt TCL}$, spatial coordinates $\vec{\cal X}$, and mass factors $(G M)_{\tt TCL}$ related to the ({\tt LCRS}), we will use the scaling for the relevant quantities in the Lunar Surface Coordinate Reference System ({\tt LSCRS}). To establish these relations, we integrate (\ref{eq:(26)}) from ${\tt T_{L0}}$ to ${\tt TCL}$, deriving the connection between the two time scales. Additionally, the spatial coordinates and mass factors are adjusted in accordance with (\ref{eq:interv-M}), resulting in:
{}
\begin{equation}
{\tt TL} = {\tt TCL}- L_{\tt L}({\tt TCL}-{\tt T}_{\tt L0}), \qquad
\vec {\cal X}_{\tt TL} = (1-L_{\tt L})\vec {\cal X}_{\tt TCL}, \qquad
(GM)_{\tt TL} = (1-L_{\tt L})(GM)_{\tt TCL},
\label{eq:LSCRS}
\end{equation}
where $ {\tt T_{L0}} $  is the initial lunar time, which, for now, we will use unspecified. 

Eqs.~(\ref{eq:(29)})–(\ref{eq:(29)-sp}) show that a purely geodetic definition of $L_L$ at the $\le 5\times 10^{-18}$ level is not yet practical; 
the dispersion from $R_{\tt MQ}$, $J_{\tt 2M}$, $C_{22}$, and tide/Love‑number variability is $\mathcal{O}(10^{-15})$.
Accordingly, and by analogy with $L_G$ for TT, we recommend treating $L_{\tt L}$ as a \emph{conventional} rate constant for TL. For early lunar timekeeping, fix $L_{\tt L}$ to a reference value $L_{\tt L}^{\rm (def)}=3.13905\times 10^{-11}$ (consistent with the selenoid‑based estimate in Eq.~(\ref{eq:(29)})), 
and realize it operationally at the reference site(s) using the best available gravity model. If a South‑Pole realization is preferred, document the realized offset $\delta L_{\tt L}$ relative to $L_{\tt L}^{\rm (def)}$ and update it as models improve.

\subsubsection{TL vs TCB}
\label{sec:TL-TCB}

To express  ${\tt TCB}$ via ${\tt TL}$, we need another constant that we call $L_{\tt M}$, which determines the rate  between ${\tt TL}$  and ${\tt TCB}$ and, similarly to (\ref{eq:constTCGTCB}),  may be formally introduced as 
{}
\begin{equation}
\Big<\frac{d{\tt TL}}{d{\tt TCB}}\Big> =1-L_{\tt M}.
\label{eq:LM}
\end{equation}

One may define the constant $L_{\tt M}$ by using the total solar system's kinetic and gravitational energy at the origin of the {\tt LCRS} and the sum of the lunar rotational, gravitational and tidal potentials at the reference surface on the Mon (similarly to the case of the constant $L_{\tt B}$, as discussed in \cite{Turyshev:2025}). Thus, keeping only the quadrupole term,  the rate $L_{\tt M}$ defined at the lunar selenoid is given as below 
{}
\begin{equation}
L_{\tt M} = \frac{1}{c^2}\Big\{ GM_{\tt S} \Big<\frac{1}{r_{\tt MS}}\Big> + GM_{\tt E} \Big<\frac{1}{r_{\tt ME}}\Big> + \big<U_{\tt MP}\big> + \big<{\textstyle\frac{1}{2}}v_{\tt M}^2\big> + {\textstyle\frac{1}{2}} \omega_{\tt M}^2R_{\tt MQ}^2 + \frac{GM_{\tt M}}{R_{\tt MQ}} ( 1 + {\textstyle\frac{1}{2}}J_{\tt 2M}) \Big\} +{\cal O}(c^{-4}; 2.11 \times 10^{-15}),
\label{eq:(nonN4)}
\end{equation}
with the error bound set by the omitted term with the tesseral harmonics $C_{22}$ of the lunar gravity field, as in (\ref{eq:(29)}), with the cumulative effect of the higher harmonics terms omitted (\ref{eq:(29)}) being on the same order, e.g., $2\times 10^{-15}$.  

Thus, the analytical definition of $L_{\tt M}$ with an accuracy below $10^{-15}$ encounters similar technical challenges—such as spatial and temporal variability at higher degrees and orders of spherical harmonics—as those discussed above for $L_{\tt L}$. This may necessitate declaring $L_{\tt M}$ as a defining constant for the {\tt LCRS}, analogous to the treatment of $L_{\tt B}$ in the {\tt GCRS}.  From (\ref{eq:(nonN4)}), the  value of the constant was found to be $L_{\tt M} =
1.485\,675\,290 \times 10^{-8}\approx 1.283\,62$~ms/d. 

Alternatively, we can use the chain of time derivatives, to establish the relationships between the constants $L_{\tt M}, L_{\tt L}$ and $L_{\tt H}$, similar to (\ref{eq:constLBLCLG}). Following this approach, with the help of (\ref{eq:(26)}), (\ref{eq:constTCGTCBH}) and (\ref{eq:LM}), we have the following expression  
{}
\begin{equation}
\Big<\frac{d{\tt TL}}{d{\tt TCB}}\Big>=\Big(\frac{d{\tt TL}}{d{\tt TCL}}\Big)\Big<\frac{d{\tt TCL}}{d{\tt TCB}}\Big> \qquad \Rightarrow \qquad (1-L_{\tt M})=(1-L_{\tt L})(1-L_{\tt H}),
\label{eq:consTM}
\end{equation}
from which the constant $L_{\tt M}$ is determined as $L_{\tt M} \simeq L_{\tt L}+L_{\tt H}-L_{\tt L}L_{\tt H}=
1.485\,675\,294\times 10^{-8}\approx 1.283\,62$~ms/d.

Note that the analytical determination of $L_{\tt M}$ below $10^{-15}$ is limited by the same geophysical uncertainties as $L_{\tt L}$. In practice, $L_{\tt M}$ should be inferred from $(L_{\tt L},L_{\tt H})$ via Eq.~(\ref{eq:consTM}), and treated as conventional for {\tt TL} standardization.

\subsection{Transformation {\tt TL} vs {\tt TDB}} \label{sec:tL-tdb}

It is useful to express the difference $({\tt TDB} - {\tt {\tt TL}})$ as a function of {\tt TDB}. This can be done by using (\ref{eq:TDBC}), (\ref{eq:LSCRS}), (\ref{eq:(30-n)}), and (\ref{eq:consTM}), yielding result  below
{}
\begin{eqnarray}
{\tt TDB}  -  {\tt TL} &= &
\frac{1-L_{\tt L}}{1-L_{\tt B}}\,
{\tt TDB}_0
 -\frac{L_{\tt B}-L_{\tt L}}{1-L_{\tt B}} \big( {\tt TDB}-{\tt T}_0\big)  +L_{\tt L}({\tt T}_0-{\tt T}_{\tt LO})+ \nonumber\\
  && \hskip -40pt +\,   
  \frac{1-L_{\tt L}}{1-L_{\tt B}}\Big\{ \frac{1}{c^2}
\int_{{\tt T_0+TDB_0}}^{\tt TDB}\Big( {\textstyle\frac{1}{2}}v_{\tt M}^2 +  \sum_{{\tt B\not= M}}{G M_{\tt B} \over r_{\tt BM}} \Big) d{\tt TDB} +
 \frac{1}{c^2} (\vec v_{\tt M}\cdot \vec {r_{\tt M}}_{\tt TDB}) +\nonumber\\
 && \hskip -2pt +\,   
  \frac{1}{c^4}\int_{{\tt T_0+TDB_0}}^{\tt TDB}\Big({\textstyle{1 \over 8}}
v_{\tt M} ^4  
+ {\textstyle{3 \over 2}} v_{\tt M} ^2 \sum_{{\tt B\not= M}}{G M_{\tt B} \over r_{\tt BM}}- 
{\textstyle{\textstyle{1 \over 2}}}\Big[\sum_{{\tt B\not= M}} {G M_{\tt B} \over r_{\tt BM}}\Big]^2\Big)d{\tt TDB}
+
 \frac{1}{c^4}\Big({\textstyle{\textstyle{1 \over 2}}}v_{\tt M} ^2 +3\sum_{\tt B \not =M}{G M_{\tt B} \over r_{\tt BM}}  \Big)(\vec v_{\tt M}  \cdot {\vec r_{\tt M}}_{\tt TDB} ) 
\Big\} +\nonumber\\
  && \hskip 0pt +\, 
 {\cal O}\Big(c^{-5};\, 6.86\times 10^{-19}\,({\tt TDB -T_0- TDB_0});\, 
 1.37\times 10^{-15}\,{\rm s}\Big).
\label{eq:(nonN2-in+)}
\end{eqnarray}

Finally,  evaluating the integral in (\ref{eq:(nonN2-in+)}) with the help of (\ref{eq:coord-tr-QQM}), we derive the following result:
{}
\begin{eqnarray}
{\tt TDB}  -  {\tt TL} &=&   \frac{1-L_{\tt M}}{1-L_{\tt B}}\,
{\tt TDB}_0 - \frac{L_{\tt B}-L_{\tt M}}{1-L_{\tt B}} \big({\tt TDB}-{\tt T}_0\big)  +L_{\tt L}({\tt T}_0-{\tt T}_{\tt LO})+\nonumber\\&+& 
P_{\tt H}({\tt TDB}) -P_{\tt H}({\tt T}_0+{\tt TDB}_0)+ \frac{1}{c^2}(\vec v_{\tt M}\cdot \vec {r_{\tt M}}_{\tt TDB})+
\frac{1}{c^4}\Big({\textstyle{\textstyle{1 \over 2}}}v_{\tt M} ^2 +3\sum_{\tt B \not =M}{G M_{\tt B} \over r_{\tt BM}}  \Big)(\vec v_{\tt M}  \cdot \vec {r_{\tt M}}_{\tt TDB} )+\nonumber\\
  &+& 
 {\cal O}\Big(c^{-5};\, 6.86\times 10^{-19}\,({\tt TDB -T_0- TDB_0});\, 
 1.37\times 10^{-15}\,{\rm s}\Big).~~
\label{eq:(nonN2+)}
\end{eqnarray}

As seen from (\ref{eq:(nonN2+)}), there is a rate difference between  ${\tt TL}$ and  ${\tt TDB}$, that is given by the combination of the constants $ (L_{\tt B}-L_{\tt M})/(1-L_{\tt B})=
6.484\,440\,414 \times 10^{-10}\simeq 56.0256~\mu$s/d, with {\tt TL} running faster than {\tt TDB}. In addition, there is also a series of small periodic terms $\propto P_{\tt H}({\tt TDB})$ and the term that depends on the lunar surface position $(\vec v_{\tt M}\cdot \vec {r_{\tt M}}_{\tt TDB})/c^2$.
 
\subsection{Practical realization of the {\tt LCRS} origin}
\label{sec:LCRS-realization}

The formal {\tt LCRS} defined above is centered on the lunar center of mass (the lunocenter), but realizing this origin at the level implied by our $5\times10^{-18}$ fractional-frequency and $0.1$ ps timing thresholds is highly nontrivial. In practice, the lunocenter is realized through a combination of dynamical lunar ephemerides and LLR data. As emphasized in \cite{Sosnica-etal-2025-ILRF}, the sparse and geographically non-uniform distribution of lunar retroreflectors leads to significant correlations between the scale and the components of the lunocenter in principal-axis coordinates, limiting the accuracy of the realized International Lunar Reference Frame (ILRF). Their combined solution yields origin and orientation errors at the several-centimeter level for the 2010--2030 period, corresponding to light travel-time uncertainties at the few $10^{-10}$ s level, which are much larger than the formal $0.1$ ps precision of our time-scale transformations.

Our analysis should therefore be understood as providing the theoretical accuracy of the {\tt TCL} and {\tt TL} transformations in an ideal reference frame. In any practical implementation, the
realized {\tt LCRS}/{\tt ILRF} origin, orientation, and scale will introduce additional errors that must be propagated into the total timing and positioning error budget. As the lunar retroreflector
network is densified and complemented by orbiter tracking and altimetry
\cite{Fienga-Rambaux-Sosnica-2024,Sosnica-etal-2025-ILRF}, the ILRF origin and orientation should improve, progressively reducing this discrepancy between the formal and realized accuracy of the
{\tt LCRS}.

\section{Transformations Between TL and TT}
\label{sec:imp}

With the introduction of the lunar timescale {\tt TL}, establishing its relation to  ({\tt TT}) is essential. In this section, we derive the $({\tt TL}-{\tt TT})$ transformation formulas required for high‑precision PNT applications in cislunar space.

\subsection{Expressing $({\tt TL} - {\tt TT})$  as a function of  {\tt TDB}}
\label{sec:tl-tt-tdb}

Using results obtained in Secs.~\ref{sec:time-E} and \ref{sec:time-M}, we can establish the relationships between {\tt TT} and {\tt TL}. For this purpose, we may use (\ref{eq:(nonN1-TCB)}) and (\ref{eq:(nonN2-in+)}) that involve the common time {\tt TDB}. Using these expressions, we can formally write:
{}
\begin{eqnarray}
{\tt TL} -   {\tt TT} &=&\frac{L_{\tt G}-L_{\tt L}}{1-L_{\tt B}}\big( {\tt TDB}-{\tt T_0}-{\tt TDB}_0\big) -L_{\tt L}({\tt T}_0-{\tt T}_{\tt LO})+\nonumber\\
  &&\hskip -50pt +\,  
  \frac{1}{1-L_{\tt B}}\bigg[ \frac{1}{c^2}\int_{\tt T_0+TDB_0}^{\tt TDB} \Big\{\Big( {\textstyle\frac{1}{2}}v_{\tt E}^2 + \sum_{{\tt B\not= E}} {G M_{\tt B} \over r_{\tt BE}} \Big)-\Big( {\textstyle\frac{1}{2}}v_{\tt M}^2 +   \sum_{{\tt B\not= M}}{G M_{\tt B} \over r_{\tt BM}}\Big)\Big\} d{\tt TDB}+
 \frac{1}{c^2}\Big( (\vec v_{\tt E}\cdot \vec {r_{\tt EM}})-(\vec v_{\tt EM}\cdot \vec {r_{\tt M}}_{\tt TDB})\Big)+
 \nonumber\\
 &&\hskip -50pt +\,  
  \frac{1}{c^4}\int_{{\tt T_0+TDB_0}}^{\tt TDB} \Big\{\Big({\textstyle{1 \over 8}}
v_{\tt E} ^4  
+ {\textstyle{3 \over 2}} v_{\tt E} ^2 \sum_{{\tt B\not= E}}{G M_{\tt B} \over r_{\tt BE}}- 
{\textstyle{\textstyle{1 \over 2}}}\Big[\sum_{{\tt B\not= E}} {G M_{\tt B} \over r_{\tt BE}}\Big]^2\Big)-
\Big({\textstyle{1 \over 8}}
v_{\tt M} ^4  
+ {\textstyle{3 \over 2}} v_{\tt M} ^2 \sum_{{\tt B\not= M}}{G M_{\tt B} \over r_{\tt BM}}- 
{\textstyle{\textstyle{1 \over 2}}}\Big[\sum_{{\tt B\not= M}} {G M_{\tt B} \over r_{\tt BM}}\Big]^2\Big)\Big\}d{\tt TDB}
+ \nonumber\\
 &+&
 \frac{1}{c^4}\Big\{\Big({\textstyle{\textstyle{1 \over 2}}}v_{\tt E} ^2 +3\sum_{\tt B \not =E}{G M_{\tt B} \over r_{\tt BE}}  \Big)(\vec v_{\tt E}\cdot \vec {r_{\tt E}}_{\tt TDB})-
\Big({\textstyle{\textstyle{1 \over 2}}}v_{\tt M} ^2 +3\sum_{\tt B \not =M}{G M_{\tt B} \over r_{\tt BM}}  \Big)(\vec v_{\tt M}  \cdot {\vec r_{\tt M}}_{\tt TDB}) \Big\}- \nonumber\\
 &&\hskip -50pt -\,  
\frac{1}{c^2}  \Big[
\int_{{\tt T_0+TDB_0}}^{\tt TDB}\Big\{ L_{\tt G}\Big({\textstyle{1\over 2}} v_{\tt E}^{2} +
\sum_{{\tt B\not= E}} {G M_{\tt B} \over r_{\tt BE}} \Big)-L_{\tt L}\Big( {\textstyle\frac{1}{2}}v_{\tt M}^2 +   \sum_{{\tt B\not= M}}{G M_{\tt B} \over r_{\tt BM}} \Big)\Big\}d{\tt TDB} +
L_{\tt G}(\vec v_{\tt E}\cdot \vec {r_{\tt E}}_{\tt TDB})-L_{\tt L}(\vec v_{\tt M}\cdot \vec {r_{\tt M}}_{\tt TDB})
\Big]\bigg]+   \nonumber\\
&+& 
 {\cal O}\Big(c^{-5};\, 6.86\times 10^{-19}\,({\tt TDB -T_0- TDB_0});\, 
 1.37\times 10^{-15}\,{\rm s}\Big),
\label{eq:(nonN2-in+2)}
\end{eqnarray}
where we used $(\vec v_{\tt E} \cdot \vec {r_{\tt E}})-(\vec v_{\tt M}\cdot \vec {r_{\tt M}})=(\vec v_{\tt E} \cdot \vec {r_{\tt EM}})-(\vec v_{\tt EM}\cdot \vec {r_{\tt M}})$, with ${r_{\tt M}}_{\tt TDB}$ being the {\tt TDB}-compatible positon of the lunar clock. The constants $ L_{\tt G}, L_{\tt C}, L_{\tt B}$  for the Earth and $L_{\tt L}, L_{\tt H}, L_{\tt M}$ for the Moon.  The constants ${\tt T}_0({\tt MJD})={\rm MJD} 43144 + 32.184~{\rm s}$ and ${\tt TDB}_0 = -65.5~\mu$s are defining constants \cite{Kaplan:2005,Petit-Luzum:2010}. The constant ${\tt T}_{\tt L0}$ has yet to be chosen.  Note that the largest term in (\ref{eq:(nonN2-in+2)}) that involves the constants multiplying the integrals, evaluated as  
$c^{-2} L_{\tt G}({\textstyle{1\over 2}} v_{\tt E}^{2} +
\sum_{{\tt B\not= E}} {G M_{\tt B}/ r_{\tt BE}})\simeq c^{-2} L_{\tt G}({\textstyle{1\over 2}} v_{\tt E}^{2} +G M_{\tt M}/ r_{\tt ME}+
G M_{\tt S}/ {\rm AU})\simeq 1.03\times 10^{-17}$ and should be kept, while the $L_{\tt L}$-term is of the order of 
$L_{\tt L}( {\textstyle\frac{1}{2}}v_{\tt M}^2 +   \sum_{{\tt B\not= M}}{G M_{\tt B} / r_{\tt BM}} )\simeq L_{\tt L}( {\textstyle\frac{1}{2}}(v_{\tt E}+v_{\tt EM})^2 +   {G M_{\tt E} / r_{\tt EM}}+{G M_{\tt S} /{\rm AU}} )\simeq 4.76\times 10^{-19}$, 
which is too small for our purposes.

\subsection{Explicit form of the constant and periodic terms}
\label{sec:AppB}

Eq.~(\ref{eq:(nonN2-in+2)}) relates  {\tt TL} and {\tt TT} with {\tt TDB} being a common time scale. Considering our target time transfer uncertainty of $0.1$~ps and the time rate uncertainty of $5.0 \times 10^{-18}=0.43$~ps/d, we can introduce simplifications.  Our objective here is to   establish a more simplified relationships  between these times scales. 

\subsubsection{The $c^{-2}$ terms}

We begin with the $c^{-2}$-terms in (\ref{eq:(nonN2-in+2)}) that involves the total energy at the Earth's orbit that is given as below:
{}
\begin{eqnarray}
\frac{1}{c^2}\Big({\textstyle\frac{1}{2}}v_{\tt E}^2 +\sum_{{\tt B}\not={\tt E}}\frac{GM_{\tt B}}{r_{\tt BE}} \Big)=
\frac{1}{c^2}\Big({\textstyle\frac{1}{2}}v_{\tt E}^2 +\frac{GM_{\tt S}}{r_{\tt SE}}+\frac{GM_{\tt M}}{r_{\tt ME}}+\sum_{{\tt B}\not={\tt E,M,S}}\frac{GM_{\tt B}}{r_{\tt BE}}\Big) +{\cal O}\Big(4.80 \times 10^{-20}\Big),
\label{eq:enE}
\end{eqnarray}
where the error bound is set by the omitted contribution for the solar quadrupole moment  $J_2 = 2.25 \times 10^{-7}$ \cite{Park:2017,MecheriMeftah:2021} in the time transformations (\ref{eq:coord-tr-T1-rec}) and shown by (\ref{w_ext-mono}).

To consider the Moon-related terms  in (\ref{eq:(nonN2-in+2)}),  it is instructive to express the {\tt BCRS}  position vector between a body {\tt B} and the Moon as $\vec r_{\tt BM}=\vec r_{\tt BE}+\vec r_{\tt EM}$, where $\vec r_{\tt BE}=\vec x_{\tt E}-\vec x_{\tt B}$ is the position vector from the body {\tt B} to the Earth, and  $\vec r_{\tt EM}=\vec x_{\tt M}-\vec x_{\tt E}$ is the Earth-Moon relative position vector, also $r_{\tt BM}\equiv|\vec x_{\tt BM}|$, $r_{\tt EM}\equiv|\vec x_{\tt EM}|$.  By treating $r_{\tt EM}/ r_{\tt BE}$ as a small parameter, we can express ${GM_{\tt B}}/{r_{\tt BM}} $ in the form of a series of tidal terms, as shown below:
{}
\begin{eqnarray}
 \frac{GM_{\tt B}}{r_{\tt BM}} &=& \frac{GM_{\tt B}}{r_{\tt BE}} - \frac{GM_{\tt B}}{r^3_{\tt BE}}(\vec r_{\tt BE}\cdot \vec r_{\tt EM}) + \sum_{\ell=2}^{N_{\tt a}}
\frac{GM_{\tt B}}{r_{\tt BE}} \Big(\frac{r_{\tt EM}}{r_{\tt BE}}\Big)^\ell P_\ell(\vec n_{\tt BE}\cdot \vec n_{\tt EM}) +{\cal O}\Big(\frac{1}{r_{\tt BE}}\Big(\frac{r_{\tt EM}}{r_{\tt BE}}\Big)^{N_{\tt a}+1} \Big),~~~
\label{eq:expand2}
\end{eqnarray}
where term with the sum is the tidal potential of external bodies at the Moon, evaluated at the Earth-Moon distance with the Sun being responsible for the dominant contribution:
{}
\begin{eqnarray}
\label{eq:(nonN2-in+2)-tid}
W^\odot_{\tt EM}
 &=&\sum_{\ell=2}^{3}
\frac{GM_{\tt S}}{r_{\tt SE}} \Big(\frac{r_{\tt EM}}{r_{\tt SE}}\Big)^\ell P_\ell(\vec n_{\tt SE}\cdot \vec n_{\tt EM}) +{\cal O}\Big(4.30\times 10^{-19}\Big),
\end{eqnarray}
where we kept the solar  octupole tidal term  $\ell =3$. The magnitude of this term was estimated to be $\simeq 1.68\times 10^{-16}$, which is small, but large enough to be part of the model.  The error bound here is set by the solar $\ell=4$ (thus, $N_a=4$) tidal contribution, evaluated to be $c^{-2}GM_{\tt S}/r_{\tt SE}(r_{\tt EM}/r_{\tt SE})^4\simeq 4.30\times 10^{-19}$. 

Using result (\ref{eq:expand2}), and representing $\vec v_{\tt M}=\vec v_{\tt E}+\vec v_{\tt EM}$, where $\vec v_{\tt E}$ is the {\tt BCRS} velocity of the Earth and $\vec v_{\tt EM}$ is the Earth-Moon relative velocity, we present the $c^{-2}$-terms in (\ref{eq:(nonN2-in+2)}) as below: 
{}
\begin{eqnarray}
\frac{1}{c^2}\Big({\textstyle\frac{1}{2}}v_{\tt M}^2+  \sum_{{\tt B}\not={\tt M}}\frac{GM_{\tt B}}{r_{\tt BM}} \Big)&=&\frac{1}{c^2}\Big\{
{\textstyle\frac{1}{2}}v_{\tt E}^2+{\textstyle\frac{1}{2}}v_{\tt EM}^2 +  \frac{GM_{\tt E}-GM_{\tt M}}{r_{\tt EM}}+\sum_{{\tt B}\not={\tt E,M,S}}\frac{GM_{\tt B}}{r_{\tt BE}}+
\frac{GM_{\tt S}}{r_{\tt SE}} +W^\odot_{\tt EM}+ \frac{d}{dt}(\vec v_{\tt E}\cdot \vec r_{\tt EM}) \Big\}+\nonumber\\
&+&
{\cal O}\Big(4.30\times 10^{-19}\Big),~~~~~
\label{eq:expra=9g}
\end{eqnarray}
where, to the required level of accuracy, the Earth's  acceleration in {\tt BCRS}, $\vec a_{\tt E}$, is given  by its Newtonian part, yielding 
{}
\begin{eqnarray}
 -\sum_{{\tt B}\not={\tt E,M}}\frac{GM_{\tt B}}{r^3_{\tt BE}}(\vec r_{\tt BE}\cdot \vec r_{\tt EM})=(\vec a_{\tt E}\cdot \vec r_{\tt EM})-\frac{GM_{\tt M}}{r_{\tt EM}}, \qquad  {\rm where}
\qquad  
 \vec a_{\tt E}=-\sum_{{\tt B}\not={\tt E}} \frac{GM_{\tt B}}{r^3_{\tt BE}}\vec r_{\tt BE}.
\label{eq:accE}
\end{eqnarray}

As a result, the group of the $c^{-2}$ terms in (\ref{eq:(nonN2-in+2)}) takes the following form:
 {}
\begin{align}
\frac{1}{c^2}\Big({\textstyle\frac{1}{2}}v_{\tt M}^2 +\sum_{{\tt B}\not={\tt M}}\frac{GM_{\tt B}}{r_{\tt BM}} \Big)
-\frac{1}{c^2}\Big({\textstyle\frac{1}{2}}v_{\tt E}^2 +\sum_{{\tt B}\not={\tt E}}\frac{GM_{\tt B}}{r_{\tt BE}}  \Big)=\frac{1}{c^2}\Big\{
{\textstyle\frac{1}{2}}v_{\tt EM}^2 +  \frac{GM_{\tt E}-2GM_{\tt M}}{r_{\tt EM}} +W^\odot_{\tt EM}+
 \frac{d}{dt}(\vec v_{\tt E}\cdot \vec r_{\tt EM}) \Big\},~~~~~
\label{eq:expra=9}
\end{align}
which is accurate to ${\cal O}\big(4.30\times 10^{-19}\big)$ set by the omitted $\ell=3$ solar tide at the Earth-Moon distance. 

\subsubsection{The $c^{-4}$ terms }

Next, we examine the group of the $c^{-4}$-terms present in under the integral sign in (\ref{eq:(nonN2-in+2)}). We again express the {\tt BCRS}  position vector between a body {\tt B} and the Moon as $\vec r_{\tt BM}=\vec r_{\tt BE}+\vec r_{\tt EM}$, and treat $r_{\tt EM}/ r_{\tt BE}$ as a small parameter, and represent $\vec v_{\tt M}=\vec v_{\tt E}+\vec v_{\tt EM}$. As a result, we estimate that the velocity-dependent term  contributes  
$c^{-4}{\textstyle\frac{1}{8}}(v_{\tt M} ^4-v_{\tt E} ^4) =c^{-4}{\textstyle\frac{1}{8}}(4 v_{\tt E}^2 (\vec v_{\tt E}\cdot \vec v_{\tt EM})+4(\vec v_{\tt E}\cdot \vec v_{\tt EM})^2+2 v^2_{\tt E} v^2_{\tt EM}+4(\vec v_{\tt E}\cdot \vec v_{\tt EM}) v_{\tt EM}^2+v^4_{\tt EM})\approx c^{-4}{\textstyle\frac{1}{2}}v_{\tt E}^2 (\vec v_{\tt E}\cdot \vec v_{\tt EM})\simeq 1.67 \times 10^{-18}$, which is too small and may be omitted. The mixed terms give 
$c^{-4}{\textstyle\frac{3}{2}}(v^2_{\tt M}  \sum_{{\tt B\not= M}}{G M_{\tt B} / r_{\tt BM}}-v^2_{\tt E}  \sum_{{\tt B\not= E}}{G M_{\tt B} / r_{\tt BE}})\approx 
c^{-4}{\textstyle\frac{3}{2}}2 (\vec v_{\tt E}  \cdot \vec v_{\tt EM})(G M_{\tt S} / r_{\tt SE})\simeq 1.00 \times 10^{-17}$, with the error term of $c^{-4}{\textstyle\frac{3}{2}} v^2_{\tt E} (G M_{\tt S} r_{\tt EM} / r^2_{\tt SE})\simeq 3.76 \times 10^{-19}$; thus, this term is above our threshold and may be kept. The last term was evaluated as 
$c^{-4}{\textstyle{\textstyle{1 \over 2}}}\big\{[\sum_{{\tt B\not= E}} {G M_{\tt B} / r_{\tt BE}}]^2- \big[\sum_{{\tt B\not= E}} {G M_{\tt B} / r_{\tt BE}}\big]^2\big\}\simeq 1.13 \times 10^{-19}$ and, thus, may be omitted. 

As a result, for the $c^{-4}$-terms present in the integrand of  (\ref{eq:(nonN2-in+2)}), we have:
{}
\begin{align}
 \frac{1}{c^4}\Big\{
\Big({\textstyle{1 \over 8}}
v_{\tt M} ^4  
+ {\textstyle{3 \over 2}} v_{\tt M} ^2 \sum_{{\tt B\not= M}}{G M_{\tt B} \over r_{\tt BM}}- 
{\textstyle{\textstyle{1 \over 2}}}\Big[\sum_{{\tt B\not= M}} {G M_{\tt B} \over r_{\tt BM}}\Big]^2\Big)
& - \Big({\textstyle{1 \over 8}}
v_{\tt E} ^4  
+ {\textstyle{3 \over 2}} v_{\tt E} ^2 \sum_{{\tt B\not= E}}{G M_{\tt B} \over r_{\tt BE}}- 
{\textstyle{\textstyle{1 \over 2}}}\Big[\sum_{{\tt B\not= E}} {G M_{\tt B} \over r_{\tt BE}}\Big]^2\Big)\Big\}=
 \nonumber\\
&=
  \frac{3}{c^4}
{G M_{\tt S} \over r_{\tt SE}}(\vec v_{\tt E}\cdot \vec v_{\tt EM})+{\cal O}\big(c^{-5}; 1.67 \times 10^{-18}\big),
\label{eq:exprtrr}
\end{align}
where the error bound is from the velocity term evaluated to be $c^{-4}{\textstyle\frac{1}{8}}(v_{\tt M} ^4-v_{\tt E} ^4)\simeq 1.67 \times 10^{-18}$.

Considering the combination of the position-dependent terms, we see that for the clocks situated on the surfaces of both Earth and the Moon, were  {\tt TT} and {\tt TL} are defined, this combination behaves as 
{}
\begin{equation}
\label{eq:mixed-c4}
 \frac{1}{c^4}\Big\{
\Big({\textstyle{\textstyle{1 \over 2}}}v_{\tt M} ^2 +3\sum_{\tt B \not =M}{G M_{\tt B} \over r_{\tt BM}}  \Big)(\vec v_{\tt M}  \cdot {\vec r_{\tt M}}_{\tt TDB}) -\Big({\textstyle{\textstyle{1 \over 2}}}v_{\tt E} ^2 +3\sum_{\tt B \not =E}{G M_{\tt B} \over r_{\tt BE}}  \Big)(\vec v_{\tt E}\cdot \vec {r_{\tt E}}_{\tt TDB})\Big\}\lesssim 2.00\times 10^{-14}\,{\rm s}+7.30\times 10^{-14}\,{\rm s},
\end{equation}
where the first value is given for a Moon-based clock with the second one is for its Earth-based analogue. Thus, for the clocks on the surface of the bodies, this combination is less than our threshold of 0.1 ps, and, thus, may be omitted.

Now, we consider the term with constants $L_{\tt G}$ and $L_{\tt L}$. First, we  evaluate the $L_{\tt G}$-term   
\begin{equation}
c^{-2} L_{\tt G}({\textstyle{1\over 2}} v_{\tt E}^{2} +
\sum_{{\tt B\not= E}} \frac{G M_{\tt B}}{r_{\tt BE}})\simeq 
 c^{-2} L_{\tt G}{\textstyle{3\over 2}} 
 \frac{G M_{\tt S}}{a_{\tt E}}\simeq  1.03\times 10^{-17}+{\cal O}(1.72\times 10^{-19}),
\end{equation}
where $a_{\tt E}$ is the semi-major axis of the Earth orbit and the error comes from the Earth orbital eccentricity ($e_{\tt E}=0.0167$) correction of $ c^{-2} L_{\tt G}{\textstyle{3\over 2}} 
 ({G M_{\tt S}}/{a_{\tt E}})e_{\tt E}\simeq 1.72\times 10^{-19}$. Although small this term is above our threshold and should be kept. The second constant-corrected term was evaluated to be  
$L_{\tt L}( {\textstyle\frac{1}{2}}v_{\tt M}^2 +   \sum_{{\tt B\not= M}}{G M_{\tt B} / r_{\tt BM}} )\simeq L_{\tt L}( {\textstyle\frac{1}{2}}(v_{\tt E}+v_{\tt EM})^2 +   {G M_{\tt E} / r_{\tt EM}}+{G M_{\tt S} /{\rm AU}} )\simeq 4.76\times 10^{-19}$, 
which is too small for our purposes. Similarly, the position-dependent terms, being evaluated on the surfaces of the Earth and the Moon, contribute $c^{-2}L_{\tt G}(\vec v_{\tt E}\cdot \vec {r_{\tt E}}_{\tt TDB})\approx c^{-2}L_{\tt G}v_{\tt E}R_{\tt E}\simeq 1.47 \times 10^{-15}$\,s and $c^{-2}L_{\tt L}(\vec v_{\tt M}\cdot \vec {r_{\tt M}}_{\tt TDB})\approx c^{-2}L_{\tt L}( v_{\tt E} + v_{\tt EM})r_{\tt MQ}\simeq 1.87 \times 10^{-17}$\,s, and, thus, both terms may be omitted. 

As a result, the constant-corrected-term in (\ref{eq:(nonN2-in+2)}) takes the form:
{}
\begin{align}
&-\frac{1}{c^2}  \Big[
\int_{{\tt T_0+TDB_0}}^{\tt TDB}\Big\{ L_{\tt G}\Big({\textstyle{1\over 2}} v_{\tt E}^{2} +
\sum_{{\tt B\not= E}} {G M_{\tt B} \over r_{\tt BE}} \Big)-L_{\tt L}\Big( {\textstyle\frac{1}{2}}v_{\tt M}^2 +   \sum_{{\tt B\not= M}}{G M_{\tt B} \over r_{\tt BM}} \Big)\Big\}d{\tt TDB} +
L_{\tt G}(\vec v_{\tt E}\cdot \vec {r_{\tt E}}_{\tt TDB})-L_{\tt L}(\vec v_{\tt M}\cdot \vec {r_{\tt M}}_{\tt TDB})
\Big]\bigg]=\nonumber\\
&=-
\frac{1}{c^2}  
\int_{{\tt T_0+TDB_0}}^{\tt TDB}\Big\{ L_{\tt G}{\textstyle{3\over 2}} 
 {G M_{\tt S} \over r_{\tt SE}} \Big\}d{\tt TDB} +{\cal O}\Big(c^{-5}; 4.76 \times 10^{-19}; 1.47 \times 10^{-15}\,{\rm s}\Big).
\label{eq:expr4-trun-mix}
\end{align}

\subsection{Expressing $({\tt TL} - {\tt TT})$  as a function of  {\tt TT}}

Collecting all the contributions remaining for $c^{-2}$ and $c^{-4}$ terms, we may present  the integrand in (\ref{eq:(nonN2-in+2)})  as below: 
{}
\begin{eqnarray}
\frac{1}{c^2}\Big({\textstyle\frac{1}{2}}v_{\tt M}^2 +\sum_{{\tt B}\not={\tt M}}\frac{GM_{\tt B}}{r_{\tt BM}} \Big)
&+&
\frac{1}{c^4}
\Big({\textstyle{1 \over 8}}
v_{\tt M} ^4  
+ {\textstyle{3 \over 2}} v_{\tt M} ^2 \sum_{{\tt B\not= M}}{G M_{\tt B} \over r_{\tt BM}}- 
{\textstyle{\textstyle{1 \over 2}}}\Big[\sum_{{\tt B\not= M}} {G M_{\tt B} \over r_{\tt BM}}\Big]^2\Big) -
\frac{L_{\tt L}}{c^2}
\Big( {\textstyle\frac{1}{2}}v_{\tt M}^2 +   \sum_{{\tt B\not= M}}{G M_{\tt B} \over r_{\tt BM}} \Big)- \nonumber \\
-\frac{1}{c^2}\Big({\textstyle\frac{1}{2}}v_{\tt E}^2 +\sum_{{\tt B}\not={\tt E}}\frac{GM_{\tt B}}{r_{\tt BE}}  \Big) &-&
\frac{1}{c^4}
\Big({\textstyle{1 \over 8}}
v_{\tt E} ^4  
+ {\textstyle{3 \over 2}} v_{\tt E} ^2 \sum_{{\tt B\not= E}}{G M_{\tt B} \over r_{\tt BE}}- 
{\textstyle{\textstyle{1 \over 2}}}\Big[\sum_{{\tt B\not= E}} {G M_{\tt B} \over r_{\tt BE}}\Big]^2\Big)+
\frac{L_{\tt G}}{c^2}
\Big({\textstyle{1\over 2}} v_{\tt E}^{2} +
\sum_{{\tt B\not= E}} {G M_{\tt B} \over r_{\tt BE}} \Big)=\nonumber\\
&&\hskip -80pt =\,
  \frac{1}{c^2} \Big\{
{\textstyle\frac{1}{2}}v_{\tt EM}^2 +  \frac{GM_{\tt E}-2GM_{\tt M}}{r_{\tt EM}} +W^\odot_{\tt EM}+
L_{\tt G}   {\textstyle{3\over 2}} 
 {G M_{\tt S} \over r_{\tt SE}}\Big\} +
  \frac{1}{c^4}  \Big\{
  3{G M_{\tt S} \over r_{\tt SE}}(\vec v_{\tt E}\cdot \vec v_{\tt EM})\Big\} +   \frac{1}{c^2}\frac{d}{dt}(\vec v_{\tt E}\cdot \vec r_{\tt EM})+\nonumber\\
 &&\hskip 48pt +\,
 {\cal O}\Big(c^{-5}; 4.76 \times 10^{-19}\Big).
\label{eq:expra=c4}
\end{eqnarray}
Substituting this result in (\ref{eq:(nonN2-in+2)}), we obtain expression for $(\tt TL-TT)$ in the following form
{}
\begin{eqnarray}
{\tt TL} -   {\tt TT} &=&\frac{L_{\tt G}-L_{\tt L}}{1-L_{\tt B}}\big( {\tt TDB}-{\tt T_0}-{\tt TDB}_0\big) -L_{\tt L}({\tt T}_0-{\tt T}_{\tt LO})-\nonumber\\
  &-& \frac{1}{c^2}\int_{{\tt T_0+TDB_0}}^{\tt TDB}  \Big\{
{\textstyle\frac{1}{2}}v_{\tt EM}^2 +  \frac{GM_{\tt E}-2GM_{\tt M}}{r_{\tt EM}} +W^\odot_{\tt EM}+L_{\tt G}   {\textstyle{3\over 2}} 
 {G M_{\tt S} \over r_{\tt SE}}\Big\} d{\tt TDB} -\frac{1}{c^2}(\vec v_{\tt EM}\cdot \vec {r}_{\tt TDB})-\nonumber\\
&-& 
  \frac{1}{c^4}\int_{{\tt T_0+TDB_0}}^{\tt TDB} \Big\{
  3{G M_{\tt S} \over r_{\tt SE}}
(\vec v_{\tt E}\cdot \vec v_{\tt EM})
\Big\}d{\tt TDB} + 
 {\cal O}\Big(c^{-5}; 4.76 \times 10^{-19} \Delta {\tt TDB}; 7.30 \times 10^{-14}\,{\rm s}\Big).
\label{eq:(nonN2-in+2)-TT-int4}
\end{eqnarray} 

Note that (\ref{eq:(nonN2-in+2)-TT-int4}) still has {\tt TDB} as the time on the right hand side. Clearly, in the $c^{-4}$ order terms, we can replace {\tt TDB} with {\tt TT}, because, as show by (\ref{eq:(nonN1-TCB)}), the difference between the two time scales is of the order of $c^{-2}$.  It turned out that we can to the same simple substitution also for the $c^{-2}$ terms. Such a substitution results in the effect of $c^{-2} \big(
{\textstyle\frac{1}{2}}v_{\tt EM}^2 +  ({GM_{\tt E}-2GM_{\tt M}})/{r_{\tt EM}} +W^\odot_{\tt EM}\big)c^{-2} \big({\textstyle{\textstyle{1 \over 2}}}v_{\tt E} ^2 +\sum_{\tt B \not =E}{G M_{\tt B} / r_{\tt BE}}  \big)\simeq 2.55\times 10^{-19}$. Similarly small value of   $2.95 \times 10^{-23} \Delta t$ is produced by changing the time for the integrand.  Finally, the factor $1/(1-L_{\tt B})$ in front of the $c^{-2}$ term in (\ref{eq:(nonN2-in+2)}), resulted in the effects of the order of $ c^{-2}  L_{\tt B}(
{\textstyle\frac{1}{2}}v_{\tt EM}^2 +  ({GM_{\tt E}-2GM_{\tt M}})/{r_{\tt EM}} +W^\odot_{\tt EM})\simeq 2.65 \times 10^{-19}$ and $ c^{-2} L_{\tt B}(\vec v_{\tt EM}\cdot \vec {\cal X}_{\tt TT})\approx c^{-2} L_{\tt B} v_{\tt EM}r_{\tt MQ}\simeq 3.06\times 10^{-16}$\,s, both of these effects are negligibly small.  

Therefore, our final expression $(\tt TL-TT)$ as a function of {\tt TT} takes the form
{}
\begin{eqnarray}
{\tt TL} -   {\tt TT} &=&\frac{L_{\tt G}-L_{\tt L}}{1-L_{\tt B}}\big( {\tt TT}-{\tt T_0}\big) -L_{\tt L}({\tt T}_0-{\tt T}_{\tt LO})-\nonumber\\
  &-&   \frac{1}{c^2}\int_{\tt T_0}^{\tt TT} \Big\{
{\textstyle\frac{1}{2}}v_{\tt EM}^2 +  \frac{GM_{\tt E}-2GM_{\tt M}}{r_{\tt EM}} +W^\odot_{\tt EM}+L_{\tt G}   {\textstyle{3\over 2}} 
 {G M_{\tt S} \over r_{\tt SE}}\Big\} d{\tt TT} -\frac{1}{c^2}(\vec v_{\tt EM}\cdot \vec {\cal X}_{\tt TT})-\nonumber\\
&-& 
  \frac{1}{c^4} \int_{\tt T_0}^{\tt TT}  \Big\{
  3{G M_{\tt S} \over r_{\tt SE}}(\vec v_{\tt E}\cdot \vec v_{\tt EM})  \Big\}d{\tt TT} + 
 {\cal O}\Big(c^{-5}; 4.76 \times 10^{-19} ({\tt TT}-{\tt TT}_0); 7.30 \times 10^{-14}\,{\rm s}\Big),
\label{eq:(nonN2-in+2)-TT-int4-fin}
\end{eqnarray} 
where the solar tidal potential at the Moon $W^\odot_{\tt EM}$ is given by (\ref{eq:(nonN2-in+2)-tid}).

Following the approach demonstrated in (\ref{eq:coord-tr-QQ}) and (\ref{eq:coord-tr-QQM}), we can present result (\ref{eq:(nonN2-in+2)-TT-int4-fin}) in a similar functional form. For that, we introduce the constant $L_{\tt EM}$ and periodic terms $P_{\tt EM}$ as below
{}
\begin{align}
 \frac{1}{c^2} \Big\{
{\textstyle\frac{1}{2}}v_{\tt EM}^2 +  \frac{GM_{\tt E}-2GM_{\tt M}}{r_{\tt EM}} +W^\odot_{\tt EM}+L_{\tt G}   {\textstyle{3\over 2}} 
 {G M_{\tt S} \over r_{\tt SE}}\Big\} &+
  \frac{1}{c^4}  \Big\{
  3{G M_{\tt S} \over r_{\tt SE}}(\vec v_{\tt E}\cdot \vec v_{\tt EM}) \Big\}=
L_{\tt EM}+\dot P_{\tt EM}(t)+{\cal O}\big(c^{-5}; 4.76\times 10^{-19}\big),
\label{eq:expra=10+}
\end{align}
where the constant rate $L_{\tt EM}\simeq L_{\tt H}-L_{\tt C}$ and the  periodic terms $\dot P_{\tt EM}(t) \simeq \dot P_{\tt H}(t)-\dot P(t)$ are given as below:
{}
\begin{eqnarray}
L_{\tt EM}&=&
\frac{1}{c^2} \Big\{
\Big<{\textstyle\frac{1}{2}}v_{\tt EM}^2 +  \frac{GM_{\tt E}-2GM_{\tt M}}{r_{\tt EM}}\Big> +
\big<W^\odot_{\tt EM}\big>
+L_{\tt G}{\textstyle{3\over 2}} G M_{\tt S}\Big< {1 \over r_{\tt SE}} \Big>\Big\},
\label{eq:expRR1+}\\
\dot P_{\tt EM}(t)&=&
\frac{1}{c^2} \Big\{{\textstyle\frac{1}{2}}v_{\tt EM}^2 +  \frac{GM_{\tt E}-2GM_{\tt M}}{r_{\tt EM}}-\Big<{\textstyle\frac{1}{2}}v_{\tt EM}^2 +  \frac{GM_{\tt E}-2GM_{\tt M}}{r_{\tt EM}}\Big>+
W^\odot_{\tt EM}-\big<W^\odot_{\tt EM}\big> \Big\}+
   \frac{3}{c^4}  {G M_{\tt S} \over r_{\tt SE}}(\vec v_{\tt E}\cdot \vec v_{\tt EM}).
\label{eq:expRR2+}
\end{eqnarray}

Result (\ref{eq:expra=10+}), together with (\ref{eq:expRR1+}) and (\ref{eq:expRR2+}), provides valuable insight into the structure of the constant term \( L_{\tt EM} \) and the periodic terms \( P_{\tt EM}(t) \). These expressions can be used to explicitly establish the structure of the series \( P_{\tt EM}(t) \). 

Finally, using   (\ref{eq:expra=10+}) in (\ref{eq:(nonN2-in+2)-TT-int4-fin}), we express $({\tt TL} -   {\tt TT})$ as a function of {\tt TT}:
{}
\begin{align}
{\tt TL} -   {\tt TT} =\frac{L_{\tt G}-L_{\tt L}-L_{\tt EM}}{1-L_{\tt B}}\big( {\tt TT}-{\tt T_0}\big) &- L_{\tt L}({\tt T}_0-{\tt T}_{\tt LO}) - \Big(P_{\tt EM}({\tt TT})-P_{\tt EM}({\tt T_0} )\Big)-\frac{1}{c^2}(\vec v_{\tt EM}\cdot \vec{\cal X}_{\tt TT})\,+\nonumber\\
&+ 
 {\cal O}\Big(c^{-5}; 4.76 \times 10^{-19} ({\tt TT}-{\tt T}_0); 7.30 \times 10^{-14}\,{\rm s}\Big),
\label{eq:(nonN2-in+2)-TT-RR}
\end{align}
 where $\vec{\cal X}_{\tt TT}$ is the {\tt TT}-compatible lunicentric position of the lunar clock.  
 
\subsection{Secular Drift Rate and Periodic Terms for $({\tt TL} - {\tt TT})$ }
\label{sec:LCRS-TT}

\subsubsection{Secular drift rate $L_{\tt EM}$}

Considering the  \(O(c^{-2})\) term in \(L_{\tt EM}\) (\ref{eq:expRR1+}), we use Moon–Earth relative speed of 
$v_{\tt EM}\approx1022$ m/s, so the kinematic dilation contributes
$c^{-2}\,\big<\tfrac12\,v_{\tt EM}^2\big>
\simeq5.81\times10^{-12},$
well above our \(5\times10^{-18}\) cutoff.
Taking $r_{\tt EM}$ to be the instantaneous Earth–Moon separation, the Newtonian monopole term at the Earth–Moon distance was estimated to contribute up to  
$c^{-2}(GM_{\tt E}-2GM_{\tt M})/r_{\tt EM}\simeq1.13\times10^{-11}$. The solar quadrupole tide \(W^\odot_{\tt EM}\) yields
$c^{-2}\big<W^\odot_{\tt EM}\big>
\simeq c^{-2}\tfrac14\,{GM_{\tt S}\,r_{\tt EM}^2}/{r_{\tt SE}^3}\simeq1.63\times10^{-14}.$
Among the \(O(c^{-4})\) terms,  the scaling term proportional to \(L_{\tt G}\), gives
$
c^{-2}\,L_{\tt G}\big(\tfrac32{GM_{\tt S}}/{r_{\tt SE}}\big)
\simeq1.03\times10^{-17}.
$
All other contributions remain below \(5\times10^{-18}\) and may be omitted.

As a result,  collecting all the contributions, the secular‐drift coefficient for the Earth–Moon system is
\begin{equation}
\label{eq:L_EM}
L_{\tt EM}
=1.709\,390\,6 \times10^{-11}
=1.4769\ \mu\mathrm{s}/\mathrm{d}.
\end{equation}

With $L_{\tt EM}\simeq L_{\tt H}-L_{\tt C}=1.4769~\mu$s/d,  the total constant rate between the clock on or near the lunar surface and its terrestrial analogue to the accepted level of accuracy is estimated to be  
{}
\begin{eqnarray}
L_{\tt B}-L_{\tt M}\simeq L_{\tt G}-L_{\tt L}-L_{\tt EM}=\big(60.2146-2.7121-1.4769\big)~\mu{\rm s/d}=56.0256~\mu{\rm s/d}.
\label{eq:(RR)}
\end{eqnarray}

Note that, if the smaller value for the lunar radius $R_{\tt M}$ is used in (\ref{eq:(29)}) instead of $R_{\tt MQ}$, the value of $L_{\tt L}$ is estimated to be $L_{\tt L}=3.140\,587\,7\times 10^{-11}\simeq 2.7135 \,\mu$s/d. With this value, the total rate in (\ref{eq:(RR)}) is $L_{\tt B}-L_{\tt M}=56.0242~\mu{\rm s/d}$. Also, if the selenoid value of $W_{\tt gM}=2821713.3~\mathrm{m}^2\mathrm{s}^{-2}$ from \cite{Martinec-Pec:1988} is used to determine  $L_{\tt L} = 3.139\,579\,5 \times 10^{-11} \simeq 2.7126~\mu\mathrm{s}/\mathrm{d}$, the value of $L_{\tt B}-L_{\tt M}=56.0251~\mu{\rm s/d}$. This dispersion highlights the need for further studies of the lunar constants. 

\subsubsection{Time‐Dependent Correction \(P_{\rm EM}(t)\)}
\label{sec:P_EM}

Now we consider the periodic term $P_{\tt EM}$, see (\ref{eq:expRR2+}). From the vis–viva relation for the Moon’s motion about the Earth–Moon barycenter given as $v^2_{\tt EM}(r)=(GM_{\tt E}+GM_{\tt M})\big({2}/{r_{\tt EM}}-{1}/{a_{\tt EM}}\big)$, with $r_{\tt EM}=a_{\tt EM}\bigl(1 - e_{\tt M}\cos E\bigr)$, with $e_{\tt M}=0.0549$ being the Moon’s orbital eccentricity With these quantities, the orbital part of  integrand in (\ref{eq:expra=10+}) reads
\[
\frac{1}{c^2}\Big\{\tfrac12\,v_{\tt EM}^2
  +\frac{GM_{\tt E}-2GM_{\tt M}}{r_{\tt EM}}\Big\}
=\frac{1}{c^2}\Big\{\frac{2GM_{\tt E} - GM_{\tt M}}{r_{\tt EM}}
  -\frac{GM_{\tt E}+GM_{\tt M}}{2a_{\tt EM}}\Big\}.
\]
Using this result in (\ref{eq:expRR2+}) and 
expanding  $r_{\tt EM}$ to first order in $e_{\tt M}$, we can write  $r_{\tt EM}=a_{\tt EM}\big(1-e_{\tt M}\cos[\omega_{\tt M}(t-t_0)] +{\cal O}(e^2_{\tt M})\big)$, where $\omega_{\tt M}$ is the Moon’s mean orbital angular rate, we have  
\begin{align}
\label{eq:orb-P-dot}
\delta \dot P_{\tt EM}(t)&=
\frac{1}{c^2} \Big\{{\textstyle\frac{1}{2}}v_{\tt EM}^2 +   \frac{GM_{\tt E}-2GM_{\tt M}}{r_{\tt EM}}-\Big<{\textstyle\frac{1}{2}}v_{\tt EM}^2 +   \frac{GM_{\tt E}-2GM_{\tt M}}{r_{\tt EM}}\Big>\Big\}=\nonumber\\
&=
\frac{1}{c^2}\big(2GM_{\tt E} - GM_{\tt M}\big)\Big(\frac{1}{r_{\tt EM}}-\frac{1}{a_{\tt EM}}\Big)\simeq 
\frac{1}{c^2}\big(2GM_{\tt E} - GM_{\tt M}\big)\frac{e_{\tt M}}{a_{\tt EM}}\cos[\omega_{\tt M}(t-t_0)]=\nonumber\\
&=1.259\,047 \times 10^{-12}\cos[\omega_{\tt M}(t-t_0)]=0.109\,\cos[\omega_{\tt M}(t-t_0)]~\mu{\rm  s/d}.
\end{align}

Integrating this result in time gives
\begin{align}
\label{eq:orb-P}
\delta  P_{\tt EM}(t)&\simeq -
\frac{1}{c^2}\big(2GM_{\tt E} - GM_{\tt M}\big)\frac{e_{\tt M}}{a_{\tt EM}\omega_{\tt M}}\sin[\omega_{\tt M}(t-t_0)]=-0.473\, \sin[\omega_{\tt M}(t-t_0)]~\mu{\rm  s}.
\end{align}

The residual solar quadrupole \(\ell=2\) tide in (\ref{eq:expRR2+}) produces the contribution of 
\[
c^{-2}\delta W^\odot_{\tt EM}(t)
\simeq \tfrac34\,\frac{GM_{\tt S}\,a^2_{\tt EM}}{c^2\,r_{\tt SE}^3}
\cos\! \big[2(\omega_{\rm syn}\,t+\varphi)\big] +{\cal O}(e_{\tt EM}),
\]
which after integration in time yields  
\[
\delta P_{\tt EM}^{({\tt S})}(t)
=-\tfrac{3}{8}\frac{GM_{\tt S} a_{\tt EM}^2}{c^2 r_{\tt SE}^3\omega_{\rm syn}}
\sin\!\big[2(\omega_{\rm syn}t+\varphi)\big]
\simeq 9.18\times10^{-9}\sin\!\big[2(\omega_{\rm syn}t+\varphi)\big]\, {\rm s},
\]
which exceeds our threshold and, thus, must be retained.  

Eq.~(\ref{eq:expRR2+}) also contains the Sun’s tidal multipoles of degree $\ell=3$ evaluated at the Moon.  Expanding this term in the synodic phase shows that the one‑way proper‑time amplitude is  
$A_{\tt S[3]}\approx3.8\times10^{-19}\,\mathrm s$, which is to small to retain.

Finally, the \(O(c^{-4})\) velocity‐cross term in (\ref{eq:expRR2+}) with the form 
$
c^{-4}3({G M_{\tt S}}/{r_{\tt SE}})(\vec v_{\tt E} \cdot \vec v_{\tt EM})\simeq 1.00\times 10^{-17}
$
integrates to
\[
P_{\tt EM}^{(\rm mix)}(t)
\simeq \frac{3GM_{\tt S}}{c^4 r_{\tt SE} }\frac{v_{\tt E}v_{\tt EM}}{\omega_{\tt M}}\,
\sin\!\big(\omega_{\tt M} t-\lambda_{\tt E}\big)
\simeq 3.77 \times10^{-12}\,\sin\!\big(\omega_{\tt M} t-\lambda_{\tt E}\big)\,\,{\rm s},
\]
which is above our threshold of 0.1 ps and, thus, large enough to be in the model.

This analysis is  indicative of the various components [resent in the overall time-series $P_{\tt EM}$. For multi‑year missions, however, all six lunar arguments and osculating variations $\{\delta a,\delta e,\delta M,\delta D,\delta F,\dots\}$ must be carried through each sinusoid—either via a full analytic re‑expansion to first order in those variations or by high‑fidelity numerical propagation + FFT—to maintain sub‑ps fidelity.  That complete osculating‑element treatment will be presented elsewhere.

\section{Proper Time in Cislunar Space}
\label{sec:pt-TT-tau}

\subsection{Relating Cislunar Proper Time and {\tt TT}}
\label{sec:pt-TT-tau-rel}

To relate the proper time, $\tau$, of an ideal clock  in cislunar space with a clock on the Earth's surface that is referenced to {\tt TT}, we use the usual chain of the time-scale transformations
{}
\begin{equation}
\frac{d\tau}{d{\tt TT}}=\frac{d\tau}{d{\tt TCL}}\frac{d{\tt TCL}}{d{\tt TL}}\frac{d{\tt TL}}{d{\tt TT}}.
\label{eq:synch-M+}
\end{equation}

With all the necessary transformations derived in preceding sections, we can now compute $({d\tau}/{d{\tt TT}})$. For convenience, we will repeat these transformations here. First, we use (\ref{eq:prop-coord-time-LCRS-TL}) that connects $\tau$ and {\tt TCL}, given as below:
{}
\begin{eqnarray}
\label{eq:prop-coord-time-LCRS-expand-L}
\frac{d\tau}{d{\tt TCL}}
&=&1
-\frac{1}{c^2}\Big\{\tfrac12\,{\cal V}^2+U_{\tt M}({\cal T},\vec {\cal X}) 
+U^*_{\rm tid}({\cal T},\vec {\cal X}) \Big\}+
{\cal O}\Big(c^{-4};\, 1.46\times 10^{-21}\Big),
\end{eqnarray}
where $U_{\tt M}({\cal T},\vec{\cal X})$ and $U^{\ast}_{\rm tid}({\cal T},\vec{\cal X})$ are the Newtonian lunar gravitational and tidal potentials, respectively.

Then, we use (\ref{eq:(26)}) that connects {\tt TCL} and {\tt TL}:
{}
\begin{equation}
\frac{d{\tt TCL}}{d{\tt TL}} = \frac{1}{1  -  L_{\tt L}} = 1 +  \frac{L_{\tt L}}{1  -  L_{\tt L}}.
\label{eq:(26)+}
\end{equation}

Finally, from (\ref{eq:(nonN2-in+2)-TT-int4-fin}), we establish rate $(d{\tt TL}/{\tt TT})$ that may be given as below:
{}
\begin{eqnarray}
\frac{d {\tt TL}}{d{\tt TT}} &=&1+\frac{L_{\tt G}-L_{\tt L}}{1-L_{\tt B}}-  \frac{1}{c^2} \Big\{
{\textstyle\frac{1}{2}}v_{\tt EM}^2 +  \frac{GM_{\tt E}-2GM_{\tt M}}{r_{\tt EM}} +W^\odot_{\tt EM}+L_{\tt G}   {\textstyle{3\over 2}} 
 {G M_{\tt S} \over r_{\tt SE}}\Big\} -\frac{1}{c^2}\frac{d}{d{\tt TT}}(\vec v_{\tt EM}\cdot \vec {\cal X}_{\tt TT})-\nonumber\\
&& \hskip 8pt -\,
  \frac{1}{c^4}   \Big\{
  3{G M_{\tt S} \over r_{\tt SE}}(\vec v_{\tt E}\cdot \vec v_{\tt EM}) \Big\} + {\cal O}\Big(c^{-5}; 4.76 \times 10^{-19}\Big).
\label{eq:(nonN2-in+2)-TT-int4-finT}
\end{eqnarray} 

As a result, substituting all these expressions (\ref{eq:prop-coord-time-LCRS-expand-L}), (\ref{eq:(26)+}), and (\ref{eq:(nonN2-in+2)-TT-int4-finT}) in the chain (\ref{eq:synch-M+}), we have
{}
\begin{eqnarray}
\frac{d\tau}{d{\tt TT}}&=&
1 + \frac{L_{\tt G}}{1-L_{\tt B}}- 
\frac{1}{c^2} \Big\{
{\textstyle\frac{1}{2}}v_{\tt EM}^2 +  \frac{GM_{\tt E}-2GM_{\tt M}}{r_{\tt EM}} +W^\odot_{\tt EM}+L_{\tt G}   {\textstyle{3\over 2}} 
 {G M_{\tt S} \over r_{\tt SE}}\Big\} -\frac{1}{c^2}\frac{d}{d{\tt TT}}(\vec v_{\tt EM}\cdot \vec {\cal X}_{\tt TT})-\nonumber\\
&-&
  \frac{1}{c^4}   \Big\{
  3{G M_{\tt S} \over r_{\tt SE}}(\vec v_{\tt E}\cdot \vec v_{\tt EM})\Big\}-
 \frac{1}{c^2}\Big\{
  \tfrac12\,{\cal  V}_{\rm }^2
+ U_{\tt M}({\cal T},\vec {\cal X})+ U^*_{\rm tid}({\cal T},\vec {\cal X})\Bigr\}+
{\cal O}\Big(c^{-5};\, 4.76\times 10^{-19}\Big).
\label{eq:synch-M+F2}
\end{eqnarray}

It is important to note that, at the stated level of accuracy all contributions in the transformations \eqref{eq:prop-coord-time-LCRS-expand-L}, \eqref{eq:(26)+}, and \eqref{eq:(nonN2-in+2)-TT-int4-finT} combine additively in (\ref{eq:synch-M+F2}), with no cross‐terms. Consequently, the small‐period variations present in each expression remain unmodulated and do not interact nonlinearly as they would generally do under Eq.~\eqref{eq:synch-M+}.

Integrating result (\ref{eq:synch-M+F2}) with respect to {\tt TT}  and reinstating the integration constants as in (\ref{eq:(nonN2-in+2)-TT-int4-fin}), yields the relation between the proper time  \(\tau\) of a cislunar clock and {\tt TT}:
{}
\begin{eqnarray}
\tau -   {\tt TT} &=&\frac{L_{\tt G}}{1-L_{\tt B}}\big( {\tt TT}-{\tt T_0}\big) -L'_{\tt L}({\tt T}_0-{\tt T}_{\tt LO})
-\frac{1}{c^2}(\vec v_{\tt EM}\cdot \vec {\cal X}_{\tt TT})-\nonumber\\
&-&
 \frac{1}{c^2}\int_{\tt T_0}^{\tt TT} \Big\{
{\textstyle\frac{1}{2}}v_{\tt EM}^2 +  \frac{GM_{\tt E}-2GM_{\tt M}}{r_{\tt EM}} +W^\odot_{\tt EM}+L_{\tt G}   {\textstyle{3\over 2}} 
 {G M_{\tt S} \over r_{\tt SE}}\Big\} d{\tt TT} -
  \frac{1}{c^4} \int_{\tt T_0}^{\tt TT}  \Big\{
  3{G M_{\tt S} \over r_{\tt SE}}(\vec v_{\tt E}\cdot \vec v_{\tt EM}) \Big\}d{\tt TT} -\nonumber\\
 &-& 
 \frac{1}{c^2} \int_{\tt T_0}^{\tt TT}\Big\{
  \tfrac12\,{\cal  V}_{\rm }^2+U_{\tt M}({\cal T},\vec {\cal X})+U^*_{\rm tid}({\cal T},\vec {\cal X})\Bigr\}d{\tt TT}
+ 
 {\cal O}\Big(c^{-5}; 4.76 \times 10^{-19} ({\tt TT}-{\tt TT}_0); 7.30 \times 10^{-14}\,{\rm s}\Big),
\label{eq:tau-TT-int}
\end{eqnarray} 

To further develop (\ref{eq:tau-TT-int}), we recognize that expression (\ref{eq:expra=10+}) together with constant rate $L_{\tt EM}$ and small periodic terms $\dot P_{\tt EM}$ introduced by (\ref{eq:expRR1+}) and (\ref{eq:expRR2+}), correspondingly, allows us to present (\ref{eq:tau-TT-int}) as below
{}
\begin{eqnarray}
\tau-{\tt TT}&=&
 \frac{L_{\tt G}-L_{\tt EM}}{1-L_{\tt B}}\big({\tt TT}-{\tt T}_0\big)-L'_{\tt L}({\tt T}_0-{\tt T}_{\tt LO})- \Big(
 P_{\tt EM}({\tt TT}) - P_{\tt EM}({\tt T}_0) \Big)-\frac{1}{c^2}(\vec v_{\tt EM}\cdot \vec {\cal X}_{\tt TT})-\nonumber\\
&-&
\frac{1}{c^2}\int_{{\tt T}_0}^{\tt TT}\Big\{
  \tfrac12\,{\cal  V}_{\rm }^2+
U_{\tt M}({\cal T},\vec {\cal X})+U^*_{\rm tid}({\cal T},\vec {\cal X})
\Bigr\}d{\tt TT}+
{\cal O}\Big(c^{-4};\, 4.76\times 10^{-19}({\tt TT}-{\tt T}_0);\, 7.30\times 10^{-14}\,{\rm s}\Big),
\label{eq:synch-M+F2Y}
\end{eqnarray}  
with $L'_{\tt L}$ being an arbitrary integration constant to be specified below.

Eq.~\eqref{eq:synch-M+F2Y} generalizes the surface‐bound synchronization law of \eqref{eq:(nonN2-in+2)-TT-RR} to any cislunar trajectory. To further  simplify this result, we again follow approach that was used in (\ref{eq:coord-tr-QQ}), (\ref{eq:coord-tr-QQM}), and (\ref{eq:expra=10+}), and introduce the constant rate $L_{\tt CL}$ and periodic terms $P_{\tt CL}(t)$ evaluated for a particular orbit of a clock in cislunar space: 
{}
\begin{align}
\frac{1}{c^2} \Big\{
  \tfrac12\,{\cal  V}_{\rm }^2 +
U_{\tt M}({\cal T},\vec {\cal X})+
U^*_{\rm tid}({\cal T},\vec{\cal X})\Bigr\}
=L_{\tt CL}+\dot P_{\tt CL}(t)+
{\cal O}\Big(c^{-4};\, 3.17\times 10^{-18}\Big),
\label{eq:synch-LP}
\end{align}  
where the constant rate $L_{\tt CL}$ and the  periodic terms $\dot P_{\tt CL}(t)$ are given as below:
{}
\begin{align}
L_{\tt CL}&=
\frac{1}{c^2} \Big< \tfrac12\,{\cal  V}_{\rm }^2 +
U_{\tt M}({\cal T},\vec {\cal X})+
U^*_{\rm tid}({\cal T},\vec{\cal X}
\Big>\Big|_{\rm orb},
\qquad
\dot P_{\tt CL}(t)=
\frac{1}{c^2} \Big\{
  \tfrac12\,{\cal  V}_{\rm }^2 +
U_{\tt M}({\cal T},\vec {\cal X})+
U^*_{\rm tid}({\cal T},\vec{\cal X})\Bigr\}-L_{\tt CL},
\label{eq:PCL}
\end{align}
where $\left<...\right>|_{\rm orb}$ denotes a long-term averaging along a particular orbit of a clock in cislunar space.

Eqs.~\eqref{eq:tau-TT-int}–\eqref{eq:PCL} split the clock’s rate in the {\tt LCRS} into a secular term $L_{\tt CL}$ and a zero–mean periodic $P_{\tt CL}(t)$, while \eqref{eq:orb-P} gives the common monthly $P_{\tt EM}(t)$ that enters the ${\tt TT}$ mapping. 
For any cislunar orbit, we evaluate the final relation \eqref{eq:tau-TT=0} by (i) computing $L_{\tt CL}$ from the appropriate kinematic and potential averages, and (ii) building $P_{\tt CL}(t)$ from the retained $c^{-2}$ harmonics (orbit–dependent). 
Explicit formulas for $L_{\tt CL}$ and $P_{\tt CL}(t)$ in the representative regimes appear in Secs.~\ref{sec:LCRS-vLLO}–\ref{sec:LCRS-NRHO}.

Taking into account that $L_{\tt CL}\simeq L_{\tt L}$ (that was estimated in Sec.~\ref{sec:tm-tdb}) and chosing $L'_{\tt L}=L_{\tt CL}$, we present (\ref{eq:synch-M+F2Y}) in the functional form similar to that of (\ref{eq:coord-tr-QQ}), (\ref{eq:coord-tr-QQM}), and (\ref{eq:expra=10+}):
{}
\begin{eqnarray}
\tau-{\tt TT}&=&
 \frac{L_{\tt G}-L_{\tt CL}-L_{\tt EM}}{1-L_{\tt B}}\big({\tt TT}-{\tt T}_0\big)-L'_{\tt L}({\tt T}_0-{\tt T}_{\tt LO})- \Big(
 P_{\tt EM}({\tt TT}) - P_{\tt EM}({\tt T}_0) \Big)-\frac{1}{c^2}(\vec v_{\tt EM}\cdot \vec {\cal X}_{\tt TT})-\nonumber\\
&-&
\Big(P_{\tt CL}({\tt TT}) - P_{\tt CL}({\tt T}_0) \Big)+
{\cal O}\Big(c^{-4};\, 3.17\times 10^{-18}({\tt TT}-{\tt T}_0);\, 7.30\times 10^{-14}\,{\rm s}\Big).
\label{eq:tau-TT=0}
\end{eqnarray} 
Note that products between the $\mathcal{O}(c^{-2})$ terms in $d\tau/d{\tt TCL}$ and constant scale factors such as $L_{\tt G}$ contribute at the level of $\lesssim 7\times10^{-21}$ and are neglected here.

Note that throughout Sec.~\ref{sec:pt-TT-tau}, the $c^{-2}$ bracket in \eqref{eq:tau-TT=0} splits into a constant (secular) part and a zero-mean periodic part $P_{\tt CL}(t)$ via \eqref{eq:PCL}. The periodic term is, in general, a \emph{sum} of harmonics driven by orbital geometry (e.g., $\omega$, $2\omega$, $3\omega$ for elliptical motion), lunar tesseral rotation sidebands, and external tides. These harmonics \emph{add linearly} and do not produce nonlinear cross-terms at the order retained here. As a result, time series of $P_{\tt CL}(t)$ naturally exhibit beating/envelope patterns when multiple nearby lines (e.g., $2\omega\pm\dot\lambda$) are present, even though the underlying model remains a linear superposition mapped to {\tt TT} via \eqref{eq:tau-TT=0}.

Result~\eqref{eq:tau-TT=0} relates the proper time of a lunar‐orbiting clock to {\tt TT}. 
To apply it for clock synchronization purposes, we need to compute the constant  $L_{\tt CL}$ and periodic terms  $\dot P_{\tt CL}(t)$ for a trajectory of interest. Below, we evaluate these quantities for five representative cis‑lunar clock locations—lunar surface, LLO,  Elliptical Lunar Frozen Orbit (ELFO), Earth–Moon \(L_1\), and NRHO.

Table~\ref{tab:representative_orbits} lists representative lunar orbital regimes, their altitude ranges, key characteristics, and orbital periods. Near‑rectilinear halo orbits (NRHOs) provide continuous polar‑region visibility, whereas low lunar orbits (LLOs) yield frequent surface passes with shorter visibility windows. Each regime imposes distinct proper‑time corrections in \eqref{eq:tau-LCRS}: LLO corrections are dominated by the lunar gravity potential with many terms contributing at significant level, while ELFO, EML1 and NRHO clocks require inclusion of many terms from the external tidal and inertial potentials.

\begin{table}[h]
  \centering
  \caption{Representative lunar orbits and their key parameters and benefits.}
\label{tab:representative_orbits}
  \begin{tabular}{l|l|l|l}
    \hline
    \parbox[t]{0.25\textwidth}{\raggedright Configuration} &
    \parbox[t]{0.135\textwidth}{\raggedright Altitude (km)} &
    \parbox[t]{0.09\textwidth}{\raggedright Period} &
    \parbox[t]{0.43\textwidth}{\raggedright Benefits / Characteristics} \\
    \hline\hline

    \parbox[t]{0.25\textwidth}{\raggedright Very Low Lunar Orbit (vLLO)} &
    \parbox[t]{0.135\textwidth}{\raggedright 10} &
    \parbox[t]{0.09\textwidth}{\raggedright 1.82\,h} &
    \parbox[t]{0.43\textwidth}{\raggedright Ultra-low altitude; highest-resolution surface access; very frequent passes; active station-keeping required.} \\

    \parbox[t]{0.25\textwidth}{\raggedright Low Lunar Orbit (LLO)} &
    \parbox[t]{0.135\textwidth}{\raggedright 100--200} &
    \parbox[t]{0.09\textwidth}{\raggedright 1.96--2.13\,h} &
    \parbox[t]{0.43\textwidth}{\raggedright High revisit frequency; short visibility windows.} \\

    \parbox[t]{0.25\textwidth}{\raggedright Polar Circular Orbit} &
    \parbox[t]{0.135\textwidth}{\raggedright 100--300} &
    \parbox[t]{0.09\textwidth}{\raggedright 1.96--2.29\,h} &
    \parbox[t]{0.43\textwidth}{\raggedright Near-global coverage; favorable lighting geometry; ideal for mapping and communications.} \\

    \parbox[t]{0.25\textwidth}{\raggedright Highly Elliptical Orbit (HEO)} &
    \parbox[t]{0.135\textwidth}{\raggedright Periapsis: 500; Apoapsis: 10\,000} &
    \parbox[t]{0.09\textwidth}{\raggedright 14.56\,h} &
    \parbox[t]{0.43\textwidth}{\raggedright Extended dwell at apoapsis; prolonged surface visibility; moderate \(\Delta V\) requirements.} \\

    \parbox[t]{0.25\textwidth}{\raggedright Elliptical Lunar Frozen Orbit (ELFO)} &
    \parbox[t]{0.135\textwidth}{\raggedright Periapsis: 1\,750; Apoapsis: 17\,400} &
    \parbox[t]{0.09\textwidth}{\raggedright $\approx 30~\mathrm{h}$} &
    \parbox[t]{0.43\textwidth}{\raggedright Long dwell at south-polar apolune; “frozen” $e$ and AOP aided by Earth perturbations; stable geometry for polar coverage; modest station-keeping $\Delta V$  \cite{RydenVolle2025_LCRNS_RefConst3_1}.} \\

    \parbox[t]{0.25\textwidth}{\raggedright Earth–Moon L1 Lagrange Point} &
    \parbox[t]{0.135\textwidth}{\raggedright Perigee: 54\,815; Apogee: 61\,245} &
    \parbox[t]{0.09\textwidth}{\raggedright 27.32\,d} &
    \parbox[t]{0.43\textwidth}{\raggedright Co-rotational with the Moon; fixed geometry in rotating frame; requires periodic station-keeping.} \\

    \parbox[t]{0.25\textwidth}{\raggedright Gateway NRHO (9:2 synodic)} &
    \parbox[t]{0.135\textwidth}{\raggedright Periapsis: 1\,630; Apoapsis: 69\,400} &
    \parbox[t]{0.09\textwidth}{\raggedright ${\approx 7.49~d}$} &
    \parbox[t]{0.43\textwidth}{\raggedright Near-rectilinear halo orbit (NRHO); minimal eclipses; continuous Earth link; low station-keeping \(\Delta V\).} \\
    \hline
  \end{tabular}
\end{table}

Here we consider several plausible clock locations including—lunar surface, vLLO, LLO, ELFO, Earth–Moon \(L_1\), and NRHO and evaluate  $L_{\tt CL}$ and period terms $\dot P_{\tt CL}(t)$ for each of them. While doing so, we will make sure to retain the terms that will allow rate estimates with accuracy better than $5\times10^{-18}$ and timing more precise than 0.1 ps.

For compact analytic development we truncate the lunar potential at degree \(\ell=9\) with Love-number variations. This level is adequate for deep cislunar regimes where tides and inertial terms dominate (e.g., Earth--Moon~L1 and NRHO), but it is generally insufficient near the Moon if the stated accuracy targets of \(\le 0.1\:\mathrm{ps}\) and \(5\times 10^{-18}\) are enforced. To make this distinction explicit, we adopt the following policy for operational realizations:
\begin{itemize}
  \item \textit{Near-surface and vLLO (\(h\lesssim 30\:\mathrm{km}\)).} Use a high-degree GRAIL-derived gravity solution with degree/order \(\ell_{\max}\gtrsim 300\) (together with the same tide/Love-number model used here). This ensures that unmodeled Newtonian-potential structure from mascons remains below the implied bound
$c^{-2} |\Delta U|\;\lesssim\;5\times 10^{-18}$ or $|\Delta U|\;\lesssim\;0.45~\mathrm{m}^{2}\,\mathrm{s}^{-2}.$ If such a field is not used, the time/frequency requirement should be relaxed accordingly and the residual bias carried in the error budget. (See Sec.~\ref{sec:LCRS-vLLO}).
  \item \textit{Low to medium-altitude LLO.} Mission designs should select \(\ell_{\max}\) by altitude and science region, verifying that the resulting \(|\Delta U|/c^{2}\) stays within the \(5\times 10^{-18}\) budget when combined with kinematic and tidal terms. (See Sec.~\ref{sec:LCRS-LLO}).
    \item \textit{Elliptical Lunar Frozen Orbits (ELFO; \(h_p\!\sim\!1{,}750~\mathrm{km}\), \(h_a\!\sim\!17{,}400~\mathrm{km}\), \(T\!\approx\!30~\mathrm{h}\)).} Adopt a GRAIL-derived field with \(\ell_{\max}\!\sim\!80\)–\(120\) together with the same tide/Love-number model as used here. High-degree lunar harmonics are strongly suppressed at apolune \(\big((R_{\tt MQ}/r)^{\ell+1}\big)\), while periselene (\(r\simeq3.5\times10^3~\mathrm{km}\)) still benefits from \(\ell\gtrsim 80\) to keep \(c^{-2}|\Delta U|\!\le\!5\times10^{-18}\) across the ellipse. Earth tides should be modeled at least through \(\ell=4\); higher solar multipoles remain below threshold for this regime. (See Sec.~\ref{sec:LCRS-ELFO}.)
  \item \textit{NRHO and Earth--Moon L1.} The \(\ell=9\) truncation of lunar gravity is sufficient for the proper-time terms retained here; operational pipelines may use higher \(\ell_{\max}\) without changing the analytic expressions. (See Secs.~\ref{sec:LCRS-L1}, \ref{sec:LCRS-NRHO}).
\end{itemize}
This policy does not modify the closed-form formulae; it only tightens the {realization} of \(U_{\tt M}\) (and thus \(L_{\tt L}\) and \(L_{\tt CL}\)) when the use case demands sub-picosecond performance near the lunar surface.

\subsection{Clock in a Very Low Lunar Orbit (vLLO)}
\label{sec:LCRS-vLLO}

\subsubsection{Relevant potential terms}

Consider a clock onboard a spacecraft in a circular polar very low lunar orbit (vLLO) at altitude \(h_{\tt vLLO}=10\,\mathrm{km}\) above the mean lunar radius \(R_{\tt MQ}\), with orbital radius of the clock is  
$r _{\tt vLLO}= R_{\tt MQ} + h_{\tt vLLO} \approx 1748$\,km, yielding orbital velocity of 
${\cal V}_{\tt vLLO}=({GM_{\tt M}}/{r_{\tt vLLO}})^\frac{1}{2}\approx 1.68\times10^3\;\mathrm{m/s}.
$
This velocity produces a special‐relativistic time dilation of  $
c^{-2}\tfrac12 {\cal V}_{\tt vLLO}^2
\simeq 1.56\times10^{-11},
$
which exceeds our \(5\times10^{-18}\) cutoff and must be retained.

The dominant gravitational redshift at the LLO orbital radius is due to the Moon’s monopole field, so that
$c^{-2}U_{\tt M0}=c^{-2}{GM_{\tt M}}/{r_{\tt vLLO}}\simeq 3.12\times10^{-11},
$
which is of the same order as the kinematic term. Also, for the chosen {vLLO} orbit, the lunar quadrupole term (see Table \ref{tab:sp-harmonics-moon}) produces contribution of the order of
\begin{align}
c^{-2}U_{\tt M[2]}=
\frac{G M_{\tt M} R_{\tt MQ}^2}{c^2 r_{\tt vLLO}^3}J_2P_{20}(\cos\theta)\lesssim 6.27\times 10^{-15},
\end{align}
which is large enough to be included in the model.
Contributions of other zonal harmonics are estimated to be
{}
\begin{align}
c^{-2}U_{\tt M[3]}=
\frac{G M_{\tt M} R_{\tt MQ}^3}{c^2 r_{\tt vLLO}^4}J_3 P_{30}(\cos\theta)\lesssim 2.60\times 10^{-16}, \qquad
c^{-2}U_{\tt M[4]}=\frac{G M_{\tt M} R_{\tt MQ}^4}{c^2 r_{\tt vLLO}^5}J_4P_{40}(\cos\theta)\lesssim 1.80\times 10^{-16}.
\end{align}

Similarly, all tesseral harmonics up to $\ell=4$ listed in Table~\ref{tab:sp-harmonics-moon} yield contributions exceeding \(5\times10^{-18}\) at {vLLO}. Thus, including a complete lunar gravity field is important for this orbit.

Tidal perturbations from the Earth and the Sun  at the quadrupole level were evaluated to be
\begin{equation}
\label{eq:tides-LLO}
c^{-2}U_{\rm tid[2]}^{*\tt (E)}=\frac{GM_{\tt E}r_{\tt vLLO}^2}{c^2r_{\tt EM}^3}P_2(\vec n_{\tt EM}\!\cdot\!\widehat{\vec {\cal X}})
\simeq 2.39\times10^{-16}, \qquad
c^{-2}U_{\rm tid [2]}^{*\tt (S)}\simeq \frac{GM_{\tt S}r_{\tt vLLO}^2}{c^2{\rm AU}^3}P_2(\vec n_{\tt SM}\!\cdot\!\widehat{\vec {\cal X}})
\simeq 1.35\times10^{-18}.
\end{equation}
The Earth $\ell=3$ tidal effect was evaluated to be $c^{-2}(GM_{\tt E}/r_{\tt EM})(r_{\tt vLLO}/{r_{\tt EM})^3}\simeq 1.09 \times 10^{-18}$.

Accordingly, retaining only terms above \(5\times10^{-18}\) gives
{}
\begin{equation}
\label{eq:tau-LCRS-LLO}
\frac{d\tau}{d{\cal T}}
=1
-\frac{1}{c^2}\Big\{
  \tfrac12\,V_{\tt vLLO}^2
 + U_{\tt M}({\cal T},\vec {\cal X}) +
 \frac{GM_{\tt E}}{r_{\tt EM}}\Big(\frac{r_{\tt vLLO}}{r_{\tt EM}}\Big)^2P_2(\vec n_{\tt EM}\!\cdot\!\widehat{\vec {\cal X}})
\Bigr\}
+ {\cal O}\Bigl(c^{-4};\,1.73\times10^{-18}\Bigr),
\end{equation}
where the error bound comes from the RMS of the solar $\ell=2$  (\ref{eq:tides-LLO}) and Earth $\ell=3$ tidal potentials. Eq.~(\ref{eq:tau-LCRS-LLO}) quantifies that in a 10\,km vLLO the kinematic and monopole gravitational terms both lie at the \(10^{-11}\) level, while Earth‐induced tides contribute at \(10^{-16}\), and all higher‐order effects are safely below our \(5\times10^{-18}\) threshold. At 10~km altitude, many lunar spherical‑harmonic terms contribute; to keep unmodeled $|{\Delta U}|/c^2$ below $5\times10^{-18}$ at all longitudes, {\it operational} models for vLLO generally require very high degree (often $\ell_{\max}\!\gtrsim\!300$), even though the \emph{illustrative} truncation in \eqref{eq:tau-LCRS-LLO} shows only the terms needed for the analytic development here. 

We can now use the form of (\ref{eq:tau-LCRS-LLO}) to determine the $L_{\tt CL}$ and $P_{\tt CL}$ for this orbit that will be used to study (\ref{eq:tau-TT=0}). 

\subsubsection{Secular drift rate \(L_{\tt CL}\)}

In direct analogy to definition of \(L_{\tt L}\) in Sec.~\ref{sec:tm-tdb}, we define the orbital‐averaged constant \(L_{\tt CL}\) for a clock in a circular vLLO by averaging the kinematic and gravitational redshifts of \eqref{eq:tau-LCRS-LLO} over many revolutions.  As a result, retaining only terms larger than $5\times10^{-18}$, the  definition of $L_{\tt CL}$  for vLLO becomes
\begin{equation}
\label{eq:LC-vLLO}
L^{\tt vLLO}_{\tt CL} =
\frac{1}{c^2}\Big\{
 \tfrac12\,V_{\tt vLLO}^2
 + \Big<U_{\tt M}({\cal T},\vec {\cal X})\Big> \Big|_{\tt vLLO}+
  \frac{GM_{\tt E}r_{\tt vLLO}^2}{4r^3_{\tt EM}}
\Big\}
+ {\cal O}\Bigl(c^{-4};\,1.73\times10^{-18}\Bigr).
\end{equation}

Limiting in (\ref{eq:LC-vLLO}), the lunar gravity potential (\ref{eq:pot_w_0sh}) only to quadrupole, $J_{\tt 2M}$,  for an equatorial vLLO one obtains    
{}
\begin{equation}
L^{\tt vLLO}_{\tt CL}
= \frac{1}{c^2}\Big\{ \tfrac12\,V_{\tt vLLO}^2
 +   \frac{GM_{\tt M}}{r_{\tt vLLO}} \big( 1 + {\textstyle\frac{1}{2}}J_{2\tt M}\big) + \frac{GM_{\tt E}r_{\tt vLLO}^2}{4r^3_{\tt EM}}\Big\}\simeq 4.6818 \times 10^{-11} =4.0451\, \mu{\rm s/d}.
\label{eq:(29)LC}
\end{equation}

Thus, in a 10 km vLLO the secular drift is overwhelmingly set by the kinematic and monopole redshifts, with harmonics and tides entering at the \(10^{-5}\)–\(10^{-6}\)\, level.

As a result, a clock on vLLO will experience a total secular rate drift with respect to {\tt TT}. Substituting the value (\ref{eq:(29)LC}) into (\ref{eq:tau-TT=0}) we determine the $\tau-{\tt TT}$ offset rate this clock as 
{}
\begin{eqnarray}
 \frac{L_{\tt G}-L^{\tt vLLO}_{\tt CL}-L_{\tt EM}}{1-L_{\tt B}}=6.33017 \times 10^{-10} =  54.6926\, \mu{\rm s/d}.
\label{eq:tau-VLLO}
\end{eqnarray} 
Compared to a surface clock (\ref{eq:(RR)}), this rate is  1.33 $\mu$s/d smaller which is due to a larger velocity for a clock at vLLO.

\subsubsection{Time‐Dependent Correction \(P_{\rm CL}(t)\)}

We can now derive the periodic proper‐time correction \(P_{\tt CL}(t)\) for a circular polar vLLO.  The only time‐varying contribution is the Earth’s quadrupole tidal potential (\ref{eq:tides-LLO}), with the orbital phase
$\theta(t)=\omega_{\tt vLLO}\,t+\varphi$, and $\omega_{\tt vLLO}={2\pi}/{T_{\tt vLLO}}\approx9.58\times10^{-4}\,\mathrm s^{-1}.$ Defining the tidal amplitude
$A_{\tt [2]} \;\equiv\; c^{-2}{GM_{\tt E}r_{\tt vLLO}^2}/{r_{\tt EM}^3} \approx 2.39\times10^{-16},$
and using the identity
$P_{2}(\cos\theta)=\tfrac12\bigl(3\cos^{2}\theta-1\bigr)
=\tfrac14+\tfrac34\cos2\theta,$
one determines \(\langle P_{2}\rangle=1/4\).  Hence expression for $\dot P_{\tt CL}$ from  \eqref{eq:PCL} simplifies to
$\dot P_{\tt CL}(t)
=A_{[2]}\big(P_{2}(\cos\theta)-\tfrac14\big)
=\tfrac34 A_{[2]}\,\cos\!\big[2(\omega_{\tt vLLO}\, t+\varphi)\big].
$
Integrating in time gives
\[
P^{\tt vLLO}_{\tt CL}(t)=-\tfrac38
\frac{A_{[2]}}{\omega_{\tt vLLO}}\,
\sin\!\big[2(\omega_{\tt vLLO}\, t+\varphi)\big]=-\tfrac38\frac{GM_{\tt E}r_{\tt vLLO}^2}{c^2r_{\tt EM}^3 \omega_{\tt vLLO}}\sin\!\big[2(\omega_{\tt vLLO}\, t+\varphi)\big]\simeq -
9.34\times10^{-14}\,\sin\!\big[2(\omega_{\tt vLLO}\, t+\varphi)\big]\;\mathrm s,
\]
meaning that the one‐way amplitude is \(9.35\times10^{-14}\,\mathrm s\) and the two‐way peak‐to‐peak excursion is
 \(\Delta P_{\tt CL}\simeq 0.19\,\mathrm{ps}.\) If a smaller value of the Moon's radius is used, this value is  \(\Delta P_{\tt CL}\simeq 0.28\,\mathrm{ps}.\) Since this exceeds our \(0.10\,\mathrm{ps}\) threshold, it must be retained.
All higher‐order lunar harmonics (\(\ell\ge4\)) and solar tides lie below \(5\times10^{-18}\) and may be omitted. The corresponding two-way peak-to-peak excursion is $\Delta P_{\tt CL}\simeq 0.19$\,ps, thus retained explicitly in the model.

Note that when relating $\tau$ to {\tt TT} via (\ref{eq:tau-TT=0}), include the common monthly term $P_{\tt EM}(t)$ from (\ref{eq:orb-P}) (one-way amplitude $0.473~\mu$s) and the geometry term $-(\mathbf v_{\tt EM}\cdot\vec {\cal X})/c^2$. 
For a circular polar vLLO, $\max|(\mathbf v_{\tt EM}\cdot\vec {\cal   X})| \sim v_{\tt EM} r_{\tt vLLO}$, giving a one-way amplitude $\sim 2.0\times10^{-8}$ s ($\sim 20$ ns), well above the sub-ps {\tt LCRS} tides and therefore to be modeled alongside $P_{\tt CL}(t)$.

\subsection{Clock in a Low Lunar Orbit (LLO)}
\label{sec:LCRS-LLO}

A representative \emph{low lunar orbit} ({\tt LLO}) is taken here to be a near-circular, near-polar orbit with altitude
\(h_{\rm LLO}\in[100,200]~\mathrm{km}\) above the lunar reference radius \(R_{\tt MQ}=1738.0~\mathrm{km}\);
the corresponding orbital radius and mean motion are
\[
r_{\tt LLO}=R_{\tt MQ}+h_{\tt LLO},\qquad
\omega_{\tt LLO}=\sqrt{\frac{GM_M}{r_{\tt LLO}^{3}}},\qquad
{\cal V}_{\tt LLO}=\sqrt{\frac{GM_M}{r_{\tt LLO}}}.
\]
For \(h_{\tt LLO}=100\,(200)~\mathrm{km}\) one finds the orbital period
\(T_{\tt LLO}=2\pi/\omega_{\tt LLO}\simeq 1.964~(2.127)~\mathrm{h}\), in agreement with the ranges summarized in Table~\ref{tab:representative_orbits}. Throughout this subsection we use the global proper-time mapping of Sec.~\ref{sec:pt-TT-tau-rel}, i.e. Eqs.~(\ref{eq:synch-LP})--(\ref{eq:tau-TT=0}), as the master relation between the spacecraft proper time \(\tau\) and {\tt TT}, specialized to the {\tt LLO} geometry.

\subsubsection{Relevant potential terms}

The proper-to-coordinate time relation in the LCRS, Eq.~(\ref{eq:prop-coord-time-LCRS-TL}), specializes for an LLO clock to
\begin{align}
\frac{d\tau}{d\cal T}
&= 1-\frac{1}{c^{2}}\Big\{
\frac{1}{2}{\cal V}_{\tt LLO}^{2}
+U_{\tt M}\big(r_{\tt LLO},\theta,\lambda\big)
+
U_{\rm tid[2]}^{*\tt (E)}
\big(r_{\tt LLO},\theta,\lambda; \ell{=}2\big)
\Big\}
+{\cal O}\!\left(c^{-4};\; \delta_{\tt LLO}\right). \label{eq:LLO-rate}
\end{align}
Here \(U_{\tt M}\) is the lunar Newtonian potential (truncated to degrees/orders that survive the \(5\times10^{-18}\) fractional threshold), and \(U_{\rm tid[2]}^{*\tt (E)}\) is the Earth–induced quadrupole tide. In explicit spherical-harmonic form (keeping the leading degree–2 tesseral terms that are important for polar LLOs),
\begin{align}
U_{\tt M}(r,\theta,\lambda) &=
\frac{GM_{\tt M}}{r}
+\frac{GM_{\tt M}R_{\tt MQ}^{2}}{r^{3}}
\Big(
J_{2\tt M}P_{20}(\cos\theta)
+2\,C_{22}\,P_{22}(\cos\theta)\cos 2\lambda
+2\,S_{22}\,P_{22}(\cos\theta)\sin 2\lambda
\Big) + \nonumber\\
&\quad
+\Big(\ell{=}3,4~\text{zonal/tesseral terms}\Big)+{\cal O}(\ell\ge 5), \label{eq:LLO-UM}
\\[4pt]
c^{-2}U_{\rm tid[2]}^{*\tt (E)}
&=\frac{GM_{\tt E}\,r^{2}}{r_{\tt EM}^{3}}\,P_{2}\!\big(\hat{\boldsymbol n}_{\tt EM}\!\cdot\!\hat{\boldsymbol {\cal X}}\big).
\label{eq:LLO-tide}
\end{align}

For \(h_{\tt LLO}\in[100,200]~\mathrm{km}\) the terms that robustly exceed the \(5\times10^{-18}\) retention threshold are:
(i) the kinematic time-dilation term \(c^{-2}{\cal V}_{\tt LLO}^{2}/2\sim(1.41\textrm{–}1.48)\times10^{-11}\);
(ii) the lunar monopole \(c^{-2}GM_{\tt M}/r_{\tt LLO}\sim(2.81\textrm{–}2.97)\times10^{-11}\);
(iii) the degree–2 zonal/tesseral contributions proportional to \(J_{2\tt M},C_{22},S_{22}\) (their instantaneous \(c^{-2}\) magnitudes lie at \(10^{-15}\)–\(10^{-14}\));
and (iv) the Earth’s quadrupole tide \(c^{-2}GM_{\tt E} r_{\tt LLO}^{2}/r_{\tt EM}^{3}\sim(2.64\textrm{–}2.93)\times10^{-16}\).
Solar \(\ell = 2\) tides and Earth \(\ell = 3\) tides remain at \(\lesssim\!2\times10^{-18}\) and can be folded into the error indicator \(\delta_{\tt LLO}\).

In a circular polar LLO the degree–2 lunar harmonics modulate \(U_{\tt M}\) at twice the orbital frequency because
\(P_{20}(\cos\theta)=\tfrac{1}{4}+\tfrac{3}{4}\cos 2\theta\) and \(P_{22}(\cos\theta)=\tfrac{3}{2}\sin^{2}\theta
=\tfrac{3}{4}(1-\cos 2\theta)\).
Thus the zonal \(J_{2\tt M}\) and the sectorials \((C_{22},S_{22})\) drive prominent \(2\omega_{\tt LLO}\) oscillations in the rate
\((d\tau/d{\cal T})\) and in the integrated timing correction \(P_{\tt CL}(t)\) defined by (\ref{eq:PCL}). 
At LLO altitudes these lunar-harmonic signatures dominate over the Earth tide in the periodic budget, while the secular budget (next paragraph) is still set by the competition of the monopole redshift and orbital kinematics.

\subsubsection{Secular drift rate \(L_{\tt CL}\)}

By definition (\ref{eq:synch-LP}), the LLO secular rate is the long-time orbital average of the bracket in \eqref{eq:LLO-rate}. Retaining terms above threshold and using \(\langle P_{2}\rangle=\tfrac14\) and \(\langle P_{20}\rangle_{\rm polar}=\tfrac14\), one obtains
\begin{align}
L_{\tt CL}^{\tt LLO}
&=\frac{1}{c^{2}}\Big\{
\frac{1}{2}{\cal V}_{\tt LLO}^{2}
+\big< U_{\tt M}\big>_{\rm orb}
+\frac{GM_E r_{\tt LLO}^{2}}{4\,r_{\tt EM}^{3}}
\Big\}
+O\!\left(c^{-4};\; \delta_{\tt LLO}\right), \label{eq:LCL-LLO}\\[3pt]
\big< U_{\tt M}\big>_{\rm orb}
&\simeq \frac{GM_{\tt M}}{r_{\tt LLO}}
+\frac{GM_{\tt M}R_{\tt MQ}^{2}}{4\,r_{\tt LLO}^{3}}\,J_{2\tt M}\qquad
(\text{circular polar LLO}).
\end{align}
Numerically, adopting the constants of Table~\ref{tab:defs}, 
\[
\begin{array}{lcl}
h_{\tt LLO}=100~\mathrm{km}:& L_{\tt CL}^{\tt LLO}=4.4521\times10^{-11},
&\Rightarrow\;\dfrac{L_{\tt G}-L_{\tt CL}^{\tt LLO}-L_{\tt EM}}{1-L_{\tt B}}
=54.8912~\mu\mathrm{s/d},
\\[2pt]
h_{\tt LLO}=150~\mathrm{km}:& L_{\tt CL}^{\tt LLO}=4.3342\times10^{-11},
&\Rightarrow\; 54.9930~\mu\mathrm{s/d},
\\[2pt]
h_{\tt LLO}=200~\mathrm{km}:& L_{\tt CL}^{\tt LLO}=4.2223\times10^{-11},
&\Rightarrow\; 55.0897~\mu\mathrm{s/d},
\end{array}
\]
where we used (\ref{eq:tau-TT=0}) to map \(\tau\) to {\tt TT}, with \(L_{\tt G}\), \(L_B\) from Table~\ref{tab:defs} and \(L_{\tt EM}=1.7093906\times10^{-11}\) from (\ref{eq:L_EM}). 
For reference, the surface realization ({\tt TL}) gives \(56.0256~\mu\mathrm{s/d}\) (\ref{eq:(RR)}), while a 10~km vLLO yields \(54.6926~\mu\mathrm{s/d}\), cf. (\ref{eq:tau-VLLO}). 

\subsubsection{Time–Dependent Correction \(P_{\tt CL}(t)\)}

The periodic part \(P_{\tt CL}(t)\) follows from (\ref{eq:PCL}) by integrating the zero-mean variations in \eqref{eq:LLO-rate}. For a circular polar LLO the dominant harmonics are at \(2\omega_{\tt LLO}\) and are contributed by:
\begin{itemize}
\item the Earth’s quadrupole tide \eqref{eq:LLO-tide} with \(\delta U^{\tt (E)}_{\ell=2}(t)=
\tfrac{3}{4}\big(GM_{\tt E} r_{\tt LLO}^{2}/r_{\tt EM}^{3}\big)\cos\!\big(2\omega_{\tt LLO}t+\varphi_{\tt E}\big)\);
\item the lunar \(J_{2\tt M}\) term \eqref{eq:LLO-UM} with \(\delta U_{J_2}(t)=
\tfrac{3}{4}\big(GM_{\tt M}R_{\tt MQ}^{2}J_{2\tt M}/r_{\tt LLO}^{3}\big)\cos\!\big(2\omega_{\tt LLO}t+\varphi_{J_2}\big)\);
\item the combined sectorials \((C_{22},S_{22})\), which produce a principal \(2\omega_{\tt LLO}\) line plus weak sidebands (sum/difference with the longitude rate); the principal-line amplitude scales with \(|C_{22}|,|S_{22}|\) exactly as the \(J_2\) line.
\end{itemize}
Integrating in time, the leading contributions may be written
\begin{align}
\delta P_{\tt CL}^{\tt (E2)}(t)
&= -\tfrac38 \frac{GM_{\tt E}\,r_{\tt LLO}^{2}}{c^{2}\,r_{\tt EM}^{3}\,\omega_{\tt LLO}}
\sin\!\big(2\omega_{\tt LLO}t+\varphi_{\tt E}\big), \label{eq:PCL-Earth2}\\[3pt]
\delta P_{\tt CL}^{(J_2)}(t)
&= - \tfrac38\frac{GM_{\tt M}R_{\tt MQ}^{2}J_{2\tt M}}{c^{2}\,r_{\tt LLO}^{3}\,\omega_{\tt LLO}}
\sin\!\big(2\omega_{\tt LLO}t+\varphi_{J_2}\big), \label{eq:PCL-J2}\\[3pt]
\delta P_{\tt CL}^{(22)}(t)
&\simeq - \tfrac38\frac{GM_{\tt M}R_{\tt MQ}^{2}}{c^{2}\,r_{\tt LLO}^{3}\,\omega_{\tt LLO}}
\,\mathcal{C}_{22}\,
\sin\!\big(2\omega_{\tt LLO}t+\varphi_{22}\big), \qquad
\mathcal{C}_{22}\equiv \sqrt{C_{22}^{2}+S_{22}^{2}}\;,
\label{eq:PCL-22}
\end{align}
where \(\varphi\)-phases encode the geometry (orbit plane, prime meridian, Earth direction). The corresponding one-way amplitudes for the Earth tide are
\[
\delta P_{\tt CL}^{\tt (E2)}
=\tfrac38 \frac{GM_{\tt E}r_{\tt LLO}^{2}}{c^{2}r_{\tt EM}^{3}\omega_{\tt LLO}}
\simeq
\begin{cases}
1.11\times10^{-13}~\mathrm{s} \;(0.111~\mathrm{ps}), & h_{\tt LLO}=100~\mathrm{km},\\
1.34\times10^{-13}~\mathrm{s} \;(0.135~\mathrm{ps}), & h_{\tt LLO}=200~\mathrm{km},
\end{cases}
\]
which exceed the 0.1~ps inclusion threshold and must be modeled. The degree–2 lunar harmonics are larger:
\[
\delta P_{\tt CL}^{(J_2)}
=\tfrac38\frac{GM_{\tt M}R_{\tt MQ}^{2}J_{2\tt M}}{c^{2}r_{\tt LLO}^{3}\omega_{\tt LLO}}
\simeq
\begin{cases}
2.28~\mathrm{ps}, & h_{\tt LLO}=100~\mathrm{km},\\
2.10~\mathrm{ps}, & h_{\tt LLO}=200~\mathrm{km},
\end{cases}
\qquad
\big|P_{\tt CL}^{(22)}\big|_{\max}\sim 0.46\textrm{–}0.50~\mathrm{ps},
\]
using \(\mathcal{C}_{22}\approx|C_{22}|\) as a representative scale. Higher-degree terms (\(\ell\ge 3\)) produce sub-dominant lines that are still \(\gtrsim 0.1~\mathrm{ps}\) at \(h\approx100~\mathrm{km}\) and should be included when a sub-ps timing budget is required near mascon-rich regions.

The analytic structure above follows the general \(\tau\)–TT mapping in (\ref{eq:tau-TT=0}). For operational realizations one whould have to:
(i) evaluate \(L_{\tt CL}^{\tt LLO}\) from \eqref{eq:LCL-LLO} using the mission’s precise gravity model; (ii) accumulate the periodic correction \(P_{\tt CL}(t)=P_{\tt CL}^{(E2)}+P_{\tt CL}^{(J_2)}+P_{\tt CL}^{(22)}+\cdots\) along the osculating orbit; and
(iii) verify that the residual unmodeled potential satisfies \(|\Delta U|/c^{2}\lesssim 5\times 10^{-18}\) (which typically implies a high-degree GRAIL field for \(h\lesssim 200~\mathrm{km}\), with degree/order chosen by altitude and theater of operation). Note that in (\ref{eq:tau-TT=0}), the largest periodic is the monthly $P_{\tt EM}$ (amplitude $0.473~\mu$s); the next is $-(\vec v_{\tt EM} \cdot \vec{\cal  X})/c^2$ with one-way amplitude 
$\sim 2.1\times10^{-8}$ s (100 km) to $2.2\times10^{-8}$ s (200 km). The {\tt LCRS} lines from $J_{\tt 2M}$ and $(C_{22},S_{22})$ then enter at the few-ps level.

In near-circular, near-polar {\tt LLO} the periodic budget is dominated by three $2\omega_{\tt LLO}$ lines. The Earth’s quadrupole tide contributes a one-way amplitude of 0.111–0.135\,ps, the lunar zonal $J_{2\tt M}$ produces a 2.10–2.28\,ps line, and the sectorials $(C_{22},S_{22})$ add a co-located principal line at $\sim$0.46–0.50\,ps. Modest eccentricity ($e\ll1$) injects additional $\omega_{\tt LLO}$ harmonics through ${\cal V}^2(t)$ and $r(t)$ with amplitudes $O\!\big(e\,{\cal V}_{\tt LLO}^2/(c^2\omega_{\tt LLO})\big)$. When mapped to {\tt TT} via \eqref{eq:tau-TT=0}, these LCRS lines are subdominant to the common monthly term $P_{\tt EM}(t)$ from Sec.~\ref{sec:P_EM} (one-way amplitude $0.473~\mu$s) and the geometry line $-(\vec v_{\tt EM} \cdot \vec {\cal X})/c^2$ (one-way amplitude $\sim$21\,ns at $h=100$\,km, scaling linearly with $r_{\tt LLO}$).

\subsection{Clock in an Elliptical Lunar Frozen Orbit (ELFO)}
\label{sec:LCRS-ELFO}

Elliptical lunar frozen orbits (ELFOs) are high‑eccentricity, near‑stable solutions in which the argument of periapsis and eccentricity exhibit slow secular evolution under the combined action of $J_{2\tt M}$ and the tesseral harmonics. Consistent with Table~\ref{tab:representative_orbits}, we adopt here the NASA Lunar Communications Relay and Navigation Systems (LCRNS) Reference Constellation (LCRNS) reference for ELFO (see \cite{RydenVolle2025_LCRNS_RefConst3_1}) with periselene $h_p=1{,}750~\mathrm{km}$ and aposelene $h_a=17{,}400~\mathrm{km}$, i.e.\ a $\sim30$\,h south‑polar apolune design used for sustained polar coverage.  For this orbit we set
\begin{equation}
\label{eq:ae-ELFO}
a=\tfrac12\,(r_p+r_a),\qquad
e=\frac{r_a-r_p}{r_a+r_p},
\end{equation}
with $r_p=R_{\tt MQ}+h_p=3{,}488~\mathrm{km}$ and $r_a=R_{\tt MQ}+h_a=19{,}138~\mathrm{km}$, yielding
\[
a=11{,}313~\mathrm{km},\qquad e=0.69168,\qquad
\omega_{\tt ELFO}=\sqrt{\frac{G M_{\tt M}}{a^3}}=5.8191\times 10^{-5}\ \mathrm{s^{-1}},
\]
so that $T=2\pi/\omega_{\tt ELFO}=29.993~\mathrm{h}$. The LCRNS reference states reported in the constellation white paper give essentially the same SMA and eccentricity (SMA $\simeq 11{,}316~\mathrm{km}$, $e\simeq0.692$) and a $\approx30$\,h period, confirming consistency of this choice. We use the global mapping of Sec.~\ref{sec:pt-TT-tau}, Eqs.~(\ref{eq:tau-TT-int})–(\ref{eq:tau-TT=0}), which relate spacecraft proper time $\tau$ to {\tt TT} via a secular drift coefficient and a zero‑mean periodic term $P_{\tt CL}(t)$. 

\subsubsection{Relevant potential terms}

Specializing the LCRS proper–to–coordinate time relation to this ELFO gives
\begin{equation}
\label{eq:tau-LCRS-ELFO}
\frac{d\tau}{d{\cal T}}
=1-\frac{1}{c^2}\Big\{
\tfrac12\,{\cal V}^2
+U_{\tt M}({\cal T},\vec{\cal X})
+\frac{G M_{\tt E}}{r_{\tt EM}}
\sum_{\ell=2}^{\ell_{\max}^{(E)}}
\Big(\frac{{\cal X}}{r_{\tt EM}}\Big)^\ell
P_{\ell}\!\big(\hat{\boldsymbol n}_{\tt EM}\!\cdot\!\widehat{\vec{\cal X}}\big)
+\frac{G M_{\tt S}}{r_{\tt SM}^3}\,{\cal X}^2
P_{2}\!\big(\hat{\boldsymbol n}_{\tt SM}\!\cdot\!\widehat{\vec{\cal X}}\big)
\Big\}
+O\!\left(c^{-4}\right).
\end{equation}
The first two terms in the braces of \eqref{eq:tau-LCRS-ELFO}—the kinematic dilation $\tfrac12{\cal V}^2$ and the lunar Newtonian potential $U_{\tt M}$—set the baseline secular offset at the $\sim\!10^{-11}$ level and provide the dominant orbital harmonics as $r$ oscillates between $r_p$ and $r_a$. For the present $(a,e)$ one finds ${\cal V}_p=\sqrt{G M_{\tt M}(2/r_p-1/a)}=1.542~\mathrm{km\,s^{-1}}$ and ${\cal V}_a=0.281~\mathrm{km\,s^{-1}}$, giving
\[
c^{-2}\tfrac12{\cal V}^2\Big|_{r_p}=1.32\times10^{-11},\qquad
c^{-2}\tfrac12{\cal V}^2\Big|_{r_a}=4.39\times10^{-13},
\]
and, for the lunar monopole,
\[
c^{-2}U_{\tt M}(r_p)=\frac{G M_{\tt M}}{c^{2}r_p}=1.56\times10^{-11},\qquad
c^{-2}U_{\tt M}(r_a)=\frac{G M_{\tt M}}{c^{2}r_a}=2.85\times10^{-12}.
\]
These survive the $5\times10^{-18}$ threshold everywhere on the ellipse and, after removing their orbit averages, generate the leading lines of $P_{\tt CL}(t)$ via (\ref{eq:PCL}) and \eqref{eq:tau-TT=0}.  

On top of the monopole, the lunar degree‑2 harmonics (zonal $J_{2{\tt M}}=-C^{\tt M}_{20}$ and sectorials $C^{\tt M}_{22},S^{\tt M}_{22}$) contribute at the $10^{-16}$–$10^{-18}$ level across the ellipse: at periselene,
\[
c^{-2}\frac{\mu_{\tt M}R_{\tt MQ}^2}{r_p^3}\,J_{2{\tt M}}\simeq7.9\times10^{-16},\qquad
c^{-2}\frac{\mu_{\tt M}R_{\tt MQ}^2}{r_p^3}\,(3C^{\tt M}_{22})\simeq2.6\times10^{-16},
\]
decreasing to $4.8\times10^{-18}$ and $1.6\times10^{-18}$ at apolune, respectively. Degree–3–4 lunar terms peak near periselene at $\sim10^{-16}$ and fall below $10^{-18}$ at apolune; we retain them to protect the periselene budget while dropping $\ell\!\ge\!5$ throughout. (Time‑variable Love‑number modulations remain below threshold for this regime; see Appendix~\ref{sec:time-M-coord}.) 

External tides grow with $r$ and thus are most important near apolune. The Earth quadrupole gives
\[
c^{-2}U^{\tt (E)}_{\rm tid[2]}=\frac{G M_{\tt E}}{c^2}\frac{r^2}{r_{\tt EM}^3}\,P_2,
\]
with a scale of $9.5\times10^{-16}$ at $r_p$ and $2.86\times10^{-14}$ at $r_a$. The solar quadrupole, $\propto G M_{\tt S} r^2/r_{\tt SM}^{3}$, is smaller (from $5.4\times10^{-18}$ at $r_p$ to $1.6\times10^{-16}$ at $r_a$) but non‑negligible in the periodic budget; higher solar multipoles are below threshold and omitted. As in the other regimes of Sec.~V, the geometric factor $P_\ell(\hat{\boldsymbol n}\!\cdot\!\widehat{\vec{\cal X}})$ injects $\omega$–/\,$2\omega$ content with slow sidereal/synodic sidebands, so $P_{\tt CL}(t)$ is a multi‑line series rather than a single sinusoid.

\subsubsection{Secular drift rate \(L_{\tt CL}\)}
\label{sec:ELFO-LCL}

Following Sec.~\ref{sec:pt-TT-tau}, the ELFO secular coefficient is the orbital average of the bracket in \eqref{eq:tau-LCRS-ELFO}:
\begin{equation}
\label{eq:LCL-ELFO}
L_{\tt CL}^{\tt ELFO}
=\frac{1}{c^{2}}
\Big\{
\tfrac12\,\big\langle{\cal V}^{2}\big\rangle
+\big\langle U_{\tt M}\big\rangle_{\rm orb}
+\frac{G M_{\tt E}}{r_{\tt EM}^{3}}\,
\big\langle r^{2}P_{2}\big\rangle_{\rm orb}
+\frac{G M_{\tt S}}{r_{\tt SM}^{3}}\,
\big\langle r^{2}P_{2}\big\rangle_{\rm orb}^{\tt (S)}
\Big\}
+O\!\left(c^{-4}\right).
\end{equation}
For a Kepler ellipse, $\langle{\cal V}^{2}\rangle=\mu_{\tt M}/a$ and $\langle \mu_{\tt M}/r\rangle=\mu_{\tt M}/a$, hence the kinematic\(+\)monopole combination contributes $(3/2)\,\mu_{\tt M}/(a c^{2})$. To leading order in $e$, $\langle r^{2}\rangle=a^{2}\big(1+\tfrac{3}{2}e^{2}\big)$ and a slow‑geometry average gives $\langle P_{2}\rangle\simeq\tfrac14$, consistent with the LLO and deep‑space cases. For the adopted $(a,e)$,
\[
L_{\tt CL}^{\tt ELFO}\;=\;\frac{1}{c^{2}}\Big(\tfrac{3}{2}\frac{\mu_{\tt M}}{a}\Big)
+\frac{1}{c^{2}}\left(\frac{G M_{\tt E}}{r_{\tt EM}^{3}}+\frac{G M_{\tt S}}{r_{\tt SM}^{3}}\right)
\frac{a^{2}}{4}\Big(1+\tfrac{3}{2}e^{2}\Big)
\;=\;7.2372\times10^{-12}\;=\;0.6253~\mu\mathrm{s/d}.
\]
Mapping to {\tt TT} via \eqref{eq:tau-TT=0} yields the resulting linear drift,
\begin{equation}
\label{eq:tau-ELFO}
\frac{L_{\tt G}-L_{\tt CL}^{\tt ELFO}-L_{\tt EM}}{1-L_{\tt B}}
= 6.7263\times10^{-10} 
= 58.1152~\mu{\rm s/d},
\end{equation}
obtained with the constants of Table~\ref{tab:defs}. This rate lies between the LLO and L1/NRHO values, as expected from the intermediate mean orbital radius and speed. 

\subsubsection{Time–Dependent Correction \(P_{\tt CL}(t)\)}
\label{sec:ELFO-PCL}

The periodic correction is the time integral of the zero‑mean part of the $c^{-2}$ bracket in \eqref{eq:tau-LCRS-ELFO}, per the definition \eqref{eq:PCL}. In an ELFO the spectrum is richer than in a circular LLO because both the radius \(r(t)\) and the argument of latitude vary. Separating the kinematic\(+\)monopole piece from the tides and using vis–viva,
\[
\Big(\tfrac12{\cal V}^{2}+U_{\tt M}\Big)(t)
=\frac{2\mu_{\tt M}}{r(t)}-\frac{\mu_{\tt M}}{2a},
\]
the zero‑mean part is $2\mu_{\tt M}\big(1/r-1/a\big)$. Expanding in $e$ gives harmonics at \(\omega_{\tt ELFO}\) (dominant), \(2\omega_{\tt ELFO}\), and \(3\omega_{\tt ELFO}\). After time‑integration the one‑way amplitudes scale as
\[
A_{\omega}^{\rm(K+M)}\approx \frac{2\mu_{\tt M}}{a c^{2}\omega_{\tt ELFO}}\,e,\quad
A_{2\omega}^{\rm(K+M)}\approx \frac{\mu_{\tt M}}{a c^{2}\omega_{\tt ELFO}}\,e^{2},\quad
A_{3\omega}^{\rm(K+M)}\approx \frac{2}{3}\frac{\mu_{\tt M}}{a c^{2}\omega_{\tt ELFO}}\,e^{3},
\]
which evaluate, for the present orbit, to
\[
A_{\omega}^{\rm(K+M)}\simeq 0.115~\mu{\rm s},\qquad
A_{2\omega}^{\rm(K+M)}\simeq 0.040~\mu{\rm s},\qquad
A_{3\omega}^{\rm(K+M)}\simeq 0.018~\mu{\rm s},
\]
so the kinematic+monopole content alone produces a visibly multi-line $P_{\tt CL}(t)$ prior to adding tides.

For the Earth quadrupole tide $U^{\tt (E)}_{\rm tid}=(G M_{\tt E}/r_{\tt EM}^{3})\,r^{2}P_{2}$, a polar–like geometry gives $P_{2}=\tfrac14+\tfrac34\cos 2\Theta$. Combining this with the $r^{2}$ modulation along the ellipse and integrating the zero‑mean part yields (to $O(e)$)
\begin{align}
\label{eq:ELFO-per}
P_{\tt CL}^{\tt (E)}(t)
&\simeq
-\frac{G M_{\tt E}a^{2}}{c^{2}r_{\tt EM}^{3}}
\bigg[
\frac{3}{8\,\omega_{\tt ELFO}}\sin\!\big(2\omega_{\tt ELFO} t+\varphi_{2}\big)
-\frac{5e}{4\,\omega_{\tt ELFO}}\sin\!\big(\omega_{\tt ELFO} t+\varphi_{1}\big)
-\frac{e}{12\,\omega_{\tt ELFO}}\sin\!\big(3\omega_{\tt ELFO} t+\varphi_{3}\big)
\bigg]
+O(e^{2}),
\end{align}
with one‑way amplitudes of $\sim64~\mathrm{ps}$ at $2\omega_{\tt ELFO}$, $\sim149~\mathrm{ps}$ at $\omega_{\tt ELFO}$, and $\sim10~\mathrm{ps}$ at $3\omega_{\tt ELFO}$. The solar quadrupole has the same form with $G M_{\tt E}/r_{\tt EM}^{3}\to G M_{\tt S}/r_{\tt SM}^{3}$; here its largest line is the \(\omega_{\tt ELFO}\) term at $\sim0.84~\mathrm{ps}$ (the $2\omega$ and $3\omega$ lines are $\lesssim0.37$\,ps and $0.06$\,ps). The lunar $J_{2\tt M}$ contributes a $2\omega_{\tt ELFO}$ line at the $\sim 1.1~\mathrm{ps}$ level for this orbit (from $\langle r^{-3}\rangle$ scaling and $1/\omega_{\tt ELFO}$ integration), while the sectorials $(C_{22},S_{22})$ add a co‑located line near \(2\omega_{\tt ELFO}\) and weak sidebands at \(2\omega_{\tt ELFO}\!\pm\!\dot\lambda\) at the $\sim0.1$\,ps scale (geometry‑dependent).  

Collecting all contributions,
\[
P_{\tt CL}(t)=P_{\tt CL}^{\rm(K+M)}(t)+P_{\tt CL}^{\tt (E)}(t)+P_{\tt CL}^{\tt (S)}(t)+P_{\tt CL}^{\tt (M)}(t),
\]
so the ELFO correction is intrinsically \emph{multi‑line}, with power at \(\omega_{\tt ELFO}\), \(2\omega_{\tt ELFO}\), and \(3\omega_{\tt ELFO}\), plus sidereal/synodic sidebands from the tesseral field and the solar tide. When relating $\tau$ to {\tt TT} via \eqref{eq:tau-TT=0}, these harmonics combine with the common monthly term $P_{\tt EM}(t)$ from Sec.~\ref{sec:P_EM} (one‑way amplitude $\simeq 0.473~\mu\mathrm{s}$) and the geometry line $-(\vec v_{\tt EM}\!\cdot\!\vec{\cal X})/c^{2}$, producing the expected slow beating rather than a single clean sinusoid. 
This geometry term $-(\boldsymbol v_{\tt EM}\!\cdot\!\boldsymbol X)/c^2$ contributes a one-way amplitude set by the ELFO apolune scale, i.e., $\sim v_{\tt EM} r_a/c^2 \simeq 0.22~\mu\mathrm{s}$ (orientation-dependent), well above the LCRS lines but below the common $P_{\tt EM}(t)$ monthly term ($0.473~\mu\mathrm{s}$).
As elsewhere in Secs.~\ref{sec:LCRS-vLLO}–\ref{sec:LCRS-NRHO}, only harmonics with instantaneous amplitude $\gtrsim\!0.1$\,ps or fractional level $\gtrsim\!5\times 10^{-18}$ need be retained explicitly; the remainder are carried in the error budget for this regime.  

\subsection{Clock at the Earth–Moon Lagrange Point L1}
\label{sec:LCRS-L1}

\subsubsection{Relevant potential terms}
\label{sec:rel-pot-LI}

The Earth–Moon Lagrange L1 point lies on the line connecting the two bodies, at a distance from the Moon of
\begin{equation}
\label{eq:rL1}
r_{\tt L1}=r_{\tt EM}\Big(\alpha-\tfrac13 \alpha^2+{\cal O}(\alpha^3)\Big), 
\quad 
{\rm where}
\quad
\alpha = \Big(\frac{\tfrac13 M_{\tt M}}{M_{\tt E}+M_{\tt M}}\Big)^\frac{1}{3}\simeq 0.1594.
\end {equation}
Being a  fixed point—not an orbit—in 
 the {\tt LCRS}, L1's position depends explicitly on the instantaneous Earth–Moon separation, which varies with the Moon’s orbital eccentricity, $e_{\tt M}=0.0549006$. To first order in $e_{\tt M}$, we can write  $r_{\tt EM}=a_{\tt EM}\big(1-e_{\tt M}\cos[\omega_{\tt M}(t-t_0)] +{\cal O}(e^2_{\tt M})\big)$, where $\omega_{\tt M}$ is the Moon’s mean orbital angular rate,
$
\omega_{\tt M}={2\pi}/{T_{\rm sid}}\approx2.66\times10^{-6}\,\mathrm{s}^{-1}$ with $T_{\rm sid}\approx27.32\,\mathrm{d}$. Therefore,  L1 is at the mean distance from the Moon of $a_{\tt L1}=\big<
r_{\tt L1}\big>\simeq a_{\tt EM}(\alpha-\tfrac13 \alpha^2)\approx5.80\times10^{7}\,\mathrm{m}.$

A clock held fixed at L1 in the {\tt LCRS} frame shares 
the Moon’s mean orbital angular rate and thus has a residual speed
$
{\cal V}_{\tt L1}=|[\vec  \omega_{\tt M}\times \vec r_{\tt L1}]|\simeq \omega_{\tt M}\,a_{\tt L1}\approx1.54\times10^{2}\,\mathrm{m/s}.
$
Although this velocity is two orders of magnitude below typical orbital velocities, its contribution to the $c^{-2}$‐term is still significant
\[
c^{-2}\,\tfrac12\,{\cal V}_{\rm L1}^2\simeq 
c^{-2}\,\tfrac12\,(\omega_{\tt M}\,a_{\tt L1})^2
\simeq 1.33\times10^{-13}.
\]

At L1 the Newtonian potential of the Moon is reduced by the larger distance, yielding contribution of 
\[
c^{-2}U_{\tt M}
=\frac{GM_{\tt M}}{c^2a_{\tt L1}}
\approx 9.43\times10^{-13}.
\]
Clearly, both corrections exceed the \(5\times10^{-18}\)  and therefore require retention of higher‐order eccentricity contributions: the kinetic‐energy perturbation through \(\mathcal{O}(e_{\mathrm{M}}^3)\) and the lunar‐gravity potential expansion through \(\mathcal{O}(e_{\mathrm{M}}^4)\).

Note that the quadrupole (\(\ell=2\)) term of the lunar gravitational field is estimated to be negligible at L1: 
 \begin{align}
c^{-2}U_{\tt M[2]}
= \frac{G M_{\tt M} R_{\tt MQ}^2}{c^2 a_{\tt L1}^3}J_2P_{20}(\cos\theta)\lesssim 1.72\times 10^{-19}.
\end{align}
Other terms in Table~\ref{tab:sp-harmonics-moon} are even smaller; therefore, only the lunar monopole term is significant.

The dominant tidal perturbations are from the Earth’s and Sun's quadrupole tidal potentials at the {\tt LCRS} are:
{}
\begin{equation}
\label{eq:E-tides}
c^{-2}U_{\rm tid[2]}^{\tt (E)}
= \frac{GM_{\tt E}\,r_{\tt L1}^2}{c^2r_{\tt EM}^3}P_2(\vec n_{\tt EM}\!\cdot\!\widehat{\vec {\cal X}})
\lesssim 2.63\times10^{-13}, \qquad
c^{-2}U_{\rm tid [2]}^{\rm (S)}
= \frac{GM_{{\tt S}}r_{\tt L1}^2}{c^2{\rm AU}^3}P_2(\vec n_{ \tt SM}\!\cdot\!\widehat{\vec {\cal X}})
\lesssim 1.49\times10^{-15}.
\end{equation}

The octupole (\(\ell=3\)) terms contribute as below 
{}
\begin{equation}
c^{-2}U_{\rm tid[3]}^{\tt (E)}
= \frac{GM_{\tt E}r_{\tt L1}^3}{c^{2}r_{\tt EM}^4}P_3(\vec n_{\tt EM}\!\cdot\!\widehat{\cal X})
\approx3.97\times10^{-14},
\qquad
c^{-2}U_{\rm tid[3]}^{\tt (S)}
= \frac{GM_{\tt S} r_{\tt L1}^3}{c^{2}{\rm AU}^4}P_3(\vec n_{\tt SM}\!\cdot\!\widehat{\cal X})
\approx5.76\times10^{-19}.
\end{equation}

One can see that while the solar $\ell=3$ tide provides a negligible contribution, the Earth $\ell=3$ tide is still strong. In fact, for a clock at L1, the Earth tides reaching the level of $4.89\times 10^{-18}$ only at  $\ell=8$. Otherwise they are larger than our threshold of $5\times 10^{-18}$. So, for L1 the Earth tidal terms must be fully included in the model up to $\ell =7$. 

Hence, retaining only terms $\gtrsim5\times10^{-18}$, the proper‐to‐coordinate time relation that must be used  at L1 is
\begin{equation}
\label{eq:tau-LCRS-L1-final}
\frac{d\tau}{d{\cal T}}
=1-\frac{1}{c^2}\Big\{
  \tfrac12\,{\cal  V}_{\rm }^2
+ \frac{GM_{\tt M}} {{\cal X}} 
+
\frac{GM_{\tt E}}{r_{\tt EM}}
\sum_{\ell=2}^{7}
\Big(\frac{\cal X}{r_{\tt EM}}\Big)^\ell
P_\ell (\vec n_{ \tt EM}\!\cdot\!\widehat{\vec {\cal X}})
+\frac{GM_{\tt S}}{r_{\tt SM}^3}{\cal X}^2P_2(\vec n_{ \tt SM}\!\cdot\!\widehat{\vec {\cal X}})\Bigr\}+
{\cal O}\Big(c^{-4};\, 3.11\times 10^{-18}\Big),
\end{equation}
where the error bound  is due to omitted $\ell=8$ Earth tidal term.

The form (\ref{eq:tau-LCRS-L1-final}) makes explicit that at L1 the residual kinetic, monopole‐gravity, and Earth‐tide contributions are each of order $10^{-13}$, while all neglected corrections lie more than four orders of magnitude below the desired accuracy. Thus, expression (\ref{eq:tau-LCRS-L1-final}) provides a unified, self‐consistent model of proper time for lunar surface, low lunar orbit, or L1 applications with frequency stability at the $5\times 10^{-18}$ level.

\subsubsection{Secular drift rate \(L_{\tt CL}\)}

In direct analogy to the definition of \(L_{\tt L}\) in Sec.~\ref{sec:tm-tdb}, we define the secular drift rate \(L_{\tt CL}\) at the Earth–Moon L1 point by averaging all time‐independent contributions in the proper‐to‐coordinate time relation (\ref{eq:tau-LCRS-L1-final}) over one synodic period.  Retaining only terms above our \(5\times10^{-18}\) threshold yields four principal contributions discussed below.

The first is the residual kinematic redshift,
$c^{-2}\,\tfrac12\,{\cal V}_{\tt L1}^2
=c^{-2}\,\tfrac12(\omega_{\tt M}\,a_{\tt L1})^2
\approx 1.33\times10^{-13}.$ The second is the lunar monopole potential,
$c^{-2}U_{\tt M}=c^{-2}\,{GM_{\tt M}}/{r_{\rm L1}}
\approx 9.43\times10^{-13},$ even without the eccentricity corrections. 

The third comprises the Earth’s tidal multipoles up to \(\ell=7\). Note that at the Earth–Moon L1 point the tide‐raising axis from the Earth (and similarly from the Sun) is exactly aligned with the radial direction $\hat{\vec{\cal X}}$, so $ (\mathbf{n}_{\tt EM}\cdot \hat{\vec{\cal X}}) = 1$, thus $P_\ell(1)=1$ for all $\ell$.
As a result, the quadrupole (\(\ell=2\)) contributes
$c^{-2}U_{\tt E[2]}=c^{-2}({GM_{\tt E}r_{\rm L1}^2}/{r_{\tt EM}^3})
\approx 2.63\times10^{-13},
$ the octupole (\(\ell=3\)):
$c^{-2}U_{\tt E[3]}=c^{-2}({GM_{\tt E} a_{\tt L1}^3}/{r_{\tt EM}^4})\approx 3.97\times10^{-14},$
the \(\ell=4\) term:
$c^{-2}U_{\tt E[4]}=c^{-2}({GM_{\tt E}a_{\tt L1}^4}/{r_{\tt EM}^5})\approx 5.99\times10^{-15}$,
the \(\ell=5\) term:
$c^{-2}U_{\tt E[5]}=c^{-2}({GM_{\tt E}a_{\tt L1}^5}/{r_{\tt EM}^6})\approx 9.04\times10^{-16}$,
the \(\ell=6\) term:
$c^{-2}U_{\tt E[6]}=c^{-2}({GM_{\tt E} a_{\tt L1}^6}/{r_{\tt EM}^7})\approx 1.36\times10^{-16},
$ and the \(\ell=7\) term:
$c^{-2}U_{\tt E[7]}=c^{-2}({GM_{\tt E}a_{\rm L1}^7}/{r_{\tt EM}^8})\approx 2.06\times10^{-17}.$
All higher‐order Earth tides (\(\ell\ge8\)) are \(<5\times10^{-18}\) and are omitted.  Clearly,  the Earth tidal terms up to $\ell=6$ would also need to include eccentricity corrections of various orders. 
The fourth contribution is the solar quadrupole tide, given as 
$c^{-2}U_{\tt S[2]}= c^{-2}({GM_{\tt S}a_{\tt L1}^2}/{r_{\tt SM}^3})\approx 1.45\times10^{-15}$, thus, also included. 

Combining these four contributions gives (since $P_\ell(1)=1)$
\begin{equation}
\label{eq:LCL-L1-tau}
L^{\tt L1}_{\tt CL}
=\frac{1}{c^2}\Bigl\{
\tfrac12\,{\cal V}_{\tt L1}^2
+ \frac{GM_{\tt M}}{r_{\tt L1}}
+ \sum_{\ell=2}^{7}\frac{GM_{\tt E}r_{\rm L1}^\ell}{r_{\tt EM}^{\ell+1}}
+ \frac{GM_{\tt S}r_{\tt L1}^2}{r_{\tt SM}^3}
\Bigr\}
\simeq  1.3827\times10^{-12}\simeq 0.1195\,\mu\mathrm{s/d}.
\end{equation}

Thus the clock at the Earth–Moon L1 point experiences a net fractional rate offset of 
$\simeq 0.1195\,\mu\mathrm{s/d}$, dominated by the lunar monopole and kinematic terms at the \(10^{-13}\) level, with Earth‐tide contributions at \(10^{-13}\)–\(10^{-17}\) and the solar tide at \(10^{-15}\).  
All omitted tidal terms with $\ell \ge 8$ lie below $4.89\times10^{-18}$.

This value (\ref{eq:LCL-L1-tau}) may be used in (\ref{eq:tau-TT=0}) to determine the secular rate drift of a clock at L1 with respect to {\tt TT}:  
{}
\begin{eqnarray}
 \frac{L_{\tt G}-L^{\tt L1}_{\tt CL}-L_{\tt EM}}{1-L_{\tt B}}=6.78452 \times 10^{-10} =  58.6182 \, \mu{\rm s/d}.
\label{eq:tau-L1}
\end{eqnarray} 
Because of the weaker gravity and smaller velocity at L1, thus smaller
$L^{\tt L1}_{\tt CL}$ (\ref{eq:LCL-L1-tau}), this result is by $2.5926\,\mu$s/d larger than for a clock positioned at the lunar surface (\ref{eq:(RR)}).

\subsubsection{Time‐Dependent Correction \(P_{\tt CL}(t)\)}

Considering kinetic and gravity terms, to first order in $e_{\tt M}$, they contribute   
\begin{align}
\label{eq:tau-LCRS-E1-pot}
\frac{1}{c^2}\Big\{
  \tfrac12\,V_{\tt L1}^2 & -\big<  \tfrac12\,V_{\tt L1}^2\big>
 + \frac{GM_{\tt M}} {{\cal X}} -\Big<\frac{GM_{\tt M}} {{\cal X}} \Big>
\Bigr\}\simeq
\frac{1}{c^2}\Big\{
  \tfrac12\,\omega^2_{\tt M}\Big(r_{\tt L1}^2-\Big<r_{\tt L1}^2\Big>\Big)
 + GM_{\tt M}\Big(\frac{1}{r_{\tt L1}}- \Big<\frac{1}{r_{\tt L1}}\Big>\Big) \Bigr\}\simeq \nonumber\\
 & \simeq
-\frac{1}{c^2}\Big(\omega^2_{\tt M}a_{\tt L1}^2 +
 \frac{GM_{\tt M}}{a_{\tt L1}}\Big) e_{\tt M}\cos[\omega_{\tt M}(t-t_0)]\simeq 6.62 \times 10^{-14} \cos[\omega_{\tt M}(t-t_0)]\simeq 5.72  \cos[n_{\tt M}(t-t_0)] ~ {\rm ns/d}.
\end{align}
Integrating this result in time, we obtain the largest periodic contribution to the clock at L1
\[
\delta P^{\tt }_{\tt CL}(t)=-\frac{1}{c^2}\Big(\omega^2_{\tt M}a_{\tt L1}^2 +
 \frac{GM_{\tt M}}{a_{\tt L1}}\Big) \frac{e_{\tt M}}{\omega_{\tt M}}\,\sin[\omega_{\tt M}(t-t_0)]\simeq -2.53\times10^{-8}\,  \sin[\omega_{\tt M}(t-t_0)]\;\mathrm s.
\]
Clearly, there will be smaller contributions with non-linear modulations due to eccentricity corrections.  Tidal terms will also provide their owns series of terms at various frequencies that must  be accounted for. 

There are also contributions from the Earth tidal gravity potential with  the largest being the  $\ell=2$ quadruple term (\ref{eq:E-tides}). Because L1 lies on the Earth–Moon line $P_2(\vec n_{\tt EM}\!\cdot\!\widehat{\vec {\cal X}})=1$, with the help of (\ref{eq:rL1}), this potential at L1 is
\[
c^{-2}U_{\rm tid[2]}^{\tt (E)}
= \frac{GM_{\tt E}\,r_{\tt L1}^2}{c^2r_{\tt EM}^3}=\frac{GM_{\tt E}\,(\alpha-\tfrac13 \alpha^2)^2}{c^2a_{\tt EM}\big(1-e_{\tt M}\cos[\omega_{\tt M}(t-t_0)] \big)}\simeq 2.63 \times 10^{-13},
\]
as in (\ref{eq:E-tides}), yielding 
\[
\dot \delta P^{\tt (E)}_{\tt CL\,\rm tid[2]}(t)\simeq \frac{GM_{\tt E}}{c^2a_{\tt EM}}\,(\alpha-\tfrac13 \alpha^2)^2e_{\tt M}\cos[\omega_{\tt M}(t-t_0)] \simeq 1.44 \times 10^{-14} \cos\big[\omega_{\tt M}(t-t_0)\big],
\]
which produces an additional
\[
 \delta P^{\tt (E)}_{\tt CL\,\rm tid[2]}(t)
=-\frac{GM_{\tt E}}{c^2a_{\tt EM}}\,(\alpha-\tfrac13 \alpha^2)^2\frac{e_{\tt M}}{\omega_{\tt M}}\sin[\omega_{\tt M}(t-t_0)]
\;\simeq\;-5.42\times10^{-9} \sin\!\big[\omega_{\tt M}(t-t_0)\big]\;\mathrm s.
\]
That $\sim5.42\,$ns amplitude is comparable to the 25.3 ns “pure‑lunar” term and must be included for sub‑ps accuracy.  

We also account for the time‐varying contribution from the Sun’s quadrupole tide.  Denoting the synodic phase by
$\theta_{\tt S}(t)=\omega_{\rm syn}\,t+\varphi$, with
$\omega_{\rm syn}={2\pi}/{T_{\rm syn}}\simeq2.46\times10^{-6}\,\mathrm s^{-1},$
with \(T_{\rm syn}\approx29.53\;\mathrm d\), we define the tidal amplitude
$A_{\tt S[2]}\equiv
c^{-2}\,({GM_{\tt S}r_{\rm L1}^2}/{r_{\tt SM}^3})
\simeq1.66\times10^{-15}$.
From  $P_{2}(\cos\theta_{\tt S})
=\tfrac12(3\cos^2\theta_{\tt S}-1)
=\tfrac14+\tfrac34\cos2\theta_{\tt S}$, we see that 
$\langle P_{2}\rangle=\tfrac14$. With this, the periodic perturbation becomes
$\dot P_{\tt CL}(t)
=A_{\tt S[2]}\bigl(P_{2}(\cos\theta_{\tt S})-\tfrac14\big)
=\tfrac34\,A_{\tt S[2]}\,\cos\!\big[2(\omega_{\rm syn}t+\varphi)\big].$
Integrating in time yields
\[
\delta P^{\tt S[2]}_{\tt CL}(t)=-\frac{3A_{\tt S[2]}}{8\omega_{\rm syn}}
\sin\!\big[2(\omega_{\rm syn}t+\varphi)\big]=-
\tfrac{3}{8}\frac{GM_{\tt S}r_{\tt L1}^2}{c^2r_{\tt SM}^3 \omega_{\rm syn}}\,\sin\!\big[2(\omega_{\rm syn} t+\varphi)\big]
\simeq-2.53\times10^{-10}\,
\sin\!\big[2(\omega_{\rm syn}t+\varphi)\big]\;\mathrm s,
\]
even before the eccentricity corrections are applied. So that the one‐way amplitude is \(2.53\times10^{-10}\,\mathrm s\) and the two‐way peak‐to‐peak excursion is 
\(\Delta P_{\tt CL}\simeq 0.51\,\mathrm{ns}.\)

All other multipoles (Earth’s \(\ell=2\)–7, higher‐order solar terms) induce periodic effects \(<5\times10^{-18}\) and may be omitted.  As \(\Delta P_{\tt CL}\) at \(L_1\) exceeds our \(0.10\,\mathrm{ps}\) goal by over $10^3$ times, this periodic correction must be retained in full.  

When related to {\tt TT}, Eq.~(\ref{eq:tau-TT=0}) adds $P_{\tt EM}(t)$ (amplitude $0.473~\mu$s) and $-(\mathbf v_{\tt EM}\cdot \vec{\cal X})/c^2$. 
Because at L1 the position $\vec {\cal X}$ is nearly radial while $\vec v_{\tt EM}$ is nearly tangential, 
the dot product is suppressed by the orbital eccentricity: $|(\mathbf v_{\tt EM} \cdot \vec{\cal  X})| \sim e_{\tt M} v_{\tt EM} a_{\tt L1}$, giving a one-way amplitude $\lesssim 3.6\times10^{-8}$ s ($\sim 36$ ns).  This is comparable to the 25.3 ns lunar monthly term and larger than the solar quadrupole line (0.253 ns).

\subsection{Clock in Near‐Rectilinear Halo Orbit (NRHO)}
\label{sec:LCRS-NRHO}

Near‐Rectilinear Halo Orbits (NRHOs) about the Moon combine a low‐altitude periapsis with a distant apoapsis near the Earth–Moon Lagrange region, yielding extreme variations in both speed and gravitational potential.  For definiteness we adopt an NRHO with  
\[
r_{p}=R_{\tt MQ}+1\,630\;\mathrm{km}\approx3.37\times10^{6}\,\mathrm m,\quad
r_{a}=R_{\tt MQ}+69\,400\;\mathrm{km}\approx7.11\times10^{7}\,\mathrm m,
\]
semi–major axis and eccentricity given as below
\begin{equation}
\label{eq:ae}
a=\tfrac12(r_{p}+r_{a})\approx3.73\times10^{7}\,\mathrm m,
\qquad 
e=\frac{r_{a}-r_{p}}{r_{a}+r_{p}}\approx0.9088.
\end{equation}

\subsubsection{Relevant potential terms}

The instantaneous orbital speed follows the vis–viva relation,
\begin{equation}
\label{eq:VV}
{\cal V}^2(r)={GM_{\tt M}\Bigl(\frac{2}{r}-\frac{1}{a}\Bigr)},
\end{equation}
so that at periapsis ${\cal V}_{p}\simeq1.667\,$km/s and at apoapsis ${\cal V}_{a}\simeq78.9\,$m/s.  The corresponding relativistic time dilation,
\[
c^{-2}\,\tfrac12\,{\cal V}_p^2\approx1.55\times10^{-11},
\qquad
c^{-2}\,\tfrac12\,{\cal V}_a^2\approx3.47\times10^{-14},
\]
exceeds our $5\times10^{-18}$ cutoff throughout the orbit and must be retained.

The lunar monopole gravitational redshift likewise dominates, with
\[
c^{-2}U_{\tt M}(r_{\rm p})=
c^{-2}\,\frac{GM_{\tt M}}{r_{p}}\approx1.62\times10^{-11},
\qquad
c^{-2}U_{\tt M}(r_{\rm a})=c^{-2}\,\frac{GM_{\tt M}}{r_{a}}\approx7.67\times10^{-13}.
\]
The quadrupole term of the Moon’s field,
\[
c^{-2}U_{\tt M[2]}=
c^{-2}\,\frac{GM_{\tt M}R_{\tt MQ}^2}{r^3}J_{\tt 2M}P_{20}(\cos\theta),
\]
is significant (up to $8.7\times10^{-16}$) only near periapsis; all higher‐degree lunar harmonics remain $\lesssim10^{-19}$ and are omitted beyond $\ell=2$, except for tesseral coefficients $C_{22},S_{22}$ at periapsis, which enter at the $10^{-16}$ level and are included.

Tidal perturbations from the Earth are dominated by its quadrupole,
\[
c^{-2}U_{\rm tid[2]}^{\tt (E)}=
c^{-2}\,\frac{GM_{\tt E}}{r_{\tt EM}^3}\,r^2P_2(\cos\theta_{\tt EM})
\qquad\Rightarrow\qquad
3.95\times10^{-13}\text{ at apoapsis},\;
8.86\times10^{-16}\text{ at periapsis},
\]
and by its higher multipoles up to $\ell=8$, all of which exceed $5\times10^{-18}$ somewhere in the orbit.  The solar quadrupole tide reaches $2.23\times10^{-15}$ at apoapsis and falls below threshold at periapsis; solar $\ell\ge3$ terms are always $\lesssim10^{-18}$ and may be dropped.

Accordingly, retaining only terms $\gtrsim5\times10^{-18}$, the proper‐to‐coordinate time relation in Gateway NRHO is
\begin{equation}
\label{eq:tau-LCRS-final-NRHO}
\frac{d\tau}{d{\cal T}}
=1 -\frac{1}{c^2}\Big\{
  \tfrac12\,{\cal  V}^2
+ U_{\tt M}({\cal T},\vec {\cal X}) 
+
\frac{GM_{\tt E}}{r_{\tt EM}}
\sum_{\ell=2}^{8}
\Big(\frac{\cal X}{r_{\tt EM}}\Big)^\ell
P_\ell\bigl(\cos\theta_{\tt EM}\bigr)
+\frac{GM_{\tt S}}{r_{\tt SM}^3}{\cal X}^2
P_2(\vec n_{ \tt SM}\!\cdot\!\widehat{\vec {\cal X}})\Bigr\}+
{\cal O}\Big(c^{-4};\, 3.17\times 10^{-18}\Big),
\end{equation}
where $U_{\tt M}({\cal T},\vec {\cal X}) $ has terms only up to $\ell=2$ and the error bound  is due to omitted $\ell=9$ Earth tidal term with the $\ell=3$ solar tide that reaches $1.09\times 10^{-18}$.

\subsubsection{Secular drift rate \(L_{\tt CL}\)}

Throughout the NRHO the clock’s instantaneous speed is given by the vis–viva relation, (\ref{eq:VV}),
so that the orbit‐average of the special‐relativistic dilation 
\[
\bigl\langle\tfrac12\,{\cal V}^2\bigr\rangle
=\frac{1}{T_{\tt NRHO}}\int_0^{T_{\tt NRHO}}\tfrac12\,{\cal V}^2\,dt
\;=\;\frac{GM_{\tt M}}{2a}
\qquad\Rightarrow\qquad
c^{-2}\,\bigl\langle\tfrac12\,{\cal V}^2\bigr\rangle
=\frac{GM_{\tt M}}{2ac^2}
=7.3083\times10^{-13}.
\]
where $T_{\tt NRHO}=2\pi/\omega_{\tt NRHO}$ with $\omega_{\tt NRHO}=\sqrt{GM_{\tt M}/a^3}\simeq 9.72 \times10^{-6}\,{\rm s}^{-1}.$

Likewise, the lunar monopole gravitational redshift averages to
\[
\langle U_{\tt M}\rangle
=\frac{1}{T_{\tt NRHO}}\int_0^{T_{\tt NRHO}}\frac{GM_{\tt M}}{r(t)}\,dt
\;=\;\frac{GM_{\tt M}}{a}
\qquad\Rightarrow\qquad
c^{-2}\,\langle U_{\tt M}\rangle
=\frac{GM_{\tt M}}{ac^2}
=1.4617\times10^{-12},
\]
which exceeds the kinematic term by a factor of two and thus dominates the secular offset.

The Earth’s quadrupole tide contributes through the mean‐square orbital radius.  Using the identity 
\(\langle r^2\rangle = a^2\bigl(1+\tfrac32e^2\bigr)\approx2.239\,a^2\) and \(\langle P_2\rangle=\tfrac14\) (which follows from averaging $P_2(\cos\theta)$ over a full orbit), one finds
\[
c^{-2}\Bigl\langle\frac{GM_{\tt E}}{r_{\tt EM}^3}\,r^2\,P_2\Bigr\rangle
=\frac{GM_{\tt E}}{r_{\tt EM}^3c^2}\tfrac14 a^2(1+\tfrac32e^2)
=6.085\times10^{-14}.
\]

The solar quadrupole tide is similarly treated,
\[
c^{-2}\Bigl\langle\frac{GM_{\tt S}}{\mathrm{AU}^3}r^2P_2\Bigr\rangle
=c^{-2}\frac{GM_{{\tt S}}}{\mathrm{AU}^3}\tfrac{1}{4}a^2(1+\tfrac32e^2)
=3.436\times10^{-16}.
\]

All other potential terms—lunar \(J_2\) and higher harmonics, Earth tides \(\ell\ge3\), and solar \(\ell\ge3\)—average below our \(5\times10^{-18}\) retention threshold and are omitted from \(L_{\tt CL}\).

Collecting these four contributions yields
\begin{eqnarray}
\label{eq:LCL-NRHO}
L^{\tt NRHO}_{\tt CL}
=\frac{1}{c^2}\Bigl\{\tfrac32
\frac{GM_{\tt M}}{a}
+\tfrac14 a^2\bigl(1+\tfrac32e^2\bigr)\Big(\frac{GM_{\tt E}}{r_{\tt EM}^3}\,
+\frac{GM_{\tt S}}{\mathrm{AU}^3}\Big)
\Bigr\}
\simeq2.2537\times10^{-12}
\;\simeq0.1947\,\mu\mathrm{s/d}.
\end{eqnarray} 

Thus, the NRHO secular drift is overwhelmingly set by the lunar monopole (\(1.46\times10^{-12}\)) and kinematic (\(7.31\times10^{-13}\)) terms, with the Earth quadrupole tide entering at the \(10^{-14}\) level and the solar tide at \(10^{-16}\).  All neglected contributions lie safely below \(5\times10^{-18}\).

Substituting result (\ref{eq:LCL-NRHO}) in (\ref{eq:tau-TT=0}), we determine  the constant rate drift of a clock on NRHO with respect to {\tt TT}:
{}
\begin{eqnarray}
 \frac{L_{\tt G}-L^{\tt NRHO}_{\tt CL}-L_{\tt EM}}{1-L_{\tt B}}=6.77581 \times 10^{-10} =  58.5431 \, \mu{\rm s/d}.
\label{eq:tau-NRHO}
\end{eqnarray} 
Thus, compared to the lunar surface clock (\ref{eq:(RR)}), the NRHO clock exhibits a larger rate offset of $2.5175\,\mu$s/d, which is because its average orbital‐energy is smaller than the combined energy at the location of a clock on the lunar surface.

\subsubsection{Time‐Dependent Correction \(P_{\tt CL}(t)\)}

The periodic correction \(P_{\tt CL}(t)\) in the NRHO is obtained by isolating, for each retained \(c^{-2}\) term in (\ref{eq:tau-LCRS-final-NRHO}), the deviation about its secular average and integrating in time.  We parametrize the orbit by the eccentric anomaly \(E\), so that
\[
r(t)=a\bigl(1 - e\cos E\bigr), 
\qquad
e\approx0.9088,\quad a\approx3.73\times10^{7}\mathrm m,
\]
and the mean motion
$\omega_{\tt NRHO}=\sqrt{{GM_{\tt M}}/{a^3}}\approx9.72\times10^{-6}\,\mathrm s^{-1},
$ and orbital period of $T_{\tt NRHO}={2\pi}/{\omega_{\tt NRHO}}\approx7.49\,\mathrm d.
$
To third order in \(e\) the principal radial expansions are
\begin{eqnarray}
\frac1r-\frac1a
&=&\frac{1}{a}\Big(e\cos E + e^2\cos2E + e^3\cos3E\Big)
+{\cal O}(e^4),\\
r^{-3}-a^{-3}
&=&\frac{1}{a^3}\Big(3e\cos E + \tfrac32e^2\cos2E + \tfrac{1}{3}e^3\cos3E\Big)
+{\cal O}(e^4),\\
r^{\ell}-a^{\ell}
&=&-\,\ell\,a^{\ell}\Big(e\cos E - \tfrac12e^2\cos2E + \tfrac13e^3\cos3E\Big)
+{\cal O}(e^4).
\end{eqnarray}

Considering kinematic and  lunar gravity monopole, we see that the combination \(\tfrac12{\cal V}^2 + U_{\tt M}\) oscillates as
\[
\dot P_{\rm km+gm}(t)
=c^{-2}GM_{\tt M}\Bigl(\frac1r-\frac1a\Bigr)
\simeq \frac{GM_{\tt M}}{c^2a}\Big(e\cos E + e^2\cos2E + e^3\cos3E\Big)+{\cal O}(e^4).
\]
Integrating this result in time gives
\[
P_{\rm km+gm}(t)
=-\frac{GM_{\tt M}}{ac^2\omega_{\tt NRHO}}
\Big(e\sin E + \tfrac12{e^2}\sin2E + \tfrac13{e^3}\sin3E\Big)+{\cal O}(e^4),
\]
with one‐way amplitudes
\[
A_{1}^{\rm km+gm}=1.37\times10^{-7}\,\mathrm s,\quad
A_{2}^{\rm km+gm}=6.21\times10^{-8}\,\mathrm s,\quad
A_{3}^{\rm km+gm}=3.77\times10^{-8}\,\mathrm s,
\]
corresponding to orbital periods \(T\), \(\tfrac12 T\) and \(\tfrac13T\) of approximately 7.48 d, 3.74 d and 2.49 d.

As for the lunar quadrupole, this \(\ell=2\) tidal term of the Moon’s field varies as \(r^{-3}\), hence
\[
P_{J_2}(t)
=-\,\frac{3GM_{\tt M}R_{\tt MQ}^2J_{2\tt M}}{a^3c^2\omega_{\tt NRHO}}
\Big(e\sin E + \tfrac{1}{2}e^2\sin2E + \tfrac{1}{3}e^3\sin3E\Big)+{\cal O}(e^4),
\]
with one‐way amplitudes
\[
A_{1}^{J_2}=1.81\times10^{-13}\,\mathrm s,\quad
A_{2}^{J_2}=8.23\times10^{-14}\,\mathrm s,\quad
A_{3}^{J_2}=4.99\times10^{-14}\,\mathrm s.
\]

Moving on to the Earth tides, we see that each multipole \(\ell\in[2,8]\) enters through \(r^\ell P_\ell(\cos\theta_{\tt EM}(t))\).  
To \(\mathcal O(e^3)\) the radial part generates harmonics at \(k\omega_{\tt NRHO}\) with amplitudes
\[
P_{\tt E[\ell],k,m}(t)
=-\,\frac{\ell GM_{\tt E}a^\ell e^k}{r_{\tt EM}^{\ell+1}c^2\omega_{\tt NRHO}}
\sin(kE)+{\cal O}(e^{k+1})
\quad(k=1,2,3),
\]
and the angular factor \(P_\ell(\cos\theta_{\tt EM})\) produces sidereal sidebands at frequencies \(k\omega_{\tt NRHO}\pm m\,\omega_{\rm M}\).  
Here, we have introduced the integer \(m\) as the order of the tesseral (longitude‑dependent) harmonic in the Fourier expansion of 
\(\,P_\ell(\cos\theta_{\tt EM}(t))\), with \(m=0,1,\dots,\ell\).  
Numerically, the dominant quadrupole (\(\ell=2\)) radial amplitudes are
\[
A_{1}^{\tt E[2]}=2.03\times10^{-8}\,\mathrm s,\quad
A_{2}^{\tt E[2]}=1.02\times10^{-8}\,\mathrm s,\quad
A_{3}^{\tt E[2]}=6.78\times10^{-9}\,\mathrm s,
\]
while the \(\ell=3\ldots8\) quadrupolar harmonics fall roughly an order of magnitude per degree, down to
\[
A_{1}^{\tt E[8]}=6.78\times10^{-14}\,\mathrm s,\quad
A_{2}^{\tt E[8]}=3.39\times10^{-14}\,\mathrm s,\quad
A_{3}^{\tt E[8]}=2.26\times10^{-14}\,\mathrm s.
\]
The primary sidereal beat for \(\ell=2\) has amplitude \(B_{\tt E[2]}\approx1.15\times10^{-8}\,\mathrm s\) at frequency \(2(\omega_{\tt NRHO}-\omega_{\tt M})\), with analogous but smaller beats for \(3\le\ell\le8\).

Finally, the solar quadrupole \(\ell=2\) tide perturbation behaves as \(r^2\), combining a pure orbital series with a synodic beat at \(2(\omega_{\tt NRHO}-\omega_{\rm syn})\).  Its one‐way radial harmonics are
\[
A_{1}^{\tt S[2]}=1.15\times10^{-10}\,\mathrm s,\quad
A_{2}^{\tt S[2]}=5.75\times10^{-11}\,\mathrm s,\quad
A_{3}^{\tt S[2]}=3.83\times10^{-11}\,\mathrm s,
\]
and the synodic beat amplitude is \(B_{\tt S[2]}\approx5.75\times10^{-11}\,\mathrm s\).

Combining all contributions yields
\begin{align}
P^{\tt NRHO}_{\tt CL}(t)
= P_{\rm km+gm}(t)
+ P_{J_2}(t)
&+ \sum_{\ell=2}^{8}\sum_{m=0}^{\ell}\sum_{k=1}^{3}
  \Big\{P_{\tt E[\ell],k,m}(t)
    +B_{\tt E[\ell],k,m}\,\sin\bigl[(k\,\omega_{\tt NRHO}\pm m\,\omega_{\tt M})t\bigr]\Big\}
+\nonumber \\
&+ P_{\tt S[2]}(t)
+ B_{\tt S[2]}\,\sin\!\bigl[2(\omega_{\tt NRHO}-\omega_{\rm syn}\,)t\bigr],
\label{eq:NRHO-per}
\end{align}
a rich multi–harmonic series at orbital harmonics \(k\omega_{\tt NRHO}\) (with \(k=1,2,3\)), sidereal sidebands, and synodic beats.  
Even the smallest retained amplitude ($2.26\times 10^{-14}$\,s $\approx 0.0226$\,ps) lies \emph{below} the 0.1\,ps threshold; 
we retain it for completeness and uniformity of the harmonic expansion. Thus,  all terms to \(\mathcal O(e^3)\) and \(\ell\le8\) must be retained for sub–ps accuracy.

\paragraph*{TT mapping:}
Beyond $P_{\tt EM}(t)$ (0.473 $\mu$s), the term $-(\mathbf v_{\tt EM} \cdot \vec {\cal X})/c^2$ can dominate near apoapsis where $r$ is largest.  A conservative bound is $\max \big|(\mathbf v_{\tt EM} \cdot \vec{\cal   X})/c^2\big| \sim v_{\tt EM} r_a/c^2 \approx 8.1\times10^{-7}$ s (0.81 $\mu$s).  Its actual amplitude depends on the apoapsis orientation; typical values for Gateway‑like NRHOs are 0.2–0.5 $\mu$s. These should be modeled together with the $k\omega_{\rm NRHO}$ harmonics listed in (\ref{eq:NRHO-per}).

\subsection{Orbit-by-orbit summary}
\label{sec:orbit-sum}
 
Table~\ref{tab:cislunar-summary} consolidates, for the representative regimes treated in Sec.~\ref{sec:pt-TT-tau}, the secular {\tt LCRS} rate $L_{\tt CL}$ from the constant/periodic split \eqref{eq:synch-LP}, the largest one-way {\tt LCRS} periodic terms obtained from \eqref{eq:PCL} as specialized in Secs.~\ref{sec:LCRS-vLLO}, \ref{sec:LCRS-LLO}, \ref{sec:LCRS-L1}, \ref{sec:LCRS-NRHO}, and the mapping to {\tt TT} via \eqref{eq:tau-TT-int}.
\begin{itemize}
\item \textit{vLLO} (Sec.~\ref{sec:LCRS-vLLO}): {\tt LCRS} tides are sub-ps; the Earth $\ell{=}2$ line at $2\omega_{\tt vLLO}$ dominates (\(\sim 0.093\) ps one-way). The {\tt TT} mapping is driven by the monthly term and the geometry term $\!-\!(\mathbf v_{\tt EM}\cdot\vec{\cal X})/c^2$ (\(\sim20\) ns). See Table~\ref{tab:cislunar-summary}.
\item \textit{LLO} (Sec.~\ref{sec:LCRS-LLO}): the lunar $J_{2\tt M}$ term at $2\omega_{\tt LLO}$ (\(\sim 2.28\) ps) dominates, with sectorials \(C_{22}\) at \(2\omega_{\tt LLO}\) at the \(\sim 0.46\)–\(0.50\) ps level and Earth $\ell{=}2$ at \(0.111\) ps; the {\tt TT} mapping adds the same monthly/geometry terms as above. See Eqs.~\eqref{eq:PCL-J2}–\eqref{eq:PCL-Earth2} and Table~\ref{tab:cislunar-summary}.
\item {ELFO (Sec.~\ref{sec:LCRS-ELFO}).}
Adopting the LCRNS reference ELFO (\(h_p=1{,}750~\mathrm{km}\), \(h_a=17{,}400~\mathrm{km}\); \(a=11{,}313~\mathrm{km}\), \(e=0.69168\), \(T=29.993~\mathrm{h}\)), the secular coefficient is \(L_{\tt CL}^{\tt ELFO}=7.237\times10^{-12}=0.625~\mu\mathrm{s/d}\) [\eqref{eq:LCL-ELFO}]. The $P_{\tt CL}(t)$ content combines $(\mathrm{K+M})$ harmonics at \(\omega,2\omega,3\omega\) with one-way amplitudes \(\{0.115,\,0.040,\,0.018\}~\mu\mathrm{s}\), Earth $\ell=2$ lines from \eqref{eq:ELFO-per} at \(\{149,\,64,\,10\}~\mathrm{ps}\), solar $\ell=2$ at \(\{0.84,\,\lesssim0.37,\,\lesssim0.06\}~\mathrm{ps}\), and lunar lines at \(2\omega\) from $J_{2\tt M}$ (\(\sim1.1~\mathrm{ps}\)) with weak $(C_{22},S_{22})$ sidebands ($\sim0.1~\mathrm{ps}$). Mapping to {\tt TT} via \eqref{eq:tau-TT=0} adds the common monthly $P_{\tt EM}(t)$ ($0.473~\mu\mathrm{s}$) and a geometry term \(-(\vec v_{\tt EM}\!\cdot\! \vec {\cal X})/c^2\) with typical one-way amplitude \(\sim0.2~\mu\mathrm{s}\).

\item \textit{EM\,L1} (Sec.~\ref{sec:LCRS-L1}): {\tt LCRS} periodic content is monthly and tidal: \(\sim 25.3\) ns (kinematic\(+\)monopole), \(\sim5.42\) ns (Earth \(\ell{=}2\)), \(\sim 0.253\) ns (solar \(\ell{=}2\)); the {\tt TT} mapping adds the common monthly term and a geometry term that is \(\lesssim 36\) ns because \(\mathbf v_{\tt EM}\!\perp\!\vec{\cal X}\) to first order. See Table~\ref{tab:cislunar-summary}.
\item \textit{NRHO} (Sec.~\ref{sec:LCRS-NRHO}): rich multi‑harmonic structure at \(k\omega_{\tt NRHO}\) with \(k{=}1,2,3\) (0.137, 0.062, 0.038 \(\mu\)s one‑way), sidereal sidebands from Earth tides and a synodic beat from the solar \(\ell{=}2\) tide, cf.\ \eqref{eq:NRHO-per}. The {\tt TT} mapping adds the monthly term and a geometry line that can reach \(\sim 0.81\,\mu\)s near apoapsis.
\end{itemize}
Across all regimes, the secular drift \((\tau-{\tt TT})\) rates follow directly from \eqref{eq:tau-TT-int} and the reported $L_{\rm CL}$ values (e.g., \(\,54.6926\,\mu\mathrm{s/d}\) in vLLO, \(\,58.6182\,\mu\mathrm{s/d}\) at L1, \(\,58.5431\,\mu\mathrm{s/d}\) in NRHO; see \eqref{eq:tau-VLLO}, \eqref{eq:tau-L1}, \eqref{eq:tau-NRHO} and Table~\ref{tab:cislunar-summary}). Also, Table~\ref{tab:retention} list all the relevant potential terms that must be kept to reach the stated accuracy. 

Finally, we note that any mission‑specific implementation must promote the orbital elements \(a, e\) (and any others) to osculating, time‑dependent quantities and re‑expand each of the above sinusoids to first order in \(\delta a(t)\) and \(\delta e(t)\), or else extract them via a high‑fidelity numerical propagation followed by a spectral (FFT) analysis.

\begin{table}[t]
\centering
\setlength{\tabcolsep}{3pt}        
\renewcommand{\arraystretch}{1.0} 
\caption{Secular and dominant periodic terms by orbit. One-way amplitudes are listed;  \emph{two-way peak-to-peak} is twice these values, see mapping via \eqref{eq:tau-TT=0}. 
The second column lists the {\tt LCRS} secular rate $L_{\tt CL}$ from the averaging defined in \eqref{eq:tau-TT-int}–\eqref{eq:PCL}. 
The third column gives the largest one-way periodic terms within the {\tt LCRS} (built from the series summarized in Secs.~\ref{sec:LCRS-vLLO}–\ref{sec:LCRS-NRHO}). 
The fourth column is the secular drift of $\tau$ versus ${\tt TT}$ from \eqref{eq:tau-TT=0}. The last column lists the largest vs.\ ${\tt TT}$ periodic terms: the common monthly $P_{\tt EM}$ from \eqref{eq:orb-P}, the geometry term $-(\mathbf v_{\tt EM}\!\cdot\!\vec{\cal X}_{\tt TT})/c^2$, and the largest {\tt LCRS} line(s) propagated through \eqref{eq:tau-TT=0}.  One-way amplitudes are shown; two-way peak-to-peak is twice these values.}
\label{tab:cislunar-summary}
\renewcommand{\arraystretch}{1.2}
\begin{tabular}{lllll}
\hline
\parbox[t]{2.2cm}{\raggedright Regime} &
\parbox[t]{2.2cm}{\raggedright $L_{\tt CL}$} &
\parbox[t]{5.2cm}{\raggedright Largest {\tt LCRS} periodic(s)} &
\parbox[t]{1.85cm}{\raggedright Secular drift: $\tau$ vs~{\tt TT}} &
\parbox[t]{5.7cm}{\raggedright Largest periodic(s): $\tau$ vs~{\tt TT}} \\
\hline

\parbox[t]{2.2cm}{\raggedright vLLO (10 km)} &
\parbox[t]{2.2cm}{\raggedright $4.6818\times10^{-11}$} &
\parbox[t]{5.2cm}{\raggedright $0.093$ ps @ $2\omega_{\tt vLLO}$ (Earth $\ell\!=\!2$); $\lesssim$\,sub-ps from lunar $J_2$, $C_{22}$} &
\parbox[t]{1.85cm}{\raggedright $54.6926~\mu$s/d} &
\parbox[t]{5.7cm}{\raggedright $0.473~\mu$s (monthly $P_{\tt EM}$); $\sim$20 ns from $-(\mathbf v_{\tt EM}\!\cdot\!\vec {\cal X})/c^2$; $0.093$ ps (Earth $\ell\!=\!2$)} \\

\parbox[t]{2.2cm}{\raggedright LLO (100 km)\footnotemark} &
\parbox[t]{2.2cm}{\raggedright $4.4521\times10^{-11}$} &
\parbox[t]{5.2cm}{\raggedright $2.28$ ps (lunar $J_2$) + $\sim$0.46–0.50 ps ($C_{22}$) @ $2\omega_{\tt LLO}$; $0.111$ ps (Earth $\ell\!=\!2$)} &
\parbox[t]{1.85cm}{\raggedright $54.8912~\mu$s/d} &
\parbox[t]{5.7cm}{\raggedright $0.473~\mu$s (monthly $P_{\tt EM}$); $\sim$21 ns from $-(\mathbf v_{\tt EM}\!\cdot\!\vec{\cal X})/c^2$; $2.28$ ps ($J_2$)} \\

\parbox[t]{2.2cm}{\raggedright ELFO (30 h; $e\!=\!0.6917$)} &
\parbox[t]{2.2cm}{\raggedright $7.2372\times10^{-12}$} &
\parbox[t]{5.2cm}{\raggedright $0.115~\mu$s, $0.040~\mu$s, $0.018~\mu$s at $k\omega_{\tt ELFO}$ ($k\!=\!1,2,3$);\; $149$ ps, $64$ ps, $10$ ps (Earth $\ell\!=\!2$);\; $\sim\!1.1$ ps ($J_{2\tt M}$)} &
\parbox[t]{1.85cm}{\raggedright $58.1152~\mu$s/d} &
\parbox[t]{5.7cm}{\raggedright $0.473~\mu$s (monthly $P_{\tt EM}$);\; $\sim\!0.1$–$0.2~\mu$s from $-(\vec v_{\tt EM}\cdot \vec {\cal X})/c^{2}$;\; $0.115~\mu$s ({\tt LCRS})} \\

\parbox[t]{2.2cm}{\raggedright Earth–Moon L1} &
\parbox[t]{2.2cm}{\raggedright $1.3827\times10^{-12}$} &
\parbox[t]{5.2cm}{\raggedright $25.3$ ns (monthly, kinematic+monopole); $5.42$ ns (Earth $\ell\!=\!2$); $0.253$ ns (solar $\ell\!=\!2$)} &
\parbox[t]{1.85cm}{\raggedright $58.6182~\mu$s/d} &
\parbox[t]{5.7cm}{\raggedright $0.473~\mu$s (monthly $P_{\tt EM}$); $\lesssim$\,36 ns from $-(\mathbf v_{\tt EM}\!\cdot\!\vec{\cal X})/c^2$ (perpendicular geometry suppresses to $\sim e_M$); $25.3$ ns ({\tt LCRS})} \\

\parbox[t]{2.2cm}{\raggedright NRHO (7.49 d; $e\!=\!0.9088$)} &
\parbox[t]{2.2cm}{\raggedright $2.2537\times10^{-12}$} &
\parbox[t]{5.2cm}{\raggedright $0.137~\mu$s, $0.062~\mu$s, $0.038~\mu$s at $k\omega_{\tt NRHO}$ ($k=1,2,3$); $\sim$20 ns (Earth $\ell\!=\!2$)} &
\parbox[t]{1.85cm}{\raggedright $58.5431~\mu$s/d} &
\parbox[t]{5.7cm}{\raggedright $0.473~\mu$s (monthly $P_{\tt EM}$); up to $0.81~\mu$s from $-(\mathbf v_{\tt EM}\!\cdot\!\vec{\cal X})/c^2$ (apoapsis-aligned); $0.137~\mu$s ({\tt LCRS})} \\
\hline
\end{tabular}
\begin{flushleft}
\vspace{-15pt}
\footnotesize \footnotetext{For 200 km LLO: $L_{\tt CL}=4.2223\times10^{-11}$ (drift $55.0897~\mu$s/d); dominant {\tt LCRS} lines are $2.10$ ps ($J_2$) and $0.135$ ps (Earth $\ell\!=\!2$).}
\end{flushleft}
\end{table}

\begin{table}[h]
\centering
\setlength{\tabcolsep}{3pt}        
\renewcommand{\arraystretch}{1.0} 
\caption{Model retention by regime (terms kept explicitly to meet the $5\times10^{-18}$ rate / 0.1 ps timing thresholds).
If $c^{-2}\Delta U$ from omitted harmonics exceeds the bound anywhere along track, raise $\ell_{\max}$ per regime.}
\label{tab:retention}
\begin{tabular}{lll}
\hline
\parbox[t]{2.7cm}{\raggedright Regime} &
\parbox[t]{6.5cm}{\raggedright Lunar field kept} &
\parbox[t]{6.5cm}{\raggedright External tides kept} \\
\hline

\parbox[t]{2.7cm}{\raggedright vLLO (10 km)} &
\parbox[t]{6.5cm}{\raggedright High-degree selenopotential; operationally $\ell_{\max}\!\gtrsim\!300$} &
\parbox[t]{6.5cm}{\raggedright Earth $\ell\!=\!2$ (dominant; $\sim$0.09--0.10 ps one-way), solar $\ell\!=\!2$; higher tides negligible} \\

\parbox[t]{2.7cm}{\raggedright LLO (100--200 km)} &
\parbox[t]{6.5cm}{\raggedright At least through degree $\ell =8$; $J_{2\tt M}$, $C_{22}$, $S_{22}$ dominate $P_{\tt CL}(t)$} &
\parbox[t]{6.5cm}{\raggedright Earth $\ell\!=\!2$ at $0.11$--$0.14$ ps; solar $\ell\!=\!2$ sub-ps} \\

\parbox[t]{2.7cm}{\raggedright ELFO (30 h)} &
\parbox[t]{6.5cm}{\raggedright $J_{2\tt M}$ at $\sim$ps, $(C_{22},S_{22})$ sidebands at $\sim$0.1 ps} &
\parbox[t]{6.5cm}{\raggedright Earth $\ell\!=\!2$ at $\{149,64,10\}$ ps; solar $\ell\!=\!2$ at $\{0.84,0.36,0.06\}$ ps} \\

\parbox[t]{2.7cm}{\raggedright L1} &
\parbox[t]{6.5cm}{\raggedright No lunar harmonics; monthly (K+M) $25.3$\,ns; Earth $\ell\!=\!2$ $5.42$\,ns; solar $\ell\!=\!2$ $0.253$ ns} &
\parbox[t]{6.5cm}{\raggedright Earth $\ell\!=\!2$--$7$ retained ($\ell\!=\!8$ $<5\times10^{-18}$); solar $\ell\!=\!2$} \\

\parbox[t]{2.7cm}{\raggedright NRHO (7.49 d)} &
\parbox[t]{6.5cm}{\raggedright (K+M) lines at $\{0.137,0.062,0.038\}\,\mu$s; weak $J_{2\tt M}$ sidebands} &
\parbox[t]{6.5cm}{\raggedright Earth $\ell\!=\!2$--$8$ retained (quadrupole dominates at $\sim$20 ns); solar $\ell\!=\!2$ sub-ns} \\
\hline
\end{tabular}
\begin{flushleft}
\vspace{-5pt}\footnotesize \emph{Notes:} K+M $=$ kinematic $+$ monopole monthly terms, see Secs.~\ref{sec:LCRS-vLLO}-\ref{sec:LCRS-NRHO} for derivations. ELFO amplitudes refer to $\{\omega,2\omega,3\omega\}$ lines. 
\end{flushleft}
\end{table}

\subsection{One-Way and Two-Way Light-Time for Earth--Moon Links}
\label{sec:lighttime}

We model the coordinate light-time in the BCRS between an emitter at $(t_1,x_1)$ and a receiver at $(t_2,x_2)$ by
\begin{equation}
\Delta t_{1\to2} \equiv t_2 - t_1 = \frac{R_{12}}{c} + \sum_{\tt B\in\{S,E,M\}} \Delta^{\rm Sh}_B + \Delta^{\rm Sag}_{(1)} + \mathcal{O}(c^{-4}),
\label{eq:lt_oneway}
\end{equation}
where $R_{12}=\|x_2-x_1\|$ is evaluated at the appropriate emission/receive times. The post‑Newtonian Shapiro delay for body $\tt B$ is
\begin{equation}
\Delta^{\rm Sh}_{\tt B} = \frac{2GM_{\tt B}}{c^3}
\ln\!\left(\frac{r_{\tt 1B}+r_{\tt 2B}+R_{12}}{r_{\tt 1B}+r_{\tt 2B}-R_{12}}\right),
\quad r_{i\tt B}=\|x_i-x_{\tt B}\|,
\label{eq:shapiro}
\end{equation}
For Earth–Moon links the Shapiro magnitudes are small but non‑negligible at our target precision: 
$\sim$20–30\,ns (Sun), $\sim$0.1–0.2\,ns (Earth), and $\sim$1–3\,ps (Moon), so each body’s \eqref{eq:shapiro} term is retained in the one‑way model \eqref{eq:lt_oneway}.

When the ground station is Earth‑fixed, the first‑order Sagnac term due to Earth's rotation is
\begin{equation}
\Delta^{\rm Sag}_{(1)} = -\frac{\boldsymbol{\Omega}_\oplus}{c^2}\cdot \left( r_2 \times r_1 \right)_{\tt GCRS},
\label{eq:sagnac}
\end{equation}
with $\boldsymbol{\Omega}_\oplus$ the Earth's rotation vector and $r_{1,2}$ the GCRS station vectors at their event times. 
Equation~\eqref{eq:lt_oneway} is the recommended one‑way model consistent with the $\le 0.1$\,ps goals and with IERS conventions; 
second‑order Sagnac and atmospheric terms may be added for specific ground realizations.

For a two‑way measurement with transmit at $t_1$ from Earth, reflection or transpond at $(t_2,x_2)$ near the Moon, 
and receive back at $t_3$ on Earth, the round‑trip light‑time (neglecting hardware delays) is
\begin{equation}
\rho \equiv t_3 - t_1 = \Delta t_{1\to2} + \Delta t_{2\to3},
\end{equation}
with $\Delta t_{2\to3}$ given by Eq.~\eqref{eq:lt_oneway} with the roles of $(1,2)$ replaced by $(2,3)$. 
Iterative solution proceeds by predicting $t_2$ from straight‑line light‑time, evaluating $\Delta^{\rm Sh}_B$ and $\Delta^{\rm Sag}_{(1)}$, and iterating until $|\delta t|<10^{-13}$\,s. This model should be used in conjunction with the proper‑to‑coordinate time transformations of Secs.~\ref{sec:imp}---\ref{sec:pt-TT-tau}. 
(Operational recipes are in \cite{Moyer:2003,Petit-Luzum:2010}.)

\section{Conclusions and Recommendations}
\label{sec:conc}

In this work we have considered high-precision relativistic time scales for cis-lunar navigation. In Section~\ref{sec:time-E} we reviewed the IAU post-Newtonian time scales for the Earth system and quantified all terms down to a fractional level of $5\times10^{-18}$ and timing precision of $0.1$ ps. Section~\ref{sec:time-M} implemented, in analytic form, the {\tt LCRS} and  {\tt TCL} prescribed by IAU Resolution II (2024) \cite{IAU:2024-Resol2}, extending the IAU {\tt BCRS}/{\tt GCRS} conventions by carrying the Moon's gravity field to degree $\ell=9$ (with Love-number variations), Earth tides to degree $\ell=8$, and inertial effects to the octupole. The resulting metric and coordinate mappings (\ref{eq:G00tr-MM})--(\ref{eq:Gabtr-MM}) and (\ref{eq:coord-time-TL})--(\ref{eq:coord-space-TL}) thus capture every secular and periodic effect of practical significance for cis-lunar timing and navigation at the adopted accuracy levels.

Based on the analysis performed in Section~\ref{sec:tm-tdb}, we note that, although the analogy with $L_{\tt G}$ suggests defining the lunar constant $L_{\tt L}$ in terms of a fixed selenopotential, in practice such a definition would be difficult to realize. Near-term lunar infrastructure will likely support only one or two primary clocks, located at specific sites (e.g., near the South Pole), with no global network to average over the selenoid. This makes it infeasible to maintain $L_{\tt L}$ with the same realization fidelity as $L_{\tt G}$, which benefits from decades of Earth-based clock data. 

Our treatment is consistent with previous analyses of lunar time scales and reference frames, while focusing on a different level of detail and scope. Thus, \cite{Kopeikin-Kaplan-2024} derived the {\tt TCL}--{\tt TCG} transformation for surface clocks
within the IAU 2000 framework, Refs.  \cite{Bourgoin-Defraigne-Meynadier-2025} and \cite{Defraigne-Meynadier-Bourgoin-2025} examined candidate lunar reference timescales and their
traceability to UTC, and  \cite{Lu-Yang-Xie-2025} provided a numerical ephemeris (LTE440) implementing the {\tt TCL} definition. In addition, \cite{Fienga-Rambaux-Sosnica-2024,Sosnica-etal-2025-ILRF} studied the definition and realization of lunar reference frames. The present work complements these by providing a unified analytic post-Newtonian realization of the {\tt BCRS}, {\tt GCRS}, and {\tt LCRS}, together with orbit-dependent error budgets that specify which terms must be retained in practice to achieve sub-picosecond timing throughout the Earth--Moon system.

We therefore, analogous to the IAU decision for \(L_{\tt G}\) in the {\tt GCRS}, recommend treating $L_{\tt L}$ as a \emph{conventional rate constant} fixed at a suitable reference value for consistency of the {\tt TL} scale, but without tying it rigidly to a fully defined selenopotential. Its operational realization should be based on the best available gravity model for the chosen reference site(s), while acknowledging that the realized potential may differ from the idealized selenoid by amounts exceeding the $5\times10^{-18}$ threshold. This approach preserves interoperability in time-scale transformations while avoiding an unachievable geodetic definition in the early phases of lunar timekeeping.

In Section~\ref{sec:imp}  we derived closed-form, analytic transformations among the six time scales of interest—{\tt TCB}, {\tt TCG}, {\tt TT}, {\tt TDB}, {\tt TCL} and {\tt TL}—truncating each series at the level dictated by modern clock and ranging stability.  In particular, we have obtained the proper‑time, $\tau$, relations (\ref{eq:synch-M+F2Y}), (\ref{eq:tau-TT=0}) that link any cis‑lunar clock to {\tt TT} through a secular drift rate and a well‑characterized set of periodic corrections.  By evaluating these expressions for four representative regimes—a 10\,km very‑low lunar orbit, a conventional low lunar orbit, the Earth–Moon L1 Lagrange point, and a near‑rectilinear halo orbit—we have demonstrated sub‑picosecond synchronization capability throughout the lunar environment. In Section~\ref{sec:lighttime} we provided an explicit one- and two-way light-time model (Shapiro and first-order Sagnac) consistent with the stated thresholds.

In Section~\ref{sec:pt-TT-tau} we evaluated those formulas in four representative regimes: a 10 km very‐low lunar orbit (vLLO), the Earth–Moon L$_1$ point, and a near‐rectilinear halo orbit (NRHO).  
Our analysis yields the secular drift rates of surface and orbiting clocks relative to terrestrial {\tt TT} with better than $5\times10^{-18}$ fractional accuracy.  For a clock on the lunar surface, the net $(\tau-{\tt TT})$ rate offset is $56.0256\,\mu\mathrm{s/d}$; for a 10 km polar orbit it is $54.6926\,\mu\mathrm{s/d}$; at L1 it is $58.6182\,\mu\mathrm{s/d}$; and in NRHO it reaches $58.5431\,\mu\mathrm{s/d}$.  The associated periodic excursions—driven by orbital eccentricity, Earth tides and solar quadrupole tides—remain below $0.1\,$ps for low orbits and below a few nanoseconds for deep cis‑lunar trajectories, in accordance with our accuracy goals.

Implementing this unified framework in both onboard and ground‐segment software will enable sub‑picosecond clock synchronization and centimeter‑level positioning across cislunar space.  We recommend that future lunar navigation architectures adopt the {\tt LCRS} as a defining standard, fix the lunar rate constant $L_{\tt L}$ by convention as was done for $L_{\tt G}$, and include spherical‑harmonic truncation through $\ell=9$ along with tidal orders through $\ell=8$.  As clock and ranging technology advance, further refinements can be made by treating orbital elements as time‐dependent and by combining high‐fidelity numerical propagation with spectral analysis to capture any residual periodic structure.

The unified post‐Newtonian framework presented here provides a single, self‐consistent basis for next‐generation lunar positioning, navigation and timing (PNT) services, quantum time‐transfer links and precision tests of gravity beyond low Earth orbit.  Its adoption will enable reliable cislunar operations, secure communication networks and fundamental‐physics experiments throughout the Earth–Moon system.

\begin{acknowledgments}

The work described here, in part, was carried out at the Jet Propulsion Laboratory, California Institute of Technology, under a contract with the National Aeronautics and Space Administration. 

\end{acknowledgments}


\begin{thebibliography}{59}
\expandafter\ifx\csname natexlab\endcsname\relax\def\natexlab#1{#1}\fi
\expandafter\ifx\csname bibnamefont\endcsname\relax
  \def\bibnamefont#1{#1}\fi
\expandafter\ifx\csname bibfnamefont\endcsname\relax
  \def\bibfnamefont#1{#1}\fi
\expandafter\ifx\csname citenamefont\endcsname\relax
  \def\citenamefont#1{#1}\fi
\expandafter\ifx\csname url\endcsname\relax
  \def\url#1{\texttt{#1}}\fi
\expandafter\ifx\csname urlprefix\endcsname\relax\def\urlprefix{URL }\fi
\providecommand{\bibinfo}[2]{#2}
\providecommand{\eprint}[2][]{\url{#2}}

\bibitem[{\citenamefont{Soffel et~al.}(2003)\citenamefont{Soffel, Klioner,
  Petit, Kopeikin, Bretagnon, Brumberg, Capitaine, Damour, Fukushima, Guinot
  et~al.}}]{Soffel-etal:2003}
\bibinfo{author}{\bibfnamefont{M.}~\bibnamefont{Soffel}},
  \bibinfo{author}{\bibfnamefont{S.~A.} \bibnamefont{Klioner}},
  \bibinfo{author}{\bibfnamefont{G.}~\bibnamefont{Petit}},
  \bibinfo{author}{\bibfnamefont{S.~M.} \bibnamefont{Kopeikin}},
  \bibinfo{author}{\bibfnamefont{P.}~\bibnamefont{Bretagnon}},
  \bibinfo{author}{\bibfnamefont{V.~A.} \bibnamefont{Brumberg}},
  \bibinfo{author}{\bibfnamefont{N.}~\bibnamefont{Capitaine}},
  \bibinfo{author}{\bibfnamefont{T.}~\bibnamefont{Damour}},
  \bibinfo{author}{\bibfnamefont{T.}~\bibnamefont{Fukushima}},
  \bibinfo{author}{\bibfnamefont{B.}~\bibnamefont{Guinot}},
  \bibnamefont{et~al.}, \bibinfo{journal}{Astron. J.}
  \textbf{\bibinfo{volume}{126}}, \bibinfo{pages}{2687} (\bibinfo{year}{2003}).

\bibitem[{\citenamefont{{Kaplan}}(2005)}]{Kaplan:2005}
\bibinfo{author}{\bibfnamefont{G.~H.} \bibnamefont{{Kaplan}}},
  \bibinfo{journal}{U.S. Naval Observatory Circulars}
  \textbf{\bibinfo{volume}{179}} (\bibinfo{year}{2005}),
  \eprint{astro-ph/0602086}.

\bibitem[{\citenamefont{{International Astronomical Union, Commission A3,
  Fundamental Standards}}(2024)}]{IAU:2024-Resol2}
\bibinfo{author}{\bibnamefont{{International Astronomical Union, Commission A3,
  Fundamental Standards}}}, \emph{\bibinfo{title}{{Resolution II: to establish
  a standard Lunar Celestial Reference System (LCRS) and Lunar Coordinate Time
  (TCL)}}}, \bibinfo{howpublished}{Resolutions for consideration by the XXXII
  General Assembly} (\bibinfo{year}{2024}), \bibinfo{note}{adopted 15 April
  2024},
  \urlprefix\url{https://iauarchive.eso.org/static/resolutions/IAU2024_Resol2_English.pdf}.

\bibitem[{\citenamefont{Kopeikin and Kaplan}(2024)}]{Kopeikin-Kaplan-2024}
\bibinfo{author}{\bibfnamefont{S.~M.} \bibnamefont{Kopeikin}} \bibnamefont{and}
  \bibinfo{author}{\bibfnamefont{G.~H.} \bibnamefont{Kaplan}},
  \bibinfo{journal}{Physical Review D} \textbf{\bibinfo{volume}{110}},
  \bibinfo{pages}{084047} (\bibinfo{year}{2024}), \eprint{2407.04862}.

\bibitem[{\citenamefont{Bourgoin et~al.}(2025)\citenamefont{Bourgoin,
  Defraigne, and Meynadier}}]{Bourgoin-Defraigne-Meynadier-2025}
\bibinfo{author}{\bibfnamefont{A.}~\bibnamefont{Bourgoin}},
  \bibinfo{author}{\bibfnamefont{P.}~\bibnamefont{Defraigne}},
  \bibnamefont{and}
  \bibinfo{author}{\bibfnamefont{F.}~\bibnamefont{Meynadier}},
  \bibinfo{journal}{arXiv e-prints}  (\bibinfo{year}{2025}),
  \bibinfo{note}{arXiv:2507.21597}, \eprint{2507.21597}.

\bibitem[{\citenamefont{Defraigne et~al.}(2025)\citenamefont{Defraigne,
  Meynadier, and Bourgoin}}]{Defraigne-Meynadier-Bourgoin-2025}
\bibinfo{author}{\bibfnamefont{P.}~\bibnamefont{Defraigne}},
  \bibinfo{author}{\bibfnamefont{F.}~\bibnamefont{Meynadier}},
  \bibnamefont{and} \bibinfo{author}{\bibfnamefont{A.}~\bibnamefont{Bourgoin}},
  \bibinfo{journal}{arXiv e-prints}  (\bibinfo{year}{2025}),
  \bibinfo{note}{arXiv:2511.02709}, \eprint{2511.02709}.

\bibitem[{\citenamefont{Lu et~al.}(2025)\citenamefont{Lu, Yang, and
  Xie}}]{Lu-Yang-Xie-2025}
\bibinfo{author}{\bibfnamefont{X.}~\bibnamefont{Lu}},
  \bibinfo{author}{\bibfnamefont{T.-N.} \bibnamefont{Yang}}, \bibnamefont{and}
  \bibinfo{author}{\bibfnamefont{Y.}~\bibnamefont{Xie}},
  \bibinfo{journal}{Astronomy and Astrophysics}  (\bibinfo{year}{2025}),
  \bibinfo{note}{in press}, \eprint{2509.18511}.

\bibitem[{\citenamefont{Fienga et~al.}(2024)\citenamefont{Fienga, Rambaux, and
  So\'snica}}]{Fienga-Rambaux-Sosnica-2024}
\bibinfo{author}{\bibfnamefont{A.}~\bibnamefont{Fienga}},
  \bibinfo{author}{\bibfnamefont{N.}~\bibnamefont{Rambaux}}, \bibnamefont{and}
  \bibinfo{author}{\bibfnamefont{K.}~\bibnamefont{So\'snica}},
  \bibinfo{journal}{arXiv e-prints}  (\bibinfo{year}{2024}),
  \bibinfo{note}{appendix of the ATLAS Concept document, ESA contract
  AO/1-10712/21/NL/CRS}, \eprint{2409.10043}.

\bibitem[{\citenamefont{So\'snica et~al.}(2025)\citenamefont{So\'snica, Fienga,
  Pavlov, Rambaux, and Zajdel}}]{Sosnica-etal-2025-ILRF}
\bibinfo{author}{\bibfnamefont{K.}~\bibnamefont{So\'snica}},
  \bibinfo{author}{\bibfnamefont{A.}~\bibnamefont{Fienga}},
  \bibinfo{author}{\bibfnamefont{D.}~\bibnamefont{Pavlov}},
  \bibinfo{author}{\bibfnamefont{N.}~\bibnamefont{Rambaux}}, \bibnamefont{and}
  \bibinfo{author}{\bibfnamefont{R.}~\bibnamefont{Zajdel}},
  \bibinfo{journal}{arXiv e-prints}  (\bibinfo{year}{2025}),
  \bibinfo{note}{arXiv:2510.15484}, \eprint{2510.15484}.

\bibitem[{\citenamefont{{Turyshev} et~al.}(2025)\citenamefont{{Turyshev},
  {Williams}, {Boggs}, and {Park}}}]{Turyshev:2025}
\bibinfo{author}{\bibfnamefont{S.~G.} \bibnamefont{{Turyshev}}},
  \bibinfo{author}{\bibfnamefont{J.~G.} \bibnamefont{{Williams}}},
  \bibinfo{author}{\bibfnamefont{D.~H.} \bibnamefont{{Boggs}}},
  \bibnamefont{and} \bibinfo{author}{\bibfnamefont{R.~S.}
  \bibnamefont{{Park}}}, \bibinfo{journal}{Astrophys. J.}
  \textbf{\bibinfo{volume}{985}}, \bibinfo{pages}{140} (\bibinfo{year}{2025}),
  \bibinfo{note}{arXiv:2406.16147 [astro-ph.EP]}.

\bibitem[{\citenamefont{Landau and Lifshitz}(1988)}]{Landau-Lifshitz:1988}
\bibinfo{author}{\bibfnamefont{L.~D.} \bibnamefont{Landau}} \bibnamefont{and}
  \bibinfo{author}{\bibfnamefont{E.~M.} \bibnamefont{Lifshitz}},
  \emph{\bibinfo{title}{{The Classical Theory of Fields {\rm (in Russian)}}}}
  (\bibinfo{publisher}{Nauka, Moscow}, \bibinfo{year}{1988}),
  \bibinfo{edition}{7th} ed.

\bibitem[{\citenamefont{Moyer}(2003)}]{Moyer:2003}
\bibinfo{author}{\bibfnamefont{T.~D.} \bibnamefont{Moyer}},
  \emph{\bibinfo{title}{{Formulation for Observed and Computed Values of Deep
  Space Network Data Types for Navigation}}}, JPL Deep-Space Communications and
  Navigation Series (\bibinfo{publisher}{Wiley-Interscience},
  \bibinfo{year}{2003}).

\bibitem[{\citenamefont{{Guinot}}(1992)}]{Guinot:1992}
\bibinfo{author}{\bibfnamefont{B.}~\bibnamefont{{Guinot}}}, in
  \emph{\bibinfo{booktitle}{Journ\'ees 1992: Syst\'emes de r\'ef\'erence
  spatio-temporels}} (\bibinfo{year}{1992}), pp. \bibinfo{pages}{12--21}.

\bibitem[{\citenamefont{{Turyshev} and
  {Toth}}(2023)}]{Turyshev-Toth:2023-grav-phase}
\bibinfo{author}{\bibfnamefont{S.~G.} \bibnamefont{{Turyshev}}}
  \bibnamefont{and} \bibinfo{author}{\bibfnamefont{V.~T.}
  \bibnamefont{{Toth}}}, \bibinfo{journal}{Phys. Rev. D}
  \textbf{\bibinfo{volume}{107}}, \bibinfo{eid}{104031} (\bibinfo{year}{2023}),
  \eprint{arXiv:2303.07270 [gr-qc]}.

\bibitem[{\citenamefont{{Groten}}(2004)}]{Groten:2004}
\bibinfo{author}{\bibfnamefont{E.}~\bibnamefont{{Groten}}},
  \bibinfo{journal}{J. of Geodesy} \textbf{\bibinfo{volume}{77}},
  \bibinfo{pages}{724} (\bibinfo{year}{2004}).

\bibitem[{\citenamefont{{Petit} and {Luzum (eds.)}}(2010)}]{Petit-Luzum:2010}
\bibinfo{author}{\bibfnamefont{G.}~\bibnamefont{{Petit}}} \bibnamefont{and}
  \bibinfo{author}{\bibfnamefont{B.}~\bibnamefont{{Luzum (eds.)}}},
  \emph{\bibinfo{title}{IERS Technical Note \#36}} (\bibinfo{publisher}{``IERS
  Conventions (2010)'', Frankfurt am Main: Verlag des Bundesamts f{\"u}r
  Kartographie und Geod{\"a}sie, 2010. 179 pp., ISBN 3-89888-989-6''},
  \bibinfo{year}{2010}),
  \urlprefix\url{http://www.iers.org/IERS/EN/Publications/TechnicalNotes/tn36.html}.

\bibitem[{\citenamefont{{Philipp} et~al.}(2020)\citenamefont{{Philipp},
  {Hackmann}, {L{\"a}mmerzahl}, and {M{\"u}ller}}}]{Philipp:2020}
\bibinfo{author}{\bibfnamefont{D.}~\bibnamefont{{Philipp}}},
  \bibinfo{author}{\bibfnamefont{E.}~\bibnamefont{{Hackmann}}},
  \bibinfo{author}{\bibfnamefont{C.}~\bibnamefont{{L{\"a}mmerzahl}}},
  \bibnamefont{and}
  \bibinfo{author}{\bibfnamefont{J.}~\bibnamefont{{M{\"u}ller}}},
  \bibinfo{journal}{\prd} \textbf{\bibinfo{volume}{101}}, \bibinfo{eid}{064032}
  (\bibinfo{year}{2020}).

\bibitem[{\citenamefont{{Luzum} et~al.}(2011)\citenamefont{{Luzum},
  {Capitaine}, {Fienga}, {Folkner}, {Fukushima}, {Hilton}, {Hohenkerk},
  {Krasinsky}, {Petit}, {Pitjeva} et~al.}}]{Luzum-etal:2011}
\bibinfo{author}{\bibfnamefont{B.}~\bibnamefont{{Luzum}}},
  \bibinfo{author}{\bibfnamefont{N.}~\bibnamefont{{Capitaine}}},
  \bibinfo{author}{\bibfnamefont{A.}~\bibnamefont{{Fienga}}},
  \bibinfo{author}{\bibfnamefont{W.}~\bibnamefont{{Folkner}}},
  \bibinfo{author}{\bibfnamefont{T.}~\bibnamefont{{Fukushima}}},
  \bibinfo{author}{\bibfnamefont{J.}~\bibnamefont{{Hilton}}},
  \bibinfo{author}{\bibfnamefont{C.}~\bibnamefont{{Hohenkerk}}},
  \bibinfo{author}{\bibfnamefont{G.}~\bibnamefont{{Krasinsky}}},
  \bibinfo{author}{\bibfnamefont{G.}~\bibnamefont{{Petit}}},
  \bibinfo{author}{\bibfnamefont{E.}~\bibnamefont{{Pitjeva}}},
  \bibnamefont{et~al.}, \bibinfo{journal}{Cel. Mech. Dyn. Astron.}
  \textbf{\bibinfo{volume}{110}}, \bibinfo{pages}{293} (\bibinfo{year}{2011}).

\bibitem[{\citenamefont{{Klioner}}(2008)}]{Klioner:2008}
\bibinfo{author}{\bibfnamefont{S.~A.} \bibnamefont{{Klioner}}},
  \bibinfo{journal}{Astron. Astrophys.} \textbf{\bibinfo{volume}{478}},
  \bibinfo{pages}{951} (\bibinfo{year}{2008}).

\bibitem[{\citenamefont{{Brumberg} and {Groten}}(2001)}]{Brumberg-Groten:2001}
\bibinfo{author}{\bibfnamefont{V.~A.} \bibnamefont{{Brumberg}}}
  \bibnamefont{and} \bibinfo{author}{\bibfnamefont{E.}~\bibnamefont{{Groten}}},
  \bibinfo{journal}{Astron. Astrophys.} \textbf{\bibinfo{volume}{367}},
  \bibinfo{pages}{1070} (\bibinfo{year}{2001}).

\bibitem[{\citenamefont{{Fukushima}}(1995)}]{Fukushima:1995}
\bibinfo{author}{\bibfnamefont{T.}~\bibnamefont{{Fukushima}}},
  \bibinfo{journal}{Astron. Astrophys.} \textbf{\bibinfo{volume}{294}},
  \bibinfo{pages}{895} (\bibinfo{year}{1995}).

\bibitem[{\citenamefont{{Irwin} and {Fukushima}}(1999)}]{Irwin-Fukushima:1999}
\bibinfo{author}{\bibfnamefont{A.~W.} \bibnamefont{{Irwin}}} \bibnamefont{and}
  \bibinfo{author}{\bibfnamefont{T.}~\bibnamefont{{Fukushima}}},
  \bibinfo{journal}{Astron. Astrophys.} \textbf{\bibinfo{volume}{348}},
  \bibinfo{pages}{642} (\bibinfo{year}{1999}).

\bibitem[{\citenamefont{{Fairhead} and
  {Bretagnon}}(1990)}]{Fairhead-Bretagnon:1990}
\bibinfo{author}{\bibfnamefont{L.}~\bibnamefont{{Fairhead}}} \bibnamefont{and}
  \bibinfo{author}{\bibfnamefont{P.}~\bibnamefont{{Bretagnon}}},
  \bibinfo{journal}{Astron. Astrophys.} \textbf{\bibinfo{volume}{229}},
  \bibinfo{pages}{240} (\bibinfo{year}{1990}).

\bibitem[{\citenamefont{{Harada} and
  {Fukushima}}(2003)}]{Harada-Fukushima:2003}
\bibinfo{author}{\bibfnamefont{W.}~\bibnamefont{{Harada}}} \bibnamefont{and}
  \bibinfo{author}{\bibfnamefont{T.}~\bibnamefont{{Fukushima}}},
  \bibinfo{journal}{Astron. J.} \textbf{\bibinfo{volume}{126}},
  \bibinfo{pages}{2557} (\bibinfo{year}{2003}).

\bibitem[{\citenamefont{IAU}(2009)}]{IAU2009ResB3}
\bibinfo{author}{\bibnamefont{IAU}}, \emph{\bibinfo{title}{Resolution B.3}}
  (\bibinfo{publisher}{Cambridge University Press}, \bibinfo{address}{New
  York}, \bibinfo{year}{2009}), vol. \bibinfo{volume}{26B} of
  \emph{\bibinfo{series}{Transactions of the IAU, Series B}},
  p.~\bibinfo{pages}{42}, \bibinfo{note}{{Proc. of the 26th General Assembly
  (Prague, Czech Republic, 2006)}}.

\bibitem[{\citenamefont{Standish}(1998)}]{Standish:DE405-1998}
\bibinfo{author}{\bibfnamefont{E.~M.} \bibnamefont{Standish}},
  \bibinfo{type}{Interoffice Memorandum} \bibinfo{number}{JPL
  IOM 312.F‑98‑048}, \bibinfo{institution}{Jet Propulsion Laboratory,
  California Institute of Technology} (\bibinfo{year}{1998}),
  \urlprefix\url{ftp://ssd.jpl.nasa.gov/pub/eph/planets/ioms/de405.iom.pdf}.

\bibitem[{\citenamefont{{Park} et~al.}(2017)\citenamefont{{Park}, {Folkner},
  {Konopliv}, {Williams}, {Smith}, and {Zuber}}}]{Park:2017}
\bibinfo{author}{\bibfnamefont{R.~S.} \bibnamefont{{Park}}},
  \bibinfo{author}{\bibfnamefont{W.~M.} \bibnamefont{{Folkner}}},
  \bibinfo{author}{\bibfnamefont{A.~S.} \bibnamefont{{Konopliv}}},
  \bibinfo{author}{\bibfnamefont{J.~G.} \bibnamefont{{Williams}}},
  \bibinfo{author}{\bibfnamefont{D.~E.} \bibnamefont{{Smith}}},
  \bibnamefont{and} \bibinfo{author}{\bibfnamefont{M.~T.}
  \bibnamefont{{Zuber}}}, \bibinfo{journal}{ApJ}
  \textbf{\bibinfo{volume}{153}}, \bibinfo{eid}{121} (\bibinfo{year}{2017}).

\bibitem[{\citenamefont{Mecheri and Meftah}(2021)}]{MecheriMeftah:2021}
\bibinfo{author}{\bibfnamefont{R.}~\bibnamefont{Mecheri}} \bibnamefont{and}
  \bibinfo{author}{\bibfnamefont{M.}~\bibnamefont{Meftah}},
  \bibinfo{journal}{MNRAS} \textbf{\bibinfo{volume}{506}},
  \bibinfo{pages}{2671} (\bibinfo{year}{2021}).

\bibitem[{\citenamefont{{Fukushima}}(2010)}]{Fukushima:2010}
\bibinfo{author}{\bibfnamefont{T.}~\bibnamefont{{Fukushima}}}, in
  \emph{\bibinfo{booktitle}{Relativity in Fundamental Astronomy: Dynamics,
  Reference Frames, and Data Analysis}}, edited by
  \bibinfo{editor}{\bibfnamefont{S.~A.} \bibnamefont{{Klioner}}},
  \bibinfo{editor}{\bibfnamefont{P.~K.} \bibnamefont{{Seidelmann}}},
  \bibnamefont{and} \bibinfo{editor}{\bibfnamefont{M.~H.}
  \bibnamefont{{Soffel}}} (\bibinfo{year}{2010}), vol. \bibinfo{volume}{261},
  pp. \bibinfo{pages}{89--94}.

\bibitem[{\citenamefont{{Bursa} and {Sima}}(1980)}]{Bursa-Sima:1980}
\bibinfo{author}{\bibfnamefont{M.}~\bibnamefont{{Bursa}}} \bibnamefont{and}
  \bibinfo{author}{\bibfnamefont{Z.}~\bibnamefont{{Sima}}},
  \bibinfo{journal}{Studia Geophysica et Geodaetica}
  \textbf{\bibinfo{volume}{24}}, \bibinfo{pages}{211} (\bibinfo{year}{1980}).

\bibitem[{\citenamefont{{Martinec} and {Pec}}(1988)}]{Martinec-Pec:1988}
\bibinfo{author}{\bibfnamefont{Z.}~\bibnamefont{{Martinec}}} \bibnamefont{and}
  \bibinfo{author}{\bibfnamefont{K.}~\bibnamefont{{Pec}}},
  \bibinfo{journal}{Earth, Moon, and Planets} \textbf{\bibinfo{volume}{43}},
  \bibinfo{pages}{21} (\bibinfo{year}{1988}).

\bibitem[{\citenamefont{{Ferrari} et~al.}(1980)\citenamefont{{Ferrari},
  {Sinclair}, {Sjogren}, {Williams}, and {Yoder}}}]{Ferrari-etal:1980}
\bibinfo{author}{\bibfnamefont{A.~J.} \bibnamefont{{Ferrari}}},
  \bibinfo{author}{\bibfnamefont{W.~S.} \bibnamefont{{Sinclair}}},
  \bibinfo{author}{\bibfnamefont{W.~L.} \bibnamefont{{Sjogren}}},
  \bibinfo{author}{\bibfnamefont{J.~G.} \bibnamefont{{Williams}}},
  \bibnamefont{and} \bibinfo{author}{\bibfnamefont{C.~F.}
  \bibnamefont{{Yoder}}}, \bibinfo{journal}{J. Geophys. Res.}
  \textbf{\bibinfo{volume}{85}}, \bibinfo{pages}{3939} (\bibinfo{year}{1980}).

\bibitem[{\citenamefont{{Ardalan} and {Karimi}}(2014)}]{Ardalan-Karimi:2014}
\bibinfo{author}{\bibfnamefont{A.~A.} \bibnamefont{{Ardalan}}}
  \bibnamefont{and} \bibinfo{author}{\bibfnamefont{R.}~\bibnamefont{{Karimi}}},
  \bibinfo{journal}{Cel. Mech. Dyn. Astron.} \textbf{\bibinfo{volume}{118}},
  \bibinfo{pages}{75} (\bibinfo{year}{2014}).

\bibitem[{\citenamefont{{Smith} et~al.}(2017)\citenamefont{{Smith}, {Zuber},
  {Neumann}, {Mazarico}, {Lemoine}, {Head III}, {Lucey}, {Aharonson},
  {Robinson}, {Sun} et~al.}}]{Smith-etal:2017}
\bibinfo{author}{\bibfnamefont{D.~E.} \bibnamefont{{Smith}}},
  \bibinfo{author}{\bibfnamefont{M.~T.} \bibnamefont{{Zuber}}},
  \bibinfo{author}{\bibfnamefont{G.~A.} \bibnamefont{{Neumann}}},
  \bibinfo{author}{\bibfnamefont{E.}~\bibnamefont{{Mazarico}}},
  \bibinfo{author}{\bibfnamefont{F.~G.} \bibnamefont{{Lemoine}}},
  \bibinfo{author}{\bibfnamefont{J.~W.} \bibnamefont{{Head III}}},
  \bibinfo{author}{\bibfnamefont{P.~G.} \bibnamefont{{Lucey}}},
  \bibinfo{author}{\bibfnamefont{O.}~\bibnamefont{{Aharonson}}},
  \bibinfo{author}{\bibfnamefont{M.~S.} \bibnamefont{{Robinson}}},
  \bibinfo{author}{\bibfnamefont{X.}~\bibnamefont{{Sun}}},
  \bibnamefont{et~al.}, \bibinfo{journal}{Icarus}
  \textbf{\bibinfo{volume}{283}}, \bibinfo{pages}{70} (\bibinfo{year}{2017}).

\bibitem[{\citenamefont{{Park} et~al.}(2021)\citenamefont{{Park}, {Folkner},
  {Williams}, and {Boggs}}}]{Park-etal:2021}
\bibinfo{author}{\bibfnamefont{R.~S.} \bibnamefont{{Park}}},
  \bibinfo{author}{\bibfnamefont{W.~M.} \bibnamefont{{Folkner}}},
  \bibinfo{author}{\bibfnamefont{J.~G.} \bibnamefont{{Williams}}},
  \bibnamefont{and} \bibinfo{author}{\bibfnamefont{D.~H.}
  \bibnamefont{{Boggs}}}, \bibinfo{journal}{Astron. J.}
  \textbf{\bibinfo{volume}{161}}, \bibinfo{eid}{105} (\bibinfo{year}{2021}).

\bibitem[{\citenamefont{{Williams} et~al.}(2014)\citenamefont{{Williams},
  {Konopliv}, {Boggs}, {Park}, {Yuan}, {Lemoine}, {Goossens}, {Mazarico},
  {Nimmo}, {Weber} et~al.}}]{Williams-etal:2014}
\bibinfo{author}{\bibfnamefont{J.~G.} \bibnamefont{{Williams}}},
  \bibinfo{author}{\bibfnamefont{A.~S.} \bibnamefont{{Konopliv}}},
  \bibinfo{author}{\bibfnamefont{D.~H.} \bibnamefont{{Boggs}}},
  \bibinfo{author}{\bibfnamefont{R.~S.} \bibnamefont{{Park}}},
  \bibinfo{author}{\bibfnamefont{D.-N.} \bibnamefont{{Yuan}}},
  \bibinfo{author}{\bibfnamefont{F.~G.} \bibnamefont{{Lemoine}}},
  \bibinfo{author}{\bibfnamefont{S.}~\bibnamefont{{Goossens}}},
  \bibinfo{author}{\bibfnamefont{E.}~\bibnamefont{{Mazarico}}},
  \bibinfo{author}{\bibfnamefont{F.}~\bibnamefont{{Nimmo}}},
  \bibinfo{author}{\bibfnamefont{R.~C.} \bibnamefont{{Weber}}},
  \bibnamefont{et~al.}, \bibinfo{journal}{J. Geophys. Res. (Planets)}
  \textbf{\bibinfo{volume}{119}}, \bibinfo{pages}{1546} (\bibinfo{year}{2014}).

\bibitem[{\citenamefont{{Konopliv} et~al.}(2013)\citenamefont{{Konopliv},
  {Park}, {Yuan}, {Asmar}, {Watkins}, {Williams}, {Fahnestock}, {Kruizinga},
  {Paik}, {Strekalov} et~al.}}]{Konopliv-etal:2013}
\bibinfo{author}{\bibfnamefont{A.~S.} \bibnamefont{{Konopliv}}},
  \bibinfo{author}{\bibfnamefont{R.~S.} \bibnamefont{{Park}}},
  \bibinfo{author}{\bibfnamefont{D.-N.} \bibnamefont{{Yuan}}},
  \bibinfo{author}{\bibfnamefont{S.~W.} \bibnamefont{{Asmar}}},
  \bibinfo{author}{\bibfnamefont{M.~M.} \bibnamefont{{Watkins}}},
  \bibinfo{author}{\bibfnamefont{J.~G.} \bibnamefont{{Williams}}},
  \bibinfo{author}{\bibfnamefont{E.}~\bibnamefont{{Fahnestock}}},
  \bibinfo{author}{\bibfnamefont{G.}~\bibnamefont{{Kruizinga}}},
  \bibinfo{author}{\bibfnamefont{M.}~\bibnamefont{{Paik}}},
  \bibinfo{author}{\bibfnamefont{D.}~\bibnamefont{{Strekalov}}},
  \bibnamefont{et~al.}, \bibinfo{journal}{J. Geophys. Res. (Planets)}
  \textbf{\bibinfo{volume}{118}}, \bibinfo{pages}{1415} (\bibinfo{year}{2013}).

\bibitem[{\citenamefont{Ryden and
  Volle}(2025)}]{RydenVolle2025_LCRNS_RefConst3_1}
\bibinfo{author}{\bibfnamefont{G.}~\bibnamefont{Ryden}} \bibnamefont{and}
  \bibinfo{author}{\bibfnamefont{M.}~\bibnamefont{Volle}}, \bibinfo{type}{White
  paper} \bibinfo{number}{03.2025-1}, \bibinfo{institution}{NASA Goddard Space
  Flight Center (GSFC)}, \bibinfo{address}{Greenbelt, MD}
  (\bibinfo{year}{2025}), \bibinfo{note}{version 3.1; publicly available via
  NASA LCRNS website and NASA Technical Reports Server},
  \urlprefix\url{https://ntrs.nasa.gov/citations/20250002698}.

\bibitem[{\citenamefont{{Turyshev}}(1996)}]{Turyshev:1996}
\bibinfo{author}{\bibfnamefont{S.~G.} \bibnamefont{{Turyshev}}},
  \emph{\bibinfo{title}{{Relativistic Navigation: A Theoretical Foundation}}},
  \bibinfo{howpublished}{Technical Report, NASA/CR-96-112596; NAS 1.26:112596;
  JPL-Publ-96-13} (\bibinfo{year}{1996}), \eprint{gr-qc/9606063}.

\bibitem[{\citenamefont{Turyshev and Toth}(2015)}]{Turyshev-Toth:2013}
\bibinfo{author}{\bibfnamefont{S.~G.} \bibnamefont{Turyshev}} \bibnamefont{and}
  \bibinfo{author}{\bibfnamefont{V.~T.} \bibnamefont{Toth}},
  \bibinfo{journal}{Int. J. Mod. Phys. D} \textbf{\bibinfo{volume}{24}},
  \bibinfo{pages}{1550039} (\bibinfo{year}{2015}), \eprint{arXiv:1304.8122
  [gr-qc]}.

\bibitem[{\citenamefont{{Moyer}}(1981{\natexlab{a}})}]{Moyer:1981-part1}
\bibinfo{author}{\bibfnamefont{T.~D.} \bibnamefont{{Moyer}}},
  \bibinfo{journal}{Celestial Mechanics} \textbf{\bibinfo{volume}{23}},
  \bibinfo{pages}{33} (\bibinfo{year}{1981}{\natexlab{a}}).

\bibitem[{\citenamefont{{Moyer}}(1981{\natexlab{b}})}]{Moyer:1981-part2}
\bibinfo{author}{\bibfnamefont{T.~D.} \bibnamefont{{Moyer}}},
  \bibinfo{journal}{Celestial Mechanics} \textbf{\bibinfo{volume}{23}},
  \bibinfo{pages}{57} (\bibinfo{year}{1981}{\natexlab{b}}).

\bibitem[{\citenamefont{Standish and Williams}(2008)}]{Standish_Williams_2008}
\bibinfo{author}{\bibfnamefont{E.~M.} \bibnamefont{Standish}} \bibnamefont{and}
  \bibinfo{author}{\bibfnamefont{J.~G.} \bibnamefont{Williams}}, in
  \emph{\bibinfo{booktitle}{Explanatory Supplement to the American Ephemeris
  and Nautical Almanac}}, edited by \bibinfo{editor}{\bibfnamefont{P.~K.}
  \bibnamefont{Seidelmann}} (\bibinfo{publisher}{Mill Valley: University
  Science Books}, \bibinfo{year}{2008}), chap.~\bibinfo{chapter}{8}.

\bibitem[{\citenamefont{{Abramowitz} and
  {Stegun}}(1965)}]{Abramovitz-Stegun:1965}
\bibinfo{author}{\bibfnamefont{M.}~\bibnamefont{{Abramowitz}}}
  \bibnamefont{and} \bibinfo{author}{\bibfnamefont{I.~A.}
  \bibnamefont{{Stegun}}}, \emph{\bibinfo{title}{Handbook of Mathematical
  Functions: with Formulas, Graphs, and Mathematical Tables.}}
  (\bibinfo{publisher}{Dover Publications, New York; revised edition},
  \bibinfo{year}{1965}).

\bibitem[{\citenamefont{{Kozai}}(1966)}]{Kozai:1966}
\bibinfo{author}{\bibfnamefont{Y.}~\bibnamefont{{Kozai}}},
  \bibinfo{journal}{Space Sci. Rev.} \textbf{\bibinfo{volume}{5}},
  \bibinfo{pages}{818} (\bibinfo{year}{1966}).

\bibitem[{\citenamefont{{Gasposchkin}}(1977)}]{Gasposchkin:1977}
\bibinfo{author}{\bibfnamefont{E.}~\bibnamefont{{Gasposchkin}}},
  \bibinfo{journal}{Philos. Trans. Royal Soc. London. Ser. A, Math. Phys. Sci.}
  \textbf{\bibinfo{volume}{284}}, \bibinfo{pages}{515} (\bibinfo{year}{1977}).

\bibitem[{\citenamefont{{Lambeck}}(1988)}]{Lambeck:1988}
\bibinfo{author}{\bibfnamefont{K.}~\bibnamefont{{Lambeck}}},
  \emph{\bibinfo{title}{Geophysical Geodesy: The Slow Deformations of the
  Earth}} (\bibinfo{publisher}{Oxford University Press, New York},
  \bibinfo{year}{1988}).

\bibitem[{\citenamefont{{NASA/JPL, a project file
  depository}}(2001)}]{JPL-EarthGrav:2021}
\bibinfo{author}{\bibnamefont{{NASA/JPL, a project file depository}}}
  (\bibinfo{year}{2001}), \bibinfo{note}{{Spherical} Harmonic Representation of
  the Gravity Field Potential},
  \urlprefix\url{https://spsweb.fltops.jpl.nasa.gov/portaldataops/mpg/MPG_Docs/Source
  Docs/gravity-SphericalHarmonics.pdf}.

\bibitem[{\citenamefont{Tapley et~al.}(1996)\citenamefont{Tapley, Watkins,
  Ries, Davis, Eanes, Poole, Rim, Schutz, Shum, Nerem et~al.}}]{Tapley-1996}
\bibinfo{author}{\bibfnamefont{B.~D.} \bibnamefont{Tapley}},
  \bibinfo{author}{\bibfnamefont{M.~M.} \bibnamefont{Watkins}},
  \bibinfo{author}{\bibfnamefont{J.~C.} \bibnamefont{Ries}},
  \bibinfo{author}{\bibfnamefont{G.~W.} \bibnamefont{Davis}},
  \bibinfo{author}{\bibfnamefont{R.~J.} \bibnamefont{Eanes}},
  \bibinfo{author}{\bibfnamefont{S.~R.} \bibnamefont{Poole}},
  \bibinfo{author}{\bibfnamefont{H.~J.} \bibnamefont{Rim}},
  \bibinfo{author}{\bibfnamefont{B.~E.} \bibnamefont{Schutz}},
  \bibinfo{author}{\bibfnamefont{C.~K.} \bibnamefont{Shum}},
  \bibinfo{author}{\bibfnamefont{R.~S.} \bibnamefont{Nerem}},
  \bibnamefont{et~al.}, \bibinfo{journal}{J. Geophys. Res.: Solid Earth}
  \textbf{\bibinfo{volume}{101}}, \bibinfo{pages}{28029}
  (\bibinfo{year}{1996}).

\bibitem[{\citenamefont{Montenbruck and Gill}(2012)}]{Montenbruck-Gill:2012}
\bibinfo{author}{\bibfnamefont{O.}~\bibnamefont{Montenbruck}} \bibnamefont{and}
  \bibinfo{author}{\bibfnamefont{E.}~\bibnamefont{Gill}},
  \emph{\bibinfo{title}{{Satellite Orbits}}} (\bibinfo{publisher}{Springer
  Berlin, Heidelberg}, \bibinfo{year}{2012}).

\bibitem[{\citenamefont{Richter and Matzner}(1983)}]{Richter-Matzner:1983}
\bibinfo{author}{\bibfnamefont{G.~W.} \bibnamefont{Richter}} \bibnamefont{and}
  \bibinfo{author}{\bibfnamefont{R.~A.} \bibnamefont{Matzner}},
  \bibinfo{journal}{Phys. Rev. D} \textbf{\bibinfo{volume}{28}},
  \bibinfo{pages}{3007} (\bibinfo{year}{1983}).

\bibitem[{\citenamefont{{Cang} et~al.}(2016)\citenamefont{{Cang}, {Guo}, {Hu},
  and {He}}}]{Cang:2016}
\bibinfo{author}{\bibfnamefont{R.}~\bibnamefont{{Cang}}},
  \bibinfo{author}{\bibfnamefont{J.}~\bibnamefont{{Guo}}},
  \bibinfo{author}{\bibfnamefont{J.}~\bibnamefont{{Hu}}}, \bibnamefont{and}
  \bibinfo{author}{\bibfnamefont{C.}~\bibnamefont{{He}}},
  \bibinfo{journal}{Astron. \& Astrophys. (Hans Publishers}
  \textbf{\bibinfo{volume}{4}}, \bibinfo{pages}{33} (\bibinfo{year}{2016}).

\bibitem[{\citenamefont{{Ashby}}(2003)}]{Ashby-2003}
\bibinfo{author}{\bibfnamefont{N.}~\bibnamefont{{Ashby}}},
  \bibinfo{journal}{Living Rev. Relativity} \textbf{\bibinfo{volume}{6}}
  (\bibinfo{year}{2003}).

\bibitem[{\citenamefont{Turyshev et~al.}(2013)\citenamefont{Turyshev, Toth, and
  Sazhin}}]{Turyshev:2012nw}
\bibinfo{author}{\bibfnamefont{S.~G.} \bibnamefont{Turyshev}},
  \bibinfo{author}{\bibfnamefont{V.~T.} \bibnamefont{Toth}}, \bibnamefont{and}
  \bibinfo{author}{\bibfnamefont{M.~V.} \bibnamefont{Sazhin}},
  \bibinfo{journal}{Phys. Rev. D} \textbf{\bibinfo{volume}{87}},
  \bibinfo{pages}{024020} (\bibinfo{year}{2013}), \eprint{arXiv:1212.0232
  [gr-qc]}.

\bibitem[{\citenamefont{Williams et~al.}(2001)\citenamefont{Williams, Boggs,
  Yoder, Ratcliff, and Dickey}}]{Williams-etal:2001}
\bibinfo{author}{\bibfnamefont{J.~G.} \bibnamefont{Williams}},
  \bibinfo{author}{\bibfnamefont{D.~H.} \bibnamefont{Boggs}},
  \bibinfo{author}{\bibfnamefont{C.~F.} \bibnamefont{Yoder}},
  \bibinfo{author}{\bibfnamefont{J.~T.} \bibnamefont{Ratcliff}},
  \bibnamefont{and} \bibinfo{author}{\bibfnamefont{J.~O.}
  \bibnamefont{Dickey}}, \bibinfo{journal}{J. Geophys. Res.}
  \textbf{\bibinfo{volume}{106}}, \bibinfo{pages}{27933}
  (\bibinfo{year}{2001}).

\bibitem[{\citenamefont{Goossens and
  Matsumoto}(2008)}]{Goossens-Matsumoto:2008}
\bibinfo{author}{\bibfnamefont{S.}~\bibnamefont{Goossens}} \bibnamefont{and}
  \bibinfo{author}{\bibfnamefont{K.}~\bibnamefont{Matsumoto}},
  \bibinfo{journal}{Geophys. Res. Lett.} \textbf{\bibinfo{volume}{35}},
  \bibinfo{pages}{L02204} (\bibinfo{year}{2008}).

\bibitem[{\citenamefont{Konopliv et~al.}(1998)\citenamefont{Konopliv, Binder,
  Hood, Kucinskas, Sjogren, and Williams}}]{Konopliv1998}
\bibinfo{author}{\bibfnamefont{A.~S.} \bibnamefont{Konopliv}},
  \bibinfo{author}{\bibfnamefont{A.~L.} \bibnamefont{Binder}},
  \bibinfo{author}{\bibfnamefont{L.~L.} \bibnamefont{Hood}},
  \bibinfo{author}{\bibfnamefont{A.~B.} \bibnamefont{Kucinskas}},
  \bibinfo{author}{\bibfnamefont{W.~L.} \bibnamefont{Sjogren}},
  \bibnamefont{and} \bibinfo{author}{\bibfnamefont{J.~G.}
  \bibnamefont{Williams}}, \bibinfo{journal}{Science}
  \textbf{\bibinfo{volume}{281}}, \bibinfo{pages}{1476} (\bibinfo{year}{1998}).

\bibitem[{\citenamefont{Lemoine et~al.}(2013)\citenamefont{Lemoine, Goossens,
  Sabaka, Nicholas, Mazarico, Rowlands, and \emph{et al.}}}]{Lemoine2013}
\bibinfo{author}{\bibfnamefont{F.~G.} \bibnamefont{Lemoine}},
  \bibinfo{author}{\bibfnamefont{S.}~\bibnamefont{Goossens}},
  \bibinfo{author}{\bibfnamefont{T.~J.} \bibnamefont{Sabaka}},
  \bibinfo{author}{\bibfnamefont{J.~B.} \bibnamefont{Nicholas}},
  \bibinfo{author}{\bibfnamefont{E.}~\bibnamefont{Mazarico}},
  \bibinfo{author}{\bibfnamefont{D.~D.} \bibnamefont{Rowlands}},
  \bibnamefont{and} \bibinfo{author}{\bibnamefont{\emph{et al.}}},
  \bibinfo{journal}{Journal of Geophysical Research: Planets}
  \textbf{\bibinfo{volume}{118}}, \bibinfo{pages}{1676} (\bibinfo{year}{2013}).

\bibitem[{\citenamefont{Lemoine et~al.}(2014)\citenamefont{Lemoine, Goossens,
  Sabaka, Nicholas, Mazarico, Rowlands et~al.}}]{Lemoine2014}
\bibinfo{author}{\bibfnamefont{F.~G.} \bibnamefont{Lemoine}},
  \bibinfo{author}{\bibfnamefont{S.}~\bibnamefont{Goossens}},
  \bibinfo{author}{\bibfnamefont{T.~J.} \bibnamefont{Sabaka}},
  \bibinfo{author}{\bibfnamefont{J.~B.} \bibnamefont{Nicholas}},
  \bibinfo{author}{\bibfnamefont{E.}~\bibnamefont{Mazarico}},
  \bibinfo{author}{\bibfnamefont{D.~D.} \bibnamefont{Rowlands}},
  \bibnamefont{et~al.}, \bibinfo{journal}{Geophysical Research Letters}
  \textbf{\bibinfo{volume}{41}}, \bibinfo{pages}{3382} (\bibinfo{year}{2014}).

\end{thebibliography}

\appendix

\section{The  IAU 2000  relativistic reference systems}
\label{sec:BCRS-GCRS}

IAU Resolution B1.3 (2000) \cite{Kaplan:2005,Soffel-etal:2003} defines two harmonic‑gauge, post‑Newtonian frames: the {\tt BCRS} at the solar‑system barycenter and the {\tt GCRS} at Earth’s center of mass. It specifies the {\tt BCRS} metric \(g_{\mu\nu}(t,\mathbf{x})\) to \(O(c^{-4})\) via the scalar and vector potentials \(w\) and \(w^i\), and similarly defines the {\tt GCRS} metric \(G_{\alpha\beta}(T,\mathbf{X})\) with potentials \(W\) and \(W^a\). Resolution B1.3 also derives the \(O(c^{-4})\) coordinate transformation \((t,\mathbf{x})\to(T,\mathbf{X})\), including the external tidal potential \(w_{\rm ext}\); Resolution B1.4 then provides explicit analytic expressions for Earth’s tidal term \(W_{\rm tidal}\) in the GCRS.
Resolution B1.5 relates Barycentric Coordinate Time ({\tt TCB}) and Geocentric Coordinate Time ({\tt TCG}) and designates Barycentric Dynamical Time ({\tt TDB}) as the practical ephemeris timescale for modern planetary and lunar ephemerides \cite{Park-etal:2021}. A detailed discussion of implementation and operational implications appears in \cite{Soffel-etal:2003}. 

Below, we summarize the IAU 2000 definitions of the {\tt BCRS} and {\tt GCRS} and then present truncated metric tensors and coordinate‑transformation laws—retaining only terms above current instrumental thresholds—to support high‑precision timing and navigation in any Earth–Moon reference frame.

\subsection{The {\tt BCRS}, as defined by IAU}
\label{sec:BCRS}

\subsubsection{Metric tensor and gravitational potentials}

The {\tt BCRS} is defined with coordinates $(ct, x^\alpha) = x^m$, where $t$ is defined as Barycentric Coordinate Time ($\tt TCB$), or $t\equiv {\tt TCB}$. The {\tt BCRS} employs the metric tensor $g_{mn}$ in barycentric coordinates $(t, \mathbf{x})$. It includes a scalar potential $w(t, \mathbf{x})$, generalizing the Newtonian potential, and a spacetime component represented by a vector potential $w^\alpha(t, \mathbf{x})$:
{}
\begin{align}
\label{BCRS_metric}
g_{00} &= 1 - {2w \over c^2} +{2w^2 \over c^4} + \OO5, 
\qquad 
g_{0\alpha} =  -{4 \over c^3} w_\alpha+\OO5, 
\qquad
g_{\alpha\beta} = \gamma_{\alpha\beta}\Big(1 + {2 \over c^2}w \Big) + \OO4,
\end{align}
where gravitational potentials  $w(t,{\bf x})$ and  $w^\alpha(t, {\bf x})$ are found from the post-Newtonian Einstein field equations 
{}
\begin{eqnarray}
\label{PN_fielda}
\Big( \Delta -  {1 \over c^2} {\partial^2 \over \partial t^2}
\Big) w &=& - 4 \pi G \sigma + \OO4,
\label{PN_fieldb}
\qquad \quad
\Delta w^\alpha = - 4 \pi G \sigma^\alpha + \OO2,
\end{eqnarray}
with $\sigma = c^{-2}(T^{00} +T^{\epsilon\epsilon})$ and $\sigma^\alpha = c^{-1}T^{0\alpha}$ representing the relativistic gravitational mass and mass current density, respectively, and where $T^{mn}$ are the components of the stress-energy tensor for the solar system bodies \cite{Turyshev:1996,Turyshev-Toth:2013}. With these equations the potentials $w$ and $w^\alpha$ are determined as follows:
{}
\begin{eqnarray}
w(t,{\bf x}) &=& G \int d^3 x' {\sigma(t,{\bf x'}) \over
 |{\bf x} - {\bf x'}|}
+ {1 \over 2c^2} G {\partial^2 \over \partial t^2} \int d^3 x' \sigma(t,{\bf x'}) |{\bf x} - {\bf x'}|, \label{eq:pot-w}
\qquad
w^\alpha(t,{\bf x}) = G \int d^3 x' {\sigma^\alpha(t,{\bf x'}) \over |{\bf x} - {\bf x'}|}, 
\label{eq:pot-wi}
\end{eqnarray}
 where the integrals are evaluated over the compact support of body ${\tt B}$ alone.
For an ensemble of $N$-bodies
{}
\begin{equation}
\label{wsum}
w(t,\vx) = \sum_{{\tt B}=1}^N w_{\tt B}(t,\vx) \, , \qquad
w^\alpha(t,\vx) = \sum_{{\tt B}=1}^N w_{\tt B}^\alpha(t,\vx),
\end{equation}
where the index ${\tt B}$ denotes the contribution from the body ${\tt B}\in [1, N]$. Note that linearity of (\ref{wsum}) does not imply that body-body interaction terms have been overlooked.

\subsubsection{{\tt BCRS} metric for N-body system}
\label{sec:mass-multipoles}

Relativistic coordinate transformations for Earth were derived in
\cite{Moyer:1981-part1,Moyer:1981-part2,Fairhead-Bretagnon:1990,Standish_Williams_2008,Brumberg-Groten:2001}
and adopted by the IAU resolutions \cite{Kaplan:2005}. However, the preceding expressions carry precision beyond what is required for current solar‐system applications.  The recommended form expresses the barycentric metric potential \(w(t,\vx)\)  in (\ref{BCRS_metric}), as follows:
{}
\begin{equation}
\label{wparts*}
w = w_0 + w_{\tt L} - {1 \over c^2} \Delta.
\end{equation}

The first term in (\ref{wparts*}), $w_0$, denotes the $\ell=0$ monopole contribution (i.e., due to spherically-symmetric part of the mass distribution) to the scalar gravitational potential $w(t,\vx)$, as given in (\ref{wsum}):
{}
\begin{equation}
\label{eq:w_0}
w_0(t,\vx) \equiv \sum_{\tt B=1}^N {G M_{\tt B} \over r_{\tt B}},
\end{equation}
with the summation is performed over all solar system bodies ${\tt B}\in [1,N]$, where ${\vec r}_{\tt B} = {\vec x} - {\vec x}_{\tt B}$ and ${\vec x}_{\tt B}$ are the {\tt {\tt BCRS}} coordinates of the center of mass of body ${\tt B}$ with $r_{\tt B} = |\vec r_{\tt B}|$.

The second term in (\ref{wparts*}), $w_{\tt L}$, includes all contributions from higher potential coefficients beyond the monopole, with $\ell \ge 1$. In the gravitational $N$-body problem, the potential coefficients of a body ${\tt B}$ are defined within its corresponding local reference system, analogous to the {\tt GCRS} for the Earth. In the vicinity of a celestial body ${\tt B}$, the potential $w_{\tt L}$ can be expressed as $w_{\tt L} = w_{\tt L,{\tt B}} + w_{\tt L,{\rm ext}}$, where $w_{\tt L,{\tt B}}$ represents the extended gravitational contribution from body ${\tt B}$, and $w_{\tt L,{\rm ext}} = \sum_{{\tt C} \neq {\tt B}} w_{\tt L,{\tt C}}$ is the contribution from other bodies in the solar system. Clearly,  in the proximity of body ${\tt B}$, its own moments are dominant and must be considered, while the contributions from external bodies are typically negligible and, for most applications,  $w_{\tt L, \rm ext}$ may be neglected. 

The last term in (\ref{wparts*}), $\Delta(t,\vx)$, is the post-Newtonian part of the gravitational potential  
{}
\begin{equation}
\Delta(t,\vx) = \sum_{\tt B=1}^N \Delta_{\tt B}(t,\vx),
\label{eq:Delta*}
\end{equation}
where individual terms $\Delta_{\tt B}(t,\vx)$, to sufficient accuracy
are given as below
{}
\begin{eqnarray}
\label{delta-A}
\Delta_{\tt B}(t,\vx) &=& {G M_{\tt B} \over r_{\tt B}}
\Big[- 2 v_{\tt B}^2 + \sum_{{\tt C} \not= {\tt B}} {G M_{\tt C} \over r_{\tt CB}}
+ {\textstyle{\textstyle{\textstyle{1 \over 2}}}}\Big((\vec n_{\tt B} \cdot \vec v_{\tt B})^2
+ (\vec r_{\tt B}\cdot \vec a_{\tt B}) \Big)\Big]
+ {2G\big(\vec v_{\tt B}\cdot[{\bf r}_{\tt B} \times {\bf S}_{\tt B}]\big) \over r_{\tt B}^3}, 
\label{eq:Da}
\end{eqnarray}
where $\ve{r}_{\tt CB} = \ve{x}_{\tt B} - \ve{x}_{\tt C}$,  $\ve{n}_{\tt B} = \ve{r}_{\tt B}/{r}_{\tt B}$  and $\ve{a}_{\tt B} = d\ve{v}_{\tt B}/dt$. 
Here, the terms with ${\bf S}_{\tt B}$ are relevant only for Jupiter ($S_{\tt J} \approx 4.50 \times 10^{38}\,{\rm m}^2{\rm s}^{-1}$kg) and Saturn ($S_{\tt S} \approx 1.42 \times 10^{38}\,{\rm m}^2{\rm s}^{-1}$kg), especially in the immediate vicinity of these planets.

Finally, for accuracy sufficient for most practical purposes, the vector potential $w^\alpha$ (\ref{wsum}), can be expressed as
{}
\begin{equation}
w^\alpha(t,\vx) = \sum_{\tt B}\Big\{ {G M_{\tt B} \over r_{\tt B}} v_{\tt B}^\alpha -{G[{\bf r}_{\tt B} \times {\bf S}_{\tt B}]^\alpha \over 2r_{\tt B}^3} \Big\},
\label{eq:vecw}
\end{equation}
where ${\bf S}_{\tt B}$ is the total angular momentum of body ${\tt B}$ and $v_{\tt B}^\alpha$ is the barycentric coordinate velocity of body ${\tt B}$.  

As a result, for most practical applications in the solar system within the modern relativistic framework, the metric tensor of the {\tt BCRS}, as outlined in (\ref{BCRS_metric}), can be expressed in a more compact form as below \cite{Soffel-etal:2003}:
{}
\begin{eqnarray}
\label{eq:BCRS-metric-00}
g_{00}(t,{\bf x})  &=&1 - {2 \over c^2} \Big(w_0(t,{\bf x}) + w_{\tt L}(t,{\bf x})\Big) + {2 \over c^4}\Big(w_0^2(t,{\bf x}) + \Delta(t,{\bf x})\Big)+{\cal O}(c^{-5}), 
\\
g_{0\alpha}(t,{\bf x})  &=&  -{4 \over c^3} w_\alpha(t,{\bf x})+{\cal O}(c^{-5}), 
\quad \,
g_{\alpha\beta}(t,{\bf x})  = \gamma_{\alpha\beta}\Big( 1 + {2w_0(t,{\bf x}) \over c^2} \Big)+{\cal O}(c^{-4}),
\label{eq:BCRS-metric-ab}
\end{eqnarray}
where the potential $w_0(t,{\bf x})$ is detailed in (\ref{eq:w_0}), and $w_{\tt L}(t,{\bf x})$ includes the expansion in terms of multipole moments representing gravitational mass and current distribution for each body. The vector potential $w^\alpha(t,{\bf x})$ is described in (\ref{eq:vecw}), and the function $\Delta(t,{\bf x})$ is outlined in (\ref{eq:Delta*})--(\ref{eq:Da}). The ${\cal O}(c^{-4})$-terms in $g_{00}$, when evaluated at the Earth,  contribute up to  $\sim 9.74\times 10^{-17}=8.42$~ps/d. The omitted ${\cal O}(c^{-5})$-terms are $\sim10^4$ times smaller. 

\subsection{The {\tt GCRS}, as defined by IAU}
\label{sec:B-GCRS}

The {\tt GCRS} is defined by the geocentric metric tensor \(G_{mn}\) in coordinates \((T,\vec X)\), where $T$ is the Geocentric Coordinate Time ($\tt TCG$) or $T\equiv \tt TCG$. The form of the metric tensor mirrors that of the {\tt BCRS} (\ref{BCRS_metric}), with the barycentric potentials replaced by the geocentric scalar and vector potentials \(W(T,\mathbf X)\) and \(W^\alpha(T,\mathbf X)\), namely  
{}
\begin{eqnarray}\label{geo_metric}
G_{00} &=& 1 - {2W \over c^2} + {2W^2 \over c^4} + \OO5, 
\qquad
G_{0\alpha} = -{4 \over c^3} W_\alpha+\OO5, 
\qquad
G_{\alpha\beta} = \gamma_{\alpha\beta}\left(1 + {2 \over c^2}W \right) + \OO4,~~~
\end{eqnarray}
with the geocentric field equations formally resemble the barycentric ones in Eq.~\eqref{PN_fielda},  but with all  variables referenced to the {\tt GCRS}. The potentials \(W\) and \(W^\alpha\) are defined as the sum of the Earth's potentials and those due to other external bodies and are given as below:
{}
\begin{equation}
W(T,{\bf X}) = W_{\tt E}(T,{\bf X}) + W_{\rm ext}(T,{\bf X}), \qquad
W^\alpha(T,{\bf X}) = W_{\tt E}^\alpha(T,{\bf X}) + W_{\rm ext}^\alpha(T,{\bf X}).
\label{eq:W-pot}
\end{equation}

The Earth’s potentials $W_{\tt E}$ and $W_{\tt E}^\alpha$ are defined similarly to $w$ and $w^\alpha$, but with quantities calculated in the {\tt GCRS} and integrals performed over the entire Earth. A spherical harmonic expansion of the post-Newtonian potential of the Earth in the {\tt GCRS}, denoted as $W_{\tt E}$, outside the Earth to sufficient accuracy can be expressed as follows \cite{Soffel-etal:2003}:
{}
\begin{eqnarray}\label{BD-WE-sphe}
W_{\tt E}(T,\vX)
&=&
{G M_{\tt E} \over R}
\Big\{ 1 + \sum_{\ell= 2}^\infty \sum_{m=0}^{\ell}
\Big( {R_{\tt E}\over R} \Big)^\ell P_{\ell m}(\cos \theta)
\Big(C^{\tt E}_{\ell m}(T,R) \cos m \phi
+ S^{\tt E}_{\ell m}(T,R) \sin m \phi \Big) \Big\} + \cO(c^{-4}),
\end{eqnarray}
where $M_{\tt E}$ and $R_{\tt E}$ are the Earth's mass and equatorial radius, respectively, while $P_{\ell k}$ are the associated Legendre-polynomials \cite{Abramovitz-Stegun:1965}. $C_{\ell m}^{\tt E}$ and $S_{\ell m}^{\tt E}$ are the post-Newtonian multipole moments. $\theta$ and $\phi$ are the polar angles corresponding to the spatial coordinates $X^\alpha(\equiv \vec X)$ of the {\tt GCRS}, and $R = |X|$.  The moments $C_{\ell m}^{\tt E}(T)$ and $S_{\ell m}^{\tt E}(T)$, which refer to the {\tt GCRS} coordinates, are associated with nearly constant potential coefficients in a terrestrial system that rotates with the Earth (i.e., those from an Earth model) through time-dependent transformations.
Note that (\ref{BD-WE-sphe}) do not include second time derivatives of the multipole moments due to negligible magnitude of the resulting contributions.
The values $C_{\ell k}$ and $S_{\ell k}$ are the spherical harmonic coefficients that characterize contributions of the gravitational field of the Earth beyond the monopole potential. Of these, $J_\ell=-C_{\ell 0}$ are the zonal harmonic coefficients. Largest among these is $J_2=1.082635854\times 10^{-3}$, with all other spherical harmonic coefficients at least a factor of $\sim10^3$ times smaller \cite{Kozai:1966,Gasposchkin:1977,Lambeck:1988,JPL-EarthGrav:2021} (see Table~\ref{tab:sp-harmonics} for details).

\begin{table*}[t!]
\vskip-15pt
\caption{Some of the Earth's spherical gravitational coefficients up to degree and order $\ell,k=4$, with $GM_{\tt E}=398\,600.4415~{\rm km}^3{\rm s}^{-2}, R_{\tt E}=6\,378.13630~{\rm km}$ \cite{Tapley-1996,Montenbruck-Gill:2012}. Also, values of some additional lower order zonal harmonics are given as
$C_{50}= 2.28\times 10^{-7}$,
$C_{60}=-5.39\times 10^{-7}$,
$C_{70}= 3.51\times 10^{-7}$,
$C_{80}= 2.03\times 10^{-7}$,
$C_{90}= 1.19\times 10^{-7}$,
$C_{10\,0}=2.48\times 10^{-7}$.
\label{tab:sp-harmonics}}
\begin{tabular}{|c| c c c c c |}\hline
$C_{\ell k}$  & $k=0$  &  1 & 2  & 3& 4 \\\hline
$\ell=0$ & +1& &  & &\\
1 & 0.00 & 0.00&  & & \\
2 & $-1.0826359\times 10^{-3}$ & 0.00& ${+1.5745}\times 10^{-6}$ &   & \\
3 & $\phantom{000}{+2.5324}\times 10^{-6}$ & $+2.1928\times 10^{-6}$ &
$\phantom{0}{+3.090}\times 10^{-7}$ & ${+1.006}\times 10^{-7}$&
 \\
4 & $\phantom{000}{+1.6193}\times 10^{-6}$ & $\phantom{0}{-5.087}\times 10^{-7}$ &
$\phantom{00}{+7.84}\times 10^{-8}$ & $\phantom{0}{+5.92}\times 10^{-8}$ &
$-3.98\times 10^{-9}$
 \\
\hline
\hline
$S_{\ell k}$  & $k=0$  &  1 & 2  & 3& 4 \\\hline
$\ell=0$ & 0.00& &  & &\\
1 & 0.00 & 0.00&  & & \\
2 & 0.00 & $\phantom{0}{+1.54}\times 10^{-9}$ & $-9.039\times 10^{-7}$ &   & \\
3 & 0.00 & $+2.680\times 10^{-7}$ &
$-2.114\times 10^{-7}$ & $+1.972\times 10^{-7}$&
 \\
4 & 0.00 & $-4.494\times 10^{-7}$ &
$+1.482\times 10^{-7}$ & $\phantom{0}{+1.20}\times 10^{-8}$ &
$+6.53\times 10^{-9}$
 \\
\hline
\end{tabular}
\end{table*}

Regarding the external potentials $W_{\rm ext}$ and $W_{\rm ext}^\alpha$ in (\ref{eq:W-pot}), it is useful to further decompose them as follows:
{}
\begin{equation}
W_{\rm ext} = W_{\rm tid} + W_{\rm iner}, \qquad
W_{\rm ext}^\alpha = W_{\rm tid}^\alpha + W_{\rm iner}^\alpha,
\label{eq:W-ext}
\end{equation}
where $W_{\rm tid}$ generalizes the Newtonian expression for the tidal potential. To sufficient accuracy, it may be given as
{}
\begin{equation}
\label{W-tidal}
W_{\rm tid}(T,\vX) =
w_{\rm ext}(\ve{x}_{\tt E} + \vX) - w_{\rm ext}(\ve{x}_{\tt E}) - \big(\vX \cdot
\vec \nabla  w_{\rm ext}(\ve{x}_{\tt E})\big)=
\sum_{\tt B\not={\tt E}}
\sum_{\ell=2}^{N}
\frac{GM_{\tt B}}{r_{\tt BE}}
\Big(\frac{ X}{r_{\tt BE}}\Big)^\ell
P_\ell\bigl(\cos\theta_{\tt BE}\bigr)+{\cal O}\Big(\frac{{ X}^N}{r^{N+1}_{\tt BE}},c^{-2}\Big),
\end{equation}
where $\vec{r}_{\tt BE}=\vec x_{\tt E}-\vec x_{\tt B}$ is the vector connecting the center of mass of body $\tt B$ with that of the Earth, with ${r}_{\tt BE}=|\vec{r}_{\tt BE}|$ and $\vec{n}_{\tt BE}=\vec{r}_{\tt BE}/{r}_{\tt BE}$, also $\widehat{\vec { X}}={\vec { X}}/{ X}$ and $\cos\theta_{\tt BE} = ( { {\vec  n}}_{\tt BE}\cdot \widehat{\vec { X}})$, with  $P_\ell\bigl(\cos\theta\bigr)$ being the Legendre polynomials.\footnote{For convenience, we  the lowest orders of the Legendre polynomials $P_\ell(x)$ that for $\ell\in[2,9]$ are  given as below \cite{Abramovitz-Stegun:1965}
\[
\begin{aligned}
P_2(x)&=\tfrac12\,(3x^2-1),\quad 
P_3(x)=\tfrac12\,(5x^3-3x),\quad
P_4(x)=\tfrac18\,(35x^4-30x^2+3),\quad
P_5(x)=\tfrac18\,(63x^5-70x^3+15x),\\
P_6(x)&=\tfrac{1}{16}\,(231x^6-315x^4+105x^2-5),
\qquad
P_7(x)=\tfrac{1}{16}\,(429x^7-693x^5+315x^3-35x),\\
P_8(x)&=\tfrac{1}{128}\,(6435x^8-12012x^6+6930x^4-1260x^2+35),~
P_9(x)=\tfrac{1}{128}\,(12155x^9-25740x^7+18018x^5-4620x^3+315x).
\end{aligned}
\]
}
Naturally, the quadratic term (i.e., $\sim {\cal O}(X^2)$) in the resulting expression for $W_{\rm tidal}$ is the dominant one.

The potentials $W_{\rm iner}$ and $W_{\rm iner}^\alpha$ are inertial contributions that are linear in $X^\alpha$. The former is primarily influenced by the interaction between Earth's non-sphericity and the external potential. In the kinematically non-rotating {\tt GCRS}, $W_{\rm iner}^\alpha$ mainly describes the Coriolis force resulting from geodesic precession. Specifically,
{}
\begin{eqnarray}\label{W-iner}
W_{\rm iner}&=& (\vec Q \cdot \vec X),
\label{W-iner-a}
\qquad 
W^\alpha_{\rm iner} = -{1\over 2}\,c^2
\big[\vec  \Omega_{\rm iner} \times \vec X\big]^\alpha.
\end{eqnarray}

The quantity  $Q^\alpha$ is associated with the 4-acceleration of the geocenter in the external gravitational field. For an idealized Earth modeled as a purely spherical, non-rotating body following a geodesic in the external field (i.e., a mass monopole), this term is zero. Consequently, the $Q^\alpha$ term arises from the coupling of Earth's higher-order multipole moments with external tidal gravitational fields. It quantifies the deviation of the GCRS origin's actual worldline from a geodesic trajectory within the external gravitational field. From  (\ref{wsum}), we determine
{}
\begin{equation}
w_{\rm ext}(t,\vx) = \sum_{{\tt B}\not= \tt E} w_{\tt B}(t,\vx),
\qquad
w^\alpha_{\rm ext}(t,\vx) = \sum_{{\tt B}\not= \tt E} w^\alpha_{\tt B}(t,\vx),
\label{eq:W-pot2}
\end{equation}
where $w_{\tt B}$ and $w_{\tt B}^\alpha$ are determined by the expressions for $w$ and $w^\alpha$, with the integrals evaluated exclusively over the volume of body {\tt B}. Introducing $\vec x_{\tt E}(t)$, $\vec v_{\tt E}(t) = d\vec x_{\tt E}/dt$, and $\vec a_{\tt E} = d\vec v_{\tt E}/dt$ as the barycentric coordinate position, velocity, and acceleration of the geocenter (the origin of the {\tt GCRS}), respectively, the Newtonian expression for $Q^\alpha$ is given by:
{}
\begin{equation}\label{}
 Q^\alpha = \frac{\partial w_{\rm ext}(\vx_{\tt E})}{\partial x^\alpha}
 -  a^\alpha_{\tt E}.
\end{equation}
Note that the magnitude the absolute value of $Q^\alpha$ due to the action of the Moon  $ Q_{\tt M}\sim4.12 \times 10^{-11}\,$m/s$^2$.

The term $W^\alpha_{\rm iner}$ in (\ref{eq:W-ext}) is a relativistic Coriolis force due to the rotation of the {\tt GCRS} relative to a dynamically non-rotating geocentric reference system. This rotation includes several components,  including the geodesic precession, \( \vOmega_{\rm GP} \),  Thomas precession, \(\vOmega_{\rm TP}\), and Lense-Thirring effect, \(\vOmega_{\rm LTP}\), as below
{}
\begin{equation}\label{Omega-iner}
\vOmega_{\rm iner} =
\vOmega_{\rm GP}
+ \vOmega_{\rm TP}
+ \vOmega_{\rm LTP},
\end{equation}
with
\begin{eqnarray}
\label{Omega-GP}
\vOmega_{\rm GP} &=& -{3\over 2 c^2}\big[ \vv_{\tt E} \times
\vec \nabla w_{\rm ext}(\ve{x}_{\tt E})\big],
\label{Omega-TP}
\qquad
\vOmega_{\rm TP} = -{1\over 2c^2}\big[ \vv_{\tt E} \times \vQ\big],
\qquad
\vOmega_{\rm LTP} = -{2\over c^2}\big[\vec \nabla \times \vw_{\rm ext}(\ve{x}_{\tt E})\big].
\label{Omega-LT}
\end{eqnarray}

The geodesic precession \(\vOmega_{\rm GP}\) arises from Earth’s barycentric velocity \(v_{\tt E}\) interacting with the gradient of the external scalar potential \(w_{\rm ext}\) at the geocenter—equivalent, at the required accuracy, to the  barycentric coordinate acceleration of the geocenter.  Its magnitude is \(\lvert\vOmega_{\rm GP}\rvert\approx\tfrac{3}{2}c^{-2}v_{\tt E}\,GM_{\tt S}/\mathrm{AU}^2\approx2.95\times10^{-15}\,\mathrm{s}^{-1}\approx1.92\) arcsec/century ($''$/cen). 

The Thomas precession \(\vOmega_{\rm TP}\) arises from the coupling of Earth’s barycentric velocity \(v_{\tt E}\) with the geodesic deviation term \(Q^\alpha\). Its magnitude is \(\lvert\vOmega_{\rm TP}\rvert\approx\tfrac12\,c^{-2}\,v_{\tt E}\,\lvert \vec Q\rvert\approx6.83\times10^{-24}\,\mathrm{s}^{-1}\approx4.44\times10^{-9}\,\text{arcsec/century}\), making it negligible compared to the geodesic precession.

The Lense–Thirring precession \(\vOmega_{\rm LTP}\) arises from the gradient of the external gravito-magnetic potential at the geocenter. For a spherically symmetric body {\tt B}, its gravito-magnetic potential in the local rest frame is
\begin{equation}
\label{eq:spin_pot}
W^\alpha_{\tt B} = -\frac{G}{2}\frac{[\vX\times\vS_{\tt B}]^\alpha}{R^3},
\end{equation}
where \(\vS_{\tt B}\) is the body’s intrinsic angular momentum. For the Earth-Moon system, the spin and motion of both the Sun and the Moon provide the largest contributions to $\vOmega_{\rm LTP}$: $\vert \vOmega_{\rm LTP} \vert \sim 1.97 \times 10^{-3}~''$/cen.

The {\tt GCRS}  spatial axes \(\mathbf X\) are defined to be kinematically non-rotating with respect to the {\tt BCRS}  axes \(\mathbf x\). However, due to geodetic precession, a locally inertial frame precesses relative to the {\tt GCRS}  at 
\(\lvert\vOmega_{\rm iner}\rvert = 1.9198''/\mathrm{century}\). 
Since the {\tt GCRS}  is not a locally inertial frame, Coriolis accelerations arising from this inertial rotation must be included in all {\tt GCRS}  dynamical equations, including those governing Earth’s satellites.

\subsubsection{Estimating magnitudes of various terms}
\label{sec:mag-terms-metr}

To assess which post‐Newtonian terms in the {\tt GCRS} metric can be neglected for Earth orbiters, we evaluate the potentials at the altitude of {\tt  GPS} satellites \(h_{\rm GPS}=20\,200\)\,km, giving
$
  r_{\tt GPS}=R_{\rm E}+h_{\tt GPS}\approx2.6578\times10^7\,\mathrm{m}.
$

We first compute Earth’s Newtonian monopole potential that yields
\begin{align}
\label{eq:W_E}
  W_{\tt E}=\frac{GM_{\rm E}}{r_{\tt GPS}}\approx1.50\times10^7\;\mathrm{m}^2/\mathrm{s}^2
  \qquad\Rightarrow\qquad
  \delta G^{\tt E}_{00}=\frac{2W_{\tt E}}{c^2}\approx3.34\times10^{-10}.
\end{align}
The combined solar and lunar tidal potentials contribututions up to more than five orders of magnitude below \(W_{\tt E}\):
\begin{align}
\label{eq:W_E-tidal}
W_{\rm tidal}
&=\sum_{{\tt B}={\tt S, M}}\frac{GM_{\tt B}}{r_{\tt BE}^3}X^2 P_2(\cos\theta_{\tt BE})\lesssim \Big(\frac{GM_{\tt S}}{\rm AU^3}+\frac{GM_{\tt M}}{r_{\tt EM}^3}\Big)r^2_{\tt GPS}\approx 88.98\;\mathrm{m}^2/\mathrm{s}^2,\nonumber \\
&\Rightarrow \quad
\delta G^{\tt tidal}_{00}  =\frac{2W_{\rm tidal}}{c^2} \simeq 1.98\times10^{-15}.
\end{align}

The Newtonian‐order dipole coefficient \(Q_i\) in the {\tt GCRS} arises solely from the coupling of Earth’s quadrupole moment $ Q_{\tt E}^{jk}$ to the external tidal field, enforcing the geocenter’s free‐fall. In the multipolar expansion one finds
{}
\begin{equation}
\label{eq:QiQij}
  Q^\alpha =\frac{\partial w_{\rm ext}(\vx_{\tt E})}{\partial x^\alpha}-a_{\tt E}^\alpha
\simeq -\frac{1}{2M_{\tt E}}\,Q_{\tt E}^{jk}\,\partial^\alpha\partial_j\partial_k\,w_{\rm ext}(x_{\tt E}),
\end{equation}
where Earth's quadrupole and external potential are given as 
\[
Q_{\tt E}^{jk}
= J_2 M_{\tt E}R_{\tt E}^2\,\mathrm{diag}(1,1,-2), \qquad
w_{\rm ext}(\mathbf x)
= \sum_{{\tt B}\neq {\tt E}}\frac{GM_{\tt B}}{|\mathbf x-\mathbf x_{\tt B}|}.
\]
For a perturber \(\tt B\) at geocentric distance \(r_{\tt B}\) and unit vector \(n_i=(x_{\tt E}^i-x_{\tt B}^i)/r_{\tt BE}\), the cubic spatial derivative of \(1/r\)  is
\[
\partial_i\partial_j\partial_k\frac1{r_{\tt B}}\Big|_{\vec x_{\tt E}}
= -\,\frac{1}{r_{\tt B}^4}\Bigl(
15\,n_i n_j n_k
-3\,(n_i\delta_{jk}+n_j\delta_{ik}+n_k\delta_{ij})
\Bigr).
\]
Contracting this with \(Q_{\tt E}^{jk}\) and carrying through the factor \(-\tfrac1{2M_{\tt E}}G\,M_{\tt B}\) expression  (\ref{eq:QiQij}) gives
\[
\mathbf Q_{\tt B}
= \frac{9G M_{\tt B}J_2R_{\tt E}^2}{2 r_{\tt BE}^4}
\Bigl(n_x(1-5n_z^2),\,n_y(1-5n_z^2),\,n_z(3-5n_z^2)\Bigr),
\]
with the magnitude of this term given as
\[
Q_{\tt B}
= \frac{9G M_{\tt B}J_2R_{\tt E}^2}{2 r_{\tt BE}^4}\sqrt{
(1-n^2_z)(1-5n_z^2)^2+n^2_z(3-5n_z^2)^2}.
\]

 Using $
J_{2\tt M}=1.08263\times10^{-3}$ and $
R_{\tt E}=6.37814\times10^6\rm\,m,
$
the lunar contribution with
$
G M_{\tt M}=4.9028\times10^{12}\rm\,m^3/s^2$ and $
r_{\tt EM}=3.84399\times10^8\rm\,m,
$
gives the prefactor value of 
\({9GM_{\tt M}J_{2\tt M}R_{\tt E}^2}/{2r_{\tt EM}^4}\approx4.46\times10^{-11}\rm\,m/s^2.\)
The maximum occurs when \(n_z=0\), giving
$
Q_{\tt M}^{\max}\approx4.46\times10^{-11}\rm\,m/s^2,
$
and for a typical lunar inclination (\(n_z\approx0.41\)),
we obtain
$Q_{\tt M}\approx
4.01\times10^{-11}\rm\,m/s^2.$
The solar term, with
$
GM_{\tt S}=1.3271244\times10^{20}\rm\,m^3/s^2,$ and $
r_{\tt E{\tt S}}=1.49598\times10^{11}\rm\,m,
$
yields
\(Q_{\tt S}\sim1.9\times10^{-14}\rm\,m/s^2.\) Hence, the total dipole coefficient
\(Q^\alpha=\sum_{\tt B}Q_{\tt B}^\alpha \approx4.01\times10^{-11}\,\mathrm{m/s^2}, $
dominated by the Moon.  
As a result, the inertial (dipole) potential is evaluated as below:
\begin{align}
\label{eq:W-iner2}
W_{\rm iner} = (\vec Q\cdot \vec X) \approx Q_i\,r_{\tt GPS}\approx1.07\times10^{-3}\,\mathrm{m^2/s^2}
\qquad
\Rightarrow \qquad
\delta G^{\tt tidal}_{00} & =\frac{2W_{\rm iner}}{c^2} \simeq 2.37\times10^{-20}.
\end{align}
Because \(W_{\rm iner}\) is $\simeq 10^5$ times smaller than \(W_{\rm tidal}\) (\ref{eq:W_E-tidal}) and purely a coordinate artifact, it is omitted from the  metric.

For the vector potentials, Earth’s spin, $S_{\tt E}\simeq 5.86\times 10^{33}\, {\rm kg\,m}^2$/s, generates the Lense–Thirring term at GPS orbit:
\[
W_{\tt E}^\alpha\sim\frac{G\,|\mathbf S_{\tt E}|}{2\,r^2_{\tt GPS}}
\approx2.77\times10^8\;\mathrm{m}^3/\mathrm{s}^3 
\qquad
\Rightarrow \qquad
\delta G^{\tt E}_{0\alpha}  =\frac{4W_{\rm E}^\alpha}{c^3} \simeq 4.11\times10^{-17}.
\]

The tidal‐vector potential in the GCRS is defined similarly to (\ref{W-tidal})  (see \cite{Soffel-etal:2003}, (28)--(29)), taking the form:
\[
W_{\rm tid}^\alpha(T,\mathbf X)
=\sum_{\tt B\neq E}\Bigl[
w_{\tt B}^\alpha(\mathbf X_{\tt E}+\mathbf X)
-\,w_{\tt B}^\alpha(\mathbf X_{\tt E})
-\,X^\beta\,\partial_\beta w_{\tt B}^\alpha(\mathbf X_{\tt E})
\Bigr],
\qquad
w_{\tt B}^\alpha(\mathbf x)=\frac{G\,M_{\tt B}\,v_{\tt B}^\alpha}{|\mathbf x-\mathbf x_{\tt B}|}.
\]
To leading (quadrupole) order in \(\mathbf X\),
$
W_{\rm tid}^\alpha
=\sum_{\tt B\not=E}{G\,M_{\tt B}\,v_{\tt B}^\alpha}{r^{-3}_{\tt BE}}X^2 P_2(\cos\theta_{\tt BE}).
$
At GPS altitude \(R_{\tt GPS}\), using lunar and solar barycentric speeds
\(\;v_{\tt M}\approx3.08\times10^4\)\,m/s and \(v_{\tt S}\approx12.71\)\,m/s, the individual contributions are
\[
\begin{aligned}
W_{\rm tid}^{\alpha}
\simeq \Big\{ \frac{G\,M_{\tt M}\,v_{\tt M}\,r_{\tt GPS}^2}
            {r_{\tt EM}^3}P_2(\cos\theta_{\tt ME})
+ \frac{G\,M_{\tt S}\,v_{\tt S}\,r_{\tt GPS}^2}
            {\mathrm{AU}^3}P_2(\cos\theta_{\tt SE})\Big\}
\lesssim 1.88\times10^6\;\mathrm{m^3/s^3}+3.56\times10^2\;\mathrm{m^3/s^3},
\end{aligned}
\]
so that the total tidal vector potential is evaluated to be
well below the \(5\times10^{-18}\) retention threshold:
\begin{align}
\label{eq:W_E-tidal-al}
W_{\rm tid}^\alpha\sim1.88\times10^6\;\mathrm{m^3/s^3}
\qquad
\Rightarrow
\qquad
\delta G_{0\alpha}^{\rm (tidal)}
=-\frac{4\,W_{\rm tid}^i}{c^3}
\simeq 2.79\times10^{-19}.
\end{align}

Taking the IAU-defined inertial (de Sitter) precession rate of   
$
\Omega_{\rm iner} = 19.2\ \mathrm{mas/yr}
\;\approx\;2.95\times10^{-15}\,\mathrm{rad/s}$, we estimate the  inertial precession as below
\[
W_{\rm iner}^\alpha \sim \frac{c^2}{4}\,\Omega_{\rm iner}\,r_{\tt GPS}
\approx 1.76\times10^{9}\;\mathrm{m^3/s^3}
\qquad
\Rightarrow \qquad
\delta G^{\rm iner}_{0\alpha}  =\frac{4W_{\rm iner}^\alpha}{c^3} \simeq 2.62\times10^{-16}.
\]
Thus at GPS height the inertial term is $\sim6.4$ times larger than the Lense–Thirring term.

Because the inertial terms enter only as a choice of coordinates (they can be set identically to zero by a small time- and-axis gauge transformation) and carry no invariant physical effect, and because their metric contributions $\delta g_{00}\lesssim10^{-21}$, $\delta g_{0i}\lesssim10^{-16}$ lie below modern measurement precision (e.g., GPS, etc.), they are formally removable and hence omitted from the working {\tt GCRS} metric. As a result, below we omit both of the inertial terms and consider only gravitational potential due to Earth and tidal potentials. 

\subsubsection{ {\tt GCRS}: Practically-relevant formulation}
\label{sec:rec-GCRS}

In practical {\tt GCRS} implementations, one includes all post-Newtonian terms up to \(\mathcal O(c^{-4})\) in \(G_{00}\), \(\mathcal O(c^{-3})\) in \(G_{0\alpha}\), and \(\mathcal O(c^{-2})\) in \(G_{\alpha\beta}\), but discards any metric perturbations smaller than \(5\times10^{-18}\).  Accordingly, the metric tensor (\ref{geo_metric})  retaining only \(\lvert\delta G_{mn}\rvert\gtrsim5\times10^{-18}\), sufficient for high-precision time-keeping applications,  becomes
{}
\begin{eqnarray}
G_{00}(T,{\bf X})  &=& 1 - \frac{2}{c^2} \Big\{W_{\tt E}(T,{\bf X}) + W_{\rm tid}(T,{\bf X})\Big\} + \frac{2}{c^4}W^2_{\rm E}(T,{\bf X}) + {\cal O}\Big(c^{-5}; 6.61\times 10^{-25}\big), 
\label{eq:G00tr}\\
G_{0\alpha}(T,{\bf X})  &=& -{2 G\over c^3}  \frac{[ {\vec S}_{\tt E}\times{\vec X}]_\alpha}{R^3}+ {\cal O}\Big(c^{-5}; 2.79\times 10^{-19}\big), 
\label{eq:G0atr}\\\
G_{\alpha\beta}(T,{\bf X})  &=& \gamma_{\alpha\beta}\Big(1 + \frac{2}{c^2} \Big\{W_{\tt E}(T,{\bf X}) + W_{\rm tid}(T,{\bf X})\Big\} \Big) + {\cal O}\Big(c^{-4}; 5.57\times 10^{-20}\Big),
\label{eq:Gabtr}
\end{eqnarray}
where $S_{\tt E}\simeq 5.86\times 10^{33}\, {\rm kg\,m}^2$/s is the Earth spin vector moment. Also, the post-Newtonian gravitational potentials $W_{\tt E}(T,{\bf X}) $ and $W_{\rm tid}(T,{\bf X})$ are given by (\ref{BD-WE-sphe}) and (\ref{W-tidal}), correspondingly. 

The error bounds in  \eqref{eq:G00tr}–\eqref{eq:Gabtr}  are due to  the dominant omitted corrections evaluated at GPS altitude, specifically:  $\delta G_{00}^{\rm (mix)}=-4c^{-4}
W_{\tt E}\,W_{\rm tid}\simeq 4c^{-4}(GM_{\tt E}/r_{\tt GPS})(GM_{\tt S}/{\rm AU}^3+Gm_{\tt M}/r_{\tt EM}^3)r_{\tt GPS}^2\simeq 6.61\times10^{-25}$, 
as given by (\ref{eq:W_E})--(\ref{eq:W_E-tidal});
$\delta G_{0\alpha}^{\rm (tid)}=-4c^{-3}W_{\rm tid}^\alpha \simeq 4c^{-3} (GM_{\tt M}/r_{\tt EM}^3)v_{\tt M} r_{\tt GPS}^2\simeq2.79\times10^{-19}$, given by (\ref{eq:W_E-tidal-al}); and $\delta G_{\alpha\beta}^{\rm (2PN)}=\gamma_{\alpha\beta}\,\tfrac32 c^{-4} W_{\tt E}^2\simeq \tfrac32c^{-4} (GM_{\tt E}/r_{\tt GPS})^2\simeq  4.78\times10^{-20}$ \cite{Richter-Matzner:1983}. Thus all neglected terms are safely below the \(5\times 10^{-18}\).
The inertial dipole  \(W_{\rm iner}\) is a coordinate artifact absorbed by the {\tt GCRS} origin choice. Also, the \(W^\alpha_{\rm iner}\) is chosen such that $G_{0\alpha}(T,{\bf X}) $ takes a particular form of (\ref{eq:G0atr}). Thus, all omitted terms are 2–8 orders of magnitude below the \(5\times 10^{-18}\) accuracy goal for GPS orbits and clocks; the inertial dipole \(W_{\rm iner}\) is a coordinate artifact absorbed by the {\tt GCRS} origin choice. Note that although we evaluated the metric components in \eqref{eq:G00tr}–\eqref{eq:Gabtr} at GPS altitude (i.e., where tidal contributions exceed those at the surface), these expressions remain valid for all Earth‐orbit regimes from LEO through GEO.

When evaluating the contributions of the {\tt GCRS} metric tensor to the proper‐time–to–{\tt TCG} transformation, \(\mathrm{d}\tau/\mathrm{d}{\tt TCG}\), at a GPS orbit \cite{Turyshev-Toth:2023-grav-phase}, the \(O(c^{-2})\) terms dominate at \(c^{-2}GM_{\tt E}/r_{\tt GPS}\sim1.67\times10^{-10}\). Contributions from \(W^2\) and \(W^\alpha\) are at most 
\(c^{-4}(GM_{\tt E}/r_{\tt GPS})^2\sim2.79\times10^{-20}\)
 and  \(c^{-4}2G S_{\tt E}v_{\tt GPS}/r^2_{\tt GPS}\sim5.31\times10^{-22}\), correspondingly, and those from the inertial potential \(W_{\rm iner}\) are \(\lesssim2.37\times10^{-20}\) (\ref{eq:W-iner2}). In fact, the metric (\ref{geo_metric}) and its truncated form \eqref{eq:G00tr}–\eqref{eq:Gabtr} may be used for time‐and‐frequency applications up to cislunar space,  satisfying the target accuracy of  \(5\times 10^{-18}\).  

\subsection{Coordinate transformations between {\tt BCRS} and {\tt GCRS}}
\label{sec:BG-RS-transform}

\subsubsection{Coordinate transformations as recommended by the IAU}
\label{Section-trans}

The metric tensors in the {\tt BCRS} and {\tt GCRS} frameworks allow for the derivation of the transformation rules between the {\tt BCRS} coordinates \( x^m \) and the {\tt GCRS} coordinates \( X^n \) using tensorial transformation principles. These transformations can be expressed in two equivalent forms: i) as \( x^m(T,\vec{X}) \) or ii) as \( X^n(t,\vec{x}) \). It is important to note that converting from one form to the other is non-trivial due to the barycentric coordinate position of the geocenter, which appears as a function of \texttt{TCG} in the first form and as a function of \texttt{TCB} in the second form.

Explicitly, for the kinematically non-rotating {\tt GCRS}, the coordinate transformations are given as below
{}
\begin{eqnarray}
T &=& t - {1 \over c^2} \Big\{ A(t) + ({\vec  v}_{\tt E} \cdot {\vec r}_{\tt E})\Big\} + {1 \over c^4}
\Big\{ B(t) +\big({\vec B}(t)\cdot \vec r_{\tt E}\big) + B_{\mu\nu}(t)r_{\tt E}^\mu r_{\tt E}^\nu + C(t,{\vec x})\Big\}
+ O(c^{-5}), \label{eq:coord-tr-T1}\\
\vec X &=&  \vec r_{\tt E} + {1 \over c^2}
\Big\{{\textstyle{\textstyle{\textstyle{1 \over 2}}}} \vec v_{\tt E}  (\vec v_{\tt E} \cdot\vec r_{\tt E}) + \vec r_{\tt E} \,w_{\rm ext}({\bf x_{\tt E}}) + \vec r_{\tt E} (\vec a_{\tt E}\cdot \vec r_{\tt E}) - {\textstyle{\textstyle{\textstyle{1 \over 2}}}}\vec a_{\tt E} r_{\tt E}^2 \Big\}+ O(c^{-4}),
 \label{eq:coord-tr-X1}
\end{eqnarray}
where $T$ = $\tt TCG$, $t$ = $\tt TCB$, ${\vec  r}_{\tt E}={\vec  x}-{\vec  x}_{\tt E}$, ${\vec  v}_{\tt E}=d{\vec  x}_{\tt E}/dt$, ${\vec  a}_{\tt E}=d^2{\vec  x}_{\tt E}/dt^2$, and 
functions $A, B, B^\mu, B^{\mu\nu}, C(t,{\vec  x})$ are
{}
\begin{eqnarray}
{d \over dt}A(t) &=& {\textstyle{\textstyle{1 \over 2}}} v_{\tt E}^2 + w_{\rm ext}({\bf x_{\tt E}}), 
 \label{eq:coord-tr-A-B1}\\
{d \over dt}B(t) &=& -{\textstyle{\textstyle{1 \over 8}}}
v_{\tt E}^4 - {\textstyle{\textstyle{3 \over 2}}} v_{\tt E}^2 w_{\rm ext}({\bf x_{\tt E} })
+ 4\big(\vec v_{\tt E}  \cdot \vec w_{\rm ext}({\bf x_{\tt E} })\big) + {\textstyle{\textstyle{\textstyle{1 \over 2}}}}w_{\rm ext}^2({\bf x_{\tt E} }),  \label{eq:coord-tr-Bt}\\
B^\mu(t) &=& -{\textstyle{\textstyle{\textstyle{1 \over 2}}}}v_{\tt E} ^2 v_{\tt E} ^\mu + 4 w_{\rm ext}^\mu({\bf
x_{\tt E} }) - 3v_{\tt E} ^\mu w_{\rm ext}({\bf x_{\tt E} }), 
 \label{eq:coord-tr-A-Bi}\\
B^{\mu\nu}(t) &=&-v_{\tt E} ^\mu Q^\nu + 2
\partial^\mu w_{\rm ext}^\nu ({\bf x_{\tt E} }) - v_{\tt E} ^\mu
\partial^\nu w_{\rm ext}({\bf x_{\tt E} })-  {\textstyle{\textstyle{\textstyle{1 \over 2}}}}\gamma^{\mu\nu}\dot{w}_{\rm ext}({\bf x_{\tt E} }), 
 \label{eq:coord-tr-A-Bij}\\
C(t,{\bf x}) &=& -{\textstyle{1 \over 10}}r_{\tt E} ^2\big(\dot{\vec  a}_{\tt E} \cdot \vec r_{\tt E} \big).
 \label{eq:coord-tr-A-C}
\end{eqnarray}

The external potential at the Earth \( w_{\rm ext}({\bf x_{\tt E} }) \)  may be represented only by the monopole contribution of the gravity field of the external bodies  \( w_{0,\texttt{ext}} \) taken at the Earth's world-line  
{}
\begin{equation}
c^{-2}w_{\rm ext}({\vec x_{\tt E} })=c^{-2}
\sum_{\tt B\not= E} {G M_{\tt B} \over r_{\tt BE}} +{\cal O}\Big(4.80 \times 10^{-20}\Big),
\label{w_ext-mono}
\end{equation}
with the summation carried out over all solar system bodies {\tt B} except the Earth, \( {\bf r}_{\tt BE} = {\bf x}_{\tt E}  - {\bf x}_{\tt B} \), with \( r_{\tt BE} = |\vec{r}_{\tt BE}| \). The error term is determined by the contribution of solar quadruple moment $J_2 = 2.25 \times 10^{-7}$ \cite{Park:2017,MecheriMeftah:2021} in (\ref{eq:W-pot2}) and (\ref{eq:spin_pot}), yielding contribution to the time transformation (\ref{eq:coord-tr-T1}) via (\ref{eq:coord-tr-A-B1}) of $ c^{-2}(GM_{\tt S}/{\rm AU}^3) J_2 R_{\tt S}^2 P_{20}(\cos\theta)\simeq 4.80 \times 10^{-20} P_{20}(\cos\theta)$, which is sufficiently small to be ignored for our purposes.  

Finally, with accuracy sufficient for most practical purposes, from (\ref{eq:vecw}), we have
{}
\begin{equation}
c^{-3}w^\alpha_{\tt ext}(t,\vx) = c^{-3}\sum_{\tt B \not =E}
{G M_{\tt B} \over r_{\tt BE}} v_{\tt B}^\alpha +{\cal O}\Big(1.04 \times 10^{-17}\Big).
\label{w_0ii}
\end{equation}
where the error term is due to the omitted contribution from the solar spin moment of $S_{\tt S} \simeq 1.8838\times 10^{41}\, {\rm kg\,m}^2/{\rm s}$ \cite{Cang:2016}, contributing effect up to $\delta w^\alpha_{\tt ext}(t,\vx)\sim c^{-3} G S_{\tt S}/2{\rm AU}^2 \simeq 1.04 \times 10^{-17} $. When this term is multiplied by $v_{\tt E}/c\simeq 9.94\times 10^{-5}$, as in (\ref{eq:coord-tr-Bt}), the results is $\sim 1.03\times 10^{-21}$ -- too small to consider for (\ref{eq:coord-tr-T1}), thus bounding (\ref{w_0ii}). 

\subsubsection{Estimating magnitudes of various terms}
\label{sec:mag-terms} 

Here we examine the magnitudes of the terms in (\ref{eq:coord-tr-T1})--(\ref{eq:coord-tr-A-C}) as they apply to {\tt GNSS}. The numerical applications will focus on time and frequency transfer involving GPS spacecraft orbiting Earth at an altitude of \( h_{\texttt{GPS}} = 20\,200 \)~km and velocity of $v_{\texttt{GPS}} \simeq 3.87 \times 10^3\)\,m/s. We consider measurement uncertainties of \( 5 \times 10^{-18} \) for frequency transfer and 0.1\,ps for time transfer (see IAU Resolutions 1.3 and 1.5 in \cite{Soffel-etal:2003}).

We begin with the expression for the time transformation (\ref{eq:coord-tr-T1}). With definition for $w_{\rm ext}({\bf x_{\tt E}})$ from (\ref{w_ext-mono}), the terms proportional to $1/c^2$ in ${d A/ dt}$ contribute  $ c^{-2}({\textstyle{\textstyle{1 \over 2}}} v_{\tt E}^2 + \sum_{\tt B\not=E} GM_{\tt B}/r_{\tt BE})\simeq1.48 \times 10^{-8}$ to the time rate $dT/ dt$. As a result, expression for ${d A(t)/ dt}$ from (\ref{eq:coord-tr-A-B1}) takes the form:
{}
\begin{eqnarray}
\frac{1}{c^2}{d \over dt}A(t) &=&\frac{1}{c^2}\Big\{ {\textstyle{\textstyle{1 \over 2}}}v_{\tt E} ^2 + \sum_{{\tt B\not= E}} {G M_{\tt B} \over r_{\tt BE}} \Big\}+{\cal O}(1.86\times 10^{-20}) \simeq 
1.48 \times 10^{-8}+{\cal O}(1.86\times 10^{-20}),
 \label{eq:coord-tr-A-B1E}
\end{eqnarray}
where the error term is determined by the contribution from the mixed potential terms, $\Delta_{\tt ext}(t,\vx)$, that were present in (\ref{eq:Da}), but omitted  in (\ref{w_ext-mono}) (see discussion in \cite{Soffel-etal:2003}.)

The position-dependent $c^{-2}$-term in (\ref{eq:coord-tr-T1})  contributes a periodic effect of $c^{-2}({\vec  v}_{\tt E} \cdot {\vec r}_{\tt E})\simeq8.81 \,\mu$s to the time transfer at the GPS altitude. Therefore, both of the $c^{-2}$-terms are significant and must be included in the model.

Terms proportional to $1/c^4$ in (\ref{eq:coord-tr-T1}) exhibit both secular and quasi-periodic behavior. Considering the term ${d B(t)/ dt}$ as given in (\ref{eq:coord-tr-Bt}), the velocity term contributes up to $v_{\tt E} ^4/8c^4 \simeq 1.22 \times 10^{-17}$ to the time rate. The second term, when evaluated for the solar potential, yields $c^{-4}(3/2)v^2_{\tt E}  GM_{\tt S}/{\rm AU} \simeq 1.46 \times 10^{-16}$. The third term, evaluated for the solar vector potential, yields $c^{-4}4v_{\tt E}  GM_{\tt S} v_{\tt S}/{\rm AU} \simeq 1.66 \times 10^{-19}$, with its total term contribution of $4\sum_{\tt B \not =E}(G M_{\tt B}/ r_{\tt BE}) (\vec v_{\tt E} \cdot \vec v_{\tt B})\sim 2.14\times10^{-19}$ and thus, is too small to be considered for high-precision timing applications. Finally, the last term contributes $c^{-4}{\textstyle\frac{1}{2}}(GM_{\tt S} /{\rm AU}+GM_{\tt M}/r_{\tt EM}+GM_{\tt J} /{\rm 4AU} )^2 \simeq 4.87 \times 10^{-17}$. Altogether, the term $d B(t)/ dt$ contributes  $\sim2.07 \times 10^{-16}$ to the time rate $(dT/ dt)$, or up to $\sim$5.2 cm in 10 days.

As a result, the entire term (\ref{eq:coord-tr-Bt}) takes the following form:
{}
\begin{eqnarray}
\frac{1}{c^4}{d \over dt}B(t) &=& \frac{1}{c^4}\Big\{-{\textstyle{1 \over 8}}
v_{\tt E} ^4 - {\textstyle{3 \over 2}} v_{\tt E} ^2 \sum_{{\tt B\not= E}}{G M_{\tt B} \over r_{\tt BE}}+ 
{\textstyle{\textstyle{1 \over 2}}}\Big[\sum_{{\tt B\not= E}} {G M_{\tt B} \over r_{\tt BE}}\Big]^2\Big\}+{\cal O}(2.14\times 10^{-19}) \simeq\nonumber \\
&\simeq&-2.07 \times 10^{-16}+{\cal O}(2.14\times 10^{-19}),
\label{eq:coord-tr-BtE}
\end{eqnarray}
where the error is set by the omitted contribution from the external vector potential in (\ref{eq:coord-tr-Bt}).

Next, considering the contribution of the $B^\mu(t)$ term as specified in (\ref{eq:coord-tr-A-Bi}), we find that its velocity-dependent term contributes up to $c^{-4}v_{\tt E} ^3 r_{\tt GPS}/2 \simeq 4.35 \times 10^{-14}$\,s to the time transfer for a  GPS spacecraft. Given that the Sun moves relative to the SSB barycenter at a speed of $v_{\tt S} \sim 12.71$ m/s, its vector potential is responsible for a time uncertainty of $c^{-4}4(GM_{\tt S} /{\rm AU})v_{\tt S} r_{\tt GPS}  \sim 1.48 \times 10^{-16}$\,s. Also, the contribution from the Jovian vector potential was evaluated to be  $\sim3.71 \times 10^{-17}$\,s, other terms are even smaller. Thus, the term with the external vector potential $4\sum_{\tt B \not =E} (G M_{\tt B} / r_{\tt BE}) (\vec v_{\tt E}\cdot \vec r_{\tt GPS} )$  may be disregarded. Considering the last term in (\ref{eq:coord-tr-A-Bi}), the presence of  the solar scalar potential was found to contribute  $c^{-4}3(GM_{\tt S}/{\rm AU}+GM_{\tt M} /r_{\tt EM}+GM_{\tt J}/{\rm 4AU})^2 v_{\tt E}  r_{\tt GPS} \sim 2.61 \times 10^{-13}$\,s to the timing uncertainty, and thus it may be included. Thus, given (\ref{w_ext-mono}) and (\ref{w_0ii}), the term $B^i(t)$ can be writen as follows:
{}
\begin{eqnarray}
\frac{1}{c^4}\big({\vec B}(t)\cdot \vec r_{\tt E} \big) &=&-\frac{1}{c^4}\Big({\textstyle{\textstyle{1 \over 2}}}v_{\tt E} ^2 +3\sum_{\tt B \not =E}{G M_{\tt B} \over r_{\tt BE}}\Big)(\vec v_{\tt E} \cdot \vec r_{\tt E} )+{\cal O}(1.91\times 10^{-16}\,{\rm s}) \simeq \nonumber \\
&\simeq&3.04 \times 10^{-13}\,{\rm s}+{\cal O}(1.91\times 10^{-16}\,{\rm s}),
 \label{eq:coord-tr-A-BiE}
\end{eqnarray}
where the error is set by the omitted contribution from the external vector potential in (\ref{eq:coord-tr-A-Bi}).  Thus, at the GPS altitude this periodic term has magnitude of 0.30 ps but when evaluated on the Earth surface it amounts to 0.07\,ps. 

The second position-dependent term with quadratic position dependence, $B^{\mu\nu}(t)$, contributes a periodic effect with magnitude of up to  $\sim7.72 \times 10^{-17}$\,s to the time difference and is too small to be considered. Similarly, the third position-dependent term \(C(t,x)\) is periodic and even smaller.  To estimate its magnitude, we take $\dot a_{\tt E}\simeq 2GM_{\tt S} v_{\tt E}/{\rm AU}^3$, then the resulting timing offset  is $c^{-4}(1/5)GM_{\tt S} v_{\tt E}R_{\tt GPS}/ {\rm AU}^3\sim5.45 \times 10^{-22}$\,s, again far below any practical threshold.

Therefore, the only \(O(c^{-4})\) contributions that must be retained are the secular/quasi‐periodic rate term $c^{-4}{d B(t)/ dt}$, which induces a fractional timing rate of \(\sim2.07\times10^{-16}\) (\ref{eq:coord-tr-BtE}), and the periodic position term \(c^{-4}(\vec B(t)\cdot\vec r_{\rm E})\), which produces a peak timing offset of \(\sim0.30\ \mathrm{ps}\) (\ref{eq:coord-tr-A-BiE}); if fractional stability at the \(\sim5\times10^{-18}\) level (or sub‐ps timing) is required, both must be included in the model.

Next, we consider the position transformation as specified by (\ref{eq:coord-tr-X1}). At altitude of  a GPS spacecraft, the first two $1/c^2$ terms in this equation contribute $c^{-2}{\textstyle{\textstyle{\textstyle{1 \over 2}}}}\vec v_{\tt E}  (\vec v_{\tt E}  \cdot\vec r_{\tt E} ) \simeq 10$ cm and $c^{-2}w_{\rm ext}\,\vec r_{\tt E} =c^{-2}(G M_{\tt S}/{\rm AU}) \vec r_{\tt E}  \simeq 20$~cm. For a ground station, the effects are $c^{-2}{\textstyle{\textstyle{\textstyle{1 \over 2}}}}\vec v_{\tt E}  (\vec v_{\tt E}  \cdot\vec r_{\tt E} ) \simeq 3.2$ cm and $c^{-2}w_{\rm ext}\,\vec r_{\tt E} =c^{-2}(G M_{\tt S}/{\rm AU})\vec r_{\tt E}  \simeq 6.3$ cm. These contributions are significant enough to be included in the model. 

The acceleration-dependent terms in (\ref{eq:coord-tr-X1}) may contribute up to $2.68 \times 10^{-6}$~m at a ground station and $4.66 \times 10^{-5}$~m at $r_{\tt GPS}$. Although these corrections are small, they prove to be significant if one aims to compare spacecraft accelerations in {\tt BCRS} and {\tt GCRS}.  The next term involves the external multipole moments.  Using the solar quadrupole moment $J_2 = 2.25 \times 10^{-7}$ \cite{Park:2017,MecheriMeftah:2021}, its  contribution to the position transformation is estimated to be $c^{-2}w_{2,{\tt S}}(t,\vx) \vec r_{\tt E}  \simeq c^{-2} (GM_{\tt S} J_2R_{\tt S}^2/{\rm AU}^3)R_{\tt GPS}\sim  1.28 \times 10^{-12}$\,m,which is negligible and therefore serves as a conservative error bound.

\subsubsection{ {\tt GCRS}: Practically-relevant formulation}
\label{sec:rec-express}

 As a result of the preceding analysis, we present the coordinate transformations between the {\tt GCRS} ($T = \tt TCG$, $\vec X$) and the {\tt BCRS} ($t = \tt TCB$, $ \vec x $) that are sufficient for modern high-precision timing and positioning applications:
{}
\begin{eqnarray}
T &=& t-c^{-2}\Big\{\int^{t}_{t_{0}}\Big({\textstyle{1\over 2}} v_{\tt E}^{2} +
\sum_{{\tt B\not= E}} {G M_{\tt B} \over r_{\tt BE}} \Big)dt + (\vec v_{\tt E}  \cdot \vec r_{\tt E} ) \Big\}-
\nonumber \\ && \hskip 6pt -\, 
   c^{-4}\Big\{\int^t_{t_0}\Big({\textstyle{1 \over 8}}
v_{\tt E} ^4  
+ {\textstyle{3 \over 2}} v_{\tt E} ^2 \sum_{{\tt B\not= E}}{G M_{\tt B} \over r_{\tt BE}}- 
{\textstyle{\textstyle{1 \over 2}}}\Big[\sum_{{\tt B\not= E}} {G M_{\tt B} \over r_{\tt BE}}\Big]^2\Big)dt
+
\Big({\textstyle{\textstyle{1 \over 2}}}v_{\tt E} ^2 +3\sum_{\tt B \not =E}{G M_{\tt B} \over r_{\tt BE}}  \Big)(\vec v_{\tt E}  \cdot \vec r_{\tt E} ) 
\Big\} +\nonumber\\
  && \hskip 6pt +\, 
 {\cal O}\Big(c^{-5};\, 2.14\times 10^{-19}(t-t_0);\, 
 1.91\times 10^{-16}\,{\rm s}\Big),
 \label{eq:coord-tr-T1-rec}\\[4pt]
\vec X &=& \vec r_{\tt E}  + c^{-2} \Big\{{\textstyle\frac{1}{2}}( \vec v_{\tt E}  \cdot\vec r_{\tt E} )\vec v_{\tt E}  +\sum_{{\tt B\not= E}}{G M_{\tt B} \over r_{\tt BE}} \vec r_{\tt E}  +(\vec a_{\tt E} \cdot \vec r_{\tt E} )\vec r_{\tt E} - {\textstyle\frac{1}{2}}r^2_{\tt E} \vec a_{\tt E} \Big\}  +
 {\cal O}\Big(c^{-4};\, 1.28\times 10^{-12}~{\rm m}\big),
\label{eq:coord-tr-Xrec}
\end{eqnarray}
where the error bounds for secular 
\(\mathcal O(2.1\times10^{-19}(t-t_{0}))\), periodic 
\(\mathcal O(1.9\times10^{-16}\,\mathrm s)\), and positional 
\(\mathcal O(1.3\times10^{-12}\,\mathrm m)\) terms 
arise from omitted external vector‐potentials (\ref{eq:coord-tr-BtE}) and (\ref{eq:coord-tr-A-BiE}), and solar \(J_2\) contributions (\ref{w_ext-mono}), respectively.   

\section{Coordinate Transformations for the Moon}
\label{sec:time-M-coord}

\subsection{Lunicentric Coordinate Reference System ({\tt LCRS})}
\label{sec:LCRS-all}

In the vicinity of the Moon, one may introduce a non-rotating coordinate system known as the Lunicentric Coordinate Reference System ({\tt LCRS}). Centered at the Moon's center of mass\footnote{A similar coordinate system is used at the Earth and is known as  the Earth-Centered Earth-Fixed (ECEF) coordinate system \cite{Ashby-2003}.}, the LCRS may be used to track orbits in the vicinity of the Moon \cite{Turyshev:2025}.  Given the fact that the {\tt BCRS} (\ref{BCRS_metric}) or (\ref{eq:BCRS-metric-00})--(\ref{eq:BCRS-metric-ab}) is a common reference system for the solar system, to define the {\tt LCRS}, we will use the same approach as we used to define {\tt GCRS} (see Sec.~\ref{sec:B-GCRS}.) Accordingly, the {\tt LCRS} is defined  by the lunicentric metric tensor ${\cal G}_{mn}$ with lunicentric coordinates $({\cal T}, \vec {\cal X})$, where $\cal T$ is the Lunicentric Coordinate Time ($\tt TCL$) or ${\cal T}\equiv \tt TCL$. The metric tensor has the same form as the {\tt BCRS} (\ref{BCRS_metric}) and {\tt GCRS} (\ref{geo_metric}) but with potentials ${\cal W}({\cal T},  \vec{\cal X})$ and ${\cal W}^\alpha({\cal T}, \vec {\cal X})$, and  may be given in the form, as below \citep{Turyshev:2012nw}:
{}
\begin{eqnarray}\label{LCRS_metric}
{\cal G}_{00} &=& 1 - {2{\cal W} \over c^2} + {2{\cal W}^2 \over c^4} + \OO5, 
\qquad
{\cal G}_{0\alpha} = -{4 \over c^3} {\cal W}_\alpha+\OO5, 
\qquad
{\cal G}_{\alpha\beta} = \gamma_{\alpha\beta}\left(1 + {2 \over c^2}{\cal W} \right) + \OO4,~~~
\end{eqnarray}
with the field equations in {\tt LCRS} formally resemble those in the {\tt BCRS}  (\ref{PN_fielda}), but  all variables referenced to the {\tt LCRS}.

The lunicentric potentials \(\mathcal W\) and \(\mathcal W^\alpha\) decompose into the Moon’s self-potentials \(\mathcal W_{\tt M},\mathcal W_{\tt M}^\alpha\) and the external tidal contributions \(\mathcal W_{\rm ext},\mathcal W_{\rm ext}^{\alpha}\) (from all solar-system bodies except the Moon), all evaluated at the {\tt LCRS} origin:
{}
\begin{equation}
{\cal W}({\cal T}, \vec {\cal X}) = {\cal W}_{\tt M}({\cal T},\vec {\cal X}) + {\cal W}_{\rm ext}({\cal T}, \vec {\cal X}), \qquad
{\cal W}^\alpha({\cal T}, \vec {\cal X}) = {\cal W}_{\tt M}^\alpha({\cal T}, \vec {\cal X}) + {\cal W}_{\rm ext}^{\alpha}({\cal T}, \vec {\cal X}).
\label{eq:W-pot-M}
\end{equation}
The self-potentials \(\mathcal W_{\tt M},\mathcal W_{\tt M}^\alpha\) are defined by the same integrals as \(w,w^\alpha\), but taken over the Moon’s mass in the {\tt LCRS}.

The Moon’s post‐Newtonian scalar gravitational potential in the {\tt LCRS}, \(\mathcal W_{\tt M}(\mathcal T,\boldsymbol{\mathcal X})\), is determined by its relativistic mass density \(\sigma_{\tt M}(\mathcal T,\mathbf x')\):
\begin{equation}
\mathcal W_{\tt M}(\mathcal T,\boldsymbol{\mathcal X})
= G \int_{V_{\rm Moon}}\frac{\sigma_{\tt M}(\mathcal T,\mathbf x')}{\lvert\boldsymbol{\mathcal X}-\mathbf x'\rvert}\,\mathrm d^3x'
+ \mathcal O(c^{-4}),
\label{eq:pot_w_0-M}
\end{equation}
where the integral extends over the Moon’s volume.  
Outside the Moon ($r>R_{\tt M}$),  ${\cal W}_{\tt M}$ admits the standard spherical harmonics expansion. At a particular location with spherical coordinates $({\cal R}\equiv|{\vec {\cal X}}|,\psi_{\tt M},\theta_{\tt M})$ (where $\psi_{\tt M}$ is the longitude and $\theta_{\tt M}$ is the colatitude, which is $0$ at the pole and ${\textstyle\frac{\pi}{2}}$ at the equator) the Moon's potential ${\cal W}_{\tt M}$ in (\ref{eq:pot_w_0-M}) is given as
{}
\begin{eqnarray}
{\cal W}_{\tt M}({\cal T}, \vec  {\cal X})&=&
\frac{GM_{\tt M}}{\cal{R}}\Big\{1+\sum_{\ell=2}^\infty\sum_{k=0}^{+\ell}\Big(\frac{r_{\tt MQ}}{\cal{R}}\Big)^\ell P_{\ell k}(\cos\theta_{\tt M})\big(C^{\tt M}_{\ell k}\cos k\psi_{\tt M}+S^{\tt M}_{\ell k}\sin k\psi_{\tt M}\big)\Big\}=\nonumber\\
&=&\frac{GM_{\tt M}}{\cal R}\Big\{1-\sum_{\ell=2}^\infty\Big(\frac{r_{\tt MQ}}{\cal R}\Big)^\ell J^{\tt M}_\ell P_{\ell0}(\cos\theta_{\tt M})+\sum_{\ell=2}^\infty\sum_{k=1}^{+\ell}\Big(\frac{r_{\tt MQ}}{\cal R}\Big)^\ell P_{\ell k}(\cos\theta_{\tt M})\big(C^{\tt M}_{\ell k}\cos k\psi_{\tt M}+S^{\tt M}_{\ell k}\sin k\psi_{\tt M}\big)\Big\},~~~
\label{eq:pot_w_0sh}
\end{eqnarray}
where $M_{\tt M}$ and $r_{\tt MQ}$ are the Moon's mass and equatorial radius, respectively, while $P_{\ell k}$ are the associated Legendre-polynomials \cite{Abramovitz-Stegun:1965}, and  $C^{{\tt M}}_{\ell k}$ and $S^{{\tt M}}_{\ell k}$ are the Moon's spherical harmonic coefficients, and ${\cal R} =|\vec {\cal{X}}|\geq R_{\tt MQ}$. The values $C^{\tt M}_{\ell k}$ and $S^{\tt M}_{\ell k}$ are the spherical harmonic coefficients\footnote{For details, see the Lunar Gravity Field: GRGM1200A at  \url{https://pgda.gsfc.nasa.gov/products/50}} that characterize contributions of the gravitational field of the Moon beyond the monopole potential. Of these, $J_\ell=-C_{\ell 0}$ are the zonal harmonic coefficients. Largest among these is $J_{\tt 2M}= -2.033\times 10^{-4}$, with all other spherical harmonic coefficients about a factor of 10 smaller \cite{Turyshev:2025} (see Table~\ref{tab:sp-harmonics-moon}).

It is also essential to account for the elastic deformation of the Moon, represented by corrections $\Delta C^{\tt M}_{\ell k}$ and $\Delta S^{\tt M}_{\ell k}$ to the lunar spherical harmonic coefficients. These corrections arise from the tidal potential induced by body $\tt B$, located at lunicentric spherical coordinates $(r_{\tt B M}, \phi_{\tt B M}, \theta_{\tt B M})$~\citep{Williams-etal:2001,Montenbruck-Gill:2012}:
\begin{equation}
\left\{\begin{matrix}
\Delta C^{\tt M}_{\ell k}\\
\Delta S^{\tt M}_{\ell k}\end{matrix}\right\}=4k_\ell^{\tt M}\frac{M_{\tt B}}{M_{\tt M}}\left(\frac{R_{\tt 0M}}{r_{\tt B M}}\right)^{\ell+1}\sqrt{\frac{(\ell+2)[(\ell-k)!]^3}{[(\ell+k)!]^3}}P_{\ell k}(\cos\theta_{\tt B M})\left\{\begin{matrix}\cos k\phi_{\tt B M}\\
\sin k\phi_{\tt B M}\end{matrix}\right\}.
\label{eq:LOVE}
\end{equation}
The lunar Love number $k_2^{\tt M} \simeq 0.025$~\citep{Goossens-Matsumoto:2008} introduces a significant time-dependent contribution to the lunar spherical harmonic coefficients $C^{\tt M}_{\ell k}$ and $S^{\tt M}_{\ell k}$. These coefficients are therefore expressed as the sums
\begin{align}
C^{{\tt M}}_{\ell k}&=C^{{\tt M0}}_{\ell k} + \Delta C^{{\tt M}}_{\ell k}\qquad {\rm and} \qquad
S^{{\tt M}}_{\ell k}=S^{{\tt M0}}_{\ell k} + \Delta S^{{\tt M}}_{\ell k},\label{eq:DS}
\end{align}
where $C^{\tt M0 }_{\ell k}$ and $S^{\tt M0 }_{\ell k}$ represent the constant (static) components of the lunar spherical harmonic field (with some of them mentioned above), and $\Delta C^{\tt M}_{\ell k}$ and $\Delta S^{\tt M}_{\ell k}$ describe the tidal variations induced by external perturbing bodies.

\begin{table*}[t!]
\vskip-15pt
\caption{Some of the Moon's unnormalized spherical-harmonic gravitational coefficients up to degree and order $\ell,k=4$, with $GM_{\tt M}=4902.800118~{\rm km}^3{\rm s}^{-2}$ and lunar equatorial radius of $R_{\tt MQ}=1738.0~{\rm km}$  \cite{Tapley-1996,Konopliv1998,Montenbruck-Gill:2012,Lemoine2013,Lemoine2014,Konopliv-etal:2013}.
\label{tab:sp-harmonics-moon}}
\begin{tabular}{|c| c c c c c |}\hline
$C_{\ell k}$  & $k=0$  &  1 & 2  & 3& 4 \\\hline
$\ell=0$ & +1& &  & &\\
1 & 0.00 & 0.00&  & & \\
2 & $+2.0330530\times 10^{-4}$ & 0.00& $+2.242\,615\times 10^{-5}$ &   & \\
3 & $-8.459\,703\times 10^{-6}$ & $+2.848\,074\times 10^{-5}$ &
$\phantom{0}\,4.840\,499\times 10^{-6}$ & $\phantom{0}\,1.711\,660\times 10^{-6}$& \\
4 &  $+5.901000\times10^{-6}$ & 0.00 &  
\(+9.754000\times10^{-7}\)&  
\(+2.387000\times10^{-7}\)& 
\(+1.118000\times10^{-7}\) \\
\hline
\hline
$S_{\ell k}$  & $k=0$  &  1 & 2  & 3& 4 \\\hline
$\ell=0$ & 0.00& &  & &\\
1 & 0.00 & 0.00&  & & \\
2 & 0.00 & 0.00 & 0.00 &   & \\
3 & 0.00 & $5.891\,555\times 10^{-6}$ &
$1.666\,142\times 10^{-6}$ & $-2.474\,276\times 10^{-7}$&
 \\
4 &  0.00 & 0.00 &  0.00 & 
\(-2.474000\times10^{-7}\) & 
\(-2.310000\times10^{-8}\) \\
\hline
\end{tabular}
\end{table*}

In the {\tt LCRS}, the external scalar and vector potentials decompose into tidal and inertial parts:
\begin{equation}
\label{eq:W-ext-M}
\mathcal W_{\rm ext}(\mathcal T,\vec{\mathcal X})
= \mathcal W_{\rm tidal}(\mathcal T,\vec{\mathcal X})
  + \mathcal W_{\rm iner}(\mathcal T,\vec{\mathcal X}),
  \qquad
\mathcal W_{\rm ext}^{\alpha}(\mathcal T,\vec{\mathcal X})
= \mathcal W_{\rm tidal}^{\alpha}(\mathcal T,\vec{\mathcal X})
  + \mathcal W_{\rm iner}^{\alpha}(\mathcal T,\vec{\mathcal X}).
\end{equation}
Here, \(\mathcal W_{\rm tidal}\) and \(\mathcal W_{\rm tidal}^{\alpha}\) generalize the Newtonian lunar tidal potential, while \(\mathcal W_{\rm iner}\) and \(\mathcal W_{\rm iner}^{\alpha}\) represent the inertial potentials arising from the non‐inertial motion of the {\tt LCRS} origin.

Insofar as the tidal potential ${\cal W}_{\rm tidal}$ is concerned, for our purposes it is sufficient to keep only its Newtonian contribution (primarily due to the Sun and the Earth) which can be given in the form similar to (\ref{W-tidal}) as below:
{}
\begin{equation}
\label{W-tidal-M}
{\cal W}_{\rm tidal}({\cal T},\vec{\cal X}) =
w_{\rm ext}(\ve{x}_{\tt M} + \vec{\cal X}) - w_{\rm ext}(\ve{x}_{\tt M}) - \big(\vec{\cal X} \cdot
\vec \nabla  w_{\rm ext}(\ve{x}_{\tt M})\big) =
\sum_{\tt B\not={\tt M}}
\sum_{\ell=2}^{N}
\frac{GM_{\tt B}}{r_{\tt BM}}
\Big(\frac{\cal X}{r_{\tt BM}}\Big)^\ell
P_\ell\bigl(\cos\theta_{\tt BM}\bigr)+{\cal O}\Big(\frac{{\cal X}^N}{r^{N+1}_{\tt BM}},c^{-2}\Big),
\end{equation}
where  \(\vec r_{\tt BM}=\vec x_{\tt M}-\vec x_{\tt B}\) is the vector connecting the center of mass of body $\tt B$ with that of the Moon, with ${r}_{\tt BM}=|\vec{r}_{\tt BM}|$ and $\vec{n}_{\tt BM}=\vec{r}_{\tt BM}/{r}_{\tt BM}$, also $\widehat{\vec {\cal X}}={\vec {\cal X}}/{ \cal X}$ and $\cos\theta_{\tt BM} = ( { {\vec  n}}_{\tt BM}\cdot \widehat{\vec {\cal X}})$, with  $P_\ell\bigl(\cos\theta\bigr)$ being the Legendre polynomials.
 
The potentials ${\cal W}_{\rm iner}$ and ${\cal W}_{\rm iner}^{\alpha}$ are inertial contributions that are linear in $\vec {\cal X}$. The former is primarily influenced by the interaction between Moon's non-sphericity and the external potential. In the kinematically non-rotating {\tt LCRS}, ${\cal W}_{\rm iner}^{\alpha}$ mainly describes the Coriolis force resulting from geodesic precession at thr {\tt LCRS}. Specifically,
{}
\begin{eqnarray}\label{W-iner-M}
{\cal W}_{\rm iner}&=& (\vec {\cal Q}\cdot \vec {\cal X}),
\label{W-iner-a-M}
\qquad 
{\cal W}^{\alpha}_{\rm iner} = - {1\over 4}\,c^2
\,\epsilon^\alpha_{\beta\epsilon}  \Omega^{*\beta}_{\rm iner} \,{\cal X}^\epsilon,
\end{eqnarray}
where ${\cal Q}^\alpha$ is the inertial dipole vector induced by the Moon’s asphericity interacting with external gravity gradients; ${\Omega}_{\rm iner}^{*\alpha}$ is the geodesic‐precession rate that generates the Coriolis‐type term in the kinematically non-rotating {\tt LCRS}.  

The quantity \(\mathcal Q^\alpha\) represents the 4-acceleration of the lunicenter relative to a geodesic in the external field.  For an ideal spherical, non-rotating Moon (a pure mass monopole), \(\mathcal Q^\alpha=0\).  In reality, \(\mathcal Q^\alpha\) arises from the coupling of the Moon’s higher‐order multipole moments to external tidal fields, and it measures the deviation of the {\tt LCRS} origin’s worldline from geodesic motion.  From (\ref{wsum}), the external {\tt BCRS} potentials read
\[
  w^*_{\rm ext}(t,\mathbf x)
    = \sum_{{\tt B}\neq{\tt M}}w_{\tt B}(t,\mathbf x),
  \quad
  w^{*\alpha}_{\rm ext}(t,\mathbf x)
    = \sum_{{\tt B}\neq{\tt M}}w^\alpha_{\tt B}(t,\mathbf x),
\]
where {\tt M} labels the Moon and each \(w_{\tt B},\,w^\alpha_{\tt B}\) is defined by the standard {\tt BCRS} integrals over body \(\tt B\).  Denoting the lunicenter’s {\tt BCRS} position, velocity, and acceleration by \(\mathbf x_{\tt M}(t)\), \(\mathbf v_{\tt M}=d\mathbf x_{\tt M}/dt\), and \(\mathbf a_{\tt M}=d\mathbf v_{\tt M}/dt\), the Newtonian expression for \(\mathcal Q^\alpha\) is given by:
{}
\begin{equation}\label{eq:q-M}
 {\cal Q}^\alpha = \frac{\partial w^*_{\rm ext}(\vx_{\tt M})}{\partial x^\alpha}
 -  a^\alpha_{\tt M}\simeq -\frac{1}{2M_{\tt M}}\,Q_{\tt M}^{jk}\,\partial^\alpha\partial_j\partial_k\,w^*_{\rm ext}(x_{\tt M}).
\end{equation}
Note that the dominant contribution to ${\cal Q}^\alpha$ comes from Earth, evaluated to be $ {\cal Q}_{\tt E}\simeq 4.53 \times 10^{-11}\,$m/s$^2$.

The term \(\mathcal W_{\rm iner}^\alpha\) in (\ref{eq:W-ext}) represents a relativistic Coriolis acceleration due to the rotation of the {\tt LCRS} relative to a dynamically non‐rotating lunicentric frame.  This rotation comprises geodesic precession \(\vOmega^*_{\rm GP}\), Thomas precession \(\vOmega^*_{\rm TP}\), and the Lense–Thirring effect \(\vOmega^*_{\rm LTP}\):
\begin{equation}\label{Omega-iner-M}
\vOmega^*_{\rm iner}
= \vOmega^*_{\rm GP}
+ \vOmega^*_{\rm TP}
+ \vOmega^*_{\rm LTP},
\end{equation}
with
\begin{eqnarray}
\label{Omega-GP-M}
\vOmega^*_{\tt GP} &=& -{3\over 2 c^2}\big[ \vv_{\tt M} \times
\vec \nabla w_{\rm ext}(\ve{x}_{\tt M})\big],
\label{Omega-TP-M}
\qquad
\vOmega^*_{\tt TP} = -{1\over 2c^2}\big[ \vv_{\tt M} \times \vec {\cal Q}\big],
\qquad
\vOmega^*_{\tt LTP} = -{2\over c^2}\big[\vec \nabla \times \vw_{\rm ext}(\ve{x}_{\tt M})\big].
\label{Omega-LT-M}
\end{eqnarray}

The geodesic precession, \(\vOmega^*_{\rm GP}\), is influenced by the lunicenter’s barycentric velocity \(v_{\tt M}\) and the gradient of the external scalar potential \(w_{\rm ext}\) at the lunicenter, which to sufficient accuracy equals the lunicenter’s barycentric acceleration.  The magnitude of this term is $ \vert \vOmega^*_{\tt GP} \vert \sim {\textstyle\frac{3}{2}} c^{-2}v_{\tt M}(G M_{\tt S}/{\rm AU}^2+G M_{\tt E}/r_{\tt EM}^2) \sim 4.44 \times 10^{-15}~{\rm s}^{-1} \sim 2.89$ $''$/cen. 

The Thomas precession at the {\tt LCRS} \(\vOmega^*_{\rm TP}\) arises from the coupling of the lunicenter’s barycentric velocity \(\vec v_{\tt M}\) with the geodesic‐deviation vector \(\mathcal Q^\alpha\), and its magnitude isestimated as \(\lvert\vOmega^*_{\rm TP}\rvert\sim\tfrac12\,c^{-2}\,v_{\tt M}\,\lvert\mathcal Q_{\tt E}\rvert\approx7.76\times10^{-24}\,\mathrm s^{-1}\approx5.05\times10^{-9}\, {''}/\mathrm{cen}\), which is negligible relative to the geodesic‐precession rate.

Lastly, the Lense–Thirring precession \(\vOmega^*_{\rm LTP}\) results from the spatial gradient of the external gravito‐magnetic potential at the lunicenter. In the {\tt LCRS}, the leading‐order vector potential for a rotating, spherically symmetric body {\tt B} is given by \eqref{eq:spin_pot} in LCRS coordinates; using \(\lvert\vec S_{\rm M}\rvert\simeq2.32\times10^{29}\,\mathrm{kg\,m}^2/\mathrm s\) for the Moon and \(\lvert\vec S_{{\tt S}}\rvert\simeq1.88\times10^{41}\,\mathrm{kg\,m}^2/\mathrm s\) for the Sun, one finds for the Earth–Moon system \(\lvert\vOmega^*_{\rm LTP}\rvert\sim c^{-2}\,(2G\,\lvert\vec S_{{\tt S}}\rvert/\mathrm{AU}^3)\approx8.3\times10^{-20}\,\mathrm s^{-1}\approx5\times10^{-5}\,{''}/\mathrm{cen}\), while the Moon’s own spin contributes at the \(\sim10^{-10}\,''\)/cen level. Therefore, \(\vOmega^*_{\rm LTP}\) is entirely negligible compared to the geodesic‐precession term.

The definition of the {\tt LCRS} specifies that its spatial coordinates \(\boldsymbol{\mathcal X}\) are kinematically non‐rotating with respect to the {\tt BCRS} axes \(\mathbf x\).  However, locally inertial frames undergo geodesic precession relative to the {\tt LCRS} at a rate 
$  \lvert\vOmega^*_{\rm GP}\rvert\approx2.9''/{\rm cen}.$ Since the {\tt LCRS} is not itself inertial, the associated Coriolis accelerations must be included in all {\tt LCRS} equations of motion, for example when modeling lunar‐satellite orbits.

\subsubsection{ {\tt LCRS}: Practically-relevant formulation}
\label{sec:rec-LCRS}

Using the same procedure that was used to derive the {\tt GCRS} metric (\ref{eq:G00tr})--(\ref{eq:Gabtr}), one can evaluate every potential contribution in the {\tt LCRS} metric tensor (\ref{LCRS_metric}).  In direct analogy with the {\tt GCRS}, we substitute the lunar self‐potential \(\mathcal W_{\tt M}\), the external tidal potential \(\mathcal W_{\rm ext}\), the vector potential \(\mathcal W^\alpha\), and the inertial corrections into the lunicentric ansatz.  We formally include all post-Newtonian terms up to  $5\times 10^{-18}$ in all the metric components:
{}
\begin{eqnarray}
{\cal G}_{00}(\mathcal T,\vec{\mathcal X})&=& 1 - \frac{2}{c^2}\Big\{{\cal W}_{\tt M}(\mathcal T,\vec{\mathcal X})+{\cal W}_{\rm tid}(\mathcal T,\vec{\mathcal X})\Big\} +\frac{2}{c^4} {\cal W}_{\tt M}^2(\mathcal T,\vec{\mathcal X}) +{\cal O}\Big(c^{-5};\, 1.04\times10^{-24}\Big), \label{eq:G00tr-MM}\\
{\cal G}_{0\alpha}(\mathcal T,\vec{\mathcal X})&=& -\frac{4}{c^3}\Big\{{G\over 2}  \frac{[ {\vec S}_{\tt M}\times{\vec X}]_\alpha}{{\cal R}^3}+\frac{GM_{\tt E}}{2r^3_{\tt EM}}v^\alpha_{\tt E}\Big(3(\vec{n}^{}_{\tt EM}\cdot\vec{\cal X})^2-\vec{\cal X}^2\Big)\Big\}+{\cal O}\Big(c^{-5};\, 2.81\times10^{-22}\Big),\label{eq:G0atr-MM}\\
{\cal G}_{\alpha\beta}(\mathcal T,\vec{\mathcal X})&=& \gamma_{\alpha\beta}\Big(1 + \frac{2}{c^2}\Big\{ {\cal W}_{\tt M}(\mathcal T,\vec{\mathcal X})+{\cal W}_{\rm tid}(\mathcal T,\vec{\mathcal X})\Big\} \Big)+{\cal O}\Big(c^{-4};\, 1.46\times10^{-21}\Big),
\label{eq:Gabtr-MM}
\end{eqnarray}
where post-Newtonian gravitational potentials ${\cal W}_{\tt M}({\cal T},\vec {\cal X}) $ and ${\cal W}_{\rm tid}({\cal T},{\vec {\cal X}})$ are given by (\ref{eq:pot_w_0sh})--(\ref{eq:DS}) and (\ref{W-tidal-M}), correspondingly.  With the stated level of accuracy, one may use only Newtonian form of these potentials.

The error bounds in (\ref{eq:G00tr-MM})--(\ref{eq:Gabtr-MM}) are due to omitted terms that were evaluated for
various orbits listed in Table~\ref{tab:representative_orbits}. To get the most conservative estimates, we will use either a circular very low lunar orbit (vLLO) with $r_{\tt vLLO}=R_{\tt MQ}+ 10\, {\rm km}= 1.748\times 10^6$\,m or the Earth-Moon L1 point with $r_{\tt L1}\approx6.13\times10^{7}\,\mathrm{m}$ (Table~\ref{tab:representative_orbits}.) We expect that at vLLO the lunar gravity will be significant, while at the E-M L1 the tidal effects may be more important. With this in mind, the terms of interest have the following magnitudes: $\delta {\cal G}_{00}^{\rm (mix)}=-4c^{-4}
{\cal W}_{\tt M}\,{\cal W}^{(\tt E)}_{\rm tid}\lesssim -4c^{-4}
(G M_{\tt M}/a_{\tt L1})(GM_{\tt E}/ r_{\tt EM}^3)a_{\tt L1}^2\simeq 9.84\times10^{-25}$ (similar to (\ref{eq:W_E})--(\ref{eq:W_E-tidal})),  
and 
$\delta {\cal G}_{0\alpha}^{\rm (tid)}=-4c^{-3}{\cal W}_{\rm tid}^{\alpha ({\tt S})} \lesssim -4c^{-3}(GM_{\tt S}/{\rm AU}^3)v_{\tt S} a_{\tt L1}^2 \simeq2.52\times10^{-22}$ (in analogy to (\ref{eq:W_E-tidal-al})), and 
$\delta {\cal G}_{\alpha\beta}^{\rm (2PN)}=\gamma_{\alpha\beta}\,\tfrac32c^{-4} {\cal W}_{\tt M}^2\simeq \tfrac32c^{-4} (G M_{\tt M}/r_{\tt vLLO})^2\simeq 1.46\times10^{-21}.$  Also, the inertial dipole \({\cal W}^*_{\rm iner}\) is a coordinate artifact absorbed by the {\tt LCRS} origin choice. The \({\cal W}^{*\alpha}_{\rm iner}\) is chosen such that ${\cal G}_{0\alpha}(T,{\bf X}) $ takes a particular form of (\ref{eq:G0atr-MM}). 

Although the {\tt LCRS} metric tensor (\ref{eq:G00tr-MM})--(\ref{eq:Gabtr-MM}) formally has the same structure as its {\tt GCRS} counterpart (\ref{eq:G00tr})--(\ref{eq:Gabtr}), it still has terms that are much smaller than  \(5\times10^{-18}\). For instance, the $c^{-4}$-order term in ${\cal G}_{00}$, is $2c^{-4} {\cal W}_{\tt M}^2 \simeq 2c^{-4}(GM_{\tt M}/r_{\tt MQ})^2\simeq 1.97 \times 10^{-21}$. The lunar Lense-Thirring term in ${\cal G}_{0\alpha}$ is only $2c^{-3}G S_{\tt M}/r_{\tt MQ}^2\simeq 3.81\times 10^{-19}$ and may be neglected. The vector tidal potential is large only at the Earth-Moon L1 point reaching $4c^{-3}(GM_{\tt E}/r_{\tt EM}^3)v_{\tt E} a_{\tt L1}^2 \simeq1.05\times10^{-16}$, while at the lunar surface it is only $4c^{-3}(GM_{\tt E}/r_{\tt EM}^3)v_{\tt E} r_{\tt MQ}^2 \simeq9.37\times10^{-20}$ and may also be omitted. 

 Although we retained these terms in (\ref{eq:G00tr-MM})--(\ref{eq:Gabtr-MM}), we will omit them as we start considering practical applications of these expressions  for which we retain only those terms whose magnitudes exceed the fractional accuracy goal of \(5\times10^{-18}\), ensuring a sufficient model for high-precision lunar timing and navigation.  
 
\subsubsection{Proper Time in Cislunar Space}
\label{sec:proper-time-LCRS}

Consider a clock moving along an arbitrary worldline \(\boldsymbol{\mathcal X}(\mathcal T)\) in {\tt LCRS}. The four‑velocity of this clock  is given as usual ${\cal U}^m = {d\mathcal X^m}/{d\mathcal T}
    = \bigl(1,\;c^{-1}\,\boldsymbol{\mathcal V}\bigr),$
where  \(\boldsymbol{\mathcal V}=d\boldsymbol{\mathcal X}/d\mathcal T\) with $\cal V=|\vec {\cal V}|$ is  clock's velocity. To quantify performance of the proper time  of this clock, $\tau$,   with respect to the coordinate time \(\mathcal T\) of the {\tt LCRS}, we consider the line element  on the clock's wordline $c^2\,d\tau^2
= \mathcal G_{mn}(\mathcal T,\boldsymbol{\mathcal X})\,d\mathcal X^m d\mathcal X^n={\cal G}_{mn}\,{\cal U}^m {\cal U}^n c^2d{\cal T}^2$. Using the {\tt LCRS}  metric tensor  \(\mathcal G_{mn}\) 
from  (\ref{eq:G00tr-MM})--(\ref{eq:Gabtr-MM}) and formally keeping all the terms through order $c^{-4}$ we have
\begin{eqnarray}
\label{eq:tau-LCRS}
\frac{d\tau}{d{\cal T}}
&=&1
-\frac{1}{c^2}\Big\{\tfrac12\,{\cal V}^2+{\cal W}_{\tt M}({\cal T},\vec {\cal X}) 
+{\cal W}_{\rm tid}({\cal T},\vec {\cal X}) \Big\}-\nonumber\\
&&\hskip 7pt -\,
\frac{1}{c^4}\Big\{\tfrac18\,{\cal V}^4
+\tfrac32\,{\cal V}^2\bigl({\cal W}_{\tt M}+{\cal W}_{\rm tid}\bigr)
-\tfrac12\bigl({\cal W}_{\tt M}+{\cal W}_{\rm tid}\bigr)^2-
\frac{2GM_{\tt E}}{r^3_{\tt EM}}\Big(3(\vec{n}^{}_{\tt EM}\cdot\vec{\cal X})^2-\vec{\cal X}^2\Big)(\vec v_{\tt E}\cdot \vec {\cal V})\Big\}+\nonumber\\
&&\hskip 7pt +\,
{\cal O}\Big(c^{-5};\, 1.04\times 10^{-24}\Big),
\end{eqnarray}
where the error bound is from the \(\mathcal G_{00}\)   metric component (\ref{eq:G00tr-MM}).
Here, the \(O(c^{-2})\) term comprises the special‑relativistic kinetic correction \(\tfrac12\mathcal V^2\) and the gravitational redshift due to the lunar monopole \(\mathcal W_{\tt M}\) and external tidal potential \(\mathcal W_{\rm tid}\).  The \(O(c^{-4})\) contributions include quartic‑velocity effects, kinetic–potential couplings, the potential‐square term, and the velocity‑dependent tidal cross term proportional to \(G M_{\tt E}/r_{\tt  EM}^3\).  All neglected terms beyond \(O(c^{-4})\) are bounded by \(\sim2\times10^{-21}\), guaranteeing sub‑picosecond accuracy for any cis‑lunar trajectory.  

We need to further ``clean''  this expression to see if the \(O(c^{-4})\) terms are needed for our purposes.  To develop the most conservative estimates, we use  a circular vLLO (Sec.~\ref{sec:LCRS-LLO}). With vLLO velocity of $v_{\tt vLLO}=\sqrt{GM_{\tt M}/r_{\tt vLLO}}\simeq 1.68 \times 10^3$ m/s,  all  \(c^{-4}\)–order terms in (\ref{eq:tau-LCRS})—including the kinetic quartic \(c^{-4}\tfrac18{\cal V}^4_{\tt vLLO}\simeq 1.23 \times 10^{-22}\), the mixed term \(c^{-4}\tfrac32{\cal V}_{\tt vLLO}^2\bigl({\cal W}_{\tt M}+{\cal W}_{\rm tid}\bigr)\simeq c^{-4}\tfrac32{\cal V}_{\tt vLLO}^2\bigl(GM_{\tt M}/r_{\tt vLLO}+{GM_{\tt E}r_{\tt vLLO}^2}/{r_{\tt EM}^3}\bigr)\simeq 1.46 \times 10^{-21}\), the potential squared \(\tfrac12{\cal W}^2/c^4\simeq c^{-4}\tfrac12\bigl(GM_{\tt M}/r_{\tt vLLO}+{GM_{\tt E}r_{\tt vLLO}^2}/{r_{\tt EM}^3}\bigr)^2\simeq 4.87 \times 10^{-22}\),  the cross term  \(c^{-4}(2GM_{\tt E}/r_{\rm EM}^3)r_{\tt vLLO}^2{v_{\tt E}{\cal V}_{\tt vLLO}}\sim2.65\times10^{-25}\)—all well below our retention threshold of \(5\times 10^{-18}\). Considering other orbits from Table~\ref{tab:representative_orbits}, we see that corresponding magnitudes of the  \(O(c^{-4})\) terms will be even smaller than for vLLO.  Therefore, these terms may be safely omitted. 

Consequently, we recast  (\ref{eq:tau-LCRS}) into a form suitable for modern timekeeping applications in cislunar space:
{}
\begin{eqnarray}
\label{eq:tau-LCRS-lim}
\frac{d\tau}{d{\cal T}}
&=&1
-\frac{1}{c^2}\Big\{\tfrac12\,{\cal V}^2+U_{\tt M}({\cal T},\vec {\cal X}) 
+U^*_{\rm tid}({\cal T},\vec {\cal X}) \Big\}+
{\cal O}\Big(c^{-4};\, 1.46\times 10^{-21}\Big),
\end{eqnarray}
where $U_{\tt M}({\cal T},\vec {\cal X})$ and $U^*_{\rm tid}({\cal T},\vec {\cal X})$ are is the Newtonian lunar gravitational and tidal potentials, correspondingly. Also,  the error bound is due to the largest omitted mixed term \(c^{-4}\tfrac32{\cal V}_{\tt vLLO}^2{\cal W}_{\tt M}\simeq c^{-4}\tfrac32{\cal V}_{\tt vLLO}^2\bigl(GM_{\tt M}/r_{\tt vLLO}\bigr)\simeq 1.46 \times 10^{-21}\).
 
\subsection{Coordinate transformations between {\tt BCRS} and {\tt LCRS}}
\label{sec:BG-RS-transform-M}

\subsubsection{Coordinate transformations based on the IAU recommendations}
\label{Section-trans-M}

In a direct analogy with the definition of the {\tt GCRS},  the metric tensors in the {\tt BCRS} and {\tt LCRS}  allow for the derivation of the transformation rules between the {\tt BCRS} coordinates \( x^m \) and the {\tt LCRS} coordinates \( {\cal X}^n \) using tensorial transformation principles. These transformations can be expressed in two equivalent forms: i) as \( x^m({\cal T},\vec{\cal X}) \) or ii) as \( {\cal X}^n(t,\vec{x}) \). It is important to note that converting from one form to the other is non-trivial due to the barycentric coordinate position of the lunicenter, which appears as a function of \texttt{TCL} in the first form and as a function of \texttt{TCB} in the second form.

Explicitly, for the kinematically non-rotating {\tt LCRS}, the coordinate transformations are given as below
{}
\begin{eqnarray}
{\cal T} &=& t - {1 \over c^2} \Big\{ {\cal A}(t) + ({\vec  v}_{\tt M} \cdot {\vec r}_{\tt M})\Big\} + {1 \over c^4}
\Big\{ {\cal B}(t) +\big({\vec {\cal B}}(t)\cdot \vec r_{\tt M}\big) + {\cal B}_{\mu\nu}(t)r_{\tt M}^\mu r_{\tt M}^\nu + {\cal C}(t,{\vec x})\Big\}
+ O(c^{-5}), \label{eq:coord-tr-T1-M}\\
\vec {\cal X} &=&  \vec r_{\tt M} + {1 \over c^2}
\Big\{{\textstyle{\textstyle{\textstyle{1 \over 2}}}} \vec v_{\tt M}  (\vec v_{\tt M} \cdot\vec r_{\tt M}) + \vec r_{\tt M} \,w^*_{\rm ext}({\bf x_{\tt M}}) + \vec r_{\tt M} (\vec a_{\tt M}\cdot \vec r_{\tt M}) - {\textstyle{\textstyle{\textstyle{1 \over 2}}}}\vec a_{\tt M} r_{\tt M}^2 \Big\}+ O(c^{-4}),
 \label{eq:coord-tr-X1-M}
\end{eqnarray}
where $\cal T$ = $\tt TCL$, $t$ = $\tt TCB$, ${\vec  r}_{\tt M}={\vec  x}-{\vec  x}_{\tt M}$, ${\vec  v}_{\tt M}=d{\vec  x}_{\tt M}/dt$, ${\vec  a}_{\tt M}=d^2{\vec  x}_{\tt M}/dt^2$, and 
functions ${\cal A}, {\cal B}, {\cal B}^\mu, {\cal B}^{\mu\nu}, {\cal C}(t,{\vec  x})$ are
{}
\begin{eqnarray}
{d \over dt}{\cal A}(t) &=& {\textstyle{\textstyle{\textstyle{1 \over 2}}}} v_{\tt M}^2 + w^*_{\rm ext}({\bf x_{\tt M}}), 
 \label{eq:coord-tr-A-B1-M}\\
{d \over dt}{\cal B}(t) &=& -{\textstyle{\textstyle{1 \over 8}}}
v_{\tt M}^4 - {\textstyle{\textstyle{3 \over 2}}} v_{\tt M}^2 w^*_{\rm ext}({\bf x_{\tt M} })
+ 4\big(\vec v_{\tt M}  \cdot {\vec w}^*_{\rm ext}({\bf x_{\tt M} })\big) + {\textstyle{\textstyle{\textstyle{1 \over 2}}}}w_{\rm ext}^{*2}({\bf x_{\tt M} }),  \label{eq:coord-tr-Bt-M}\\
{\cal B}^\mu(t) &=& -{\textstyle{\textstyle{\textstyle{1 \over 2}}}}v_{\tt M} ^2 v_{\tt M} ^\mu + 4 w_{\rm ext}^{*\mu}({\bf
x_{\tt M} }) - 3v_{\tt M} ^\mu w^*_{\rm ext}({\bf x_{\tt M} }), 
 \label{eq:coord-tr-A-Bi-M}\\
{\cal B}^{\mu\nu}(t) &=&-v_{\tt M} ^\mu {\cal Q}^\nu + 2
\partial^\mu w_{\rm ext}^{*\nu} ({\bf x_{\tt M} }) - v_{\tt M} ^\mu \partial^\nu w^*_{\rm ext}({\bf x_{\tt M} })-  {\textstyle{\textstyle{\textstyle{1 \over 2}}}}\gamma^{\mu\nu}\dot{w}^*_{\rm ext}({\bf x_{\tt M} }), 
 \label{eq:coord-tr-A-Bij-M}\\
{\cal C}(t,{\bf x}) &=& -{\textstyle{1 \over 10}}r_{\tt M} ^2\big(\dot{\vec  a}_{\tt M} \cdot \vec r_{\tt M} \big).
 \label{eq:coord-tr-A-C-M}
\end{eqnarray}

The external potential at the Moon \( w^*_{\rm ext}({\bf x_{\tt M} }) \) may be represented only by the monopole contribution of the gravity field of the external bodies  \( w_{0,\texttt{ext}} \) taken at the Moon's world-line  
{}
\begin{equation}
c^{-2}w^*_{\rm ext}({\vec x_{\tt M} })=
\sum_{\tt B\not= M} {G M_{\tt B} \over r_{\tt BM}} +{\cal O}(4.80 \times 10^{-20}),
\label{w_ext-mono-M}
\end{equation}
with the summation carried out over all solar system bodies {\tt B} except the Moon, \( {\bf r}_{\tt BM} = {\bf x}_{\tt M}  - {\bf x}_{\tt B} \), with \( r_{\tt BM} = |\vec{r}_{\tt BM}| \). The error term is determined by the contribution of solar quadruple moment $J_2 = 2.25 \times 10^{-7}$ \cite{Park:2017,MecheriMeftah:2021} in (\ref{eq:W-pot2}), yielding $c^{-2}(GM_{\tt S}/{\rm AU}^3) J_2 R_{\tt S}^2 P_{20}(\cos\theta)\simeq 4.80 \times 10^{-20} P_{20}(\cos\theta)$.

Finally, with accuracy sufficient for most practical purposes, from (\ref{eq:vecw}), we have
{}
\begin{equation}
c^{-3}w^{*\alpha}_{\tt ext}(t,\vx) = \sum_{\tt B \not =M}
{G M_{\tt B} \over r_{\tt BM}} v_{\tt B}^\alpha+{\cal O}(1.04 \times 10^{-17}),
\label{w_0ii-M}
\end{equation}
where the error term is due to the omitted term with the solar spin moment of $S_{\tt S} \simeq 1.8838\times 10^{41}\, {\rm kg\,m}^2/{\rm s}$ \cite{Cang:2016}, which results in the effect on the order of 
$c^{-3} G S_{\tt S}/2{\rm AU}^2 \simeq 1.04 \times 10^{-17} $.

This formulation will ensure an uncertainty of $\lesssim 5 \times 10^{-18}$ in the time rate, and for quasi-periodic terms,  $\lesssim 5 \times 10^{-18}$ in the rate amplitude and 0.1 ps in the phase amplitude for locations beyond a few solar radii from the Sun. The same level of uncertainty applies to the transformation between \texttt{TCB} and \texttt{TCL} for locations within $r\simeq 60,000$~km of the Moon. However, inaccuracies in astronomical quantities may lead to larger errors in these calculations \cite{Soffel-etal:2003}.

\subsubsection{Estimating magnitudes of various terms}
\label{sec:mag-terms-M}

Here we will examine the magnitudes of the terms in (\ref{eq:coord-tr-T1-M})--(\ref{eq:coord-tr-A-C-M}) as they apply to lunar orbiters at various orbits.  The numerical applications will focus on time and frequency transfer involving a spacecraft at Earth-Moon Lagrange point (L1) that is at the distance of \(a_{\texttt{L1}} = 58\,018\)~km from the center of the Moon (Sec.~\ref{sec:rel-pot-LI}). We consider measurement uncertainties of \( 5 \times 10^{-18} \) for frequency transfer and 0.1 ps for time transfer. 

We begin with the expression for the time transformation (\ref{eq:coord-tr-T1-M}). Taking the Moon's velocity around the Earth to be $v_{\tt EM}=1022$ m/s, we have the Moon's barycentric velocity of $\vec v_{\tt M}=\vec v_{\tt E}+\vec v_{\tt EM}$, then, with definition for $w^*_{\rm ext}({\bf x_{\tt E}})$ from (\ref{w_ext-mono-M}), we estimate the magnitude of the terms proportional to $1/c^2$ in ${d {\cal A}/ dt}$ to see that they contribute  $ c^{-2}({\textstyle{\textstyle{1 \over 2}}} v_{\tt M}^2 + \sum_{\tt B\not=M} GM_{\tt B}/r_{\tt BM})\simeq1.52 \times 10^{-8}$ to the time rate $d{\cal T}/ dt$.  As a result, expression for ${d {\cal A}(t)/ dt}$ from (\ref{eq:coord-tr-A-B1-M}) takes the form:
{}
\begin{eqnarray}
\frac{1}{c^2}{d \over dt}{\cal A}(t) &=&\frac{1}{c^2}\Big\{ {\textstyle{\textstyle{\textstyle{1 \over 2}}}}v_{\tt M} ^2 + \sum_{{\tt B\not= M}} {G M_{\tt B} \over r_{\tt BM}} \Big\}+{\cal O}(1.86\times 10^{-20}) \simeq 
1.52 \times 10^{-8}+{\cal O}(1.86\times 10^{-20}),
 \label{eq:coord-tr-A-B1M}
\end{eqnarray}
where the error term is determined by the contribution from the mixed potential terms, $\Delta_{\tt ext}(t,\vx)$, that were present in (\ref{eq:Da}), but omitted  in (\ref{w_ext-mono}), as discussed in \cite{Soffel-etal:2003}.

The position-dependent $c^{-2}$-term in (\ref{eq:coord-tr-T1-M})  contributes a periodic effect of $c^{-2}({\vec  v}_{\tt M} \cdot {\vec r}_{\tt L1})\simeq 19.88 \,\mu$s to the time transfer at the Earth-Moon L1. Therefore, both of the $c^{-2}$-terms are significant and must be included in the model.

Terms proportional to $1/c^4$ in (\ref{eq:coord-tr-T1-M}) exhibit both secular and quasi-periodic behavior. Considering the term ${d {\cal B}(t)/ dt}$ as given in (\ref{eq:coord-tr-Bt-M}), the velocity term contributes up to $v_{\tt M} ^4/8c^4 \simeq 1.32 \times 10^{-17}$ to the time rate. The second term, when evaluated for the solar potential, yields $c^{-4}(3/2)v^2_{\tt M}  (GM_{\tt S}/{\rm AU} +GM_{\tt M}/r_{\tt EM}  + GM_{\tt J}/{\rm 4 AU})\simeq 1.57 \times 10^{-16}$. The third term, evaluated for the solar vector potential, yields $c^{-4}4v_{\tt M}  GM_{\tt S} v_{\tt S}/{\rm AU} \simeq 1.72 \times 10^{-19}$, with its total term contribution of $ c^{-4}4\big(\vec v_{\tt M}  \cdot {\vec w}^*_{\rm ext}({\bf x_{\tt M} })\big)=c^{-4}4\sum_{\tt B \not =M} (G M_{\tt B}/ r_{\tt BM}) (\vec v_{\tt M} \cdot \vec v_{\tt B})\sim 6.86\times10^{-19}$, too small to be considered for high-precision timing applications. Finally, the last term contributes $\sim c^{-4}{\textstyle\frac{1}{2}}(GM_{\tt S} /{\rm AU} +GM_{\tt E}/r_{\tt EM}+GM_{\tt J} /{\rm 4AU} )^2 \simeq 4.89 \times 10^{-17}$. Altogether, the term $d {\cal B}(t)/ dt$ contributes  $\sim1.22\times 10^{-16}$ to the time rate $(d{\cal T}/ dt)$, or up to $\sim$2.84 cm in 10 days.

As a result, the entire term (\ref{eq:coord-tr-Bt-M}) takes the following form:
{}
\begin{eqnarray}
\frac{1}{c^4}{d \over dt}{\cal B}(t) &=& \frac{1}{c^4}\Big\{-{\textstyle{1 \over 8}}
v_{\tt M} ^4 - {\textstyle{3 \over 2}} v_{\tt M} ^2 \sum_{{\tt B\not= M}}{G M_{\tt B} \over r_{\tt BM}}+ 
{\textstyle{\textstyle{1 \over 2}}}\Big[\sum_{{\tt B\not= M}} {G M_{\tt B} \over r_{\tt BM}}\Big]^2\Big\}+{\cal O}(6.86\times 10^{-19}) \simeq\nonumber \\
&\simeq&-1.22 \times 10^{-16}+{\cal O}(6.86\times 10^{-19}),
\label{eq:coord-tr-BtE-M}
\end{eqnarray}
where the error is set by the omitted contribution from the external vector potential in (\ref{eq:coord-tr-Bt-M}).

Next, considering the contribution of the ${\cal B}^\mu(t)$ term as specified in (\ref{eq:coord-tr-A-Bi-M}), we find that its velocity-dependent term contributes up to $c^{-4}v_{\tt M} ^3 a_{\tt L1}/2 \simeq 1.05 \times 10^{-13}$\,s to the time transfer for a  spacecraft at Earth-Moon L1 (Sec.~\ref{sec:rel-pot-LI}). 
The contribution of the term with the external vector potential was evaluated to be $c^{-4}4(GM_{\tt S} v_{\tt S} /{\rm AU}+GM_{\tt E} v_{\tt E}/r_{\tt EM}+GM_{\tt J} v_{\tt J}/{\rm 4AU})a_{\tt L1}  \sim 1.30 \times 10^{-15}$\,s, which is too small to be included in the model. Thus, the entire term with the external vector potential $4\sum_{\tt B \not =M} (G M_{\tt B} / r_{\tt BM}) (\vec v_{\tt B}\cdot \vec r_{\tt M} )$  may be disregarded. Considering the last term in (\ref{eq:coord-tr-A-Bi-M}), the presence of  the solar scalar potential was found to contribute $c^{-4}3(GM_{\tt S}/{\rm AU}+GM_{\tt E} /r_{\tt EM}+GM_{\tt J}/{\rm 4AU})^2 v_{\tt M} a_{\tt L1} \sim 5.90 \times 10^{-13}$\,s  to the timing uncertainty, and thus it may be included. Consequently, given (\ref{w_ext-mono-M}) and (\ref{w_0ii-M}), the term ${\cal B}^i(t)$ can be written as follows:
{}
\begin{eqnarray}
\frac{1}{c^4}\big({\vec {\cal B}}(t)\cdot \vec r_{\tt M} \big) &=&-\frac{1}{c^4}\Big({\textstyle{\textstyle{1 \over 2}}}v_{\tt M} ^2 +3\sum_{\tt B \not =M}{G M_{\tt B} \over r_{\tt BM}}\Big)(\vec v_{\tt M} \cdot \vec r_{\tt M} )
+{\cal O}(1.37\times 10^{-15}\,{\rm s}) \simeq \nonumber \\
&\simeq&6.95 \times 10^{-13}\,{\rm s}+{\cal O}(1.30\times 10^{-15}\,{\rm s}),
 \label{eq:coord-tr-A-BiE-M}
\end{eqnarray}
where the error is set by the omitted contribution from the external vector potential in (\ref{eq:coord-tr-A-Bi-M}).  Thus, at the Earth-Moon L1 distance this periodic term has magnitude of  $\simeq 0.70$ ps; when evaluated on the Moon's surface it is $0.02$\,ps. Accordingly, we will drop this term from the discussions and treat it as an error bound in the timing expression (\ref{eq:coord-tr-T1-M}). 

The second position-dependent term with quadratic position dependence, ${\cal B}^{\mu\nu}(t)$, contributes a periodic effect with magnitude of up to  $\sim4.25 \times 10^{-17}$\,s to the time difference and is too small to be considered. Similarly, the contribution of the third position-dependent term, ${\cal C}(t,x)$, is also periodic and small. To estimate its magnitude we take $\dot a_{\tt M}\simeq 2GM_{\tt S} v_{\tt M}/{\rm AU}^3$, than this term may amount to $c^{-4}(1/5)GM_{\tt S} v_{\tt M}a_{\tt L1}/ {\rm AU}^3\sim6.90 \times 10^{-21}$\,s in the time difference at Earth-Moon L1 orbit, and is also much too small to be practically important.

Therefore, only one term of the order of $c^{-4}$, specifically, ${d {\cal B}(t)/ dt}$, is not0.1 ps negligible in modern-day timing applications and may each reach an amplitude of $\sim 1.22 \times 10^{-16}$ in time rate in geostationary orbit. As a result, this term will be included in the model accurate to $\sim5.0 \times 10^{-18}$, or better, is required.

Next, we consider the position transformation as specified by (\ref{eq:coord-tr-X1-M}). At altitude of a Earth-Moon L1 spacecraft, the first two $1/c^2$ terms in this equation contribute $c^{-2}{\textstyle{\textstyle{\textstyle{1 \over 2}}}}\vec v_{\tt M}  (\vec v_{\tt M}  \cdot\vec r_{\tt M} ) \simeq 30.29$ cm and $c^{-2}w^*_{\rm ext}\,\vec r_{\tt M} =c^{-2}(G M_{\tt S}/{\rm AU}) \vec r_{\tt M}  \simeq 57.73$~cm. For a  station on the lunar surface, the effects are $c^{-2}{\textstyle{\textstyle{\textstyle{1 \over 2}}}}\vec v_{\tt M}  (\vec v_{\tt M}  \cdot\vec r_{\tt M} ) \simeq 0.95$ cm and $c^{-2}w^*_{\rm ext}\,\vec r_{\tt M} =c^{-2}(G M_{\tt S}/{\rm AU})\vec r_{\tt M}  \simeq 1.61$\,cm. These contributions are significant enough to be included in the model. 

The acceleration-dependent terms in (\ref{eq:coord-tr-X1-M}) may contribute up to $2.90 \times 10^{-7}$~m to station position and $3.61 \times 10^{-4}$~m to an L1 observer. Although these corrections are small, they prove to be significant if one aims to compare spacecraft accelerations in {\tt BCRS} and {\tt LCRS}.  The next term involves the contribution of external multipole moments to (\ref{eq:coord-tr-X1-M}). Based on the value of the solar quadrupole moment, $J_2 = 2.25 \times 10^{-7}$ \cite{Park:2017,MecheriMeftah:2021}, the contribution of the solar $J_2$ to the position transformation is estimated even at the Earth-Moon L1 point to be $c^{-2}w^*_{2,{\tt S}}(t,\vx) \vec r_{\tt M}  \simeq c^{-2} (GM_{\tt S} J_2R_{\tt S}^2/{\rm AU}^3)a_{\tt L1}\sim  2.78 \times 10^{-12}$\,m, and, as such, is totally negligible for our purposes and will serve as an error bound.

\subsubsection{Practical coordinate transformations for  {\tt LCRS}}
\label{sec:rec-express-M}

As a result of the order-of-magnitude considerations above, similar to (\ref{eq:coord-tr-T1-rec})--(\ref{eq:coord-tr-Xrec}), we  present the practically-relevant form of coordinate transformations between the {\tt LCRS} (${\cal T} = \tt TCL$, $\vec {\cal X}$) and the {\tt BCRS} ($t = \tt TCB$, $ \vec x $)  that are suffuicient for modern high-precision PNT applications in cislunar space:
{}
\begin{eqnarray}
{\cal T} &=& t-c^{-2}\Big\{\int^{t}_{t_{0}}\Big({\textstyle{1\over 2}} v_{\tt M}^{2} +
\sum_{{\tt B\not= M}} {G M_{\tt B} \over r_{\tt BM}} \Big)dt + (\vec v_{\tt M}  \cdot \vec r_{\tt M} ) \Big\}-
\nonumber \\ && \hskip 6pt -\, 
   c^{-4}\Big\{\int^t_{t_0}\Big({\textstyle{1 \over 8}}
v_{\tt M} ^4  
+ {\textstyle{3 \over 2}} v_{\tt M} ^2 \sum_{{\tt B\not= M}}{G M_{\tt B} \over r_{\tt BM}}- 
{\textstyle{\textstyle{1 \over 2}}}\Big[\sum_{{\tt B\not= M}} {G M_{\tt B} \over r_{\tt BM}}\Big]^2\Big)dt
+
\Big({\textstyle{\textstyle{1 \over 2}}}v_{\tt M} ^2 +3\sum_{\tt B \not =M}{G M_{\tt B} \over r_{\tt BM}}  \Big)(\vec v_{\tt M}  \cdot \vec r_{\tt M} ) 
\Big\} +\nonumber\\
  && \hskip 6pt +\, 
 {\cal O}\Big(c^{-5};\, 6.86\times 10^{-19}\,(t-t_0);\, 
 1.37\times 10^{-15}\,{\rm s}\Big),
 \label{eq:coord-tr-T1-rec-M}\\[4pt]
\vec {\cal X} &=& \vec r_{\tt M}  + c^{-2} \Big\{{\textstyle\frac{1}{2}}( \vec v_{\tt M}  \cdot\vec r_{\tt M} )\vec v_{\tt M}  +\sum_{{\tt B\not= M}}{G M_{\tt B} \over r_{\tt BM}} \vec r_{\tt M}  +(\vec a_{\tt M} \cdot \vec r_{\tt M} )\vec r_{\tt M} - {\textstyle\frac{1}{2}}r^2_{\tt M} \vec a_{\tt M} \Big\}  +
 {\cal O}\Big(c^{-4};\, 2.94\times 10^{-12}~{\rm m}\Big),
\label{eq:coord-tr-Xrec-M}
\end{eqnarray}
where $\vec r_{\tt M} \equiv \vec x - \vec x_{\tt M}(t)$ with $\vec x_{\tt M}$  and $\vec v_{\tt M}=d\vec x_{\tt M}/dt$ being the Moon's position and velocity vectors  in the {\tt BCRS}. Also, the error in the time transformation is set by the omitted contribution of the external vector potential in (\ref{eq:coord-tr-Bt-M}) and (\ref{eq:coord-tr-A-Bi-M}), yielding (\ref{eq:coord-tr-BtE-M});  the error in the position transformation is due to omitted contribution of the solar quadrupole moment to (\ref{w_ext-mono-M}), which is clearly impractical for our purposes. 

\end{document}